\documentclass[epj]{svjour}
\usepackage{graphicx,wrapfig,lipsum}
\usepackage{amssymb,amsmath}
\usepackage{subeqnarray}
\usepackage{hyperref}

\newcommand{\greeksym}[1]{{\usefont{U}{psy}{m}{n}#1}}
\newcommand{\ugamma}{\mbox{\scriptsize\greeksym{g}}}
\newcommand{\umu}{\mbox{\greeksym{m}}}
\newcommand{\udelta}{\mbox{\greeksym{d}}}

\newcommand{\upi}{\mbox{\greeksym{p}}}
\newcommand{\uSigma}{\mbox{\greeksym{S}}}
\newcommand{\uLambda}{\mbox{\greeksym{L}}}
\newcommand{\uXi}{\mbox{\greeksym{X}}}

\newcommand{\uphi}{\mbox{\greeksym{f}}}
\newcommand{\uOmega}{\mbox{\greeksym{W}}}

\newcommand{\pLambda}{\mbox{\footnotesize\greeksym{L}}} 
\newcommand{\sLambda}{\mbox{\scriptsize\greeksym{L}}}
\newcommand{\sSigma}{\mbox{\scriptsize\greeksym{S}}}

\newcommand{\ie}{{\it i.e.\/}}
\newcommand{\eg}{{\it e.g.\/}}
\newcommand{\etal}{{\it et.al.\/}}
\newcommand{\loccit}{{\it loc.cit.\/}}
\newcommand{\muB}{\ensuremath{\umu_{\rm B}}} 

\newcommand{\nc}{\newcommand}   
\nc{\req}[1]{Eq.\,(\ref{#1})}    \nc{\reqp}[1]{Eq.\,(\ref{#1}) on page \pageref{#1}}     
\nc{\rf}[1]{Fig.~\ref{#1}}   \nc{\rfp}[1]{Fig.~\ref{#1} on page \pageref{#1}}     
\nc{\rt}[1]{table~\ref{#1}}   \nc{\rtp}[1]{table~\ref{#1} on page \pageref{#1}}     
\nc{\Th}{\ensuremath{T_\mathrm{H}\,}}
\nc{\pp}{\ensuremath{pp\ }}
\nc{\pA}{\ensuremath{p A\ }}
\nc{\hAA}{\ensuremath{AA\ }}
\nc{\D}{\mathrm{d}}
\nc{\E}{\mathrm{e}}
\def\beq{\begin{equation}}
\def\eeq{\end{equation}}

\begin{document}
\title{Melting Hadrons, Boiling Quarks}
\author{Johann Rafelski\inst{1}\inst{2}}
%
%
\institute{CERN-PH/TH, 1211 Geneva 23, Switzerland\and
Department of Physics, The University of Arizona 
Tucson, Arizona, 85721, USA }
\date{Submitted: August 11, 2015 / Print date: \today}
%
\abstract{In the context of the Hagedorn temperature half-centenary I describe our understanding of the hot phases of hadronic matter both below and above the Hagedorn temperature. The first part of the review addresses many frequently posed questions about properties of hadronic matter in different phases, phase transition and the exploration  of quark-gluon plasma (QGP). The historical context of the discovery of QGP is shown and the role of strangeness and strange antibaryon signature of QGP illustrated.  In the second part I discuss the corresponding theoretical ideas and show how experimental results can be used to  describe the  properties of QGP at hadronization. The material of this review is complemented by  two  early and unpublished reports containing the  prediction of the different forms of hadron matter, and of the formation of QGP in relativistic heavy ion collisions, including the discussion of  strangeness, and in particular strange antibaryon signature of QGP. 
\vskip -11.2cm \phantom{.}\hfill {\bf arXiv:1508.03260} 13 Aug 2015 and  PREPRINT {\bf CERN-PH-TH-2015-194}\\
\vskip 10.2cm
\PACS{{24.10.Pa}{Thermal and statistical models} \and
      {25.75.-q}{Relativistic heavy-ion collisions} \and 
      {21.65.Qr}{Quark matter} \and
      {12.38.Mh}{Quark-gluon plasma}
     } 
\vskip 0.5cm  
} 
\maketitle

\section{Introduction}

The year 1964/65 saw the rise of several new ideas which in the following 50 years shaped the  discoveries in fundamental subatomic physics:
\begin{enumerate}
\item The  Hagedorn  temperature \Th; later recognized as the melting point of hadrons into
\item Quarks as building blocks of hadrons;   and,
\item The Higgs particle and field escape from the  Goldstone theorem, allowing the  understanding of  weak interactions,   the source of inertial mass of the  elementary particles.
\end{enumerate}

The topic in this paper is  Hagedorn  temperature \Th\ and the strong interaction phenomena near to \Th. I present  an overview of 50 years of effort with emphasis on:\\   
a) Hot nuclear and hadronic matter;\\
b) Critical behavior near \Th;\\ 
c) Quark-gluon plasma (QGP);\\ 
d) Relativistic heavy ion (RHI) collisions\footnote{We refer to atomic nuclei which are heavier than the $\alpha$-particle as \lq heavy  ions\rq.};\\ 
e) The  hadronization process of QGP;\\ 
f) Abundant production of strangeness flavor.\\ 
This  presentation  connects   and extends a recent retrospective work, Ref.~\cite{HagedornBook}:  {\it Melting Hadrons, Boiling Quarks; From Hagedorn temperature to  ultra-relativistic heavy-ion  collisions at CERN; with a tribute to Rolf Hagedorn}. This report complements prior summaries of our  work: 1986~\cite{Koch:1986ud}, 1991~\cite{Eggers:1990dm},1996~\cite{Rafelski:1996hf}, 2000~\cite{Letessier:2000ay}, 2002~\cite{Letessier:2002gp}, 2008~\cite{Letessier:2005qe}.

A report  on \lq Melting Hadrons, Boiling Quarks and \Th\rq\  relates strongly to quantum chromodynamics (QCD), the  theory of quarks and gluons, the building blocks of hadrons, and its lattice numerical solutions;  QCD is the  quantum (Q) theory of color-charged  (C)   quark  and gluon dynamics (D); for numerical study the  space-time continuum is  discretized on a \lq lattice\rq. 

Telling the story of how we learned that  strong interactions are a gauge theory  involving  two types of particles, quarks and gluons, and the working of the lattice numerical method would entirely change  the contents of this article, and be beyond the  expertise of the author.  I recommend instead  the book by Weinberg~\cite{Weinberg:QFT}, which also shows the  historical path to QCD.  The best sources of the QCD relation to the topic of this article  are: (a) the book by Kohsuke Yagi and Tetsuo Hatsuda~\cite{Yagi:QGP} as well as, (b)  the now 15 year old monograph by Letessier and the author~\cite{Letessier:2002gp}. We often refer to lattice-QCD method to present QCD properties of interest in this article. There are  books and  many reviews  on  lattice implementation of gauge theories of interacting fields, also specific to hot-lattice-QCD method. At the time of writing I do not have a favorite to recommend. 

Immediately in the following Subsection~\ref{introwhy}  the famous {\it Why?} is addressed. After that I turn to answering  the {\it How?} question in  Subsection~\ref{introhow}, and include a few reminiscences about the accelerator race in   Subsection~\ref{introrace}. I close this Introduction with Subsection \ref{introformat}   where the organization and contents of this  review will be explained. 


\subsection{What are the conceptual challenges of the\\ QGP/RHI collisions research program?}\label{introwhy}
 
Our conviction that we achieved in laboratory  experiments the conditions required for melting (we can also say, dissolution) of hadrons into a soup of boiling quarks and gluons became firmer in the past 15-20 years. Now we can ask, what are the \lq applications\rq\ of the quark-gluon plasma physics? Here is a short wish list:
\vskip 0.15cm 
\noindent 1) Nucleons  dominate the mass of matter by a factor 1000. The mass of the three  \lq elementary\rq\ quarks found in nucleons is about 50 times  smaller than the nucleon mass.  Whatever compresses and keeps the quarks  within the nucleon  volume is thus the source of nearly all of mass of matter. This clarifies that the Higgs field provides the   mass scale to all particles that we view today as elementary. Therefore only  a small \%-sized fraction of the mass of matter originates {\em directly} in the Higgs field; see Section~\ref{MassMatter} for further  discussion.  The  question:  {\it What is  mass?\/}  can be  studied by melting hadrons into quarks in RHI collisions. 
\vskip 0.15cm 
\noindent 2)  Quarks are kept inside hadrons by the \lq vacuum\rq\ properties which abhor the color charge of quarks. This explanation of 1) means that there must be at least two different forms of the modern \ae ther that we call \lq vacuum\rq: the world around us, and the holes in it that are called hadrons. The question: {\it Can we form  arbitrarily   big  holes  filled with almost free  quarks and gluons?\/} was and remains the  existential issue for laboratory study of hot matter made of quarks and gluons, the QGP. Aficionados of the  lattice-QCD should take note that  the presentation of two phases of matter in  numerical simulations  does {\em not} answer this question as the lattice method studies the entire Universe, showing hadron properties at low temperature, and QGP properties at high temperature.
\vskip 0.15cm 
\noindent 3) We all agree that QGP was the primordial Big-Bang stuff that filled the Universe before \lq normal\rq\ matter formed. Thus any laboratory exploration  of the QGP properties solidifies our models of the Big Bang and allows us to ask these questions: {\it What are   the properties  of the primordial matter content of the Universe?\/} and  {\it How does  \lq normal\rq\ matter   formation in early Universe work?\/} 
\vskip 0.15cm 
\noindent 4) {\it What is flavor?} In elementary particle collisions, we deal with  a few, and in most cases only one, pair of newly created 2nd, or 3rd flavor family of particles at a time. A new situation arises in  the QGP formed in relativistic heavy ion collisions. QGP includes a large number of particles from the second family: the strange quarks  and   also, the yet heavier charmed quarks; and from  the third family at the LHC we expect an appreciable abundance of  bottom quarks.  The novel ability to study a large number of these 2nd and 3rd generation particles offers a new opportunity to approach in an experiment  the riddle of flavor.
\vskip 0.15cm 
\noindent 5) In relativistic heavy ion collisions the kinetic energy of ions feeds the growth of quark population. These quarks  ultimately turn into final state material particles. This means that  we study  experimentally the mechanisms leading to the conversion of the colliding ion kinetic energy into mass of matter. One can wonder aloud if this  sheds some light on the reverse process: {\it Is it possible to  convert   matter into energy in the laboratory?\/} 

The last two points show the potential of  \lq applications\rq\ of QGP physics to change   both  our understanding of, and our place in the world. For the present we keep these questions  in  mind. This review will  address  all the other challenges listed under  points 1), 2), and 3) above; however, see  also  thoughts along comparable foundational  lines  presented  in Subsections~\ref{EinsteinAether2} and \ref{QuarkUniverseLab}.

\subsection{From melting hadrons to boiling quarks}\label{introhow}

With the hindsight of 50 years I believe that Hagedorn's effort to interpret  particle multiplicity data  has led  to the recognition of the opportunity  to study  quark deconfinement at high temperature. This is the topic of the book\,\cite{HagedornBook} {\it Melting Hadrons,\,Boiling Quarks; From Hagedorn temperature to  ultra-relativistic heavy-ion  collisions at CERN; with a tribute to Rolf Hagedorn\/} published at Springer Open, \ie\ available for free on-line. This article should be seen as a companion addressing more recent developments, and setting a contemporary context for this book. 

{\it How did we get here?} There were two critical milestones:\\[0.15cm]
\noindent I) The first  milestone occurred in 1964--1965, when Hagedorn, working to resolve discrepancies of the statistical particle production model with the   \pp  reaction data, produced his \lq\lq distinguishable particles\rq\rq\  insight. Due to a twist of history, the initial research work was archived without publication and has only become available to a wider public recently;  that is, 50 years later, see Chapter 19 in \cite{HagedornBook} and Ref.\cite{CernCourierTh}.  Hagedorn   went on to interpret the observation he made. Within  a few months, in Fall 1964, he created the Statistical Bootstrap Model (SBM)~\cite{Hagedorn:1965st}, showing how the large diversity of strongly interacting particles could arise; Steven Frautschi~\cite{Frautschi:1971ij} coined in 1971 the name \lq Statistical Bootstrap Model\rq.\\[0.15cm]
\noindent II) The second milestone occurred in the  late 70s and early 80s   when we spearheaded the development of an experimental program to study  \lq melted\rq\ hadrons and the \lq boiling\rq\ quark-gluon plasma  phase of matter. The intense theoretical and experimental work on the thermal properties of strongly interacting matter, and the confirmation of a new quark-gluon plasma paradigm started in 1977    when the SBM mutated to  become a model for  melting  nuclear matter.  This development motivated the experimental exploration in the collisions of heavy nuclei at relativistic energies of the phases of matter in conditions close to those last seen in the early Universe. I refer to Hagedorn's account of these developments for further details~Chapter 25 {\it loc.cit.\/} and Ref.\cite{Hagedorn:1984hz}.  We return to this time period in Subsection~\ref{RHI_QGP}.

At the beginning of this new field of research in the late 70s,  quark confinement was a  mystery for many of my colleagues; gluons mediating the strong color force were neither discovered nor widely accepted, especially not among my nuclear physics peers, and QCD vacuum structure was just finishing kindergarten. The discussion of a new phase of deconfined quark-gluon matter was therefore in many eyes not consistent with established wisdom and certainly too ambitious for the time. 

Similarly, the special situation  of Hagedorn deserves remembering: early on Hagedorn's research was undermined by outright personal hostility; how could Hagedorn dare to introduce thermal physics into the field governed by particles and fields?
However, one should also take note of the spirit of academic tolerance at CERN.   Hagedorn   advanced through the ranks along with his critics, and  his presence attracted other like-minded researchers,  who were welcome in the CERN Theory Division, creating a bridgehead towards  the new field of RHI collisions when the opportunity appeared on  the horizon.

In those days, the field of RHI collisions was in other ways  rocky terrain:\\[0.15cm]
\noindent 1) RHI collisions required the use of atomic nuclei at highest energy. This required cooperation between experimental nuclear and particle physicists. Their culture, background, and experience differed. A similar situation prevailed  within the domain of theoretical physics, where an interdisciplinary program of research merging the  three traditional physics domains had to form. While ideas of thermal and  statistical physics needed to be applied, very  few subatomic physicists, who usually deal with individual particles, were prepared to deal with many body questions.  There were also several practical issues: In which (particle, nuclear, stat-phys) journal can one  publish and who could be the reviewers (other than direct competitors)? To whom to apply for funding? Which conference to contribute to? \\[0.15cm]
\noindent 2) The small group of scientists who practiced RHI collisions were divided on many important questions. In regard to what happens in relativistic collision of nuclei the situation was most articulate: a) One group believed that nuclei (baryons)  pass through each other with  a new phase of matter formed in a somewhat heated projectile and/or target. This picture required detection systems of very different character than the systems required by, both:  b) those who believed that in RHI collisions energy would be consumed by a shock wave compression  of nuclear matter crashing into the center of momentum frame; and c) a third group who argued that up to top  CERN-SPS ($\sqrt{(s_{NN})}\simeq {\cal O}(20)$\,GeV)  collision energy    a high temperature, relatively low baryon density quark matter fireball will be formed.  The last   case turned out to be closest to  results obtained at CERN-SPS and  at Brookhaven National Laboratory (BNL) RHI Collider (RHIC).

From outside, we were ridiculed as being speculative; from within we were in state of uncertainty about the fate of colliding matter and the  kinetic energy it carried, with disagreements that ranged across theory vs. experiment, and  particle vs. nuclear physics. In this situation, \lq QGP formation in  RHI collisions\rq\  was a field of research that could have easily fizzled out.  However, in Europe there was CERN, and in the US there was strong institutional support. Early on it was realized that RHI collisions required large experiments  uniting much more human  expertise and manpower as compared  to the prior nuclear and even some particle physics projects. Thus work had to be centralized in a few \lq pan-continental\rq\  facilities. This meant that expertise from a few laboratories would need to be  united in   a third location where  prior investments  would help limit the preparation time and cost.  

\begin{figure*} 
\centering\resizebox{\textwidth}{!}{%
\includegraphics{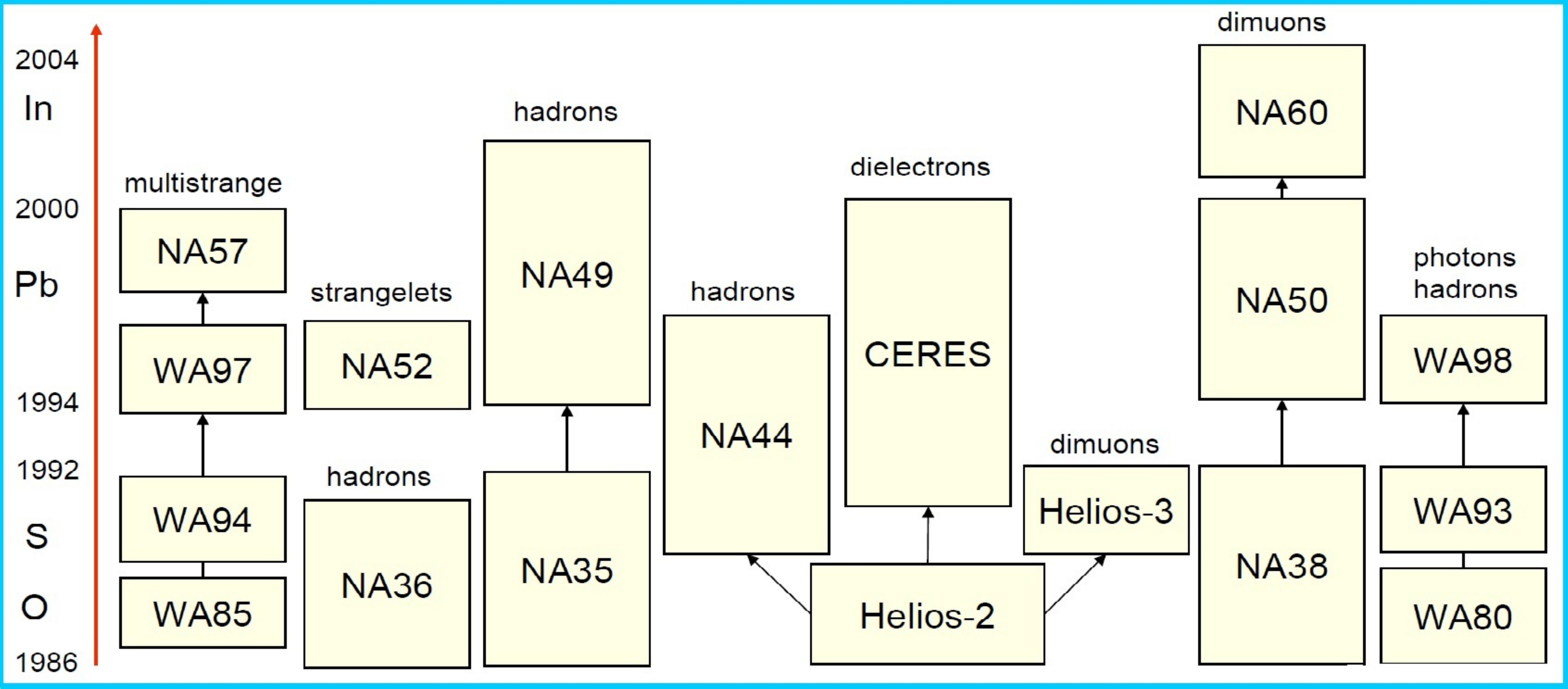}
}
\caption[]{Time line  of the CERN-SPS RHI program: on the left axis we see year and ion beam available, S=Sulfur, Pb=lead, In=Indium) as a \lq function\rq\ of the experimental code. The primary observables are indicated next to each square; arrows connecting the squares indicate that the prior equipment and group, both  in updated format, continued. Source: CERN release February 2000 modified by the  author.}\label{CERNSPS}
\end{figure*}

\subsection{The  accelerator race for quark matter}\label{introrace}
These considerations meant that in Europe the QGP formation in RHI collisions research program found its home at CERN. The CERN site benefited from being a multi-accelerator laboratory  with a large pool of engineering expertise and where some of the necessary experimental equipment already existed, thanks to prior related particle physics efforts.

The  CERN program took off by the late 80s. The time line of the many  CERN RHI experiments through the beginning of this millennium is shown in \rf{CERNSPS}; the representation is based on a similar CERN document from the year 2000. The  experiments WA85, NA35, HELIOS-2, NA38  were built largely from instrumental components from  prior particle physics detectors.  Other experiments and/or experimental components were contributed by US and European laboratories. These include  the heavy ion source and its preaccelerator complex, required for heavy ion insertion into CERN beam lines.

When the CERN SPS program faded out early in this millennium, the resources were focused on the LHC-ion collider operations, and in the US, the  RHIC  came on-line.  As this is written, the SPS fixed target program experiences a second life; the experiment NA61, built with large input from the NA49 equipment, is searching for the onset of QGP formation, see Subsection~\ref{HornThreshhold}.

The success of the SPS research program at CERN has strongly supported the continuation of the RHI collision program.  The Large Hadron Collider  (LHC) was designed to accept highest energy counter propagating heavy ion beams opening the study of a new domain of collision energy. LHC-ion operation allows us to exceed the top RHIC energy range by more than an order of magnitude. In preparation for LHC-ion operations, in the mid-90s the SPS groups  founded  a new collider collaboration, and have built one of the four LHC experiments dedicated to the study of RHI collisions. Two other experiments also participate in the LHC-ion research program which we will introduce  in  Subsection~\ref{Subsec:Experiments}.

In parallel to CERN there was a decisive move in the same direction in the US. The roots of the US relativistic heavy ion program predate the interest of CERN by nearly a decade.  In 1975, the Berkeley SuperHILAC, a low energy heavy ion accelerator  was  linked to the Bevatron,  an antique particle accelerator at the time, yet capable of accelerating the injected ions to relativistic energies with the Lorentz factor above two.  The system of accelerators was called the Bevalac. It offered beams of ions which were used in study of properties of compressed nuclear matter, conditions believed to be similar to those seen within collapsing neutron stars.

As interest in the study of quark matter grew by 1980 the Bevalac scientists formulated the future Variable Energy Nuclear Synchrotron (VENUS) heavy ion facility. Representing the Heavy Ion Program at Berkeley Howell Pugh~\cite{Pugh81} opened in October 1980 the \lq Quark Matter 1\rq\ conference at GSI in Germany making this comment
\begin{quote}
\lq\lq $20\, A\mathrm{\,GeV}<E<1000\,A$\,GeV\ldots LBL's VENUS proposal. In view of the long lead time in VENUS construction it would be extremely valuable to proceed with the necessary modifications to accelerate light nuclei at CERN\ldots the rich environment of sophisticated detectors would be hard to reproduce elsewhere.\rq\rq
\end{quote}
It is clear from the context that CERN was in these remarks synonymous with CERN-ISR, a collider. Within following  two years the incoming CERN Director Herwig Schopper closed ISR and created an alternative, the SPS heavy ion program capable of using the heaviest ions.

However, Pugh's remarks created  in the minds of all concerned in the US a question: was there a place in the US, other than LBL, with capabilities similar to CERN?  When Berkley moved to define the research program for an ultrarelativistic heavy ion collider in 1983, another candidate laboratory was waiting in the  wings: The Brookhaven National Laboratory (BNL) had a  completed  project with 4 experimental halls. This was to be the $pp$ collider ISABELLE, now mothballed having been scooped by CERN's bet on the S$p\bar p$S collider in the race to discover the W and Z weak interaction mesons. If ISABELLE were modified to be a RHI Collider (RHIC), it was thought that it could be completed within a few years, offering the US a capability comparable to that expected by Pugh  at CERN.

\begin{figure*}
\centering\resizebox{0.99\textwidth}{!}{%
\includegraphics{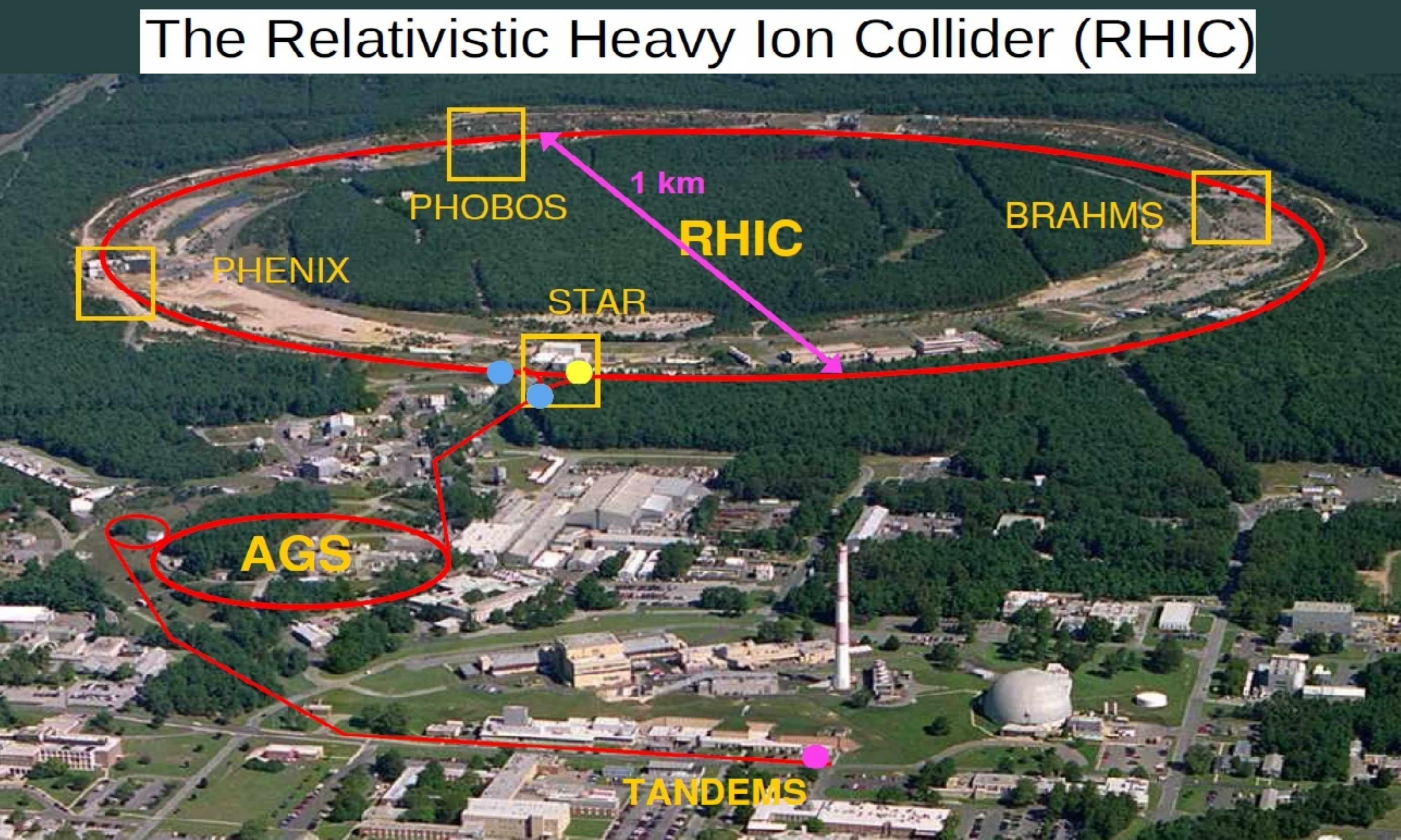}
}
\caption[]{The Brookhaven National Laboratory heavy ion accelerator complex. Creative Commons picture modified by the author.}\label{BNLRHIC}
\end{figure*}
This evaluation prompted a major investment decision by the US Department of Energy to create a new relativistic heavy ion research center at BNL shown in \rf{BNLRHIC}, a plan that would be cementing the US leadership role in the field of heavy ions. In a first step, already existing tandems able to create low energy heavy ion  beams were connected by a transfer line to the already existing AGS proton synchrotron adapted to accelerate these ions. In this step a system similar to the Berkeley Bevalac was formed, while beam energies about 7 times greater than those seen at Bevalac could be reached.

The AGS-ion system performed experiments with fixed targets, serving as a training ground for the next generation of experimentalists. During this time, another transfer line was built connecting AGS to the ISABELLE project tunnel, in which the RHIC was installed. The initial RHIC experiments are shown around their  ring  locations in \rf{BNLRHIC}: STAR, PHENIX, PHOBOS and BRAHMS, see also Subsection~\ref{QGPdiscovered}. The first data taking at RHIC began in Summer 2001, about 10 years later than many  had hoped for back in 1984.

I recall vividly that when in 1984 we were told at a meeting at BNL that RHIC was to operate by 1990, a colleague  working at the Bevalac asked, why not 1988?  So a big question remains today: why in the end was it 2001? In seeking an answer we should note that while the RHIC project took 17 years to travel from the first decision to first beam, SPS took 11 years (Pb  beam capability). However, SPS was an already built, functional accelerator. Moreover, RHIC development was hindered by the need to  move  heavy ion activities from LBL to BNL, by the adaptation of ISABELLE design to fit RHIC needs, and by typical funding constraints. As this work progressed nobody rushed. I think this was so since at BNL the opinion prevailed that RHIC was invulnerable,  a dream machine not to be beaten in the race to discover the new phase of matter. Hereto I note that nobody back then could tell what the  energy threshold for QGP formation in the very heavy ion collisions would be. The theoretical presumption that this threshold was above the energy produced at SPS turned out to be false. 

Because data taking for the RHIC beam did not happen until 2001, the priority in the field of heavy ions that the US pioneered in a decisive way  at Berkeley in the early 70s passed on to CERN where a  large experimental program at SPS was developed, and as it is clear today, the energy threshold for QGP formation in Pb-Pb collisions was within SPS reach, see Section~\ref{QGPdiscovered}. It  is important to remember that  CERN moved on to develop the relativistic heavy ion research program under the leadership of Herwig Schopper. Schopper, against great odds, bet his  reputation on Heavy Ions to become  one of the pillars of CERN's future. This decision was strongly supported by many national nuclear physics laboratories in Europe, where in my opinion the most important was the support offered by the GSI and the continued development of relativistic heavy ion physics by one of GSI directors, Rudolph Bock.

To conclude the remarks about where we came from and where we are now: a new fundamental set of science arguments about the formation of quark-gluon plasma and deep-rooted institutional support carried the field forward.   CERN was in a unique position to embark on RHI research by having not only the accelerators, engineering expertise, and  research  equipment,  but mainly  due to Hagedorn, also the scientific expertise on the ground, for more detail consult Ref.\cite{HagedornBook}. In the US a major new experimental facility, RHIC at BNL, was developed. With the construction of LHC at CERN a new RHI collision energy domain was opened. The experimental programs at SPS, RHIC and LHC-ion continue today.

\subsection{Format of this review}\label{introformat}
More than 35 years into the QGP endeavor I can say with conviction that the majority of nuclear and particle physicists and the  near totality  of the  large sub-group  directly  involved with the relativistic heavy ion collision research agree that a new form of matter, the (deconfined)  quark-gluon plasma phase has been discovered. The discovery announced at CERN in the year 2000, see Subsection~\ref{QGPdiscovered}, has been confirmed  both at RHIC and by the recent results obtained at LHC. This  review has, therefore, as its primary objective, the  presentation of the part of this  announcement that lives on, see Subsection~\ref{subsec:expStra}, and  how more recent results are addressing these  questions: {\it What are the properties of hot hadron matter? How does it turn  into  QGP, and how does QGP turn back into normal matter?\/} These are to be  the topics addressed in  the  second half of this  review.

There are literally   thousands of research papers in this field today; thus this report cannot aim to be inclusive of all work in the field. We follow the example of John  A. Wheeler. Addressing in his late age a large audience of physicists, he showed one  transparency with one line, \lq\lq What is the  question?\rq\rq\.  In this spirit, this review begins with a series of questions, and  answers,  aiming  to find the answer to:   {\it Which question is THE  question today?} A few issues we raise are truly fundamental present day  challenges. Many provide an opportunity to recognize the  state of the art, both in theory  and experiment.   Some questions are  historical in character and will kick off a debate with other witnesses with a different set of personal memories.

These introductory questions are grouped into three separate sections: first come the theoretical concepts on the hadron side of hot hadronic matter, Section~\ref{HadronSide}; next, concepts on the quark side, Section~\ref{QuarkSide}; and third,  the experimental \lq side\rq\ Section~\ref{QGPinLab} about   RHI collisions. Some of the  questions formulated in Sections~\ref{HadronSide}, \ref{QuarkSide}, and \ref{QGPinLab} introduce  topics that  this review addresses in later sections in depth. The roles of strangeness enhancement and strange antibaryon signature of QGP  are  highlighted. 

We follow this discussion by addressing the near future of the QGP and RHI collision research in the context of this review centered around the strong interactions and hadron-quarks phase.
In   Section~\ref{RHItodayQuestions} I present   several conceptual RHI topics that both are under present active study,  and which will help determine  which direction the field will move on in the  coming decade. Section~\ref{RHItoday} shows the current experimental research program  that address these questions. Assuming that this effort is successful, I propose in Section~\ref{fullwhy} the next generation of physics challenges. The topics discussed are very  subjective; other authors will certainly see other directions and propose other challenges of their interest.

In Section~\ref{meltHad} we deepen the discussion of the origins and the  contents of the theoretical ideas that have led Hagedorn to invent the theoretical foundations leading  on to \Th\ and melting hadrons. The technical discussion is brief and serves as  an introduction  to    Ref.\cite{appenA} which is published for the first time as addendum to this review. Section~\ref{meltHad} ends with a discussion, Subsection~\ref{connectLQCDSBM}, of how the present day lattice-QCD studies test and  verify the theory of hot nuclear matter based on SBM.

Selected theoretical topics related to the study of QGP hadronization  are introduced in the following: In Section~\ref{HadronizationModel} we describe the numerical analysis tool within the Statistical Hadronization Model (SHM); that is, the SHARE suite of computer programs and its parameters. We introduce practical items such as triggered centrality events and rapidity  volume $dV/dy$, resonance decays, particle number fluctuations, which all enter into the   RHIC and LHC data analysis.  

Section~\ref{AnalysisHadronization}  presents the results of the SHM analysis with emphasis put on bulk properties of the fireball;  Subsection~\ref{subsec:Bulk}, addresses SPS and RHIC prior to LHC, while in Subsection~\ref{subsec:BulkLHC}  it is shown how  hadron production can be used to determine the properties of  QGP and how the threshold energy for QGP formation is determined. The results of  RHIC and  LHC are compared and the  universality of QGP  hadronization  across a vast range of energy and fireball sizes described. Subsection~\ref{sec:earlier} explains, in terms of evaluation by example of prior work, why the prior two subsections address solely the  SHARE-model results. In Subsection~\ref{subsec:EvalSHM} the relevance of LHC results to QGP physics is described, and further lattice-QCD relations to these results pointed out. 
  
The final Section~\ref{Comments} does not attempt a summary which in case of a review would mean presenting a review of a review. Instead, a  few  characteristic objectives and  results of this review are highlighted.

An integral part of this  review are  two previously unpublished technical papers which are for the first time in print as an addendum to this review, one from 1980 (Ref.\cite{appenA}) and another from 1983 (Ref.\cite{appenB}).  These two are just a tip of an iceberg; there are many other  unpublished papers by many  authors hidden in conference volumes. There is already a  {\em published} work reprint volume~\cite{Rafelski:2003zz} in which the pivotal works describing QGP theoretical foundations are reproduced; however, the much less accessible and often equally interesting  unpublished  work is at this juncture in time  practically out of sight. This was one of the  reasons leading to the presentation of  Ref.\cite{HagedornBook}. These two papers were selected from this volume and are shown here unabridged. They best complement the contents of this review, providing technical detail not repeated here, while also offering a historical perspective. Beside the key results and/or discussion they also show  the rapid shift in the understanding that manifested itself within a short span of two years. 

Ref.\cite{appenA} presents   \lq\lq Extreme States of Nuclear Matter\rq\rq\ from the  {\it Workshop on Future Relativistic Heavy Ion Experiments\/} held 7-10 October 1980. This small gathering convened by Rudolph Bock and Reinhard Stock is now considered to be the first of  the  \lq\lq Quark Matter\rq\rq\  series {\it \ie\ \/} QM80  conference. Most of this report is a summary of the  theory of hot hadron gas based on Hagedorn's Statistical Bootstrap Model (SBM). The  key new insight in this work was that in RHI collisions the production of  particles rather than the compression of existent matter  was the determining factor.  The had\-ron gas phase study was  complemented by a detailed QGP model  presented as a large, hot, interacting quark-gluon bag.  The phase boundary between these two phases  characterized by Hagedorn temperature \Th\ was evaluated in quantitative manner. It was  shown how the consideration of different collision energies allows us to explore the phase boundary.  This 1980 paper ends with the description of strangeness flavor as the observable of QGP. Strange antibaryons are introduced  as a signature of  quark-gluon plasma.

Ref.\cite{appenB} presents   \lq\lq Strangeness and Phase Changes in Hot Hadronic Matter\rq\rq\  from the {\it Sixth High Energy Heavy Ion Study}, Berkeley, 28 June -- 1 July 1983. The meeting, which had a strong QGP scientific component, played an important role in  the  plans to develop a dedicated relativistic heavy  ion collider (RHIC). In this lecture  I  summarize and update in qualitative terms the technical phase transition consideration seen in   Ref.\cite{appenA}, before turning to the physics of strangeness in hot hadron and quark matter.  The process of strangeness production is presented as being a consequence of dynamical collision processes both among hadrons and in QGP, and the dominance of  gluon-fusion processes in QGP is described. The  role of strangeness in QGP search experiments is presented. For a more extensive historical recount  see Ref.\cite{Rafelski:2007ti}.

%
%
%
\section{The Concepts: Theory Hadron Side}\label{HadronSide}

\subsection{What is the Statistical Bootstrap Model (SBM)?}\label{SBMdef}
Considering that the interactions between   hadronic particles are well characterized by resonant scattering, see Subsection~\ref{HRGsec}, we can describe the gas of interacting hadrons as a mix of all possible particles  and their resonances \lq $i$\rq.  This motivates us to  consider the case of  a gas comprising several types of particles of mass $m_i$, enclosed in a heat bath at temperature $T$,  where the  individual populations \lq $i$\rq\ are unconstrained in their number, that is like photons in a black box   adapting abundance to what is required for the ambient $T$. The nonrelativistic limit of the partition function this gas takes the form
\begin{equation}\label{2chap2eq3.2}
\ln Z = \sum_i\ln Z_i = V\left(\frac{T}{2\pi}\right)^{3/2}\sum_i m_i^{3/2}\E^{-m_i/T}\;,
\end{equation}  
where the momentum integral was carried out and the  sum \lq $i$\rq\ includes all particles of differing parity, spin, isospin, baryon number, strangeness etc. Since each  state is counted, there is no degeneracy factor. 

It is convenient to introduce the mass spectrum $\rho(m)$, where
\begin{equation}\label{2chap2eq3.3}
\rho(m)\D m =\mbox{number of\,\lq $i$\rq\,hadron states in $\{m,m\!+\!\D m\}$}\;.
\end{equation}
Thus we have
\begin{equation}\label{2chap2eq3.4}
Z(T,V) = \exp\left[V\left(\frac{T}{2\pi}\right)^{3/2} \int_0^\infty \rho(m)m^{3/2}\E^{-m/T}\D m\right]\;.
\end{equation} 
On the other hand, a hadronic fireball comprising many components seen on the left in \rf{SBMimage}, when compressed to its natural volume  $V\to V_0$, is itself a highly excited hadron, a resonance that we must include in \req{2chap2eq3.4}. This is what Hagedorn realized in 1964~\cite{Hagedorn:1965st}. This observation leads to an  integral equation for $\rho(m)$ when  we close the \lq bootstrap\rq\ loop that   emerges.

\begin{figure}[t]
\centering\resizebox{0.45\textwidth}{!}{%
\includegraphics{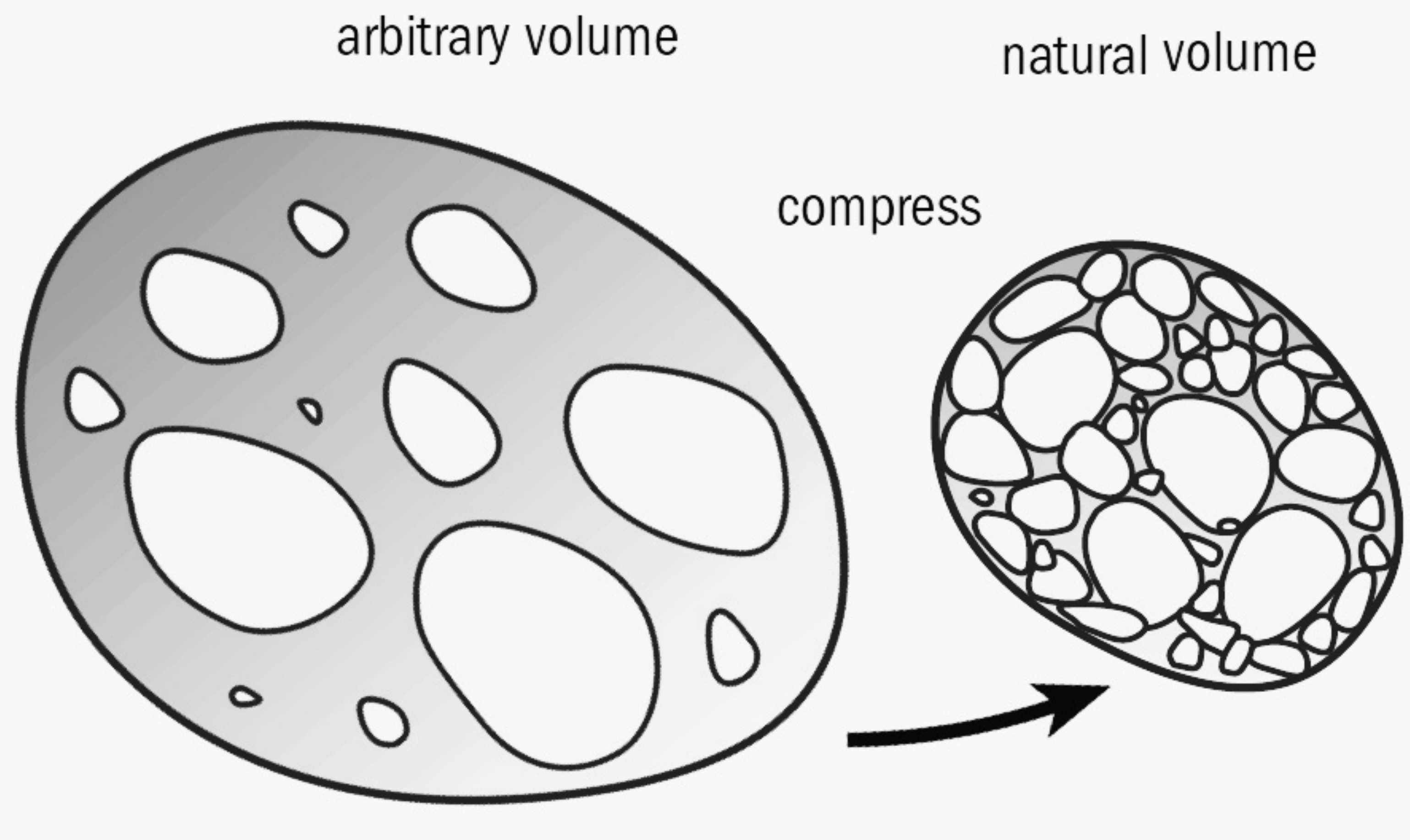}
}
\caption[]{Illustration of the Statistical Bootstrap Model idea: a volume comprising  a gas of fireballs when compressed to natural volume is itself again a fireball. Drawing from Ref.\cite{Ericson:2003ya} modified for this review.}\label{SBMimage}
\end{figure}
 
Frautschi~\cite{Frautschi:1971ij}  transcribed Hagedorn's grand canonical formulation  into microcanonical format. The microcanonical bootstrap equation  reads in invariant Yellin~\cite{Yellin:1973nj} notation
\begin{equation}\label{2chap2eq3.11}
\begin{array}{rl}
H\tau(p^2) = &H\sum_{m_{\rm in}}\delta_0(p^2-m_{\rm in}^2)\\[0.3cm]
           + &\sum_{n=2}^\infty\frac{1}{n!}\int\delta^4\bigg(p-\sum_{i=1}^np_i\bigg)\prod_{i=1}^nH\tau(p_i^2)\D^4p_i\;,
\end{array}
\end{equation}
where $H$ is a universal constant  assuring that \req{2chap2eq3.11} is dimensionless; $\tau(p^2)$ on the left-hand side of \req{2chap2eq3.11} is the   fireball mass spectrum with the mass $m=\sqrt{p_\mu p^\mu}$ which we are seeking to model. The right-hand side of \req{2chap2eq3.11}  expresses that the fireball is either just one  input particle of a given mass $m_{\rm in}$, or else composed of several (two or more) particles $i$  having mass spectra $\tau(p_i^2)$, and 
\begin{equation}
\tau(m^2)\D m^2\equiv \rho(m)\D m,
\end{equation} 
A solution to \req{2chap2eq3.11} has naturally an exponential form
\beq\label{eqepi1}
\rho(m)\propto m^{-a}\exp (m/\Th).
\eeq  
 
The appearance of the exponentially growing mass spectrum, \req{eqepi1}, is a key SBM result. One of the important consequences is that the number  of different hadron states grows so rapidly that  practically every strongly interacting particle found in the fireball is distinguishable. Hagedorn realized that the distinguishability of hadron states was an essential input in order to reconcile statistical hadron multiplicities with experimental data. Despite his own initial rejection of a  draft paper, see Chapters 18 and 19 {\it loc.cit.\/}, this insight was the birth  of the theory of hot hadronic matter as it produced the next step, a model~\cite{Hagedorn:1984hz}.

SBM solutions provide  a wealth of information including the  magnitude of the power index $a$ seen in \req{eqepi1}. Frautschi, Hamer,  Carlitz~\cite{Frautschi:1971ij,Hamer:1971zj,Frautschi:1973xd,Hamer:1974ua,Hamer:1974kc,Carlitz:1972uf}  studied solutions to \req{2chap2eq3.11} analytically and  numerically and  by 1975 drew important conclusions:
\begin{itemize}
\item Fireballs would predominantly decay into two fragments, one heavy and one light.
\item By iterating their  bootstrap equation with realistic input, they found numerically $\Th\approx 140$~MeV and $a=2.9\pm 0.1$, which ruled out the previously adopted approximate value~\cite{Hagedorn:1965st,Hagedorn:1967dia} $a=5/2$ .
\item Each imposed conservation law implemented by fixing a quantum number, e.g., baryon number $\rho(B,m)$, in the mass spectrum, increases the value of $a$ by 1/2.
\end{itemize}
Werner Nahm independently  obtained $a= 3$~\cite{Nahm:1972zc}. Further refinement was possible. In  Ref.\cite{appenA},   a SBM with compressible finite-size  hadrons is introduced where one must  consistently replace Eq.\,(29) by   Eq.\,(30). This  leads to a finite energy  density already for a  model which produces $a=3$ with incompressible hadrons. 

\begin{table}
\centering
\caption[]{Thermodynamic quantities assuming exponential  form of hadron mass spectrum  with preexponential index $a$, \req{eqepi1}; results from Ref.\cite{Rafelski:1979cia}}\label{3chap1tab3.1}
\setlength\tabcolsep{7pt}
\begin{tabular}{@{}|l ||c | c| c| l|@{}}
\hline
\phantom{\Large$\frac{1}{1}$\hspace{-4pt}}$a$ & $P$  & $\varepsilon$ & $\udelta\varepsilon/\varepsilon$
\\\hline\hline
\phantom{\Large$\frac{1}{1}$\hspace{-5pt}}1/2 & $C/\varDelta T^2$ &   $C/\varDelta T^3$ & $C+C\varDelta T$ 
\\\hline
\phantom{\Large$\frac{1}{1}$\hspace{-5pt}}1 & $C/\varDelta T^{3/2}$ &   $C/\varDelta T^{5/2}$ & $C+C\varDelta T^{3/4}$ 
\\\hline
\phantom{\Large$\frac{1}{1}$\hspace{-5pt}}3/2 & $C/\varDelta T$ &   $C/\varDelta T^2$ & $C+C\varDelta T^{1/2}$ 
\\\hline
\phantom{\Large$\frac{1}{1}$\hspace{-5pt}}2 & $C/\varDelta T^{1/2}$   & $C/\varDelta T^{3/2}$ & $C+C\varDelta T^{1/4}$ 
\\\hline
\phantom{\Large$\frac{1}{1}$\hspace{-5pt}}5/2 & $C\ln(T_0/\varDelta T)$   & $C/\varDelta T$ & $C$ 
\\\hline
\phantom{\Large$\frac{1}{1}$\hspace{-5pt}}3 & $P_0-C\varDelta T^{1/2}$   & $C/\varDelta T^{1/2}$ & $C/\varDelta T^{1/4}$ 
\\\hline
\phantom{\Large$\frac{1}{1}$\hspace{-5pt}}7/2 & $P_0-C\varDelta T$ &  $\varepsilon_0$ & $C/\varDelta T^{1/2}$ 
\\\hline
\phantom{\Large$\frac{1}{1}$\hspace{-5pt}}4 & $P_0-C\varDelta T^{3/2}$   & $\varepsilon_0-C\varDelta T^{1/2}$ & $C/\varDelta T^{3/4}$ 
\\\hline
\end{tabular}
\end{table}

For any $\rho(m)$ with  a given value of $a$,  \req{eqepi1}, it is easy to understand the  behavior near to \Th. Inserting \req{eqepi1} into the relativistic form of \req{2chap2eq3.2},  see Chapter 23 {\it loc.cit.}, allows the evaluation  near   critical condition, $\Th-T\equiv \varDelta T\to 0$ of the physical properties such as shown in  Table~\ref{3chap1tab3.1}:  pressure $P$,  energy density $\varepsilon$, and other  physical properties, as example  the mean relative fluctuations $\udelta\varepsilon/\varepsilon$ of $\varepsilon$ are shown,  for $a=1/2,2/2,\ldots, 8/2$. We see that, as $T\rightarrow \Th$ ($\varDelta T\rightarrow 0$), the energy density diverges for $a\le 3$.  

In view of the entries shown  in Table~\ref{3chap1tab3.1} an important further result can be obtained  using these leading order terms for all cases of $a$ considered:  the speed of sound at which the small density perturbations   propagate   
\begin{equation}\label{vsound}
c_s^2:= \displaystyle \frac{dP}{d\varepsilon}\propto \Delta T\to 0\;.
\end{equation} 
This universal for all $a$ result is due to the exponential mass spectrum of hadron matter studied here.  $ c_s^2 \to 0$ at \Th\  harbors an interesting new definition of the phase boundary in the context of lattice-QCD. A nonzero but small value $c_s^2$ should arise from the  subleading terms contributing to $P$ and $\varepsilon$ not shown in Table~\ref{3chap1tab3.1}. The way singular properties  work, it could be that the $c_s^2=0$ point exists.  The insight that the sound velocity  vanishes at \Th\ is known since 1978, see Ref.\cite{Rafelski:1979cia}. An \lq almost\rq\ rediscovery of this  result is seen in Sections~3.5 and 8.7 of Ref.\cite{Satz:2012zza}.

The above discussion shows both the ideas that led to the invention of SBM, and how SBM can evolve with our understanding of the strongly interacting matter, becoming more adapted to the physical properties of the  elementary \lq input\rq\ particles. Further potential refinements  include  introducing strange quark related scale into characterization of the hadron volume,  making baryons more compressible as compared to mesons. These improvements could generate a highly realistic shape of the mass spectrum,  connecting SBM more closely to the numerical study  of QCD in lattice approach. We will return to SBM, and the mass spectrum, and describe the method of finding a solution of \req{2chap2eq3.11} in Section~\ref{meltHad}.

\subsection{What is the Hagedorn temperature \Th?} \label{Thdef}
Hagedorn temperature is  the parameter entering the exponential mass spectrum  \req{eqepi1}. It is measured  by  fitting to data the exponential shape of the hadron mass spectrum. The experimental mass spectrum is discrete; hence a smoothing procedure is often adopted to fit the shape \req{SMOO} to data. In technical detail one usually  follows the method of Hagedorn (see Chapter 20 {\it loc.cit.\/} and Ref.\cite{Hagedorn:1967dia}), applying  a   Gaussian distribution with a width of 200 MeV for all hadron mass states. However,   the  accessible experimental distribution  allows  fixing \Th\   uniquely {\em only} if we know the value of the preexponential power \lq $a$\rq.

The fit procedure is encumbered   by the   singularity for $m\to 0$.  Hagedorn proposed a regularized form of~\req{eqepi1} 
\begin{equation}\label{SMOO}
\displaystyle{ \rho(m) =c\,\frac{e^{m/\Th}}{(m_0^2+m^2)^{\scriptstyle a/2}}}\;.
\end {equation}
In fits to experimental data all three parameters $\Th,m_0,c$ must be varied and allowed to find their best value. In 1967 Hagedorn   fixed $m_0=0.5$\,GeV as he was working in the limit $m> m_0$, and he also fixed $a=2.5$ appropriate for his initial SBM approach~\cite{Hagedorn:1967dia}. The introduction of a fitted value $m_0$ is necessary to improve the  characterization of the hadron mass spectrum for low values of $m$, especially when  a range of possible values for $a$ is considered.

\begin{table}
\caption{\protect\small Parameters of \req{SMOO} fitted  for a prescribed preexponential power  $a$. Results  from Ref.\cite{Tounsi:1994tu}.
\protect\label{tableMss}}{
\[\begin{array}{|l||c|c|c|c|}
\hline
\phantom{\displaystyle\frac{1}{2}}    \!a\!\phantom{\displaystyle\frac{1}{2}}  &     \phantom{\displaystyle\frac{1}{2}}   \!c[\mathrm{GeV}^{a-1}]\!\phantom{\displaystyle\frac{1}{2}} &    \phantom{\displaystyle\frac{1}{2}}    \!m_0[\mathrm{GeV}]\! \phantom{\displaystyle\frac{1}{2}}&   \phantom{\displaystyle\frac{1}{2}}    \!\Th[\mathrm{MeV}]\!\phantom{\displaystyle\frac{1}{2}}& 
\phantom{\displaystyle\frac{1}{2}}    \!\Th[10^{12}\cdot \mathrm{K}]\!\phantom{\displaystyle\frac{1}{2}}\\ 
\hline\hline
2. 5\phantom{\displaystyle\frac{1}{2}} &\  0. 83479\  & \    0. 6346\  & \     165.36\ & \ 1.9189 \ \\[0.1cm]
3.  \phantom{\displaystyle\frac{1}{2}} & 0. 69885 &    0. 66068 &    157.60 &   1.8289\\[0.1cm]
3.5 \phantom{\displaystyle\frac{1}{2}} & 0. 58627 &    0. 68006 &    150.55 &   1.7471\\[0.1cm]
4.  \phantom{\displaystyle\frac{1}{2}} & 0. 49266 &    0. 69512 &    144.11 &   1.6723\\[0.1cm]
5. \phantom{\displaystyle\frac{1}{2}}  & 0. 34968 &    0. 71738 &    132.79 &   1.5410\\[0.1cm]
6. \phantom{\displaystyle\frac{1}{2}}  & 0. 24601 &    0. 73668 &    123.41 &   1.4321\\[0.1cm]
 \hline
\end{array}\]
}\end{table}

The fits to experimental mass spectrum shown in Table \ref{tableMss} are from  1994~\cite{Tounsi:1994tu} and thus  include a smaller set of hadron states than   is available today. However, these results are stable since the new hadronic states found  are  at high mass.  We see in  Table~\ref{tableMss} that as the preexponential power law $a$ increases, the fitted value of $\Th$ decreases. The  value of $c$ for $a=2.5$   corresponds to $c=2.64\times 10^4\,\mathrm{MeV}^{3/2}$, in excellent agreement to the value obtained by Hagedorn in 1967. In \rf{3chap3CUPfig1} the  case $a=3$ is illustrated and  compared to the result of the 1967 fit by Hagedorn and the experimental smoothed spectrum. All fits for different $a$ were found  at nearly equal and convincing  confidence level as can be inferred from \rf{3chap3CUPfig1}.   
 
Even cursory inspection of Table~\ref{tableMss} suggests that the  value of  \Th\ that plays an important role in physics of  RHI  collisions depends on the understanding of the  value of  $a$. This  is the reason that we discussed the different cases in depth in previous subsection~\ref{SBMdef}. The preexponential power value $a=2.5$ in \req{SMOO} corresponds to Hagedorn's original preferred value;  the value $a=3$ was adopted by the mid-70s following extensive study of the SBM as described. However, results seen in Table~\ref{3chap1tab3.1} and  Ref.\cite{appenA} imply $a\ge 7/2$.

This is so since for $a<7/2$   we expect $\Th$ to be a maximum temperature, for which we see in Table~\ref{3chap1tab3.1} a divergence in energy density. Based on study of the statistical bootstrap model of nuclear matter with conserved baryon number  and compressible hadrons presented in  Ref.\cite{appenA}, I believe that $3.5\le a\le 4$. A yet greater value $a\ge 4$ should emerge if in addition strangeness and charge are  introduced as a distinct conserved degree of freedom -- in any  consistently  formulated SBM with canonically  conserved   quantum numbers   one unique value of \Th\ will emerge for the mass spectrum, that is $\rho(m,b,S,\ldots)\propto \exp(m/\Th)$ for any value of $b, S, Q, \ldots$ \ie\  the same \Th\ for mesons and baryons. Only the preexponential function can depend  on $b, S, Q,\ldots$  An example for this is provided by  the  SBM model of Beitel, Gallmeister and Greiner~\cite{Beitel:2014kza}. Using a conserved discrete quantum numbers approach, explicit fits lead to the same (within 1 MeV) value of \Th\ for mesons and baryons~\cite{Beitel:2014kza}.

\begin{figure}
\centering\resizebox{0.44\textwidth}{!}{%
\includegraphics{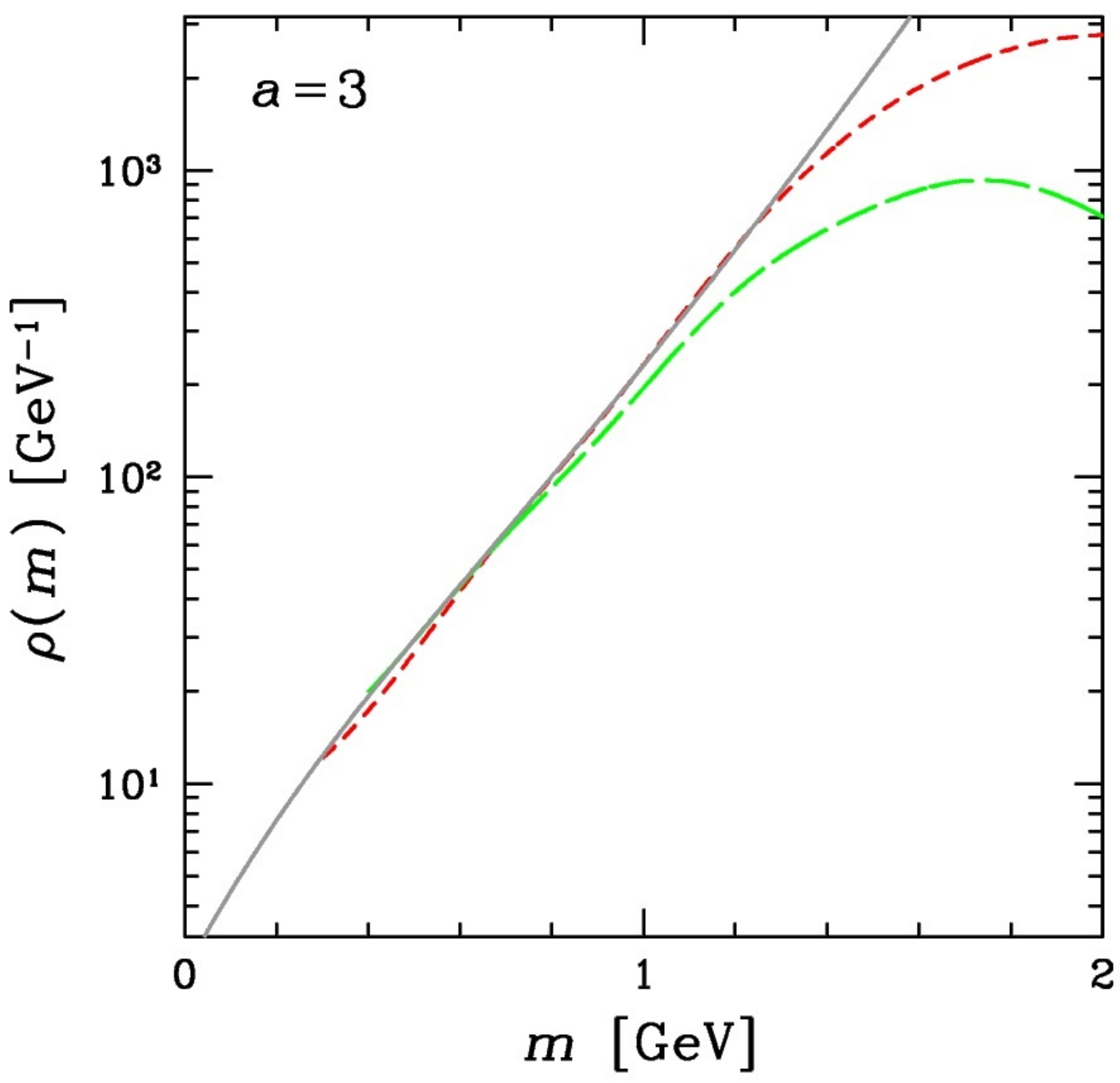}
}
\caption[]{The experimental mass spectrum (solid line), the fit (short dashed), compared to 1967 fit of Hagedorn (long dashed): The case  $a=3$ is shown,  for parameters see table \ref{tableMss}. Figure  from Ref.\cite{Ericson:2003ya} with results obtained in \cite{LetRafMass} modified for this review.}\label{3chap3CUPfig1}
\end{figure}

These results of~Ref.\cite{Beitel:2014kza} are seen   in \rf{hagSpect}: the top frame for mesons and the bottom frame for baryons.  Two different fits are shown characterized by a model parameter   $R$ which, though different from $H$ seen in Eq.\,(15) in Ref.\cite{appenA}, plays a similar role. Thus the  two results   bracket  the value of \Th\  from above (blue, $\Th\simeq 162$\,MeV) and from below (red, $\Th\simeq 145$\,MeV) in agreement with   typical empirical results seen in  table~\ref{tableMss}. 
\begin{figure}
\centering\resizebox{0.44\textwidth}{!}{%
\includegraphics{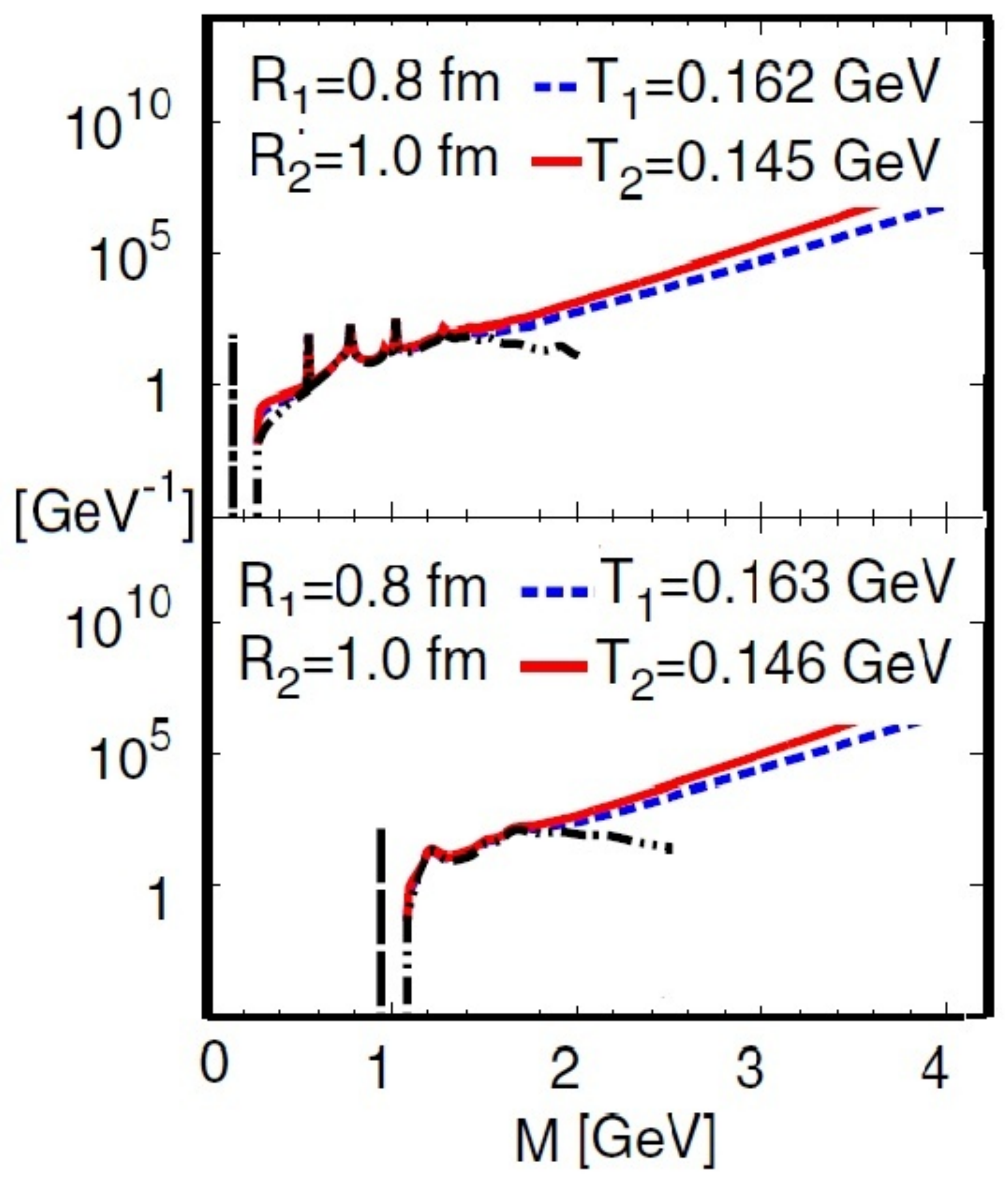}
}
\caption[]{Meson- (top) and baryon- (bottom) mass spectra $\rho(M)$ (particles per GeV): dashed line the experimental spectrum including discrete states. Two different fits are shown, see test. Figure from Ref.\cite{Beitel:2014kza}  modified for this review \label{hagSpect}}
\end{figure}

We further see  in \rf{hagSpect} that a noticeably different number of $M>2$\,GeV states can be expected depending on the value of \Th, even if the  resonances for $M<1.7$\,GeV are equally well fitted in both cases. Thus it would seem that the value of \Th\ can be fixed more precisely in the future when more hadronic resonances are known. However,  for $M\simeq 3$\,GeV there are about  $10^5$ different meson or baryon  states per GeV. This means that states  of this mass  are on average separated by 10\,eV in energy. On the other hand,  their natural width  is at least $10^6$ larger. Thus there is little if any hope to experimentally resolve such \lq Hagedorn\rq\ states. Hence we cannot expect to determine, based on experimental mass spectrum, the  value of \Th\ more precisely than it is already done today. However, there are  other approaches  to measure the value of \Th. For example, we address at the end of Subsection~\ref{WhatIsQGP} why the behavior of lattice-QCD determined speed of sound suggests that    $\Th\simeq 145$\,MeV.

To summarize, our  current understanding is that  Hagedorn temperature has a value still needing an improved determination, 
\begin{equation}\label{Thvalue}
140\le \Th\le 155 \,\mathrm{MeV}\qquad \Th\simeq (1.7\pm 0.1)\times 10^{12} \,\mathrm{K}. 
\end{equation}
\Th\  is the maximum temperature at which matter can exist in its usual form. \Th\ is  not a maximum temperature in the Universe. The value of \Th\ which we evaluate in the study of hadron mass spectra is, as we return to discuss in Section~\ref{WhatIsQGP},  the melting point of hadrons  dissolving into the quark-gluon plasma  (QGP), a liquid phase made of Debye-screened color-ionic quarks and gluons.   A further heating of the quark-gluon plasma  \lq liquid\rq\   can and will continue. A similar transformation can occur already at  a lower temperature at a finite baryon density. 

Indeed, there are two well studied ways to obtain deconfinement:  a) high temperature; and b) high baryon density. In both cases the trick is that the number of particles per unit volume is increased.
\begin{itemize}
\item[a)]
In absence of all matter (zero net baryon number corresponding to baryochemical potential $\muB\to 0$), in full thermal equilibrium temperature alone  controls the abundance of particles as we  already saw in the context of SBM. The result of importance to this review is that confinement  is shown to dissolve in the study of QCD by Polyakov~\cite{Polyakov:1978vu}, and this has been also argued early on and independently in the context of lattice-QCD~\cite{Susskind:1979up}. 
\item[b)]
At nuclear (baryon)  densities an order of magnitude greater than the prevailing nuclear  density in large nuclei, this transformation probably can occur  near to,  or even at, zero temperature; for further quantitative discussion see   Ref.\cite{appenA}.  This is the context in which asymptotically free quark matter was proposed in the context of neutron star physics~\cite{Collins:1974ky}.  
\end{itemize}
Cabibbo and Parisi~\cite{Cabibbo:1975ig} were first to recognize that these two distinct limits are smoothly  connected  and that the  phase boundary could be  a smooth line in the  $\muB,T$  plane. Their qualitative remarks did not address a method to form, or to explore, the phase boundary  connecting these limits. The  understanding of high baryon density matter properties in the limit $T\to 0$ is a separate vibrant  research topic which will not be further discussed here~\cite{Barrois:1977xd,Alford:1997zt,Alford:1998mk,Alford:2007xm}.  Our primary interest is the domain in which the effects of temperature dominate, in this sense the limit of small $\muB\ll T$.

\subsection{Are there several possible values of ${\boldmath{\Th}}$?} \label{Thdef2}
The singularity of the  SBM at \Th\ is a unique  singular point of the model.  If and when within SBM  we implement distinguishability of mesons from baryons, and/or  of strange and nonstrange hadrons, all these families of  particles would have a mass spectrum with a common value of \Th. No matter how complex are the so-called  SBM \lq input\rq\  states, upon Laplace transform they always lead  to one  singular point, see Subsection~\ref{SBMexample}. In subsequent projection of the generating SBM function onto individual families of hadrons one common exponential is found for  all. On the  other hand, it is evident from the  formalism that when extracting from the  common expression the specific forms of the mass spectrum for different particle families, the preexponential function must  vary  from family to family.  In  concrete terms this means that we must fit the individual mass spectra with common \Th\ but particle family  dependent values of $a$ and dimensioned parameter  $c,m_0$ seen in \rt{tableMss}, or any  other assumed preexponential function.

There are several recent phenomenological studies of the hadron mass spectrum claiming to relate to SBM of Hagedorn, and the approaches taken are often disappointing.
The frequently seen   defects are: i) Assumption of $a=2.5$ along  with the Hagedorn 1964-67 model, a value   obsolete since 1971 when $a=3$ and higher was recognized; and ii) Choosing to change \Th\ for different particle families, {\it e.g.\/} baryons and mesons or strange/nonstrange hadrons instead of modifying the preexponential function for different particle families. iii) A third technical  problem is that an integrated (`accumulated') mass spectrum is considered, 
\begin{equation}\label{IntRho}
R(M)=\int_0^M\rho(m)dm \;.
\end {equation}

While the Hagedorn-type approach requires smoothing of the spectrum, adopting an effective Gaussian width for all hadrons, the integrated spectrum \req{IntRho} allows one to address directly the step function arising from integrating the discrete hadron mass spectrum, \ie\  avoiding the  Hagedorn smoothing. One could think that the Hagedorn smoothing process loses information that is now available in the new approach, \req{IntRho}. However, it also could be that a  greater information loss comes from the consideration of the integrated `signal'.   This situation is not uncommon when considering any integrated signal function.

The Krakow group Ref.\cite{Broniowski:2000bj,Broniowski:2004yh} was first to consider   the integrated  mass spectrum \req{IntRho}. They also break the large set of hadron resonances into different classes,  e.g. non-strange/strange hadrons, or  mesons/baryons. However, they  chose same preexponential fit function and  varied  \Th\  between particle families. The  fitted value  of \Th\ was found to be strongly varying  in dependence on supplementary hypotheses made about the procedure, with the value of \Th\  changing by 100's MeV, possibly  showing the  inconsistency of procedure aggravated by the  loss of signal information.  

Ref.\cite{Cleymans:2011fx} fixes $m_0=0.5$\,GeV  at $a=2.5$, \ie\  Hagedorn's 1968 parameter choices. Applying the Krakow  method approach, this fit produces with present day data $\Th=174$\,MeV. We keep in mind that the assumed value of $a$  is incompatible with SBM, while the assumption of a relatively small $m_0=0.5$\,GeV is forcing a relatively large value of  \Th, compare here also the dependence of \Th\ on $a$ seen in \rt{tableMss}. Another similar work is Ref.\cite{Cohen:2011cr}, which seeing poor phenomenological results that emerge from an {\em inconsistent} application of Hagedorn SBM, criticizes unjustly  the current widely accepted Hagedorn approach and Hagedorn temperature. For reasons already described, we do not share in any of the views presented in this work.

However, we note two  studies~\cite{Biro:2005iu,Lo:2015cca}  of differentiated (meson vs. baryon) hadron mass spectrum done in the way that we consider correct: using a common singularity, that is one and the same exponential \Th, but \lq family\rq\ dependent preexponential functions obtained in projection on the appropriate quantum number. It should be noted that the hadronic volume $V_h$ enters any reduction of the  mass spectrum by the  projection method, see  Ref.\cite{appenA}, where volume effect for strangeness is shown. 

Biro and Pe\-shier~\cite{Biro:2005iu} search for \Th\ within nonextensive thermodynamics. They consider  two different values of $a$ for mesons and  baryons (somewhat on the low side), and in their Fig.\,2  the two fits show a common value of \Th\  around  150--170\,MeV.   A very recent lattice motivated effort assumes  differing shape of the preexponential function for different families of particles~\cite{Lo:2015cca}, and  uses a common, but assumed, {\em not fitted},  value of \Th.

Arguably, the most important recent step forward in regard to improving the Hagedorn mass spectrum analysis is the realization first made by  Majumder and M\"uller~\cite{Majumder:2010ik} that one can infer important information about the hadron mass spectrum from  lattice-QCD numerical results. However, this first effort also assumed  $a=2.5$ without a good reason. Moreover, use of asymptotic expansions of the Bessel functions  introduced errors, preventing  a comparison of these results with those seen in \rt{tableMss}. 

To close let us emphasize that  phenomenological approach in which one forces same preexponential function and fits different values of \Th\ for different families of particles is at least within the SBM framework blatantly wrong. A more  general argument indicating that this is always wrong could be also made: the only universal natural constant governing phase boundary is the value of \Th, the preexponential function, which varies depending on how we split up the hadron particle family -- projection of baryon  number (meson, baryon), and strangeness, are two natural choices.

\subsection{What is hadron resonance gas\,(HRG)?} \label{HRGsec}
We are seeking a description of the phase of matter made of individual hadrons. One would be  tempted to think that  the SBM provides a valid framework. However, we already know from discussion above that the  experimental realities limit the  ability to fix the parameters of this model; specifically, we do not know  \Th\ precisely.

In the present day  laboratory   experiments one therefore approaches the situation differently. We employ all  experimentally known hadrons  as explicit partial fractions in the hadronic gas: this is   what in general is called the  hadron resonance gas\,(HRG),  a gas represented by the non-averaged, discrete sum partial contributions, corresponding to the discrete format of $\rho(m)$ as known empirically. 

The emphasis here is on \lq resonances\rq\ gas, reminding us that all hadrons, stable and unstable,  must be included. In his writings Hagedorn went to great length to justify how the inclusion of  unstable hadrons, \ie\  resonances, accounts for the dominant part of the interaction between all hadronic particles. His argument was based on work of  Belenky (also spelled Belenkij)~\cite{Belenky:1956}, but the intuitive content is simple: if and when reaction cross sections are dominated by resonant scattering, we can view resonances as being all the time present along with the scattering particles in order to characterize the state of the physical system. This idea works well for strong interactions since the S-matrix of all reactions is pushed to its unitarity limit. 

To illustrate the situation, let us imagine a hadron system at \lq low\rq\ $T\simeq \Th/5$ and at zero baryon density; this is in essence  a gas made of the three types of pions, $\upi^{(+,-,0)}$. In order to account for dominant interactions between pions we   include their scattering resonances as individual contributing fractions. Given that these particles have considerably higher mass   compared to that of two pions, their number is relatively small. 

But as  we warm up our  hadron gas, for  $T> \Th/5$ resonance contribution becomes more noticeable and in turn their scattering with pions requires inclusion of other  resonances and so on. As we reach $T\lesssim  \Th$  in the heat-up process, Hagedorn's distinguishable particle limit applies: very many {\em different} resonances are present such that this hot gas develops properties of classical numbered-ball system, see Chapter 19 \loccit. 

All heavy resonances ultimately decay, the process creating  pions observed experimentally. This yield is well ahead of what one  would expect from a pure pion gas. Moreover, spectra of particles born in resonance decays differ from what one could expect without resonances.  As a witness of the early Hagedorn work from before 1964,   Maria Fidecaro of CERN told  me recently, I paraphrase \lq\lq when Hagedorn produced his first pion yields, there were many too few,  and with a wrong momentum spectrum\rq\rq. As we know, Hagedorn did not let himself be  discouraged by this initial difficulty.  

The introduction of HRG can be tested  theoretically  by comparing HRG properties with lattice-QCD. In \rf{P_HRG_BorsanyLQCD} we show the pressure presented in  Ref.\cite{Borsanyi:2013bia}. We indeed see a   good agreement of lattice-QCD results obtained for  $T\lesssim  \Th$ with HRG,   within the lattice-QCD uncertainties. In this way  we have ab-initio confirmation that Hagedorn's ideas of using particles and their resonances to describe a strongly interacting hadron gas is correct, confirmed by more fundamental theoretical ideas involving quarks, gluons, QCD.
\begin{figure}
\centering\resizebox{0.45\textwidth}{!}{%
\includegraphics{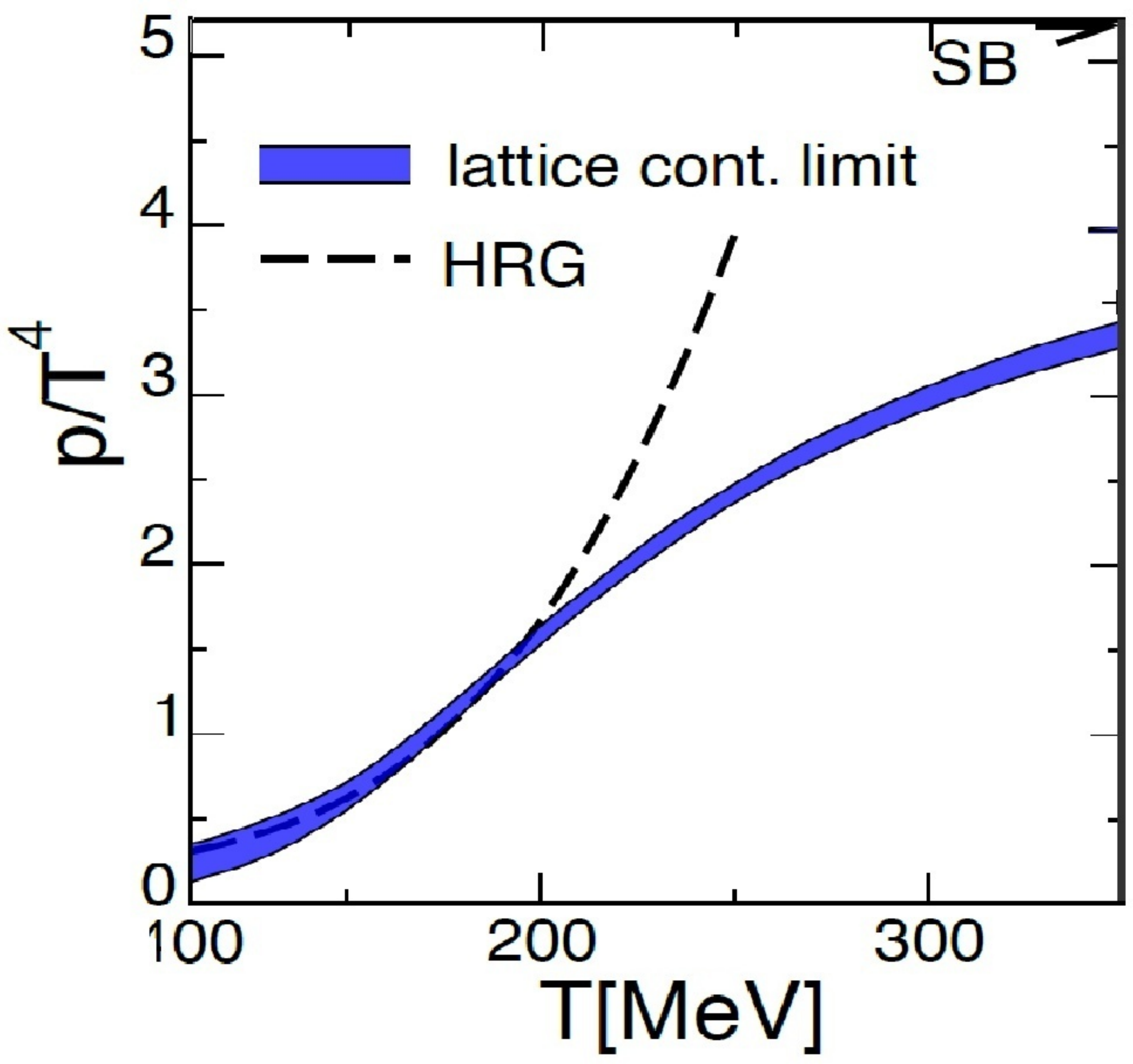}
}
\caption[]{Pressure $P/T^4$ of QCD matter evaluated in lattice approach (includes 2+1 flavors and gluons) compared with their result for the HRG pressure, as function of $T$. The  upper limit of $P/T^4$ is the free Stephan-Boltzmann (SB) quark-gluon pressure with three flavors of quarks in the  relativistic limit $T\gg$\,strange quark mass. Figure from Ref.~\cite{Borsanyi:2013bia} modified for this  review}\label{P_HRG_BorsanyLQCD}
\end{figure}

Results seen in \rf{P_HRG_BorsanyLQCD} comparing pressure of lattice-QCD with HRG show that, as temperature decreases  towards and below  \Th, the color charge of  quarks and gluons  literally freezes, and for   $T\lesssim\Th$ the properties of strongly interacting matter should be  fully characterized by a HRG. Quoting Redlich and Satz~\cite{Redlich:2015bga}:
\begin{quote}
\lq\lq The crucial question thus is,  if the equation of state of hadronic matter introduced  by Hagedorn can describe the corresponding results obtained from  QCD within lattice approach.\rq\rq\  and they continue: \lq\lq There is a clear coincidence of the  Hagedorn resonance model results and the lattice data on the equation of states. All bulk thermodynamical observables  are very strongly changing with temperature when approaching the  deconfinement transition.  This behavior  is well  understood in the Hagedorn  model  as being due to the contribution of resonances. \ldots resonances are indeed the essential  degrees of freedom near deconfinement. Thus, on the thermodynamical level, modeling hadronic interactions by formation and excitation of resonances, as introduced by Hagedorn, is an excellent approximation of strong interactions.\rq\rq\   
\end{quote}

\subsection{What does lattice-QCD tell us about HRG\\ 
and about the emergence of equilibrium ?}\label{hadronMatter}
The thermal pressure reported in \rf{P_HRG_BorsanyLQCD} is the quantity least sensitive to missing high mass resonances which are nonrelativistic and thus contribute  little to pressure. Thus the agreement we see in \rf{P_HRG_BorsanyLQCD} is testing: a) the principles of Hagedorn's  HRG ideas; and b) consistency with the  part of the hadron mass spectrum already known, see \rf{3chap3CUPfig1}. A more thorough study is presented in Subsection~\ref{connectLQCDSBM}, describing the  compensating effect for pressure of finite hadron size and missing high mass states in HRG, which than produces good fit  to energy density.

Lattice-QCD results apply to a fully thermally equilibrated system filling all space-time. This in principle is true  only   in  the early Universe. After hadrons are born at $T\lesssim  \Th$, the  Universe cools in expansion and evolves, with the expansion time constant governed by the magnitude of the (applicable to this period) Hubble parameter; one finds~\cite{Letessier:2002gp,Rafelski:2013obw}  $\tau_q\propto 25\cdot \mu\mathrm{s}$ at \Th,  see also Subsection~\ref{QuarkUniverseLab}. The  value of $\tau_q$ is long   on hadron scale. A  full thermal equilibration of all HRG particle components can be expected in the early  Universe.

Considering the early Universe conditions, it is possible  and indeed {\em necessary} to interpret the  lattice-QCD results in terms of a coexistence era  of  hadrons and QGP. This picture is usually associated with a 1st order phase transition, see Kapusta and  Csernai~\cite{Csernai:1992tj} where one finds separate spatial domains of quarks and hadrons. However, as one can see modeling the more experimentally accessible smooth transition of hydrogen gas to hydrogen plasma, this type of consideration applies  in analogy also to any smooth phase transformation. The difference is that for smooth  transformation, the  coexistence means that the mixing of the two phases is complete at microscopic level; no domain formation occurs. However, the physical properties of the mixed system like in the 1st order transition case are obtained in a superposition of fractional gas components.

\begin{figure}[t]
\centering\resizebox{0.45\textwidth}{!}{%
\includegraphics{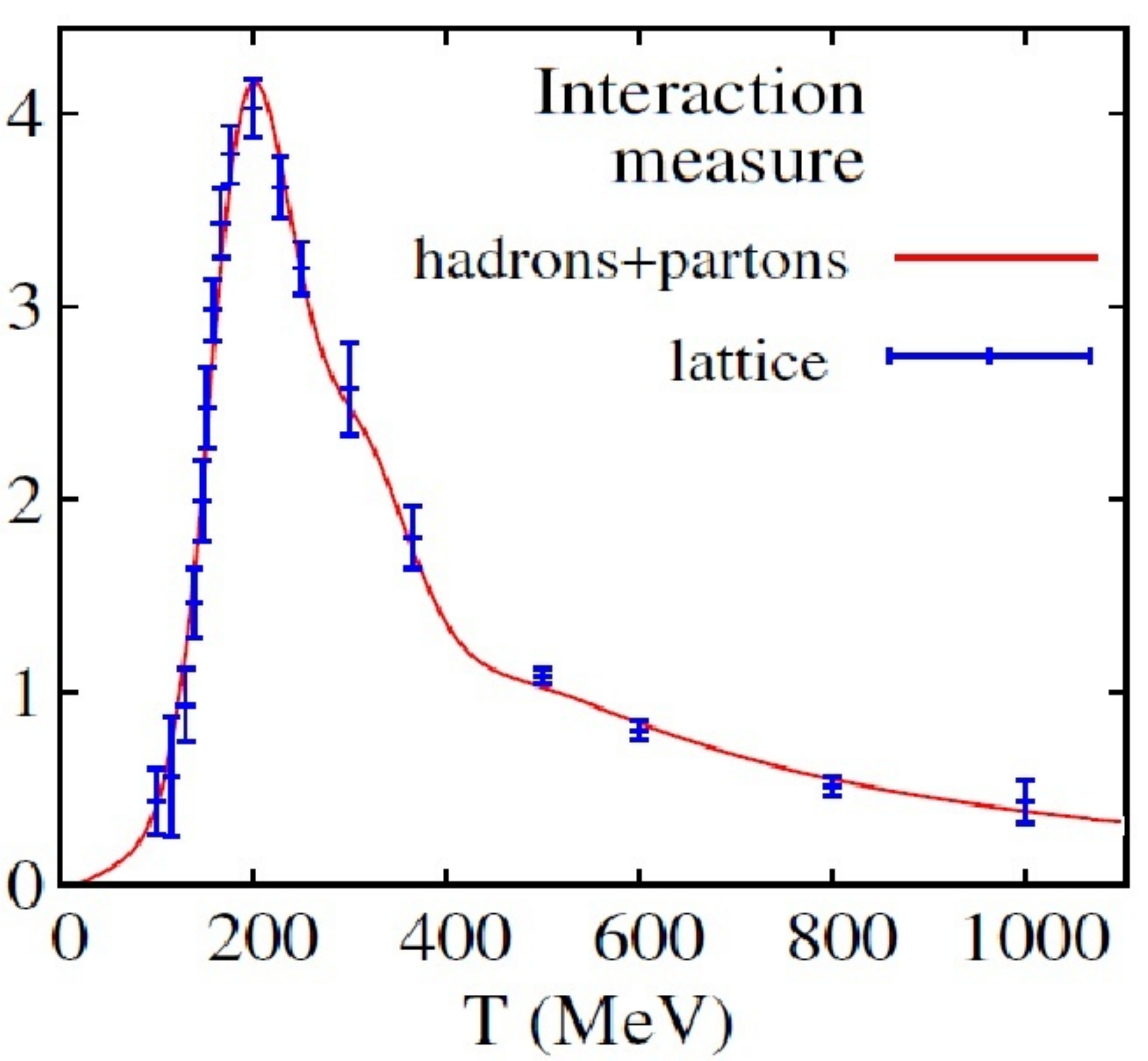}
}
\caption[]{The Interaction measure $(\varepsilon-3P)/T^4$ within mixed parton-hadron model, model fitted to match the lattice data of  Ref.\cite{Borsanyi:2013bia}. Figure from Ref.\cite{Biro:2014sfa} modified for this review}\label{BiroIm}
\end{figure}

The recent analysis of lattice-QCD results of Biro and  Jakovac~\cite{Biro:2014sfa} proceeds   in terms of a perfect microscopic mix of partons and hadrons. One should take note that as soon as QCD-partons appear, in such a picture color deconfinement is present. In Figures~10  and 11 in \cite{Biro:2014sfa} the  appearance of partons for $T>140$\,MeV is noted. Moreover, this  model is able to describe precisely the  interaction measure
\begin{equation}\label{interact_measure}
I_\mathrm{m}\equiv \displaystyle\frac{\varepsilon-3P}{T^4}
\end{equation} 
as shown in \rf{BiroIm}. $I_\mathrm{m}$ is a dimensionless quantity that  depends on the  scale  invariance violation in QCD.  We note the  maximum value of $I_\mathrm{m}\simeq 4.2$ in  \rf{BiroIm}, a value    which reappears  in the   hadronization   fit   in \rf{fig:TraceHad}, Subsection~\ref{subsec:BulkLHC}, where we  see for a few classes of collisions the same value $I_\mathrm{m}\simeq 4.6\pm 0.2$.

Is this  agreement between a hadronization fit  and lattice $I_\mathrm{m}$  an accident? The question is  open  since a priori this agreement has to be considered allowing for the rapid dynamical evolution occurring in laboratory experiments, a situation differing vastly from the lattice simulation of static properties. The dynamical situation   is also more complex and one cannot expect that the matter content of the fireball is a parton-hadron ideal mix. The rapid expansion could and should mean that the parton system evolves without having time to enter equilibrium mixing with hadrons, this is normally called super-cooling in the context of a 1st order phase transition, but in  context of a mix of partons and hadrons~\cite{Biro:2014sfa},  these ideas should also apply: as the parton phase evolves to lower temperature, the yet nonexistent hadrons  will need to form. 

To be specific,  consider a dense hadron phase created in RHI collisions with a size   $R_h\propto 5$\,fm and  a $T\simeq 400$ MeV, where the fit of Ref.\cite{Biro:2014sfa} suggest small if any  presence of hadrons. Exploding into space this parton domain dilutes  at, or even above, the speed of sound  in the transverse direction and even faster into the  longitudinal direction. For relativistic matter the  speed of sound \req{vsound} approaches $c_s=c/\sqrt{3}$, see  \rf{c2s2013} and becomes small only  near to \Th. Within time  $\tau_h\propto 10^{-22}s$ a volume dilution by a factor 50 and more can be   expected. 

It is likely  that this expansion is too fast to allow hadron population to develop from   the  parton domain. What this means is that for both  the lattice-QGP interpreted as parton-hadron mix, and for a HRG formed in laboratory, the reaction time is too short  to allow  development of a multi-structure hadron  abundance equilibrated state,  which one refers to as \lq chemical\rq\ equilibrated hadron gas, see here the early studies in  Refs.\cite{Mekjian:1978zz,Montvay:1978sv,Koch:1984tz}. 


To conclude: lattice results allow various interpretations, and HRG is a consistent simple approximation for $T\lesssim 145$\,MeV. More complex models which  include  coexistence of partons and hadrons manage a good fit to all lattice results, including  the hard to get interaction measure $I_\mathrm{m}$. Such models in turn can be used in developing dynamical model of the QGP fireball explosion. One can argue that the laboratory QGP   cannot be close to  the  full chemical equilibrium; a kinetic  computation will be needed to assess how the properties of   parton-hadron phase evolve given a characteristic lifespan of about   $\tau_h\propto 10^{-22}s$. Such a study may be  capable of justifying accurately specific hadronization models.

\begin{figure}[t]
\centering\resizebox{0.45\textwidth}{!}{%
\includegraphics{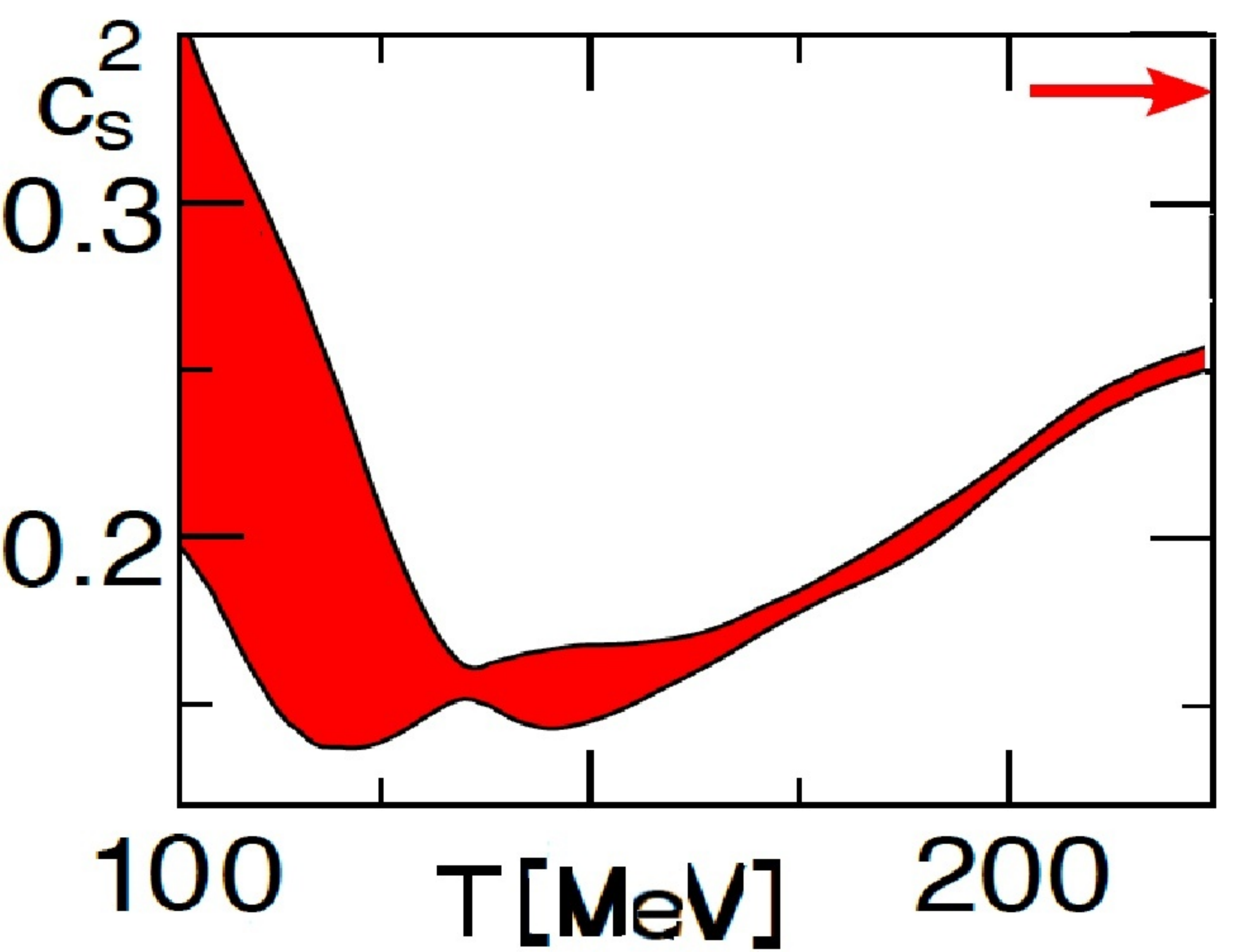}
}
\caption[]{The square of speed of sound $c_s^2$ as function of temperature $T$, the relativistic limit is indicated by  an arrow. Figure from Ref.\cite{Borsanyi:2013bia} modified for this review \label{c2s2013}}
\end{figure}

\subsection{What does lattice-QCD tell us about \Th?}\label{latticeTh}
We will see in Subsection~\ref{WhatIsQGP}  that  we do have two different lattice results showing identical behavior at $T\in \{150\pm25\}$\,MeV. This suggests that it should be possible to obtain a narrow range of \Th. Looking at  \rf{P_HRG_BorsanyLQCD},  some see \Th\ at 140--145 MeV, others as high as 170 MeV. Such disparity can arise when using eyesight to evaluate \rf{P_HRG_BorsanyLQCD} without applying a valid criterion. In fact such a criterion is available if we believe in exponential mass spectrum.

When presenting critical properties  of SBM  \rt{3chap1tab3.1} we reported that sound velocity \req{vsound} has the unique property  $c_s\to 0$ for $T\to\Th$. What governs this result is solely the exponential mass spectrum, and this result holds  in leading order {\em irrespective of the value of the power index $a$}. Thus a surprisingly simple SBM-related criterion for the value of \Th\ is that there  $c_s\to 0$.  Moreover, $c_s$ is available in lattice-QCD computation; \rf{c2s2013} shows $c_s^2$ as function of $T$, adapted from   Ref.\cite{Borsanyi:2013bia}. There is a noticeable domain where  $c_s$ is relatively small. 

In \rf{c2s2013} the bands show the computational uncertainty. To understand better the value of  \Th\  we  follow the  drop of  $c_s^2$ when temperature increases, and when $c_s$ begins to increase that is presumably,  in the context of lattice-QCD,   when the plasma material is mostly made of deconfined and progressively more mobile quarks and gluons. As temperature rises further, we expect to reach  the speed of sound  limit of ultra relativistic matter $c_s^2\to  1/3$, indicated in \rf{c2s2013} by an arrow. This  upper limit, $c_s^2\le 1/3$ arises according to \req{vsound}  as long as  the constraint $\varepsilon -3P\to 0$ from above at high $T$ applies; that is $I_\mathrm{m}>0$ and $I_\mathrm{m}\to 0$ at high $T$.

The behavior of the lattice result-bands in \rf{c2s2013} suggests hadron dominance below $T=125$ MeV, and  quark dominance above $T=150$ MeV. This is a decisively more narrow range compared to the wider one seen in the fit in which a mixed   parton-hadron phase  was used to describe lattice results~\cite{Biro:2014sfa}; see discussion in Subsection~\ref{hadronMatter}.

The shape of  $c_s^2$ in \rf{c2s2013} suggests that $\Th=138\pm 12$ MeV. There are many  ramifications of such a low value, as is discussed in the context of   hadronization model in the following Subsection~\ref{SHMsec}.

\subsection{What\,is\,the\,statistical\,hadronization\,model\,(SHM)?} \label{SHMsec}
The pivotal point leading on from the last subsection is that  in view of \rf{P_HRG_BorsanyLQCD} we can say that HRG  for  $T\lesssim \Th\simeq 150$\,MeV  works well  at a precision level that rivals the numerical precision of lattice-QCD results.   This result {\em justifies} the  method  of data analysis that we call  Statistical Hadronization Model (SHM).  SHM  was invented to characterize  how a  blob of primordial matter that we call QGP  falls apart into individual hadrons. At zero baryon density this \lq hadronization\rq\ process is expected  to occur near if not exactly at \Th. The SHM  relies on the hypothesis that a hot fireball made of building blocks of future hadrons    populates all available phase space cells proportional to their respective size, without regard to any additional interaction strength governing the process. 

The  model is presented in depth in Section~\ref{HadronizationModel}. Here we would like to place emphasis on the fact that the  agreement of lattice-QCD results with the HRG provides today a firm theoretical foundation for the use of the SHM, and it sets up the high  degree of  precision at which SHM can be trusted.

Many  argue that  Koppe~\cite{Koppe:1949zza,Koppe:1949zzb}, and later, independently, Fermi~\cite{Fermi:1950jd} with improvements made by Pomeranchuk~\cite{Pomeranchuk:1951ey}, invented SHM in its microcanonical format; this is the so called Fermi-model,  and that Hagedorn~\cite{Hagedorn:1965st,Hagedorn:1968zz}  used these ideas in computing within grand canonical formulation. However, in all these approaches the  particles emitted were not newly formed; they were seen as already being  the constituents of the fireball.  Such models therefore are what we today call freeze-out models. 

The difference between QGP  hadronization and freeze-out models is that a priori  we do not know if right at the time of QGP hadronization particles will be born into a condition that allows free-streaming and thus evolve in hadron form to the  freeze-out condition.  In a freeze-out model all particles that ultimately free-stream to a detector are not emergent from a   fireball but are already present. The fact that the  freeze-out condition must be established in a study of particle interactions was in the early days of the  Koppe-Fermi model of no relevance since the experimental outcome was governed by the phase space and microcanonical constraints as Hagedorn explained in his  very  vivid account \lq\lq The long way to the Statistical Bootstrap Model\rq\rq, Chapter 17 \loccit.

In the Koppe-Fermi-model,  as of the instant of their formation, all hadrons are free-streaming. This is also Hagedorn's fireball pot with boiling matter. This reaction view was formed before two different phases of hadronic matter were recognized. With the introduction of a second primordial phase  a new picture emerges: there are no hadrons to begin with. In this case in a first step quarks freeze into hadrons at or near \Th, and in a second step at $T<\Th$ hadrons  decouple into free-streaming particles. It is possible that \Th\ is low enough so that when the quark freezing into hadrons occurs,  hadrons are immediately free-streaming; that is $T\simeq \Th$, in which case one would expect abundances of observed individual particles to be constrained by the properties of QGP, and not of the HRG.  

On the other hand if in the QGP hadronization a dense phase of hadron matter  should  form, this will assure both chemical and thermal equilibrium of later free-streaming hadrons as was clearly  explained in 1985~\cite{Koch:1985hk}: \lq\lq Why the Hadronic Gas Description of Hadronic Reactions Works: The Example of Strange Hadrons\rq\rq. It is argued that the  way parton deconfinement manifests itself is to allow a short lived small dynamical system to reach nearly full thermal and chemical equilibrium. 

The analysis of the  experimental data within the SHM allows us to determine the degree of equilibration for different collision systems. The situation can be very different in \pp and \hAA collisions and depend on both collision energy and the size $A$ of atomic nuclei, and the related variable describing the  variable classes, the participant number $N_\mathrm{part}$, see  Subsection~\ref{ssec:particip}. Study of strangeness which is not present in initial RHI  states allows us to address the  equilibration question in a quantitative way as was noted already 30 years ago~\cite{Koch:1985hk}. We return to the SHM strangeness results in Section~\ref{AnalysisHadronization} demonstrating the absence of chemical equilibrium in the final state, and  the   presence of (near) chemical equilibrium in QGP formed at LHC, see \rf{fig:strangeness}  and \ref{fig:sSratio}.  

One cannot say it strongly enough: the transient presence of the primordial phase of matter means that  there are two different possible scenarios describing  production of hadrons  in RHI collisions:\\
{\bf a)} A dense fireball disintegrates into hadrons. There can be two temporally separate physical phenomena: the re\-com\-bi\-nant-evaporative hadronization of the fireball made of quarks and gluons  forming a HRG; this is followed by  freeze-out; that is, the beginning of free-streaming   of the newly created particles.\\
{\bf b)} The quark fireball expands significantly before converting into hadrons,  reaching a low density before hadronization. As a result, some features of  hadrons upon production are already free-streaming: i) The hadronization temperature may be low enough to freeze-out particle abundance (chemical freeze-out at hadronization), yet elastic scattering can still occur and as result momentum distribution will evolve (kinetic non-equilibrium at hadronization). ii) At a yet lower temperature domain, hadrons would be born truly free-streaming and both chemical and kinetic freeze-out conditions would be the same. This condition has been proposed for SPS yields and spectra in the year 2000  by  Torrieri~\cite{Torrieri:2000xi}, and named \lq single freeze-out\rq\ in a later study of  RHIC  results~\cite{Broniowski:2001uk,Baran:2003nm}.

\subsection{Why  value  of \Th\ matters to SHM analysis?} \label{SHMsecTh}
What exactly happens in RHI collisions in regard to particle production depends to a large degree on the value of the chemical freeze-out temperature\footnote{We omit subscript for all different \lq temperatures\rq\ under consideration -- other than \Th\ -- making the meaning   clear  in the text contents.} $T\le \Th$. The value of \Th\ as determined from mass spectrum of hadrons  depends on the value of the preexponential power index $a$, see table~\ref{tableMss}. The lower is \Th, the  lower the value of $T$ must be. Since the value  of $T$ controls the  density of particles, as seen in e.g. in \req{2chap2eq3.2}, the less dense would be the HRG phase that can be formed. Therefore, the  lower is  \Th\ the  more likely that  particles boiled off in the hadronization process  emerge without rescattering, at least without the rescattering that changes one type of particle into another \ie\ \lq chemical\rq\ free streaming. In such a situation  in chemical abundance analysis we expect to find $T\simeq \Th$.
 
The SHM analysis of particle production allows us to determine both the statistical  parameters including the value of $T$  characterizing the hadron phase space,  as well as the extensive (e.g. volume) and intensive  (e.g. baryon density) physical properties of the fireball source. These  govern  the outcome of the experiment on the hadron side, and thus can be measured employing experimental data on  hadron production as we show in Section~\ref{AnalysisHadronization}. 

The faster is the hadronization process, the more information is retained  about the QGP fireball in the hadronic populations we study. For this reason there is a long-lasting discussion in regard to how fast or, one often says, sudden is the breakup of QGP into hadrons. Sudden hadronization means that the time between QGP breakup and  chemical freeze-out is short as compared to the time needed to change abundances of particles in scattering of hadrons.

Among the source (fireball) observables we note the nearly conserved, in the hadronization process, entropy content, and the strangeness content, counted in terms of the emerging multiplicities of hadronic particles. The physical relevance  of these quantities is that they  originate, e.g. considering entropy or strangeness  yield, at an earlier fireball evolution stage as compared to the hadronization process itself; since entropy can only increase, this provides a simple and transparent example how in hadron abundances which express total entropy content there can be memory of the  the initial state dynamics. 

Physical bulk properties such as the conserved (baryon number), and  almost conserved (strangeness pair yields, entropy yield) can be measured  independent of how fast the hadronization process is, and independent of the  complexity of  the evolution during the eventual period in  time while the fireball cools from  \Th\ to chemical freeze-out $T$. We do not know how   the bulk energy density $\varepsilon$ and pressure $P$ at hadronization after scaling with $T^4$ evolve in time to freeze-out point, and even more interesting is how $I_m$  \req{interact_measure} evolves. This can be a topic of future  study. 
 
Once scattering processes came into discussion,  the  concept of dynamical models of freeze-out of particles could be  addressed. The review  of Koch \etal~\cite{Koch:1986ud} comprises many original research results and includes for the first time the consideration of dynamical QGP fireball evolution into free-streaming hadrons and an implementation of SHM in a format that we could today call SHM with sudden hadronization. In parallel it was recognized that the experimentally observed particle abundances allow the determination of physical  properties of the source. This insight is introduced in  Ref.\cite{appenB}, Fig.\,3 where we  see  how the ratio K$^+$/K$^-$ allows the evaluation of  the baryochemical potential \muB; this  is  stated explicitly in pertinent discussion. Moreover, in the following  Fig.\,4  the comparison is made between  abundance of final state $\overline{\uLambda}/\uLambda$ particle ratio  emerging from equilibrated HRG with abundance  expected in direct evaporation of the quark-fireball an effect that we attribute today  to chemical nonequilibrium with enhanced phase space abundance.

Discussion of how sudden the hadronization process is reaches back to  the  1986  microscopic model description  of strange (antibaryon) formation by Koch, M\"uller and the author~\cite{Koch:1986ud} and the application of hadron afterburner. Using these ideas in 1991, SHM model saw its first humble   application in the study of strange (anti)baryons~\cite{Rafelski:1991rh}. Strange baryon and antibaryon  abundances were interpreted assuming a fast hadronization of QGP -- fast meaning that their relative yields are  little changed in the  following evolution.  For the past 30 years the  comparison of data with   the sudden hadronization concept has never led to an inconsistency. Several theoretical  studies support the sudden hadronization approach, a sample of works includes  Refs.\cite{Csorgo:1994dd,Csernai:1995zn,Biro:1998dm,Rafelski:2000by,Keranen:2002sw}. Till further  notice we must presume that   the case has been made. 

Over the past 35 years a simple and naive thermal model of particle production has resurfaced multiple times, reminiscent of the work of Hagedorn from the  early-60s. Hadron yields emerge from a fully equilibrated hadron fireball at a given $T,V$ and to account for baryon content at low collision energies one  adds \muB. As Hagedorn found  out, the price of simplicity is that the yields can differ from experiment by a factor two or more. His effort to resolve this riddle gave us SBM. 

However, in the context of experimental results that need attention,  one seeks to understand  systematic behavior across yields varying by many  orders of magnitude as parameters (collision energy, impact parameter) of RHI collision change. So if a simple model practically \lq works\rq, for many the  case is closed. However, one finds in such a simple model study the value of chemical freeze-out $T$   well above \Th. This is so since in fitting abundant strange antibaryons there are two possible solutions: either  a $T\gg \Th$, or   $T\lesssim\Th$  with chemical  nonequilibrium.  A model with  $T\gg \Th$ for the price of getting strange antibaryons right creates other contradictions, one of which is discussed in Subsection~\ref{subsec:EvalSHM}.

\begin{figure*}
\centering\resizebox{0.87\textwidth}{!}{%
\includegraphics{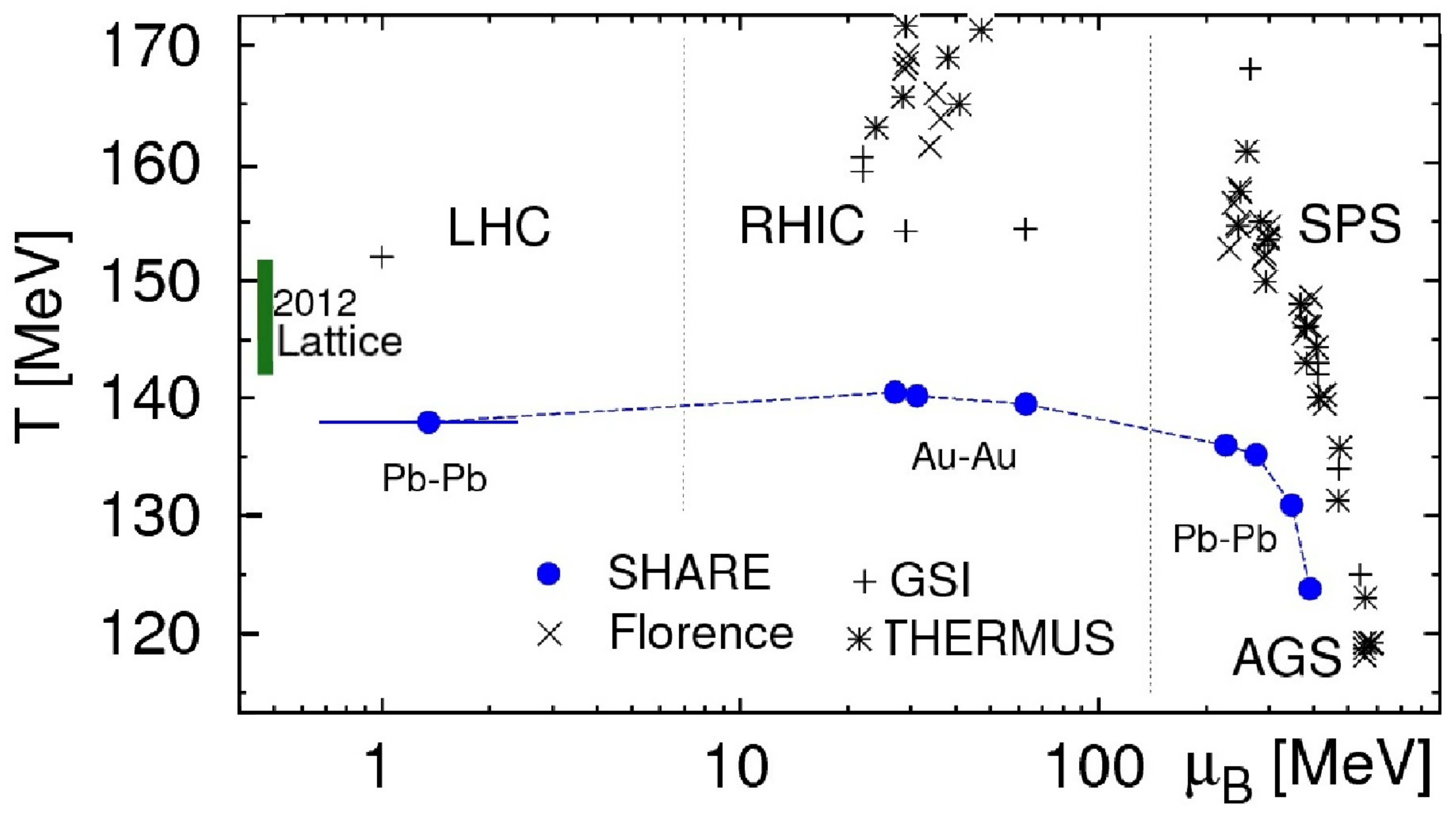}
}
\caption{\label{fig:phasediagram} $T,\muB$  diagram showing current lattice value of critical temperature $T_c$ (bar on left~\cite{Borsanyi:2012rr,Borsanyi:2013bia}, the SHM-SHARE  results (full circles)~\cite{Letessier:2005qe,Petran:2013lja,Petran:2013qla,Petran:2013dea,Petran:2011aa,Rafelski:2009jr,Rafelski:2009gu,Letessier:1998sz} and results of other groups~\cite{BraunMunzinger:1994xr,Andronic:2005yp,Becattini:2010sk,Manninen:2008mg,Becattini:2003wp,Cleymans:2005xv,Abelev:2008ab,Abelev:2012wca,Abelev:2013vea}. Figure from Ref.\cite{Petran:2013lja} modified for this review}
\end{figure*}

How comparison of chemical freeze-out  $T$ with \Th works is shown in~\rf{fig:phasediagram}.  The  bar near to the temperature axis displays the range    $\Th =147\pm5\,\mathrm{MeV}$~\cite{Borsanyi:2012rr,Borsanyi:2013bia}. The symbols show the results of hadronization analysis in the $T$--$\muB$ plane as compiled in Ref.\cite{Petran:2013lja} for results involving  most (as available) central collisions and heaviest nuclei.  The  solid circles are   results obtained using the full SHM parameter set~\cite{Letessier:2005qe,Petran:2013lja,Petran:2013qla,Petran:2013dea,Petran:2011aa,Rafelski:2009jr,Rafelski:2009gu,Letessier:1998sz}.  The  SHARE LHC  freeze-out temperature is clearly below the   lattice critical temperature range. The results of other groups are obtained with simplified parameter sets:  marked   GSI~\cite{BraunMunzinger:1994xr,Andronic:2005yp}, Florence~\cite{Becattini:2010sk,Manninen:2008mg,Becattini:2003wp},   THERMUS~\cite{Cleymans:2005xv},   STAR~\cite{Abelev:2008ab} and   ALICE~\cite{Abelev:2012wca,Abelev:2013vea}.  These results show the chemical freeze-out temperature $T$ in general well above the lattice \Th. This means that these restricted SHM studies are incompatible with lattice calculations, since chemical hadron decoupling should not occur inside the QGP domain.

\subsection{How is SHM analysis of data performed?} \label{SHMsecProc}
Here the procedure steps are described which  need technical implementation presented in  Section~\ref{HadronizationModel}. 

{\bf Data:}  The experiment provides,  within a well defined collision class, see Subsection~\ref{ssec:particip}, spectral  yields of many  particles. For the  SHM analysis we focus  on integrated spectra, the  particle number-yields. The reason that such data are chosen for study is that  particle yields are independent of  local matter velocity in the fireball which imposes spectra deformation  akin to the  Doppler shift.  However, if the $p_\bot$ coverage is not full, an extrapolation of spectra needs to be made that introduces the same uncertainty into the  study. Therefore it is important to achieve experimentally as large as possible $p_\bot$ coverage in order to minimize extrapolation  errors on particles yields considered. 

{\bf Evaluation:}
In first step we evaluate, given an assumed SHM parameter set, the phase space size for all and every  particle fraction that could be  in principle measured, including resonances. This complete set is necessary since the  observed particle set includes particles arising from a sequel chain of resonance decays. These decays are implemented and we obtain the relative phase space size of all potential particle yields. 

{\bf Optional:} Especially should hadronization $T$ be at a relatively large value, the primary  particle populations can undergo modifications in subsequent  scattering. However, since $T<\Th$, a large $T$ requires an even larger \Th\ which shows importance of knowing \Th.  If $T$ is large, a further evolution of hadrons can be treated with   hadron \lq after-burners\rq\ taking the system from \Th\ to  $T$. Since in our  analysis the  value of hadronization $T$  is small, we do not address this stage further here; see however Refs.\cite{Karpenko:2012yf,Botvina:2014lga}.

{\bf Iteration:} The particle yields obtained from phase space evaluation represent the  SHM  parameter set assumed. A comparison of this predicted yield with observed yields allows the formation of a value parameter such as 
\begin{equation}\label{chi1value}
\chi^2=\sum_i(\mathrm{theory-data})^2/\mathrm{FWHM}^2,
\end{equation}
where FWHM is the error in the data, evaluated as \lq Full Width at Half Maximum\rq\ of the data set. In an iterative approach minimizing $\chi^2$ a best set of parameters is found.

{\bf Constraints:} There may be significant  constraints; an example is the required balance of $\bar s=s$ as strangeness is produced in pairs and strangeness changing weak decays  have  no time to operate~\cite{Letessier:1993hi}. Such constraints can be implemented most effectively by constraints in the iteration steps; the iterative steps do not need till the very end to conserve \eg\ strangeness.

{\bf Bulk properties:} When our iteration has converged, we have obtained all primary particle yields; those that are measured, and all others that are, in essence, extrapolations from known to unknown. It is evident that we can use all these yields in order to compute the bulk properties of the fireball source, where the statement is exact for the  conserved quantities such as net baryon number (baryons less antibaryons) and approximate for quantities where kinetic models show little modification of the value during  hadronization. An example here is the number of strange quark pairs or entropy. 

{\bf Discussion:} The best fit is characterized by a value  function, typically  $\chi^2$ \req{chi1value}. Depending on the  complexity of the model, and the accuracy of the  inherent physics picture,  we can arrive at either  a  well converged fit, or at a poor one where  $\chi^2$  normalized by  degrees of freedom (dof) is significantly  above unity. Since the objective of the SHM is the  description of the  data, for  the case of a bad  $\chi^2$ one must seek a more complex model. The  question  about analysis degeneracy also arises: are there two different SHM model variants that  achieve in a systematic way  as a function of reaction energy and/or collision parameters always a success? Should degeneracy be  suspected, one  must  attempt to break degeneracy by  looking at specific experimental observables, as was argued in Ref.\cite{Rafelski:2002ga}.

We perform  SHM analysis of all \lq elementary\rq\ hadrons  produced -- that is we exclude composite light nuclei and antinuclei that in their tiny abundances may have  a different production history;  we will allow the data to decide what are the  necessary model characteristics. We find that for all the  data we study and report on in Section~\ref{AnalysisHadronization}, the result is strongly consistent with the parameter set and values associated with  chemical non-equilibrium. In any  case,   we obtain a  deeper look into the history of the expanding QGP fireball and QGP properties at chemical freeze-out temperature  $T<\Th$  and,   we   argue that QGP was formed. In a study of the  bulk fireball  properties   a precise description of {\em all relevant} particle yields is needed.  Detailed results of SHM analysis are presented in Section~\ref{AnalysisHadronization}.
 

\section{The Concepts: Theory Quark Side}\label{QuarkSide}
\subsection{Are quarks and gluons \lq real\rq\ particles?}\label{quarksBags}
The   question to be addressed in our context is: {\em How can quarks and gluons be  real particles and yet we fail to produce them?\/} The fractional electrical charge of quarks is  a  strong characteristic  feature and  therefore the literature is full of false discoveries. Similarly, the understanding and  explanation of quark confinement has many  twists and turns, and some of the arguments though on first sight  contradictory are saying one and the  same thing. Our present understanding   requires the introduction of  a new paradigm,  a new conceptual context how in comparison to the other interactions the  outcome of strong interactions is different.
 
A clear statement  is seen  in the September 28, 1979 lecture by T.D. Lee~\cite{Lee:1979nz} and the  argument is also presented in T.D. Lee's textbook~\cite{Lee:1981}:  at zero temperature quarks can only appear  within a bound state with other quarks  as a result of  transport properties of the vacuum state, and {\em NOT} as a consequence of the enslaving nature of  inter-quark forces. However, indirectly QCD  forces  provide the vacuum structure, hence quarks are enslaved by the same QCD forces that also  provide the  quark-quark interaction.   Even so the conceptual difference is clear: we can liberate quarks by changing  the  nature of the vacuum, the modern day \ae ther, melting its confining structure.

The quark confinement paradigm is seen as an expression of the incompatibility  of quark and gluon color-electrical fields with the vacuum structure. This insight was inherent in the work by Ken Wilson~\cite{Wilson:1974sk} which was the backdrop against which an effective picture of hadronic structure, the \lq bag model\rq\ was created in 1974/1975~\cite{Chodos:1974je,DeGrand:1975cf,Johnson:1975zp,Thomas:1982kv}. Each hadronic particle is a bubble~\cite{Chodos:1974je}. Below \Th, with their color field lines expelled from the vacuum,  quarks can only exist in colorless cluster states:  baryons $qqq$ (and antibaryons $\overline q \overline q\overline q$) and mesons $q\overline q$ as illustrated in \rf{Bagimage}. 

\begin{figure}
\centering\resizebox{0.45\textwidth}{!}{%
\includegraphics{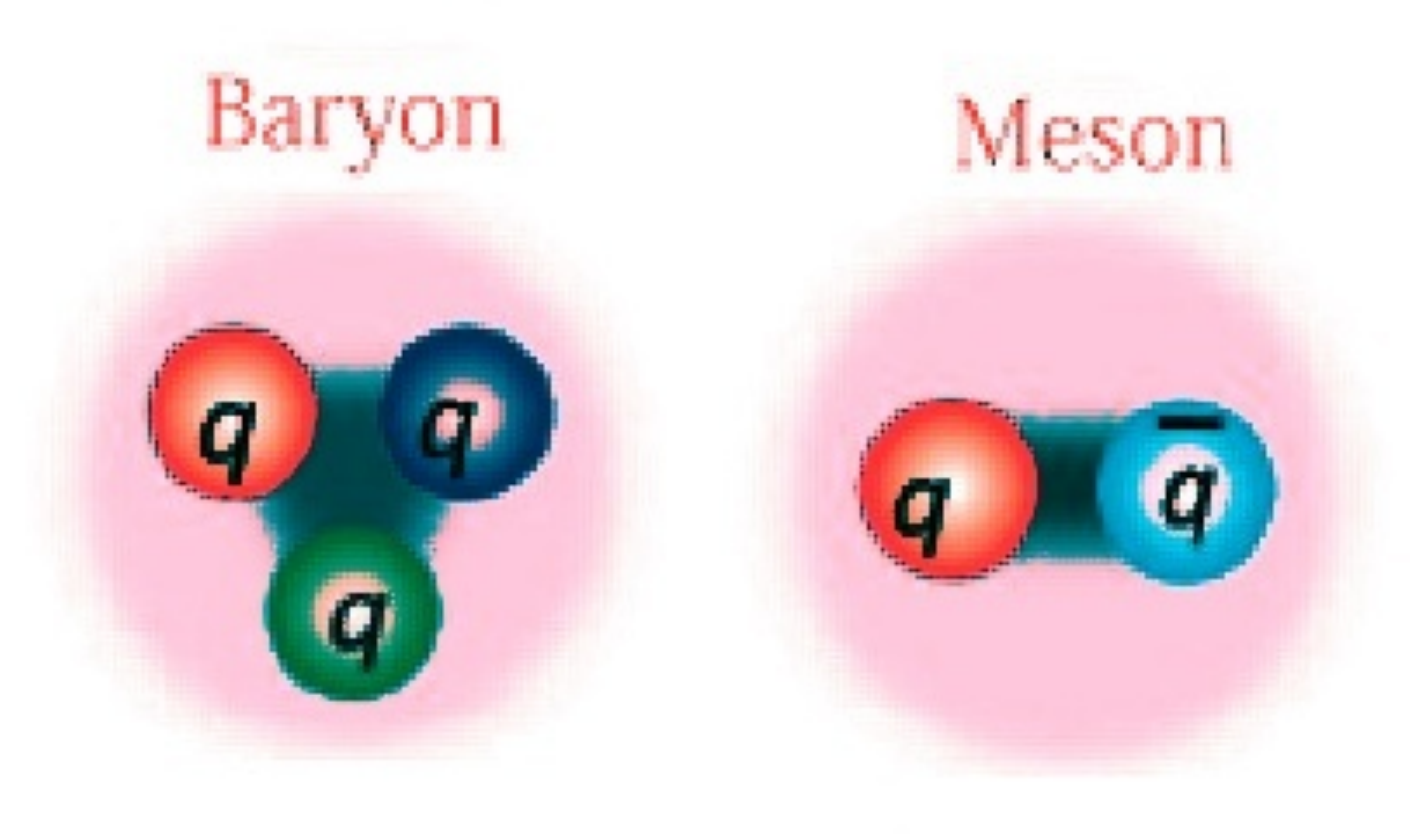}
}
\caption[]{Illustration of the quark bag model colorless states:  baryons $qqq$ and mesons $q\overline q$. The range of the quantum wave function of quarks, the  hadronic radius is  indicated as a (pink) cloud, the color electrical field lines connect individual quarks.}\label{Bagimage}
\end{figure}

These are bubbles with the electric field lines contained in a small space domain, and the color-magnetic (spin-spin hyperfine) interactions contributing the details of the hadron  spectrum~\cite{DeGrand:1975cf}.  This implementation of quark confinement is the so-called (MIT) quark-bag model. By imposing boundary conditions between the two vacuua,  quark-hadron wave functions in a localized bound state were obtained; for a succinct review see Johnson~\cite{Johnson:1975zp}. The later developments which  address the chiral symmetry are summarized  in 1982 by Thomas~\cite{Thomas:1982kv}, completing the model. 

The quark-bag model works akin to the localization of quantum states in an infinite square-well potential.  A   new ingredient is that the domain occupied by quarks and/or their chromo-electrical fields has a higher energy density called   bag constant $\cal B$: the deconfined state is the state of higher energy compared to the conventional confining vacuum state. In our context an additional finding is  important:   even for small physical  systems comprising three quarks and/or quark-antiquark pairs once strangeness is correctly accounted for, only  the volume energy density $\cal B$ without a  \lq\lq surface energy\rq\rq\ is present. This was shown by an unconstrained  hadron spectrum model study~\cite{Aerts:1984vv,Aerts:1985xv}. This result confirms the two vacuum state hypothesis as the correct picture of quark confinement, with non-analytical structure difference at $T=0$ akin to what is expected in a phase transition situation.

The reason that in the bag model  the color-magnetic hyperfine interaction dominates the color-electric interaction  is due to local color neutrality of hadrons made of light quarks; the quark wave-function of all light quarks fill   the entire bag volume in same way, hence if the global state is colorless so is the color charge density in the  bag. However, the  situation changes when considering the heavy charm $c$, or bottom $b$,  quarks  and antiquarks. Their mass scale dominates, and their semi-relativistic  wave functions are localized. The color field lines connecting the charges are, however, confined. When we place  heavy quarks relatively far apart, the  field lines are, according to the above, squeezed into a cigar-like shape, see top of \rf{Stringimage}. 

\begin{figure}[t]
\centering\resizebox{0.45\textwidth}{!}{%
\includegraphics{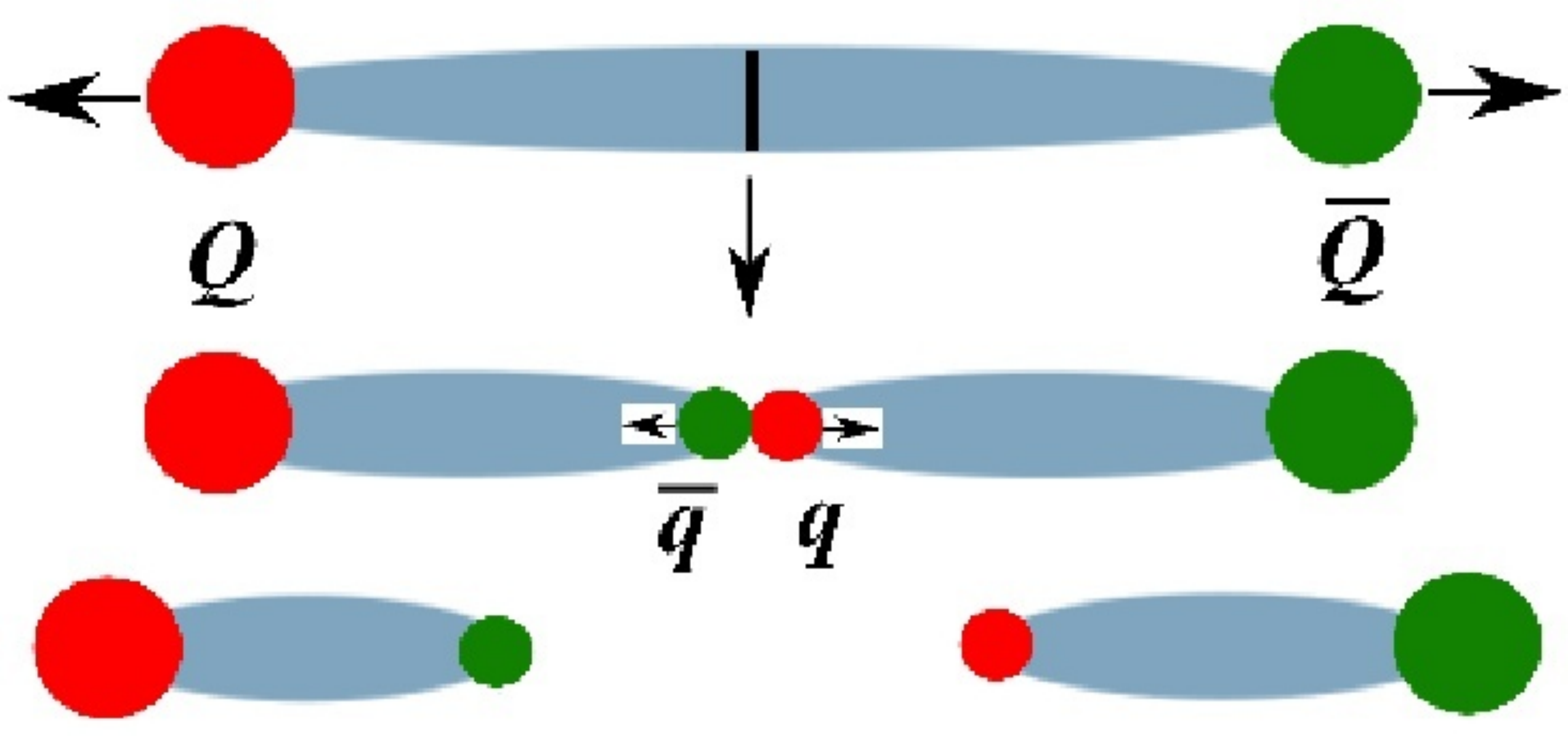}
}
\caption[]{Illustration of the heavy quark $Q=c,b$ and antiquark $\overline{Q}=\bar c, \bar b$ connected by a color field string. As  $Q\overline Q$ separate, a pair of light quarks $q\bar q$ caps the  broken field-string ends.  }\label{Stringimage}
\end{figure}

The field occupied volume grows linearly with the size of the long axis of the cigar. Thus  heavy quarks interact when pulled apart by a nearly linear potential, but only when the  ambient  temperature $T<\Th$. One can expect that at some point the field line connection snaps, producing a quark-antiquark pair. This means that when we pull on a heavy quark,  a colorless heavy-meson escapes from the colorless bound state, and another colorless heavy-antimeson is also produced; this sequence is shown from top to bottom in \rf{Stringimage}. The field lines connecting the  quark to its color-charge source are called a \lq QCD string\rq. The energy per length of the string, the string tension, is nearly   1\,GeV/fm.  This value includes the modification of the vacuum introduced by the color field lines.

For $T>\Th$ the field lines can spread out and mix with thermally produced light quarks. However, unlike light  hadrons which  melt at  \Th, the heavy $Q\bar Q$ mesons (often referred as \lq onium states, like in charmonium $c\bar c$) may remain bound, albeit with different strength for $T> \Th$. Such heavy quark clustering in QGP has been of profound interest: it impacts the pattern of production of heavy particles in QGP hadronization~\cite{Thews:2000rj,Schroedter:2000ek}. Furthermore, this is a more accessible   model of what happens to light quarks in close vicinity of \Th, where considerable clustering before and during hadronization must occur. 

The shape of the heavy quark potential, and thus the stability of \lq onium states can be studied as a function of  quark separation, and  of the temperature, in the framework of lattice-QCD,  showing how  the  properties of the heavy quark potential change when deconfinement sets in for $T>\Th$~\cite{Rothkopf:2011db,Bazavov:2012bq}.

To conclude, quarks and gluons are real particles and can, for example, roam  freely  above the vacuum melting point, \ie\ above Hagedorn temperature \Th. This understanding of confinement  allows us to view  the quark-gluon plasma as a domain in space in which confining vacuum structure is dissolved, and  chromo-electric field lines can exist. We will return to discuss further ramifications of the QCD vacuum structure in Subsections~\ref{EinsteinAether1} and~\ref{EinsteinAether2}.

\subsection{Why do  we care about  lattice-QCD?}\label{WhatIsLQCD} 
The understanding  of quark confinement as a confinement of the color-electrical field lines and characterization of hadrons as quark bags suggests as a further   question: how can there be around us, everywhere,  a vacuum structure that expels color-electric field lines?  Is there a lattice-QCD  based computation showing color field lines confinement? Unfortunately, there seems to be no answer available. Lattice-QCD produces values of static observables, and not interpretation of confinement in terms of  moving quarks and dynamics of the color-electric field lines.
 
So why care about lattice-QCD?  For the  purpose of this  article  lattice-QCD upon convergence is  the ultimate authority, resolving in an unassailable way all questions pertinent to the properties of interacting  quarks and  gluons,   described within the  framework of QCD. The word lattice reminds us how continuous  space-time is represented in a discrete numerical implementation on the most powerful computers of the world.

The  reason that  we trust lattice-QCD is that it is not a model but a solution of what we think is the foundational characterization of the hadron world. Like in other theories, the  parameters of the theory are the measured properties of observed particles. In case of QED we use the Coulomb force interaction strength at large distance, $\alpha=e^2/\hbar c=1/137$. In QCD the magnitude of the  strength of the interaction $\alpha_s=g^2/\hbar c$ is provided in terms of a scale, typically a mass that the lattice approach captures precisely; a value of $\alpha_s$ at large distance cannot be measured given the confinement paradigm.

There are serious issues that  have impacted the capability  of the lattice-QCD in the past. One is the problem of Fermi-statistics which is not easily addressed by classical computers. Another is  that the properties we wanted to learn about depend in a decisive way on the inclusion of quark flavors, and require  accurate value of the mass of the strange quark; the properties of QCD at finite $T$ are very  finely  tuned. Another complication is that in view of today's achievable lattice point and  given the quark-, and related hadron-,  scales, a lattice must be much more finely spaced than was believed necessary 30 years ago. Serious advances in numerical and theoretical methods were needed, see \eg\ Refs.\cite{Yagi:QGP,Satz:2012zza}.

Lattice capability is limited by  how finely  spaced  lattice points  in terms of their separation must be  so that over typical hadron volume sufficient number is found. Therefore, even the largest  lattice  implemented at present  cannot \lq see\rq\ any spatial structure that is larger than a few proton diameters, where for me: few=2. The  rest of the Universe is, in the lattice approach,  a periodic repetition of the same elementary  cell. 

The  reason that lattice at finite temperature cannot replace models in any foreseeable future is  the  time evolution: temperature and time are related in the theoretical formulation. Therefore considering hadrons in a heat bath  we are restricted to consideration of a thermal equilibrium system. When we include temperature,  nobody  knows how to include  time in lattice-QCD, let alone the  question of time sequence that  has not been so far implemented at $T=0$. Thus  all we can hope for in hot-lattice-QCD is what  we see in this article, possibly much refined in understanding of internal structure,  correlations, transport coefficient evaluation, and achieved computational precision. 

After this description some may wonder why we should  bother with lattice-QCD at all, given on one hand its limitations in scope, and on another the enormous cost rivaling the  experimental effort in terms of manpower and computer equipment. The answer is simple; lattice-QCD provides what  model builders need, a reference point  where models of reality meet with solutions of theory describing the reality. 

We have already  by example shown how this works. In the  previous Section~\ref{HadronSide}  we connected in several different ways the    value of \Th\ to lattice results. It seems clear that the interplay of lattice, with experimental data  and with models   can fix \Th\ with a sufficiently  small error. A further similar  situation is addressed in the following Subsection~\ref{WhatIsQGP} where we seek to interpret the  lattice results on hot QCD and to understand the properties of  the new phase of matter, quark-gluon plasma.

\subsection{What is  quark-gluon plasma?}\label{WhatIsQGP} 
An artist's view of Quark-Gluon Plasma (QGP), \rf{QGPpic},   shows several  quarks and \lq springy\rq\ gluons --  in an image similar to \rf{Bagimage}.  It is common to represent gluons by springs, a   historical metaphor from  times when we viewed gluons as creating a force that grew at a distance so as to be able to permanently  keep  quarks confined. Our views of confinement evolved, but  springs remain in gluon illustrations. In principle these springs are also colored: there are 9 bi-color combination, and excluding the  \lq white\rq\  case we have 8 bi-colors of gluons. As this is hard to illustrate, these springs are gray. The domain of space  comprising   quarks and gluons is colored to indicate that we expect this to be a much different space domain from the surroundings. 

\begin{figure}[t]
\centering\resizebox{0.45\textwidth}{!}{%
\includegraphics{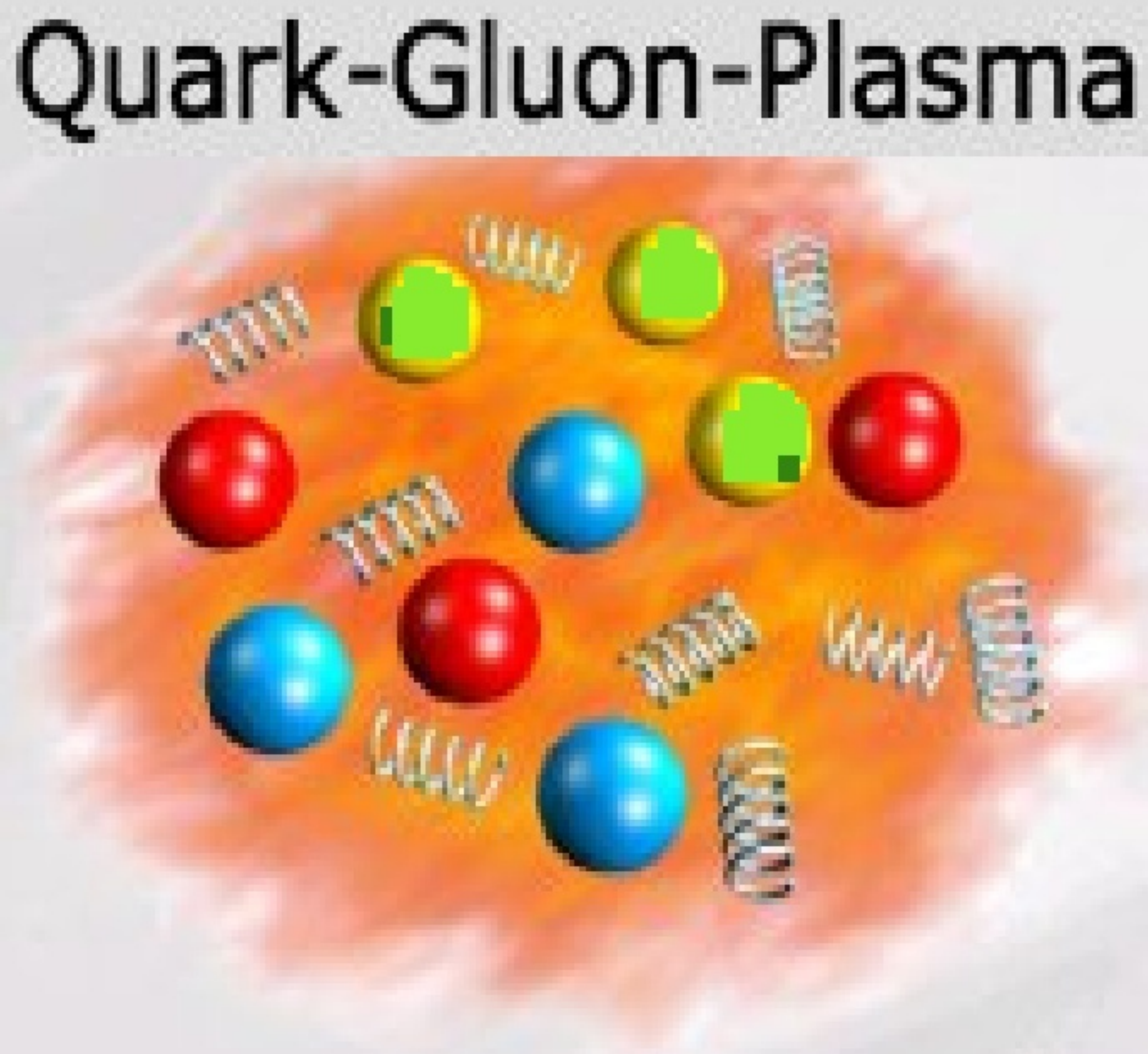}
}
\caption[]{Illustration of quark-gluon plasma (QGP) comprising several red, green, blue (RGB) colored quarks and springy gluons in a modified vacuum state. }\label{QGPpic}
\end{figure}
  
In a nutshell, QGP  in the contemporary use of the language is an interacting localized assembly of quarks and gluons at thermal (kinetic) and (close to) chemical (abundance) equilibrium. The  word \lq plasma\rq\  signals that  free color charges are allowed. Since the temperature is above \Th\  and thus  above the scale of light  quark $u,d$-mass, the pressure exhibits the  relativistic Stefan-Boltzmann format, 
\begin{equation}\label{SBLaw}
P =\left(g^{*}_{\mathrm{B}}+\frac 7 8 g^{*}_{\mathrm{F}}\right) \frac{(\pi T)^4}{90\pi^2 }  
 + g^{**}_{\mathrm{F}}\!\! \left(\! \frac{(\pi T)^2\mu^2}{ 24\pi^2}+ \frac{ \mu^4}{ 48\pi^2}\!\right)\;.
\end{equation}
The stars next to degeneracy $g$ for Bosons B, and Fermions F, indicate that these quantities are to be modified by the QCD interaction    which affects this degeneracy significantly, and differently for B, F and also fort the two terms  $^{*}$F vs. $^{**}$F. 

In \req{SBLaw} the traditional Stefan-Boltzmann $T^4$ terms, and the zero temperature  limit quark-chemical potential $\mu^4$ term are well known and  also easy to obtain by integrating the Bose/Fermi gas  expressions in the respective limit.  The  ideal (QCD interaction $\alpha_s=0$) relativistic hot quark gas at finite $T$ including the $T^2\mu^2$ term in explicit analytical form of the expression  was for the first time presented  by Harrington and Yildiz in 1974~\cite{Harrington:1974fc} in a work which has the telling title \lq\lq High-Density Phase Transitions in Gauge Theories\rq\rq.

Regarding the degeneracy factors for the ideal gases: The Boson term generalizes the usual Stefan-Boltzmann expression by an added factor $8_c$ for color degeneracy of  gluons: 
\begin{equation}
g_\mathrm{B}=2_s\times 8_c=16.
\end{equation} 
The corresponding $T^4$ Fermi (quark) Stefan-Boltzmann term  differs  by  the  well known factor 7/8 for each degree of freedom. We count particles and antiparticles as degrees of freedom: \begin{equation}
g_\mathrm{F}=2_s\times2_p\times 3_c\times (2+1)_f  =31.5,
\end{equation}  
where indices stand for: $s$=spin (=2), $p$-particle and antiparticle (=2),  $c$=color (=3, or =8), $f$-flavor: 2 flavors $q=u,d$  always satisfy $m_q\ll T$ and one flavor (strangeness $s$) at phase boundary  satisfies $m_s\lesssim T$ and turns into a light flavor at high temperatures. To make sure this situation is remembered we write $(2+1)_f$.

The analytical and relatively simple form of the first order  in ${\cal O}(\alpha_s)$ thermal QCD perturbative correction results are given in analytical format in the work of Chin in 1979~\cite{Chin:1978gj}, and result in the following   degeneracy:
\begin{equation}
\begin{array}{rl}\label{QGPgas}
g_{*}\equiv& g^{*}_{\mathrm{B}}+\frac 7 8 g^{*}_{\mathrm{F}}=  2_s\times 8_c\left(1-\frac{15\alpha_s}{4\pi}\right)\\[0.2cm]
&+
\frac 7 8 \,2_s\times2_p\times 3_c\times (2+1)_f\left(1-\frac{50\alpha_s}{21\pi}\right)\\[0.35cm]
g^{**}_{\mathrm{F}}=& 2_s\times2_p\times 3_c\times (2+1)_f\left(1-\frac{2\alpha_s}{\pi}\right).
\end{array}
\end{equation} 

$\alpha_s$ is the QCD energy scale dependent coupling constant. In the domain of $T$ we consider     $\alpha_s\simeq 0.5$,  but it is rapidly decreasing with $T$. To some extent this is why  in \rf{gPres},  showing the lattice-QCD results for pressure, we see a relatively rapid rise of $g_{*}$ as a function of $T$, towards the indicated limit $g=47.5$ of a free gas, horizontal dashed line. In \rf{gPres}   three results also depict the   path to the current understanding of the  value of \Th. The initial  results  (triangles)were presented by  Bazavov 2009~\cite{Bazavov:2009zn}; this work did not well describe the \lq low\rq\ temperature domain where the value of \Th\ is determined. This is the origin of the urban legend that   $\Th\simeq 190$ MeV, and  a lot of confusion.
\begin{figure}
\centering\resizebox{0.45\textwidth}{!}{%
\includegraphics{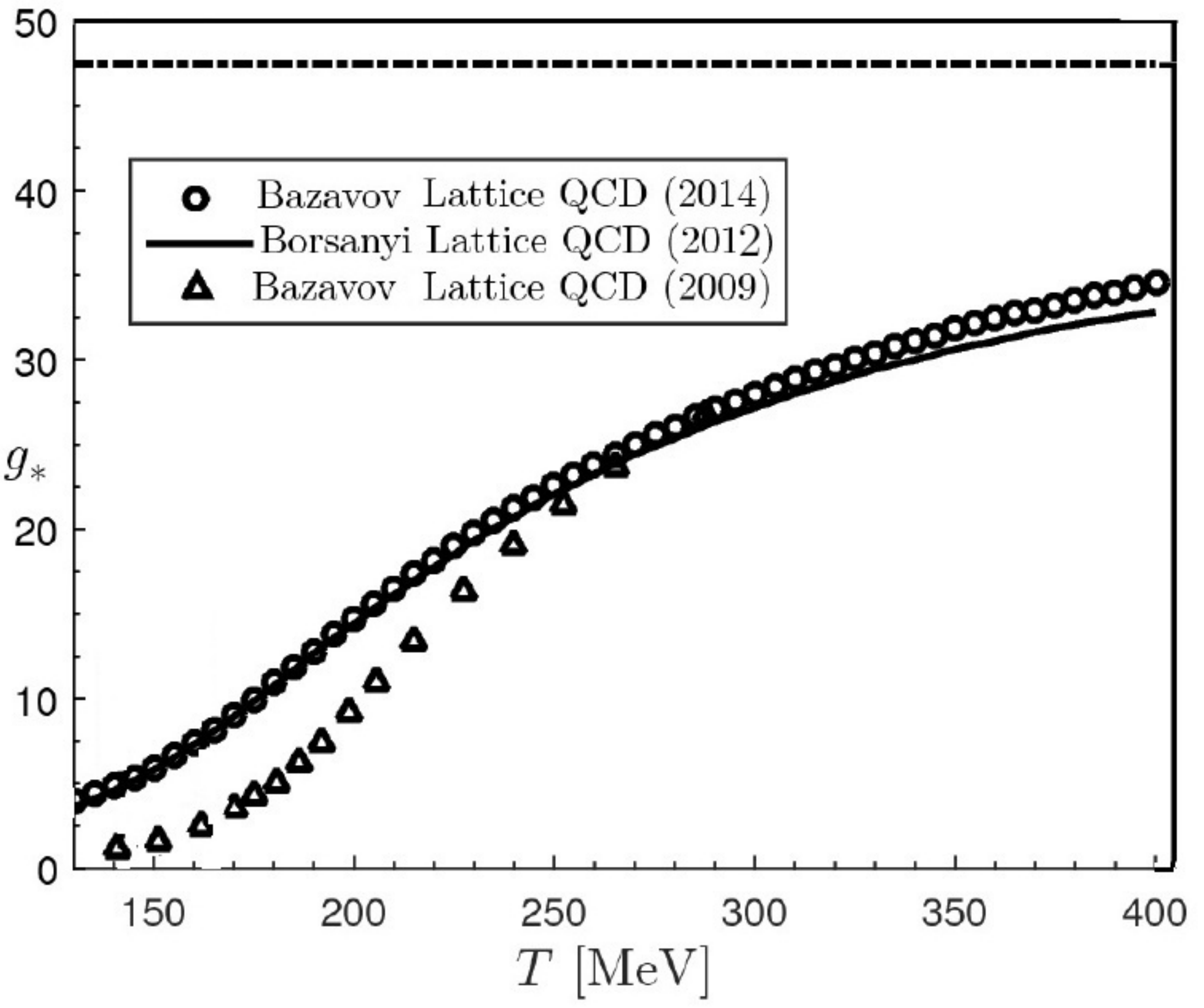}
}
\caption[]{The number of degrees of freedom $g_*$ as function of temperature $T$. The solid line includes the effect of QCD interactions  as obtained within the framework of lattice QCD  by Borsanyi \etal\  beginning in 2012  and published in 2014 (see text).  Horizontal dashed line: $g=47.5$ for free quark-gluon  gas}\label{gPres}
\end{figure}

The solid line in \rf{P_HRG_BorsanyLQCD} shows Borsanyi \etal\ 2012~\cite{Borsanyi:2012rr} results presented at the Quark Matter 2012 meeting, with later formal paper comprising the same results~\cite{Borsanyi:2013bia}; these are the same results  as we see in  \rf{P_HRG_BorsanyLQCD} connecting at low $T$ with HRG. These results of the Wuppertal-Budapest group  at first contradicted the  earlier and highly cited result of~\cite{Bazavov:2009zn}. However, agreement between both lattice groups was restored by the revised results of Bazavov 2014 [HotQCD Collaboration]~\cite{Bazavov:2014pvz}.

Let us also remember that the low value of \Th\ we obtained at the end of Subsection~\ref{latticeTh} is due to  the difference seen below in  \rf{gPres} between the results of 2009, and those reported a few years later, through 2014. This difference is highly relevant and shows that  hadrons melt into quarks near to $\Th=138\pm 12$ MeV corresponding to $4<a<4.5$ see \rt{tableMss}.

As the results of lattice-QCD became   reliable in the  high $T> 300$\,MeV domain a decade ago, it became apparent  that an accurate understanding of $g_*$ emerges~\cite{Letessier:2003uj} by taking ${\cal O}(\alpha_s)$ corrections literally and evaluating the behavior $ \alpha_s(T)$. A more modern study of the behavior of thermal QCD and its comparison with lattice QCD is available~\cite{Andersen:2014dua,Strickland:2014zka,Haque:2014rua}. The thermal QCD 
explains the  difference  between the asymptotic value  $g=47.5$ and lattice results which we see in \rf{gPres} to be significant at the highest $T=400$ considered. In fact thermal quarks are never asymptotically free; asymptotic freedom for hot QCD matter quarks suffers from logarithmic behavior.  $ \alpha_s(T)$ drops  slowly and  even at the thermal end of the standard model  $T\to 150,000$\,MeV the  QCD interaction remains relevant and   $g_*/g\simeq  0.9$. This of course is also true for very high density cold QCD matter, a small disappointment when considering the qualitative ideas seen in the work of  Collins and Perry \cite{Collins:1974ky}.

We can conclude by looking at high $T$ domains of all these results that the state of strongly interacting matter at $T\simeq 4 \Th$ is composed of the expected number of nearly free quarks and gluons, and the count of these particles in thermal-QCD and   lattice-QCD agree. We can say that  QGP emerges to be {\em the phase of strongly interacting matter which manifests its physical properties in terms of nearly free dynamics of practically massless  gluons and quarks.} The \lq practically massless\rq\ is inserted also for gluons as we must remember that in dense plasma matter all color charged particles including gluons acquire an effective in medium mass.

It seems that today we are in control of the  hot QCD matter, but what properties characterize  this QGP that differ in a decisive way  from more \lq normal\rq\ hadron matter? It seems that the safest approach in a theoretical review is to rely on theoretical insights. As the results of lattice-QCD demonstrate,  the \lq\lq quark-gluon plasma\rq\rq\  is a phase of matter comprising color charged particles (gluons and quarks) that can move nearly freely so as to create ambient pressure close to the  Stefan-Boltzmann limit and  whose motion freezes into hadrons across a narrow temperature domain characteristic of the Hagedorn temperature \Th. The properties of QGP that we check for are thus:
\begin{enumerate} 
\item Kinetic equilibrium -- allowing a meaningful definition of temperature;
\item Dominance by effectively massless particles  assuring that $P\propto T^4$;
\item Both quarks in their large number,  and gluons, must be present in conditions near chemical (yield) equilibrium with their color charge \lq open\rq\ so that the count of their number produces the correctly modified Stefan-Boltzmann constant of QCD.
\end{enumerate}

\subsection{How did the  name QGP  come into use?}\label{QGPstory} 
In this article we use practically always the words Quark-Gluon Plasma and the acronym QGP to describe the phase of matter made of deconfined quarks and gluons interacting according to (thermal) QCD and described in numerical lattice simulations with ever increasing accuracy. However, even today there is a second equivalent name; the series of conferences devoted to the  study of  quark-gluon plasma formation in laboratory calls itself \lq\lq Quark Matter\rq\rq. In 1987,   L\'eon Van Hove (former scientific director general of CERN) wrote a report entitled \lq\lq Theoretical prediction of a new state of matter, the \lq\lq quark-gluon plasma\rq\rq\ (also called \lq\lq quark matter\rq\rq)\rq\rq\ ~\cite{VanHove1987} establishing the common meaning of these two terms. 

When using Quark Matter we can be misunderstood to refer to zero-temperature  limit. That is  why QGP seems the preferred term. However,  to begin, QGP actually meant something  else. This is not unusual; quite often in physics in the naming of  an important new insight older terms are reused. This phenomenon reaches back to antiquity:   the early ancient Greek word \lq Chaos\rq\  at first meant \lq emptiness\rq. The science of that day concluded that emptiness would contain disorder, and the  word  mutated in its meaning into the present day use. 

At first  QGP  denoted  a  parton gas  in the context of \pp collisions; Hagedorn  attributes this to Bjorken 1969, but I could not find in the one paper Hagedorn cited the explicit mention of \lq QGP\rq\ (see Chapter 25 in~\cite{HagedornBook}). Shuryak in 1978~\cite{Shuryak:1978ij}   used  \lq QGP\rq\ in his publication title addressing partons in \pp collisions, thus using the language in the old fashion. 

Soon after \lq\lq QGP\rq\rq\  appears in another publication title, in July 1979 work by Kalashnikov and Klimov~\cite{Kalashnikov:1979dp}, now describing the strongly interacting quark-gluon thermal equilibrium matter. This work  did {\em not} invent   what the authors called QGP. They were, perhaps inadvertently,  connecting with the term  used by others in another context giving it the contemporary   meaning.  The  results of Kalashnikov-Klimov agree with our \req{QGPgas} attributed to a year earlier, July 1978, work of S.A. Chin~\cite{Chin:1978gj} presented under the title \lq\lq Hot Quark Matter". This  work  (despite the title) included hot gluons and their interaction with quarks and with themselves.

But QGP in its new meaning already had deeper roots. Quark-star models~\cite{Ivanenko:1965dg} appear as soon as quarks are proposed; \lq after\rq\ gluons join quarks~\cite{Fritzsch:1973pi}, within a year  
\begin{itemize}
\item
Peter Carruthers in 1973/74~\cite{Carruthers:74}   recognized that dense quark matter would be a quite \lq bizarre\rq\ plasma and he explores its many body  aspects.  His paper has priority but is also hard to obtain, published in a new journal that did not last. 
\item  A theory of thermal quark matter is that of Harrington and Yidliz 1974~\cite{Harrington:1974fc}, but has no discussion of the role of gauge interaction in quantitative terms. This paper is little known in the  field of RHI collisions yet it lays the foundation for the celebrated work by Linde on electroweak symmetry restoration in the early Universe~\cite{Linde:1978px}. There is a remarkable bifurcation in the literature: those who study the hot Universe and its early stages use the same physics as those who explore the properties of hot quarks and gluons; yet the cross-citations between the  two groups are sparse.
\item Collins and Perry 1975~\cite{Collins:1974ky} in \lq\lq Superdense Matter: Neutrons or Asymptotically  Free Quarks\rq\rq\  propose  that  high density nuclear matter turns into quark matter due to weakness of asymptotically free QCD. Compared to Harrington and Yidliz this is a step back to a zero-temperature environment, yet also a step forward as   the argument that interaction could be sufficiently weak  to view the dense matter as a Fermi gas of quarks is explicitly made.
\end{itemize}
Following this there are  a  few, at times parallel developments -- but this is not the place to present a full  history of the field. However  fragmentary, let me mention instead those papers I remember best:
\begin{itemize}
\item Freedman and McLerran 1976/77~\cite{Freedman:1976ub}  who address  the thermodynamic potential of an interacting relativistic quark gas. 
\item Shuryak 1977/78~\cite{Shuryak:1977ut}, writes about  \lq\lq Theory  of Hadron Plasma\rq\rq\    developing  the  properties of QGP in the framework of QCD.  
\item Kapusta 1978/79~\cite{Kapusta:1979fh} which work completes  \lq\lq Quantum Chromodynamics at High Temperature\rq\rq.  
\item Chin 1978~\cite{Chin:1978gj} synthesized all these results and was the first to provide the full analytical first order $\alpha_s$ corrections as seen in \req{QGPgas}.
\end{itemize}
 
However, in none  of the early thermal QCD  work is the  acronym \lq QGP\rq, or spelled out \lq Quark-Gluon Plasma\rq\rq\  introduced. So where did  Kalashnikov-Klimov~\cite{Kalashnikov:1979dp} get the  idea to use it? I can speculate that seeing the work by Shuryak on \lq\lq Theory  of Hadron Plasma\rq\rq\ they  borrowed the term from  another  Shuryak paper~\cite{Shuryak:1978ij} where he  used  \lq QGP\rq\ in his title addressing partons in \pp collisions. Indeed, in an aberration of credit  Shuryak's \pp parton work is cited in  \hAA QGP  context, clearly in recognition of the  use of the QGP acronym in the title, while Shuryak's \lq true\rq\  QGP paper, \lq\lq Theory  of Hadron Plasma\rq\rq\ is often not cited in this context. In his 1980 review Shuryak~\cite{Shuryak:1980tp} is almost shifting to QGP nomenclature, addressing  \lq QCD Plasma\rq\ and also uses  in the text \lq Quark Plasma\rq, omitting to mention  \lq gluons\rq\ which  are not  established experimentally for a few more years. In this he echoes the approach of  others in this period. 

Having said all the above, it is clear that  when  \lq QGP\rq\ is mentioned as the theory of both hot quarks and hot gluons, we should remember Kalashnikov-Klimov~\cite{Kalashnikov:1979dp}  for as I said, the  probably  inadvertent introduction of this name into its contemporary use.

\section{Quark-Gluon Plasma in Laboratory}\label{QGPinLab}
\subsection{How did RHI collisions and  QGP  come together?} \label{RHI_QGP}
\begin{figure}[h]
\centering\resizebox{0.48\textwidth}{!}{%
\includegraphics{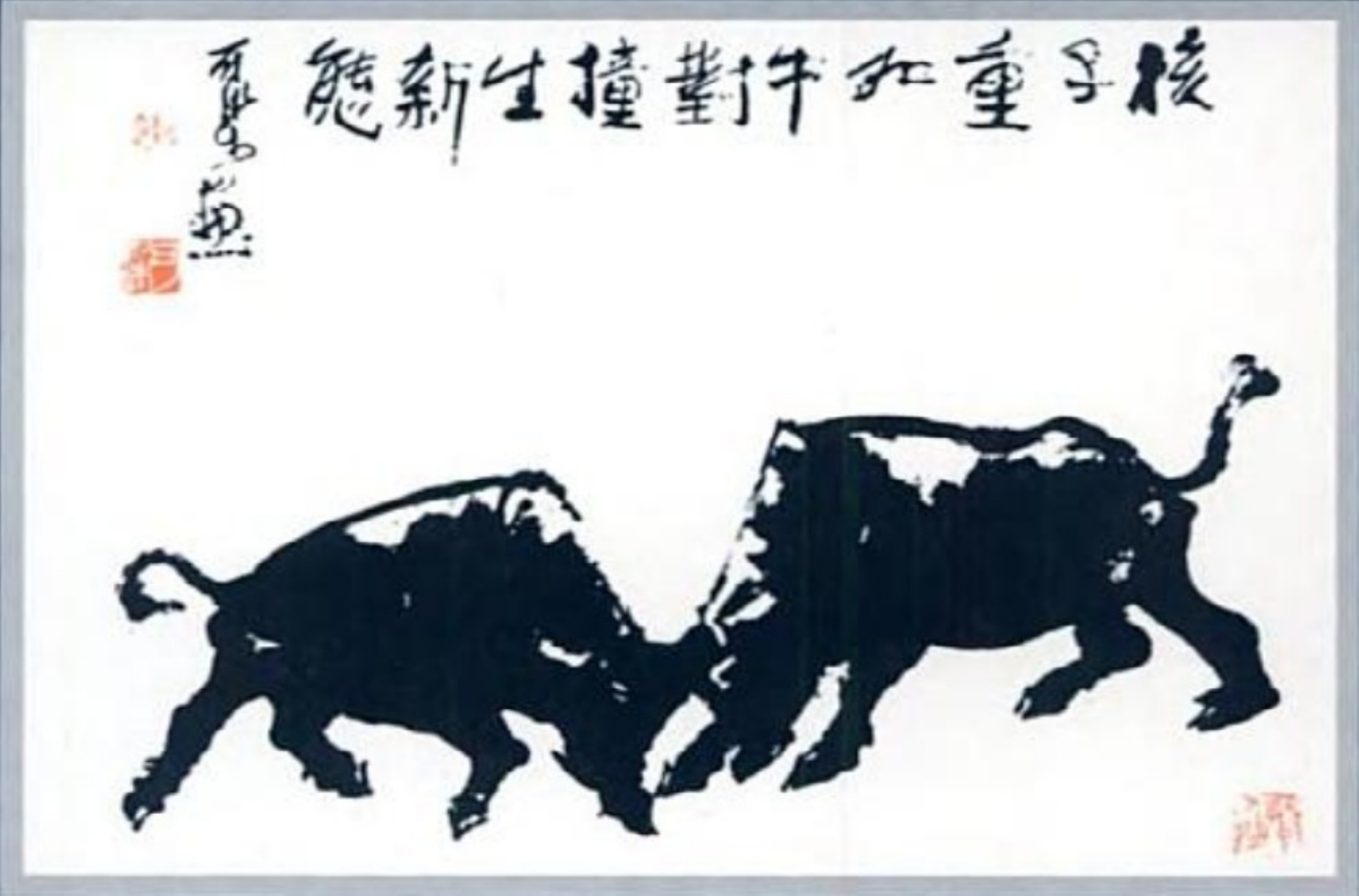}
}
\caption[]{Ink painting masterpiece 1986: \lq\lq Nuclei as Heavy as Bulls, Through Collision Generate New States of Matter\rq\rq\  by Li Keran, reproduced from open source works of T.D.Lee.}\label{keran-bulls}
\end{figure}
The  artistic representation of RHI collisions is seen in \rf{keran-bulls} -- two fighting bulls. The ink   masterpiece was created in 1986 by Li Keran and has been around the field of heavy ions for the past 30 years, a symbol of nascent symbiosis between science and art, and also a symbol of great friendship between   T.D. Lee and Li Keran. The bulls are the  heavy ions, and the  art depicts the paradigm of heavy-bull (ion) collisions.

So how did the bulls aka heavy ions connect to QGP?  In October 1980, I remarked in a citation\footnote{The present day format requirement means that these   words are now found in the text of Ref.\cite{appenA}, end of 3rd paragraph below Eq.\,(61),  so that each of the two references can be cited and hyperlinked as a separate citation item.}   \lq\lq The possible formation of quark-gluon plasma in nuclear collisions was first discussed {\em quantitatively} by S.A. Chin: Phys. Lett. B {\bf 78}, 552 (1978); see also N. Cabibbo, G. Parisi: Phys. Lett. B {\bf 59}, 67 (1975)\rq\rq. Let me refine this:
\begin{itemize}
\item[a)]
The pioneering insight of the work by Cabibbo and Parisi~\cite{Cabibbo:1975ig} is: i) to recognize the need to modify SBM to include melting of hadrons and ii) in a qualitative drawing to recognize that both high temperature and baryon density allow a phase transformation process. However, there is   no  mention direct or indirect in this work about  \lq bull\rq\ collision. 
\item[b)]
The paper by Chin~\cite{Chin:1978gj}, of July 1978,  in its Ref.\,[7] grants the origin of the idea connecting RHI  with QGP to  Chapline  and Kerman~\cite{Chapline:1978kg}, an  unpublished manu\-script entitled \lq\lq On the possibility of making quark matter in nuclear collisions\rq\rq\   of March 1978.  This paper clearly  states the connection of QGP and RHI collisions  that Chin explores in a quantitative fashion recomputing the  QCD thermodynamic potential,  and  enclosing particles, quarks and gluons, in the  bag-like structure, see Section~\ref{quarksBags}.
\end{itemize}

The preprint of Chapline  and Kerman is  available on-line at MIT~\cite{Chapline:1978kg}. It is a qualitative, mostly conceptual idea paper, a  continuation of an earlier effort by Chapline and others in 1974~\cite{Chapline:1974zf} where we read (abstract): 
\begin{quote}
It is suggested that very hot and dense nuclear matter may be formed in a transient state in \lq\lq head-on\rq\rq\ collisions of very energetic heavy ions with medium and heavy nuclei. A study of the particles emitted in these collisions should give clues as to the nature of dense hot nuclear matter.\end{quote}

At the time of the initial Chapline effort in 1974  it was too early for a mention of quark matter and heavy ions in together. Indeed, at  the Bear Mountain~\cite{BearMountain} workshop in Fall 1974  the physics of the forthcoming RHI collisions was discussed in a retreat motivated by  Lee-Wick~\cite{Lee:1974ma} matter, a proposed new state of nuclear matter. These authors claim:
\begin{quote}
\ldots the state \ldots inside a very heavy nucleus can become the minimum-energy state, at least within the tree approximation; in such a state, the \lq\lq effective\rq\rq\ nucleon mass inside the nucleus may be much lower than the normal value.
\end{quote}

In presenting this work,  the  preeminent theorists T.D. Lee and G.C. Wick  extended  an open invitation to explore in relativistic heavy  ion collisions the new exotic state of dense nuclear matter. This work generated exciting scientific prospects for  the  BEVELAC accelerator  complex at  Berkley. We keep in mind that there is no mention of quark matter in any document related to BEVELAC~\cite{Grazyna}, nor at the Bear Mountain workshop~\cite{BearMountain}. However, the ensuing experimental search for the Lee-Wick  nuclear matter generated the experimental expertise and equipment  needed to plan and perform experiments in search of  quark-gluon plasma~\cite{Hans}. And, ultimately,  T.D. Lee  will turn  to recognize QGP as the new form of hot nuclear matter resulting, among other things, in the very beautiful painting by Li Keran, \rf{keran-bulls}.

Now back to the March 1978 Chapline-Kerman  manu\-script: why was it never published? There are a few possible answers: a) It is very qualitative; b) In the 5y run up period 1973--1978 the  field of RHI collisions was dominated by other  physics such as   Lee-Wick.  In fact at the time  quarks were not part of nuclear physics which \lq owned\rq\ the field of heavy  ions. Judging by personal experience I am not really surprised that Chapline-Kerman  work was not published.   Planck was dead for 30 years\footnote{Many credit Planck with fostering an atmosphere of openness and tolerance as a publisher; certainly he did not hesitate to take responsibility for printing Einstein miraculous 1905 papers.}. It is regrettable that once Chapline-Kerman   ran into   resistance they did not pursue  the publication, or/and further development of their idea; instead, 
\begin{itemize}
\item[a)]
A year later, Kerman  (working   with   Chin  who gave him the  credit for the QGP-RHI connection idea in his paper),  presents  strangelets~\cite{Chin:1979yb}, cold drops of quark matter containing a large strangeness content. 
\item[b)]
And a few years later, Chapline~\cite{Chapline:1986sw}  gives credit for the quark-matter connection to RHI collisions both to  Chapline-Kerman~\cite{Chapline:1978kg} work, and the  work of Anishetty,  Koehler, and McLerran of 1980~\cite{Anishetty:1980zp}.  Anishetty \etal\ claim  in their abstract 
\begin{quote}
\ldots two hot fireballs are formed. These fireballs would have rapidities close to the rapidities of the original nuclei. We discuss the possible formation of hot, dense quark plasmas in the fireballs.
\end{quote}
\end{itemize}
That  Anishetty,  Koehler, and McLerran view of RHI collision dynamics is in direct conflict with the   effort of Hagedorn to describe particle production in \pp collisions which at the time was being adapted to the \hAA case and presented e.g. in the  QM1-report~\cite{Hagedorn:1981sb}.

Anishetty \etal\ created the false paradigm that QGP was not produced centrally (as in center of momentum), a point that was corrected  a few years later in 1982/83 in the renowned paper of J.D. Bjorken~\cite{Bjorken:1982qr}. He  obtained  an analytical, one dimensional, solution of relativistic hydrodynamics that could be  interpreted for the case of the RHI collision as description at asymptotically high energy of the collision events. If so, the RHI collision  outcome would be  a trail of energy connecting the two nuclei that  naturally qualifies to be the QGP. While this replaced the Anishetty,  Koehler, and McLerran \lq\lq cooking nuclei, nothing in-between\rq\rq\  picture, this new  asymptotic energy idea  also distracted from the laboratory situation of the period which had to deal with  realistic, rather than asymptotic collision energies. 

In that formative period I wrote  papers  which argued that the  hot, dense QGP  fireball would be formed due to hadron  inelasticity stopping some or even all of nuclear matter in the center of momentum frame (CM). However, my referees literally said I was delusional. As history has shown (compare Chapline and Kerman) referees are not always useful. The long paper on the topic of forming QGP at central rapidity was first published 20 years later in the memorial volume dedicated to my collaborator on this project, Michael Danos~\cite{Danos:2000mv}. 

Here it is good to remember that  the CERN-SPS discovery story relies on the formation of a baryon-rich QGP in the CM frame of reference \ie\ at  \lq central rapidity\rq. RHIC is in transition domain in energy, and LHC energy scale, finally and  30 years later, is near to the Bjorken \lq scaling\rq\ limit. The word scaling is used, as we should in a rather wide range of rapidity  observe the  same state of hot  QGP, a claim still awaiting an experiment.

To close the topic, some  regrets: an \lq idea\rq\ paper equivalent to Ref.\cite{Chapline:1978kg}  introducing the bootstrap model of hot finite sized hadron matter and transformation into QGP in RHI collisions could have been written by Hagedorn and myself in late 1977. Hagedorn, however, desired a working model. After 10 months  of telling the world about our work, and much further effort in Summer 1978  we wrote  with I. Montvay   a  99 page long paper~\cite{Hagedorn:1978kc}, as well as  a few months later a much evolved shorter version\cite{Rafelski:1979cia}.

Only in the Spring of 1980  was  Hagedorn sure we understood the  SBM and the hadron melting into QGP in RHI. Of course we were looking at central rapidity \ie\ CM system, quite different from the  work of Anishetty \etal~\cite{Anishetty:1980zp}. Hagedorn explains the time line of our and related work in his 1984 review~\cite{Hagedorn:1984hz}. His later point of view is succinctly represented in a letter of September 1995, \rf{Hagedorn1995}, where he says\footnote{German Original: {\it \ldots werde ich noch den eindeutigen Nachweis der Existenz des QGP erleben? Ich bin sowieso davon \"uberzeugt denn wohin soll der Phasen\"ubergang (den es doch sicher gibt) sonst f\"uhren?\/}}: \lq\lq \ldots can I hope to witness a proof of existence  of QG plasma? I am in any case convinced of its existence, where else could the phase transition (which with certainty is present) lead?\ldots\rq\rq.

\begin{figure}[ht]
\centering\resizebox{0.44\textwidth}{!}{%
\includegraphics{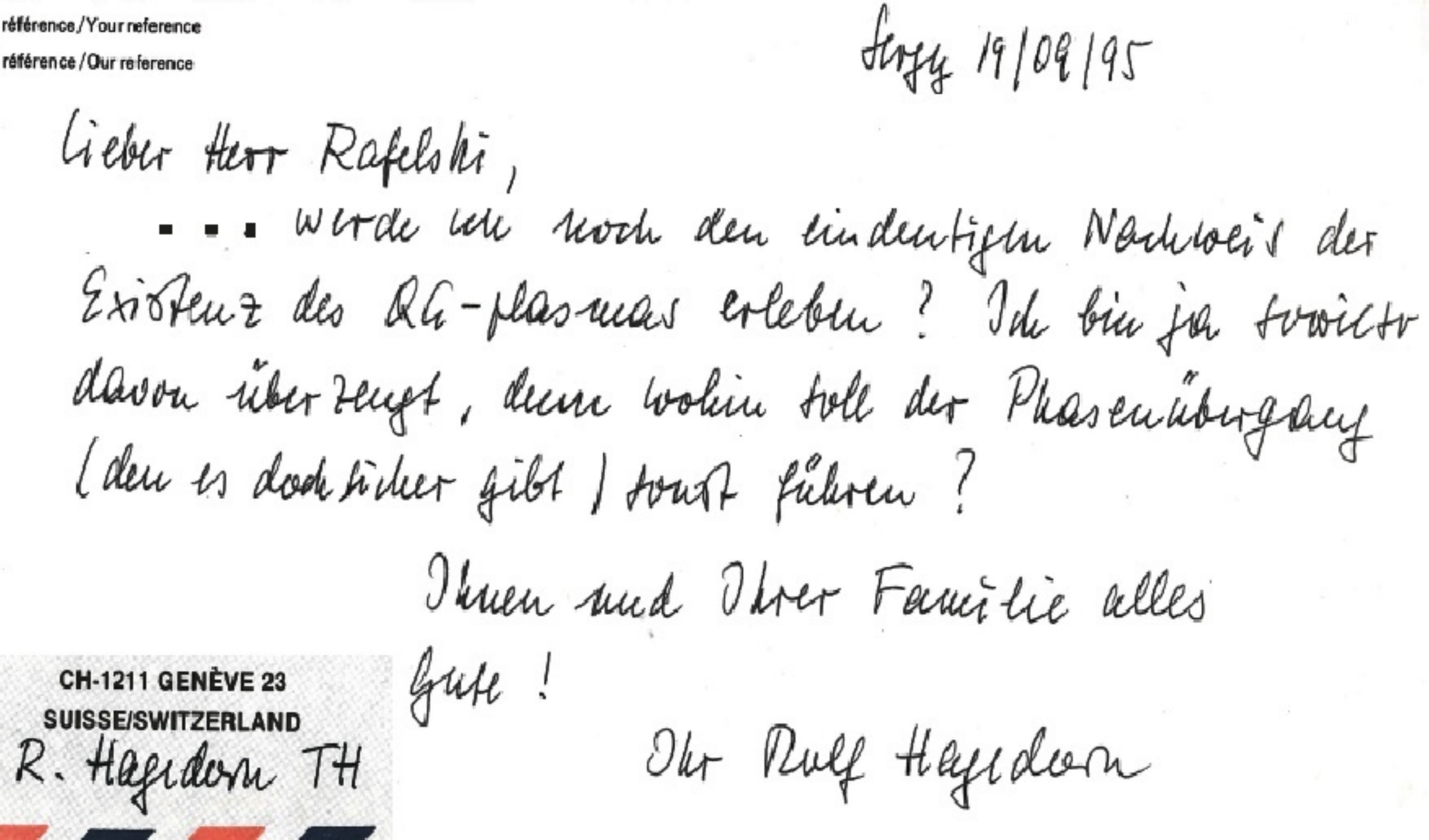}
}
\caption[]{Hagedorn in September 1995 awaiting QGP discovery, see text.}\label{Hagedorn1995}
\end{figure}


\subsection{When  and where was QGP discovered?}\label{QGPdiscovered}
\begin{figure}
\centering
\resizebox{0.48\textwidth}{!}{%
\includegraphics{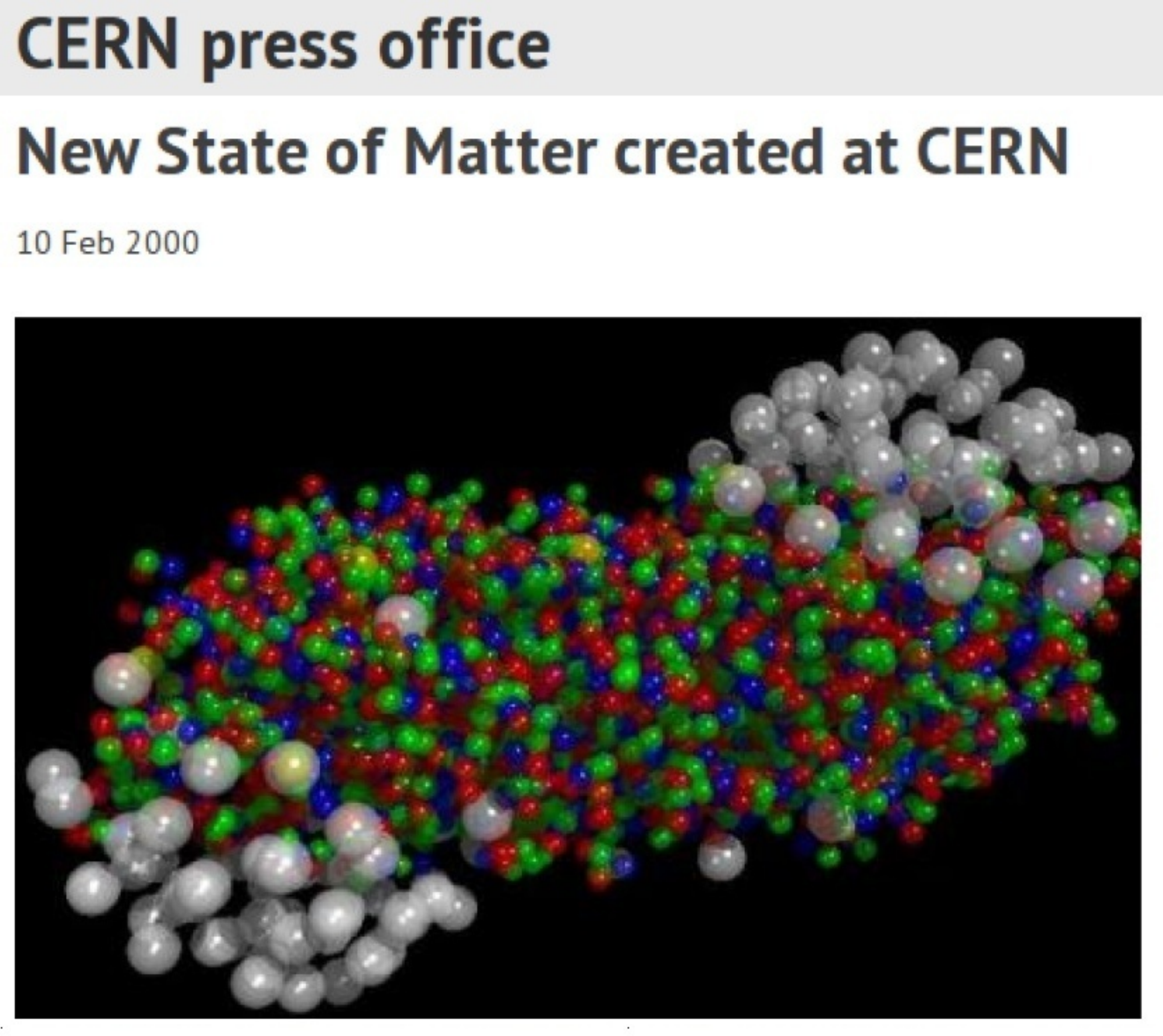}}
\caption[]{The press release text: \lq\lq At a special seminar on 10 February 2000, spokespersons from the experiments on CERN's Heavy Ion program presented compelling evidence for the existence of a new state of matter in which quarks, instead of being bound up into more complex particles such as protons and neutrons, are liberated to roam freely.\rq\rq\ }\label{CERNExtremeFig}
\end{figure}
Both CERN and BNL have held press conferences describing their experimental work.   In \rf{CERNExtremeFig} a screen shot shows how CERN advertised its position in February 2000 to a wider public~\cite{NewCERN+HIPress}. The document for scientists agreed to by those representing the seven CERN experiments (see the time line of CERN-SPS experiments in \rf{CERNSPS}) provided at the event read: \begin{quote}
\lq\lq The year 1994 marked the beginning of the CERN lead beam program. A beam of 33 TeV (or 160 GeV per nucleon) lead ions from the SPS now extends the CERN relativistic heavy ion program, started in the mid eighties, to the heaviest naturally occurring nuclei. A run with lead beam of 40 GeV per nucleon in fall of 1999 complemented the program towards lower energies. Seven large experiments participate in the lead beam program, measuring many different aspects of lead-lead and lead-gold collision events: NA44, NA45/CERES, NA49, NA50, NA52/NEWMASS, WA97/NA57, and WA98. \ldots\\
 Physicists have long thought that a new state of matter could be reached if the short range repulsive forces between nucleons could be overcome and if squeezed nucleons would merge into one another. Present theoretical ideas provide a more precise picture for this new state of matter: it should be a quark-gluon plasma (QGP), in which quarks and gluons, the fundamental constituents of matter, are no longer confined within the dimensions of the nucleon, but free to move around over a volume in which a high enough temperature and/or density prevails. \ldots (explicative in original:) {\bf A common assessment of the collected data leads us to conclude that we now have compelling evidence that a new state of matter has indeed been created, \ldots. The new state of matter found in heavy ion collisions at the SPS features many of the characteristics of the theoretically predicted quark-gluon plasma.}\ldots
In spite of its many facets the resulting picture is simple: the two colliding nuclei deposit energy into the reaction zone which materializes in the form of quarks and gluons which strongly interact with each other. This early, very dense state (energy density about 3--4 GeV/fm$^3$, mean particle momenta corresponding to $T$\,$\approx$\,240 MeV) suppresses the formation of charmonia, enhances strangeness and begins to drive the expansion of the fireball.\ldots\rq\rq\ 
\end{quote}
BNL presented  the following comment~\cite{BNL_Resp}
\begin{quote}
The CERN results are quite encouraging, says Tom Ludlam, Brookhaven's Deputy Associate Director for High-Energy and Nuclear Physics. \lq\lq These results set the stage for the definitive round of experiments at RHIC in which the quark-gluon plasma will be directly observed, opening up a vast landscape for discovery regarding the nature and origins of matter.\rq\rq

Brookhaven's Director John Marburger congratulated CERN scientists on their achievement, stating that \lq\lq piecing together even this indirect evidence of the quark-gluon plasma is a tour de force. The CERN teams have pressed their capabilities to the limit to extract these tantalizing glimpses into a new domain of matter.\rq\rq
\end{quote}
Dr. Marburger was evidently expecting a better \lq direct evidence\rq\ to ultimately emerge. Let us look at what this may be:  The  turn of BNL to announce its  QGP arrived 5 years later.  At the April 2005 meeting of the American Physical Society, held  in Tampa, Florida a press conference took place  on Monday, April 18, 9:00 local time. The public announcement of this event was made April 4, 2005: 
\begin{quote} 
EVIDENCE FOR A NEW TYPE OF NUCLEAR MATTER 
At the Relativistic Heavy Ion Collider (RHIC) at Brookhaven National Lab (BNL), two beams of gold atoms are smashed together, the goal being to recreate the conditions thought to have prevailed in the universe only a few microseconds after the big bang, so that novel forms of nuclear matter can be studied. At this press conference, RHIC scientists will sum up all they have learned from several years of observing the world’s most energetic collisions of atomic nuclei. The four experimental groups operating at RHIC will present a consolidated, surprising, exciting new interpretation of their data. Speakers will include: Dennis Kovar, Associate Director, Office of Nuclear Physics, U.S. Department of Energy's Office of Science; Sam Aronson, Associate Laboratory Director for High Energy and Nuclear Physics, Brookhaven National Laboratory. Also on hand to discuss RHIC results and implications will be: Praveen Chaudhari, Director, Brookhaven National Laboratory; representatives of the four experimental collaborations at the Relativistic Heavy Ion Collider; and several theoretical physicists.
\end{quote}
The participants  at the press conference each obtained  a \lq\lq Hunting for Quark-Gluon Plasma\rq\rq\  report, of which the cover  in \rf{RHICover}  shows the four BNL experiments operating at the time: BRAHMS, PHOBOS, PHENIX, and  STAR, which reported on the QGP physical properties that have been discovered in the first three years of RHIC operations. These four experimental reports  were later published  in an issue of Nuclear Physics A~\cite{Arsene:2004faB,Adcox:2004mhB,Back:2004jeB,Adams:2005dqB}.  

\begin{figure}
\centering\resizebox{0.45\textwidth}{!}{%
\includegraphics{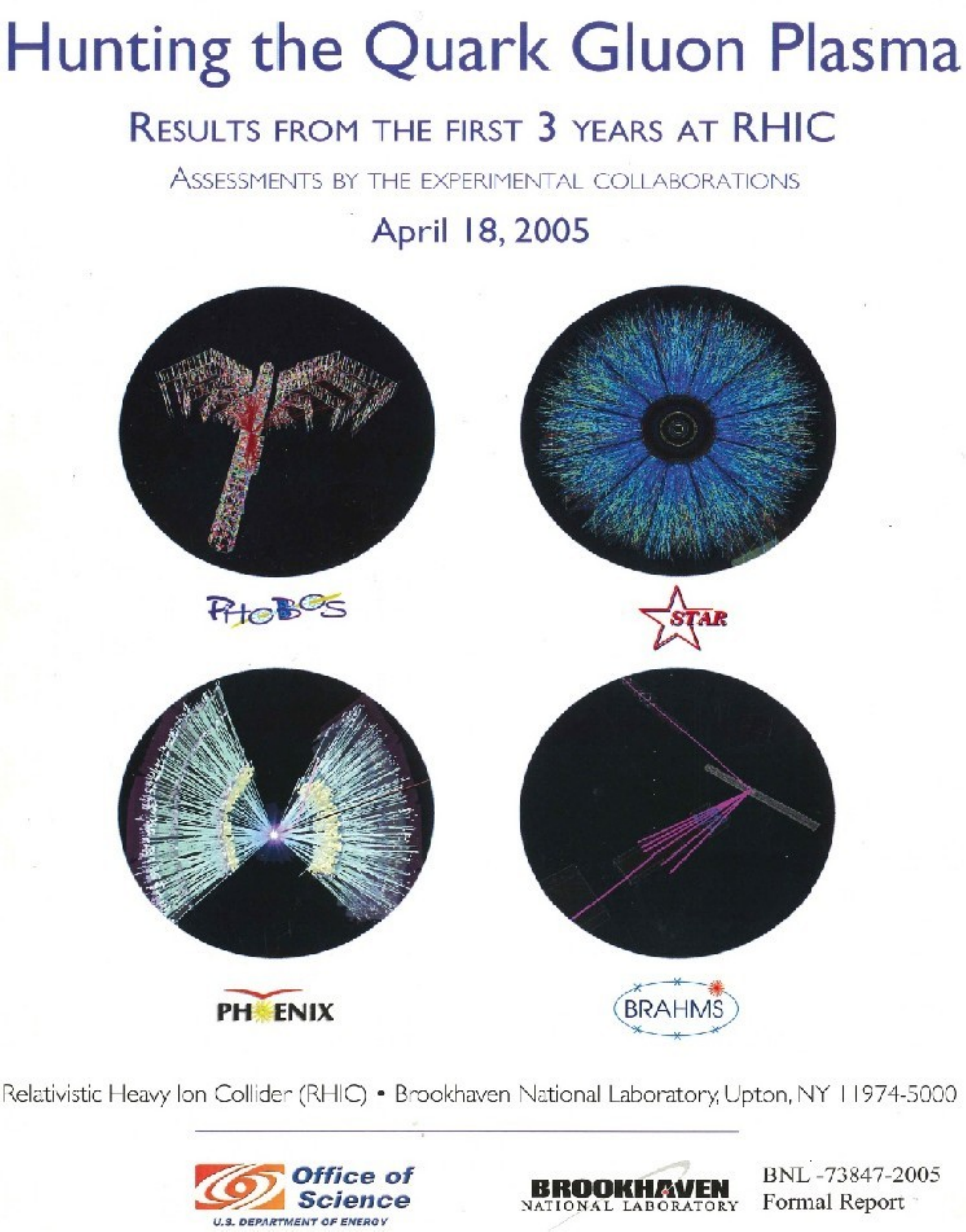}}
\caption[]{The cover of the   BNL-73847-2005  Formal Report   prepared by the Brookhaven National Laboratory, on occasion of the  RHIC experimental program press conference  April 2005. The cover identified the four RHIC experiments.
}\label{RHICover}
\end{figure}

The 10 year anniversary was relived at the 2015  RHIC \& AGS Users' Meeting, June 9-12, which included a special celebration session  \lq\lq The Perfect Liquid at RHIC: 10 Years of Discovery\rq\rq.  Berndt M\"uller, the 2015  Brookhaven' Associate Laboratory Director for Nuclear and Particle Physics is quoted as follows~\cite{BNL10years}:
\begin{quote}
\lq\lq RHIC lets us look back at matter as it existed throughout our universe at the dawn of time, before QGP cooled and formed matter as we know it, \ldots The discovery of the perfect liquid was a turning point in physics, and now, 10 years later, RHIC has revealed a wealth of information about this remarkable substance, which we now know to be a QGP, and is more capable than ever of measuring its most subtle and fundamental properties.\rq\rq
\end{quote}

An uninvolved scientist  will ask: \lq\lq Why is  the flow property of QGP: a) Direct evidence of QGP and  b) Worth full scientific attention 15 years after the new phase of matter was announced for the first time?\rq\rq\ Berndt  M\"uller   answers for this article:
\begin{quote}
Nuclear matter at \lq room temperature\rq\ is known to behave like a superfluid. When heated the nuclear fluid evaporates and turns into a dilute gas of nucleons and, upon further heating, a gas of baryons and mesons (hadrons). But then something new happens; at \Th\ hadrons melt and the gas turns  back into a liquid.  Not just any kind of liquid. At RHIC we have  shown that this is the most perfect liquid ever observed in any laboratory experiment at any scale. The new phase of matter consisting of dissolved hadrons exhibits less resistance to flow than any other substance known. The experiments at RHIC have a decade ago  shown that the Universe at its beginning was uniformly filled with a new type of material, a super-liquid, which once Universe cooled below \Th\   evaporated into a gas of hadrons.\\[-0.15cm]

Detailed measurements over the past decade have shown that this liquid is a quark-gluon plasma; \ie\  matter in which quarks, antiquarks and gluons flow independently. There remain very important questions we need to address: What makes the interacting quark-gluon plasma such a nearly perfect liquid? How exactly does  the transition to confined quarks work?  Are there conditions under which the transition becomes  discontinuous first-order phase transition? Today we are  ready to address these questions. We are eagerly awaiting new results from the upgraded STAR and PHENIX experiments at RHIC.
\end{quote}

\subsection{How did the SPS-QGP announcement\\  withstand  the test of time?}\label{subsec:expStra}
It  is impossible to present in extensive manner in this review all the physics results that  have driven the SPS announcement, and I will not even venture into the grounds of the RHIC announcement. I will focus here instead on what I consider my special expertise, the strangeness signature of QGP.  The events accompanying the discovery and development of strangeness signature of QGP more than 30 years ago have been reported~\cite{Rafelski:2007ti}, and the first extensive literature mention of strangeness signature of QGP from 1980 is  found in Ref.\cite{appenA}. 

So, what exactly is this signature? 
The situation  is  illustrated in  \rf{MultistrangFig} and described in detail in Ref.\cite{appenB}. In the center of the figure we see thermal QCD based strangeness production processes. This thermal production dominates the production occurring in first collision of the colliding nuclei. This is unlike heavier flavors where the mass threshold $2m_Q\gg T,\ Q=c,b$. Strange quark pairs: $s$ and antiquarks $\bar s$, are found  produced  in processes dominated by gluon fusion~\cite{Rafelski:1982pu}. Processes based on light quark collisions contribute fewer  $s\bar s$-pairs by nearly a factor 10~\cite{Biro:1981zi}. When $T\ge m_s$ the chemical equilibrium abundance of strangeness in QGP   is  similar in abundance to the other  light $u$ and $d$ quarks~\cite{appenA}.

\begin{figure}
\centering\resizebox{0.42\textwidth}{!}{%
\includegraphics{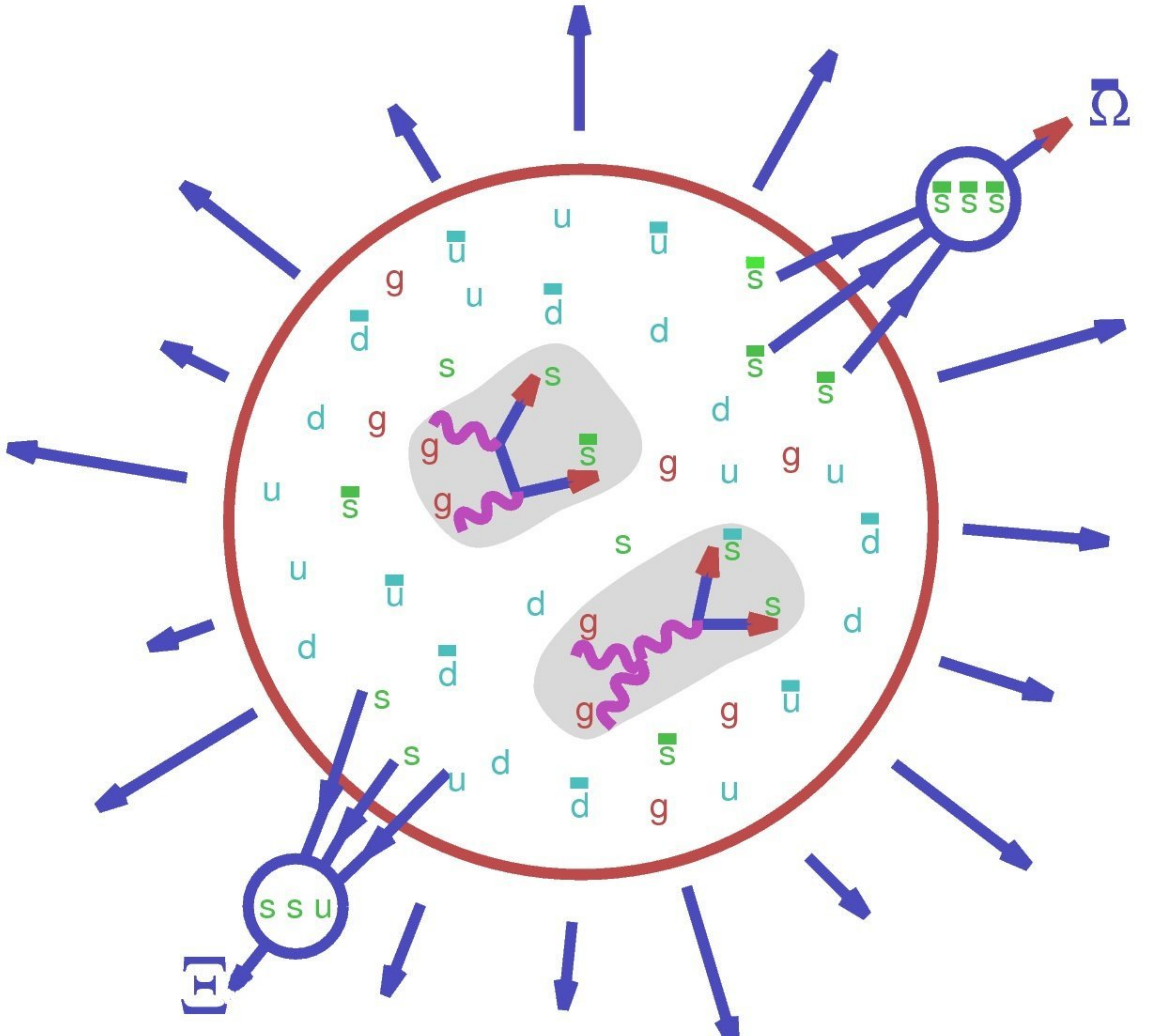}
}
\caption[]{Multistrange (anti)baryons as signature  of QGP, see text for further discussion}\label{MultistrangFig}
\end{figure}

Even for the  gluon fusion processes enough lifespan of QGP is needed to reach the large abundance of strange quark pairs in chemical equilibrium. The lifespan of the QGP fireball  increases as the collision volume increases and/or the  energy increases.
Since the gluon fusion $GG\to s\bar s$ dominates quark flavor conversion $q\bar q \to s\bar s$  the abundance of strangeness is   signature of the formation of a thermal gluon medium.

Of course we need to ask, how come there is a gluon medium at SPS energy scale?  In the  cascade evolution model one  finds that  gluons are in general the first to equilibrate in their number and momentum distribution. Equilibration means entropy $S$ production, a topic of separate importance as $S$ production is proceeding   in temporal sequence other hadronic observables of QGP, and how entropy  is produced remains today an unresolved question, see Subsection~\ref{whatQGP}.

The  gluon based processes are driving the equilibration of  quarks and antiquarks; first light $q=u,d$, next the  slightly massive $s$  and   also some  thermal evolution of charm is possible. Strangeness evolves along with the light ($u,\bar u, d, \bar d$) quarks and gluons $G$ until  the time of hadronization, when these particles seed the formation of hadrons observed in the experiment. In QGP, $s$ and  $\bar s$ can move freely and their large QGP abundance  leads to unexpectedly large yields of particles with a large $s$ and $\bar s$ content~\cite{Rafelski:1982ii,Rafelski:1982bi}, as is illustrated  exterior of the QGP domain in \rf{MultistrangFig}. 

A signature of anything requires a rather background free environment, and a good control of anything that is there as no signature is background free. There are ways two other than QGP to make strange antibaryons:\\[0.1cm]
I) Direct production of  complex multistrange (anti)baryons  is less probable for two reasons:
\begin{enumerate}
\item When new particles are produced in a color  string breaking process, strangeness is known to be produced less often by a factor 3  compared to lighter quarks. 
\item The generation of multistrange content requires   multiple such suppressed steps.
\end{enumerate}
Thus the conclusion  is that with increasing strangeness content the production by string processes of strange hadrons is progressively more suppressed.\\[0.1cm]
II) Hadron-hadron collisions can redistribute strangeness into multistrange hadrons. Detailed kinetic model study  shows that the hadron-reaction based production of multistrange hadrons is rather slow and requires time that exceeds collision time of RHI collisions significantly. This means that both $\uXi, \overline\uXi $ and $\uOmega,\overline\uOmega$ are in their abundance signatures of QGP formation and hadronization, for further details see Refs.\cite{appenB,Koch:1984tz,Koch:1986ud}.

\begin{figure}
\centering\resizebox{0.44\textwidth}{!}{%
\includegraphics{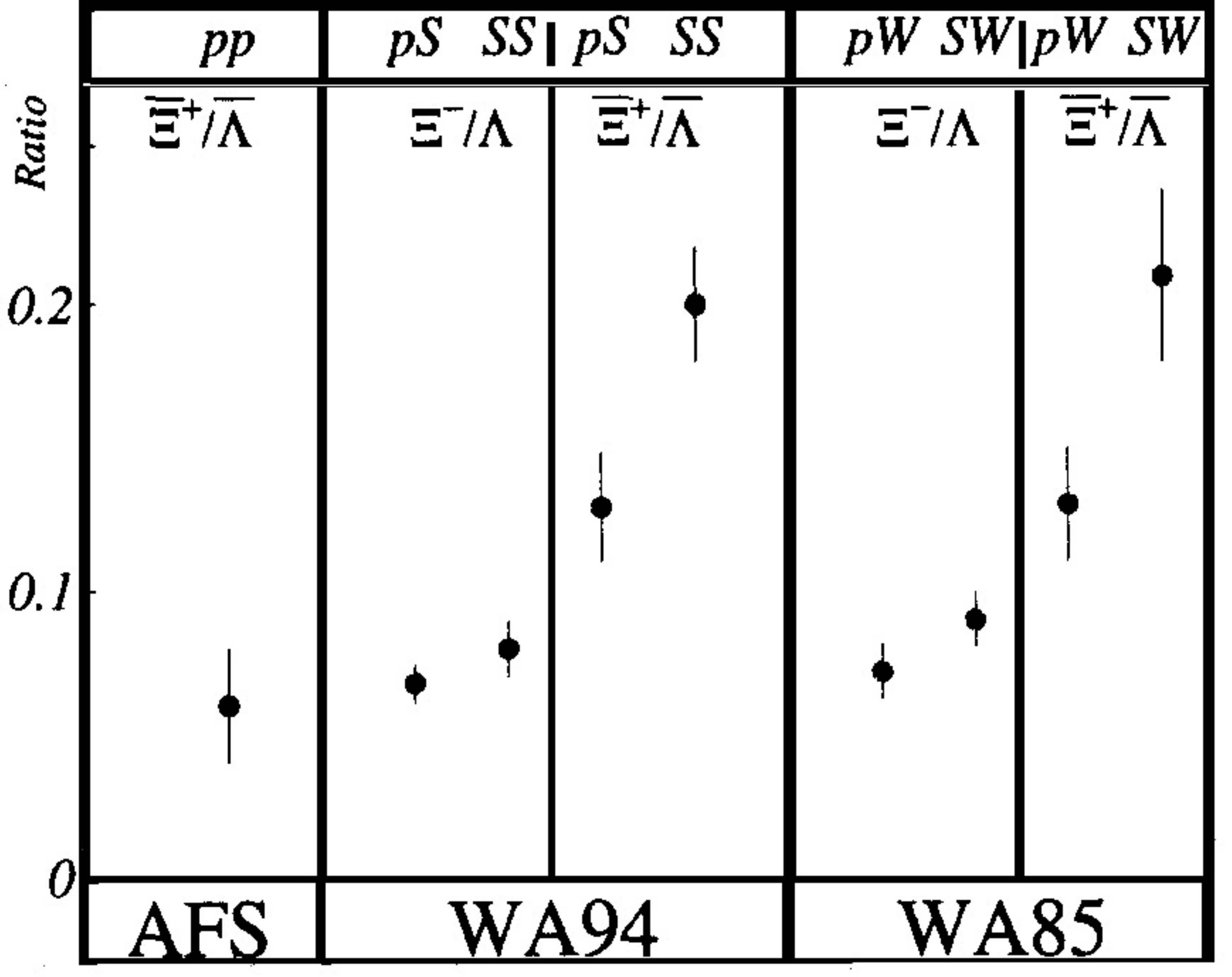}
}
\caption[]{Results obtained at the CERN-SPS  $\uOmega'$-spectrometer for $\uXi/\uLambda$-ratio in fixed target  S-S and  S-Pb at 200\,$A$\,GeV/$c$; results from  the compilation presented in Ref.\cite{Antinori:1997nn} adapted for this report}\label{SigLamCERNFig}
\end{figure}

L\'eon Van Hove, the former DG (1976-1980), characterized the strange  antibaryon signature after hearing the reports~\cite{Rafelski:1982ii,Rafelski:1982bi} as follows~\cite{VanHove:1982qz}: 
\begin{quote}
{\em In the \lq\lq Signals for Plasma\rq\rq\  section:} \ldots implying (production of) an abnormally large antihyperon to antinucleon ratio when plasma hadronizes. The qualitative nature of this prediction is attractive, all the more so that no similar effect is expected in the absence of plasma formation.
\end{quote}
Given this opinion of the \lq man in charge\rq, strange antibaryons became the intellectual cornerstone  of the experimental strangeness program  carried out at the CERN SPS, see \rf{CERNSPS}. Thus it was no accident that SPS research program included as a large part the  exploration of the  predicted strange (anti)baryon enhancement. We see this on  left in \rf{CERNSPS} noting that \lq hadrons\rq\ include of course (multi)strange hadrons and strange antibaryons.

In \hAA collisions at the CERN-SPS  $\uOmega'$-spectrometer,  the production of higher strangeness content baryons and antibaryons was   compared to lower strangeness content particles,  $\uXi/\uLambda$ and  $\bar\uXi/\bar \uLambda$. These early SPS experiments published  in 1997 clearly confirmed the QGP prediction in a systematic fashion, as we see in the 1997 compilation of the pertinent experimental WA85 and WA94 results by Antinori~\cite{Antinori:1997nn}, see \rf{SigLamCERNFig}. Given the systematic multiple observable 3 s.d. agreement of experiment with the   model predictions,  I saw this result as first  and  clear experimental evidence of QGP obtained by  the  experiment-line WA85 and WA94  designed to discover QGP.

In these experiments WA85 and WA94 (see \rf{CERNSPS}) the sulfur ions (S) at 200\,$A$\,GeV hit stationary laboratory targets, S, W (tungsten), respectively, with reference date from  \pp (AFS-ISR experiment at CERN) and $p$ on S shown for comparison. The  $\uXi/\uLambda$ and  $\bar\uXi/\bar \uLambda$ ratio enhancement rises with the size of the reaction volume measured in terms of target $A$,  and is larger for antimatter as compared to matter particles. Looking at \rf{SigLamCERNFig}, the effect is  systematic, showing the QGP predicted pattern~\cite{appenA,appenB,Koch:1984tz,Koch:1986ud}.

The \lq enhancement\rq\ results obtained by the  same group now working in CERN North Area   for  the top SPS  energy Pb (lead) beam of 156\,$A$\,GeV as published in 1999 by Andersen [NA57 Collaboration]~\cite{Andersen:1999ym} is  shown in \rf{EOSTPCFig}. On the right hadrons made  only of quarks and antiquarks that are created in the collision are shown. On the left some of the hadron valence quarks from matter can be brought into the reaction volume. 

The enhancement in production of higher strangeness content baryons and antibaryons in \hAA collisions  increases  with the particle strangeness content. To arrive at this result, the \lq raw\rq\ \hAA  yields are compared with reference  $pp$-, \pA-reaction results and presented per number of  \lq participants\rq\  $\langle N_\mathrm{part} \rangle$  obtained from geometric models of reaction based on energy and particle flows. We will discuss this in Subsection~\ref{ssec:particip}. The number of collision participants for all data presented  in \rf{EOSTPCFig} is large, greater than 100, a point to remember in further discussion. 

We see that production of hadrons made entirely from newly created quarks are up to 20 times more abundant in \hAA -reactions when compared to \pA reference measurement. This enhancement falls with decreasing strangeness content and increasing contents of the valence quarks which are  brought into collision. These reference results at yield ratio \lq 1\rq\ provide the dominant error measure. The pattern of enhancement follows the  QGP prediction and is now   at a level greater than 10 s.d. There is no known  explanation of these results other than QGP. This is also the largest \lq medium\rq\ effect observed in  RHI collision experiments.

\begin{figure}
\centering\resizebox{0.43\textwidth}{!}{%
\includegraphics{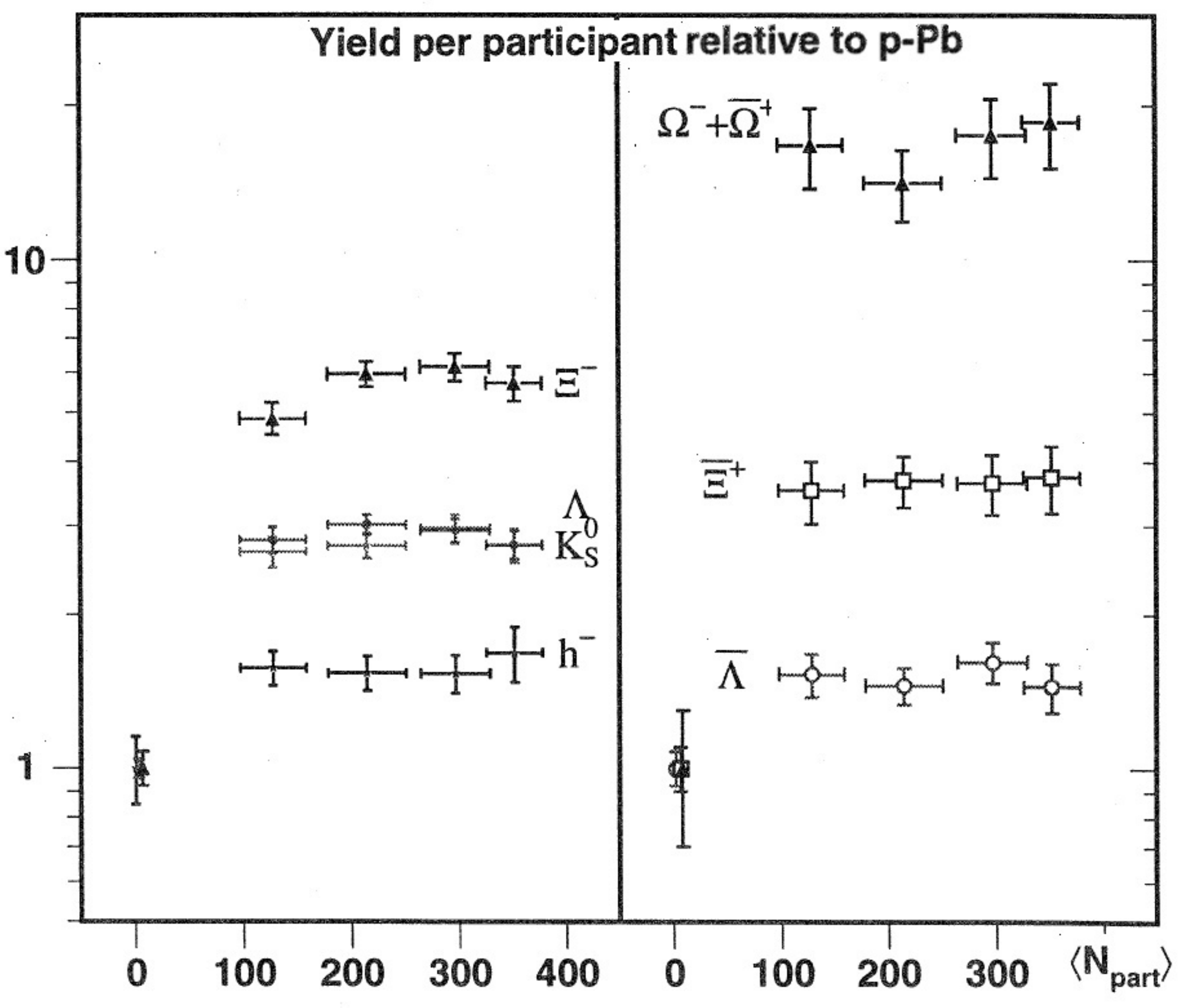}
}
\caption[]{Results  obtained by the CERN-SPS  NA57 experiment (former $\uOmega'$-spectrometer WA85 and WA94 team) for multi-strangeness enhancement at mid-rapidity  $|y_\mathrm{CM}|\,<\,0.5$ in fixed target  Pb-Pb collisions at 158\,$A$\,GeV/$c$  as a function of the mean number of participants $\langle N_\mathrm{part} \rangle$, from  Ref.\cite{Andersen:1999ym}.}\label{EOSTPCFig}
\end{figure}

These discoveries are now all more than 15 years old. They have been  confirmed by further results obtained at SPS, at RHIC, and at the LHC.  The present day experimental summary   is shown in the figure \rf{14AliceEnhFig}.  We see   results obtained by the collaborations:\\
{\bf SPS:}    NA57   for collision energy $\sqrt{s_\mathrm{NN}}=17.2$\,GeV\\ (lighter open symbols);\\
{\bf RHIC:}   STAR   for collision energy $\sqrt{s_\mathrm{NN}}=200$\,GeV\\  (darker open symbols);\\  
{\bf LHC:}  Alice   for collision energy $\sqrt{s_\mathrm{NN}}=2760$\,GeV\\ (filled symbols).\\
These results span a range of collision energies that differ by a factor 160 and yet they  are remarkably similar. 

Comparing the results of \rf{14AliceEnhFig} with those seen in \rf{EOSTPCFig} we note that  $\langle N_\mathrm{part} \rangle$ is now on a logarithmic scale:   the results of \rf{EOSTPCFig} which show that the enhancement is volume independent are in \rf{14AliceEnhFig}  compressed to a relatively small domain on the right in both panels. The   SPS-NA57 results  in \rf{14AliceEnhFig}  are in agreement with the  1999 \lq high\rq\ participant number results shown in \rf{EOSTPCFig}. 

The rise of enhancement which we see in \rf{14AliceEnhFig} as a function of the number of participants $2\, <\,\langle N_\mathrm{part} \rangle\, <\,80$  reflects on the rise of strangeness content in QGP to its chemical equilibrium abundance with an increase in volume and thus lifespan of QGP fireball. It is not surprising that the enhancement at SPS is larger than that seen at RHIC and LHC, considering that the reference yields  play an important role in this comparison. Especially the  high energy LHC \pp reactions should begin to create space domains that resemble QGP and nearly achieve the degree of chemical strangeness equilibration that could erase the enhancement effect entirely.

The study of the $\uphi(s\bar s)$ abundance and enhancement  corroborates these findings~\cite{Abelev:2008zk}. The importance in the present context is that while $\uphi(s\bar s)$ by its strangeness   connects to $\uXi^-(ssd),\overline{\uXi}$, $\uphi(s\bar s)$  is a net-strangeness free particle. Therefore if it follows the pattern of enhancement established for $\uXi,\overline{\uXi}$, this confirms strangeness as being the quantity that causes the effect. For some of my colleagues, these year 2008 results were the decisive turning point to differentiate the strangeness effect from the effect associated with the source volume described in the closing discussion of Ref.\cite{appenA}. Those reading more contemporary literature should note that this volume source effect has been rediscovered three times since, and at some point in time was called \lq\lq canonical suppression\rq\rq.

The reader should also consult   Subsection~\ref{subsec:Bulk}, where it is shown that QGP formation  threshold  for Pb--Pb collisions is found at about 1/4 of the    156\,$A$\,GeV projectile energy, and that the properties of physical QGP fireball formed at SPS are just the  same, up to volume size, when SPS results are compared to RHIC, and with today data from LHC. Today, seen across energy, participant number, and type of hadron considered, there cannot be any doubt that the source of enhancement is the mobility of quarks in the  fireball, with the specific strangeness content showing gluon based processes.

Recall, in February 2000 in the snap of the QGP announcement event, the highly influential Director of BNL, Jeff Marburger\footnote{
Jeff Marburger  was a long term Presidential Science Advisor, President of Stony Brook campus of the NY State University  System, Director of BNL.}  called these NA57 results and other CERN-ion experimental results, I paraphrase the earlier year 2000 precise quote: \lq\lq pieced together indirect glimpse of QGP\rq\rq. Today I would respond to this assessment as follows:  the NA57 results seen in \rf{EOSTPCFig} and confirmed in past 15 years of work, see \rf{14AliceEnhFig} are a direct, full panoramic sight of QGP, as good as one will ever obtain.  There is nothing more direct, spectacular, and convincing that we have seen as evidence of QGP formation in RHI collision experiments.

\begin{figure*}
\centering\resizebox{0.69\textwidth}{!}{%
\includegraphics{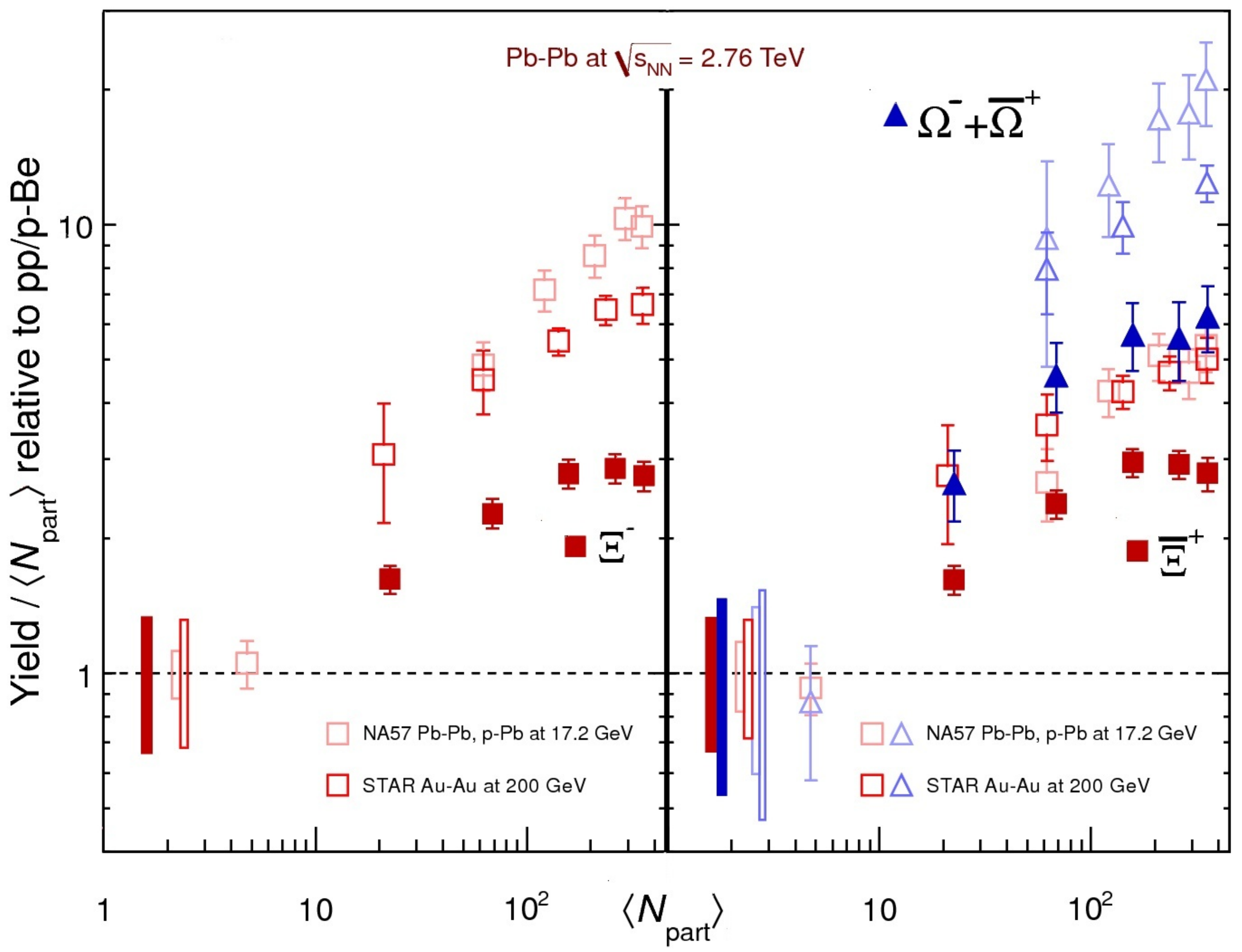}
}
\caption[]{Enhancements of $\uXi^-,\, \overline{\uXi}^+,\, \uOmega^-+\overline{\uOmega}^+$ in the rapidity range $|y_\mathrm{CM}| <  0.5$ as a function of the mean number of participants $\langle N_\mathrm{part} \rangle$: LHC-ALICE: full symbols; RHIC-STAR and SPS-NA57: open symbols. The LHC reference data use interpolated in energy   \pp reference values.
Results at the dashed line (at unity) indicate statistical and systematic uncertainties on the \pp or $p$Be (at SPS) reference. Error bars on the data points represent the corresponding uncertainties for all the heavy-ion measurements. Results presented and compiled in  Ref.\cite{ABELEV:2013zaa}.}\label{14AliceEnhFig}
\end{figure*}
\section{The RHI physics questions of today}\label{RHItodayQuestions}
\subsection{How is energy and matter stopped?}\label{RHIstopping}
We arrange to collide at very high, relativistic energies, two nuclei such as lead (Pb) or gold (Au), having   each  about 12 fm   diameter. In the rest frame of one of the two nuclei we are looking at the other Lorentz-contracted nucleus. The Lorentz contraction factor is  large and thus what an observer traveling along with each nucleus sees approaching is a thin, ultra dense matter pancake. As this pancake penetrates into the other nucleus, there are many reactions that occur, slowing down projectile matter. 

For sufficiently high initial  energy the collision occurs at the speed of light $c$ despite the loss of motion energy. Hence each  observer  comoving with with each of the nuclei records  the interaction time $\tau $ that  a pancake needs to traverse the other  nucleus. The geometric collision time thus is $c\tau^0=12$fm  as measured by  an observer comoving with one of the nuclei. Thus if you are interested like   Anishetty \etal~\cite{Anishetty:1980zp} in hot projectile and target nuclei there is no doubt this is one of the outcomes of the  collision.

An observer in  the center of momentum (CM) frame can determine the fly-by time that  two  nuclei need to pass  each other should they  miss to hit: this is $\tau ^0/\gamma$, where $\gamma$ is the Lorentz-factor  of each of the nuclei with respect to CM frame.  This time is, in general, very short and even if nuclei were to touch in such short time very little could happen. The situation changes if we model this  like a collision of the two bulls of Li Keran and T.D. Lee. Once some of the  energy (and  baryon number) of two nuclei has slowed down to rest in CM, the clocks of both `slowed' bulls tick nearly at the same speed as the  clock comoving with CM frame -- for the  stopped energy and baryon number the lifespan of the  fireball is again quite large.

But how do we stop the bulls or at least some of their energy? The answer certainly  depends on the  energy regime. The lower is the  energy of the bulls, the  less we need to worry; the pancakes are not thin and one can try to make parton-collision cascade to describe  the physics case, see e.g.    Geiger-Sriwastava~\cite{Geiger:1997ie,Geiger:1992si,Geiger:1992ac}  for SPS energy  range. The use of these methods for RHIC or even LHC  energies looks less convincing~\cite{Geiger:1994he}.

To put the  problem in perspective, we need a way to concentrate entropy so that a thermal state can rapidly arise. Beginning with the work of Bjorken~\cite{Bjorken:1982qr}  a  formation time is introduced, which is more than an order of magnitude shorter compared to $\tau^0$. It is hard to find tangible experimental evidence which compels a choice such as 0.5fm/c, and theory models describing this stage  are not fully  convincing. A model aims to explain how as a function of collision energy and centrality the easy to observe final entropy (hadron multiplicity) content arises. For some related effort see  review work of the Werner-group~\cite{Drescher:2000ha}  and  Iancu-Venugopalan~\cite{Iancu:2003xm}. 

To summarize, in the \lq low\rq\ energy  regime of SPS we can try to build a parton cascade model to capture  the  essence of heavy-ion collision dynamics~\cite{Geiger:1992si,Geiger:1992ac}.  The understanding of the initial \lq formation\rq\  of QGP as a function of collision energy and the understanding of the mechanism that describe energy and baryon number stopping  remains one of the fundamental challenges of the ongoing theoretical and experimental research program.   

\subsection{How and what happens, allowing QGP creation?}\label{whatQGP} 
In the previous Subsection~\ref{RHIstopping} we addressed the  question how the  energy and baryon number is extracted from fast moving nuclei. In this section the added challenge is, how is the  entropy produced that we find in the  fireball? While in some solutions of the initial state formation in RHI collisions these two topics are confounded, these are two different  issues: stopping   precedes and  is not the  same as abundant entropy  production.

For many  the   mechanism of fast, abundant entropy formation is associated with  the breaking of color bonds, the melting of vacuum structure, and the deconfinement of quarks and gluons. How exactly this should work has never been shown: Among the first to address a parton based entropy production quantitatively within a kinetic collision model  was Klaus Geiger~\cite{Geiger:1992si,Geiger:1992ac} who built computer cascade models at parton level, and studied thermalization as a collision based process.  

In order to understand   the QGP formation process a solution of this riddle is necessary. There is more to entropy production:  it controls the kinetic energy conversion into material particles. The contemporary wisdom how to describe the  situation  distinguishes  several  reaction steps in  RHI collision:\\
\indent 1) Formation of  the primary fireball; a momentum equi\-partitioned partonic  phase comprising in a limited space-time domain, speaking in terms of orders of magnitude, almost the final state entropy;\\ 
\indent 2) The cooking of the energy content of the hot matter fireball  towards the particle  yield (chemical) equilibrium in a hot perturbative QGP phase;\\
\indent 3) Expansion and evaporation cooling towards  the temperature phase boundary;\\
\indent 4) Hadronization; that is, combination of  effective and strongly interacting $u,d,s, \bar u,\bar d$ and $ \bar s$ quarks  and   anti-quarks   into the final state  hadrons, with  the yield probability weighted by accessible phase space.\\
It is the  first step that harbors a mystery.

The current textbook wisdom is that entropy production requires the  immersion of the quantum system in a classical environment. Such an environment is not so readily available for a RHI collision system that has a lifespan of below $10^{-22}$\,s and a size less than 1/10,000 of atomic size.  For a year 2011  review  on entropy production during the different stages of RHI collision see Ref.\cite{Muller:2011ra}. The search for a fast entropy generating  mechanism  continues, see for example Ref.\cite{Kurkela:2014tea}.  
 
So what could be  a mechanism of rapid entropy  formation? Consider the spontaneous pair production in presence  of a strong field: the stronger is the field the greater is the  rate of field conversion into particles. One finds that  when the field strength is such that it is capable of accelerating particles with a unit strength  critical  acceleration, the speed of field decay into particle pairs is such that a field filled state makes no sense as it decays  too fast~\cite{Labun:2010wf}. For this reason  there is an effective limit to the strength of the field, and forces capable to accelerate particles at critical limit  turn the field filled space  into a gas of particles. 

The conversion of energy stored in   fields into particles, often referred to, in the QCD context, as the breaking of color strings, must be an irreversible process. Yet the  textbook wisdom will assign to the time evolution pure quantum properties, and in consequence, while the complexity  of the state evolves, it  remains \lq unobserved\rq\ and thus a pure state with vanishing entropy content. Intuitively,  this  makes little sense. Thus the riddle of entropy production in RHI collisions which involve an encounter of two pure quantum states and turns rapidly  into state of large entropy carried by  many particles maybe  related to our poor formulation of quantum processes for unstable critical field filled  states  decaying into numerous pairs.

However, the  situation may also call for a more fundamental revision of the laws of physics. The  reason  is that our  understanding  is based in experience, and we really do not have  much  experience with critical acceleration conditions. When we study  acceleration phenomena  on a microscopic scale, usually these are very small, even in principle zero. However, in  RHI  collisions when we stop partons in the central rapidity region we encounter the critical acceleration,   an acceleration that in natural units is unity and which further  signals a drastic change in the way fields and particles behave. The framework of physical laws which is based on present experience may not be  sufficiently complete to deal with this situation and we will need to increase the  pool of our  experience by  performing many  experiments involving critical forces.

To conclude: a) The measurement of entropy production is   relatively straightforward as all   entropy produced   at the end is found in newly  produced particles; b) The QGP formation  presents an efficient   mechanism  for the  conversion of the kinetic  energy of the colliding nuclei  into particles in a process that is not understood despite many years of effort;  c) Exploration and understanding of the  principles that lead to the abundant formation of entropy in the  process of QGP  formation in RHI collisions  harbors  potential opportunity to expand the  horizons of knowledge.

\subsection{Nonequilibrium in fireball hadronization}\label{subsec:nonequilWhy}
Heavy flavor production cross sections, in lowest order in coupling constant, scale according to $\sigma\propto \alpha_s^2/m^2$.  Considering a smaller (running) coupling, and a  much larger mass of \eg\ heavy quarks $c,b$, we obtain a significant reduction in the speed of  thermal QGP production reactions. For charm and bottom, contribution for  thermal production depends on the profile of temperature but  is very likely negligible, and for charm it is at the level of a few percent.  Conversely,  light quarks equilibrate rather rapidly compared to the even more strongly self coupled gluons and in general can be assumed to follow and define QGP matter properties.

Heavy quark yields originate in  the pre-thermal parton dynamics. However, heavy quarks may acquire through elastic collisions a momentum distribution characteristic of the medium, providing an image of the collective dynamics of the dense quark matter flow. Moreover, the  question of yield evolution arises, in particular with regard to annihilation of heavy flavor in QGP evolution.

Our  \lq boiling\rq\  QGP fireball is not immersed in a bath. It is expanding or, rather, exploding into empty space  at a high speed. This assures that the entropy $S \propto V$ is not decreasing, but increasing, in consideration of internal collisions  which describe  the bulk viscosity. The  thermal energy content is not conserved since the sum of the kinetic energy of expanding motion, and thermal energy,  is conserved. Since the positive internal pressure of QGP accelerates the expansion into empty  space, an explosion, the thermal energy  content decreases and the fireball   cools rapidly.

In this dynamical evolution  quark flavors undergo chemical freeze-out. The heavier the quark, the  earlier the  abundance  freeze-out should occur. Charm is produced in the first collisions in the formative stage of QGP. The coupling to thermal environment is weak. As ambient temperature drops the charm quark phase space given its mass drops rapidly. The quantum Fermi phase space distribution which maximizes the entropy at fixed particle number is~\cite{Letessier:1993qa,Letessier:1993qb} 
\begin{align}\label{fermiDist}
n_\mathrm{F} (t)=&\displaystyle\frac 1 {\gamma^{-1}(t)  \E^{(E \mp \mu )/T(t)}+1}, \\
{{d^6N_\mathrm{F}}\over {d^3pd^3x}} =&{g\over (2\pi)^3}n_\mathrm{F},\notag
\quad E =\sqrt{p^2+m ^2}\;, 
\end{align}
where  $g$ is the statistical degeneracy, and the  chemical nonequilibrium fugacity (phase space occupancy) $\gamma (t)$  is the same for particles and antiparticles while the chemical potential  $\mu $ describes  particle-antiparticle asymmetry, and changes sign as indicated. Our $\mu$ is \lq relativistic\rq\ chemical potential. In the nonrelativistic limit $\mu\equiv m+\mu_\mathrm{nr}$ such that $m$ implicit in $E$ cancels out for particle  but turns to a $2m/T$ suppression for the antiparticles.  Note that independent of the values of all parameters, $n_\mathrm{F} \le 1$ as required.  

The  integral of the  distribution \req{fermiDist} provides the particle yield. When addressing SHARE phase space  properties in Subsection~\ref{YieldFluct}, we will inspect the  more exact result, here we consider  the Boltzmann   nonrelativistic limit suitable for heavy quarks ($c,b$)
\begin{align}\label{fermiY}
N = &\frac{g V T^3}{2\pi^2} \gamma \E^{\pm\mu/T}  x^2 K_2(x)\;,\quad x=\frac{m}{T}\\[0.2cm]
 &\to gV  (mT/2\pi)^{3/2} \gamma \E^{-(m\mp\mu)/T} \label{charmY}
\end{align}
$T(t)$ is time dependent because the system cools. Let us look at the  case $\mu=0$, appropriate for physics at LHC and, in the context of present discussion, also  a good approximation at RHIC.    

Considering charm abundance, in QGP chemical equilibrium $\gamma^\mathrm{QGP} (t)\to 1$. However, we recall that charm froze out  shortly after first collisions. Therefore  the value of  $\gamma^\mathrm{QGP}_c (t)$ in \req{charmY} is established by  need to preserve the  total charm pair number $N_c=Const$. The exponential factor $m/T$ changes from about 2  to 8 near to hadronization. Thus for prescribed yields at LHC  and RHIC it is likely that  $\gamma^\mathrm{QGP}_c(t)> 1$. More generally  there is nobody who disagrees with the need to have   $\gamma^\mathrm{QGP}_c\ne 1$.  $\gamma^\mathrm{QGP}_c= 1$ is an accidental condition. We have established that  charm, and  for the  very same reason, bottom flavor, cannot be  expected to emerge in chemical equilibrium abundance  at hadronization.

A QGP filled volume at high $T$ cooks up  a  high content of strangeness pairs, in essence as many as there are of each light  flavor $u,d$; in plasma strangeness suppression disappears; the Wroblewski suppression factor~\cite{Wroblewski:1985sz} (see also next subsection) is therefore close to unity.  As plasma evolves and cools at some relatively low temperature the yield of strangeness freezes-out, just like it did for  charm (and  bottom) at higher value of $T$. 

In  earlier discussion  we have   assumed that in QGP strangeness will follow the evolution in its pair abundance, and always be in chemical equilibrium in the fireball. This tacit assumption is not supported by kinetic theory for $T<T_s\simeq 180$ MeV; however for such low value of $T$ the systematic error of perturbative QCD is large, thus  we really  do not know  where approximately strangeness pair yield freezes out. We must introduce a pair fugacity parameter   aside of charm also for strangeness and we now have  $\gamma_{b,c,s}^\mathrm{QGP} \ne 1$. The phase space size of strangeness on the hadron side is smaller so once strangeness emerges one must expect that a relatively  large value  could be  measured.

So what about $\gamma_{u,d}$? If the  evolution as a function of $T$ of the fireball properties is smooth as lattice computation suggests, then the strongly  coupled light  quarks and gluons are defining the  QGP  properties and,  remain in equilibrium: \lq{\em \ldots really?\/}\rq\  The flaw in this argument is that  only quarks  define final hadrons. Thus   gluons transform  into quark pairs feeding additional mesons and baryons in that way and helping preserve entropy content. Thus gluon dissolution into additional hadrons assures that  the light  quark phase  space  occupancies as measured in terms of observed hadron abundances should show $\gamma_{u,d}^\mathrm{HG}>\gamma_{u,d}^\mathrm{QGP}>1$.
 
The introduction hadron-side of phase space occupancy $\gamma_s$~\cite{Rafelski:1991rh} and later $\gamma_{u,d}$~\cite{Letessier:1998sz} into the study of hadron production in the  statistical hadronization approach has been challenged. However there was no scientific case,  challenges were driven solely by an  intuitive  argument  that  in  RHI collisions at sufficiently  high reaction energy  aside of thermal, also chemical equilibrium is reached. One of the objectives of this  review is to explain why this intuition is wrong when QGP is formed. 

Note further that  there is a difference between an assumption and the demonstration of a result. All know that to make a proof  one generally  tries to show a contrary behavior and  arrives at a contradiction: in this case  one starts with  $\gamma_{s,u,d}\ne 1$  and shows that results are right only for  $\gamma_{s,u,d}\to 1$. However, we will see in Section~\ref{AnalysisHadronization} that  results are right when   $\gamma_{s,u,d}\ne 1$ and we show by  example in Subsection~\ref{sec:earlier} how the urban legend \lq chemical equilibrium works\rq\  formed  relying on a set of errors and/or omissions.
 
The question about chemical non/equilibrium conditions  has  to be  resolved so that consensus can emerge about the properties of the hadronizing QGP drop, and the mechanisms and processes that govern the  hadronization process. 


\section{How is the experimental study of QGP continuing today?}\label{RHItoday}
Today RHI collisions and QGP is a research field that has grown to be a large fraction of nuclear science research programs on several continents. A full account of methods, ongoing experiments, scheduled runs, future plans including the development of new experimental facilities is a separate review that this author cannot write. The question how to balance presenting \lq nothing\rq, with \lq everything\rq, is never satisfactorily soluble. The  selection of the following  few topics is  made in support  of  a few highlights of greatest importance to this review.

\subsection{Short survey  of recent QGP probes \& results}\label{QGPobservables}
A short   list of contemporary  QGP probes and results includes:\\[-0.15cm] 

\noindent{\bf Strangeness and other soft hadrons}

This cornerstone observable of  QGP is a topic of personal expertise of the author and is addressed elsewhere and at length  in these pages. The following is   a brief summary: Strangeness, the lightest unstable quark flavor, appears in \pp collisions with an abundance that  is about a factor 2.5--3 below that of each light  quark  flavor; this is the mentioned \lq Wroblewski\rq\ ratio~\cite{Wroblewski:1985sz}. It is natural to expect that  in a larger physical \hAA collision system additional scattering opportunity among all particles creates a more democratic abundance  with $u,d,s$ quarks being available in nearly equal abundance.   However, this initial simple hypothesis, see Ref.\cite{appenA}, needed to be refined with actual kinetic theory evaluation; see Ref.\cite{appenB}, in consideration of the  short time available and demonstration that quark collisions were too slow~\cite{Biro:1981zi} to achieve this goal. 

It was shown that the large abundance of strangeness depends on gluon reactions mechanism; thus the \lq gluon\rq\ particle component in quark gluon plasma is directly involved~\cite{Rafelski:1982pu}, see Ref.\cite{appenA}. The high   strangeness density in QGP and `democratic' abundance at nearly the  same level also implies that the production of (anti)baryons with multiple strangeness content is abundant, see  \rf{MultistrangFig}, which attracted  experimental interest, see Subsection~\ref{subsec:expStra}.  The observation of strange hadrons involves the identification of non-strange hadrons and thus a full characterization of all particles emitted is possible. This in turn creates an opportunity to understand the properties of the QGP at time of hadronization, see Subsection~\ref{subsec:Bulk}. \\[-0.15cm]

\noindent{\bf Hard hadrons: jet quenching}

With increasing energy, like in $pp$, also in \hAA collisions hard parton back-scattering must occur, with a rate described by the perturbative QCD~\cite{Muller:2002fa,Majumder:2010qh}. Such hard partons are observed in back-to-back jets, that is two jet-like assembly of particles into which the hard parton hadronizes. These jets are created within the primordial medium. If geometrically such a pair is produced near to the edge of colliding matter, one of the jet-partons can escape and the  balancing momentum of the immersed jet-parton tells us how it travels across the entire nuclear domain, in essence traversing QGP that has evolved in the collision. The energy of such a parton can be partially or completely  dissipated, \lq thermalized\rq\ within the QGP distance traveled. Since at the production point a second high energy quark (parton) was produced, we can deduce from  the \lq jet\rq\ asymmetry that the dense matter we form in RHI collisions is very opaque, and with some effort we can quantify the strength of such an interaction. This establishes the strength of interaction of a parton at given energy with the QGP medium. \\[-0.15cm]

\noindent{\bf Direct Photons}

Hot electromagnetic charge plasma radiates both photons and  virtual photons, dileptons~\cite{Bhattacharya:2015ada,Ryblewski:2015hea}. The hotter   is the plasma, the  greater is the radiation yield; thus we  hope for a large  early QGP stage  contribution. Electromagnetic probes emerge from the reaction zone without noticeable loss. The  yield  is the  integral over the  history of QGP evolution, and the measured uncorrected yield is polluted by contributions from the  ensuing hadron  decays. 

On the other hand, at first glancce  photons are  the   ideal probes of the primordial QGP period if one can  control the background photons from the decay of strongly  interacting particles such  as $\pi^0\to \gamma \gamma$ which in general are   dominant\footnote{Note that $\gamma$ when used as a symbol for photons is not to be confounded with parallel use of $\gamma$ as a fugacity, meaning is always clear in the  context.}. Recognition  of the signal as direct QGP photon  depends on a very precise understanding of the background.

At the highest collision energy the  initial QGP temperature increases and thus direct photons should be  more abundant. In \rf{directGamma}  we see the first still at the time of writing preliminary result from the  Alice experiment at LHC. The  yield shown  is \lq direct\rq; that is, after the indirect  photon part has  been removed. The removal procedure appears reliable as for large $p_\bot$ scaled \pp yields match the outcome. At small $p_\bot$, we see a very strong excess above the scaled \pp yields. The $p_\bot$ is high enough to believe that the origin   are direct QGP photons, and not collective charge acceleration-radiation phenomena.

\begin{figure*}
\centering\resizebox{0.75\textwidth}{!}{%
\includegraphics{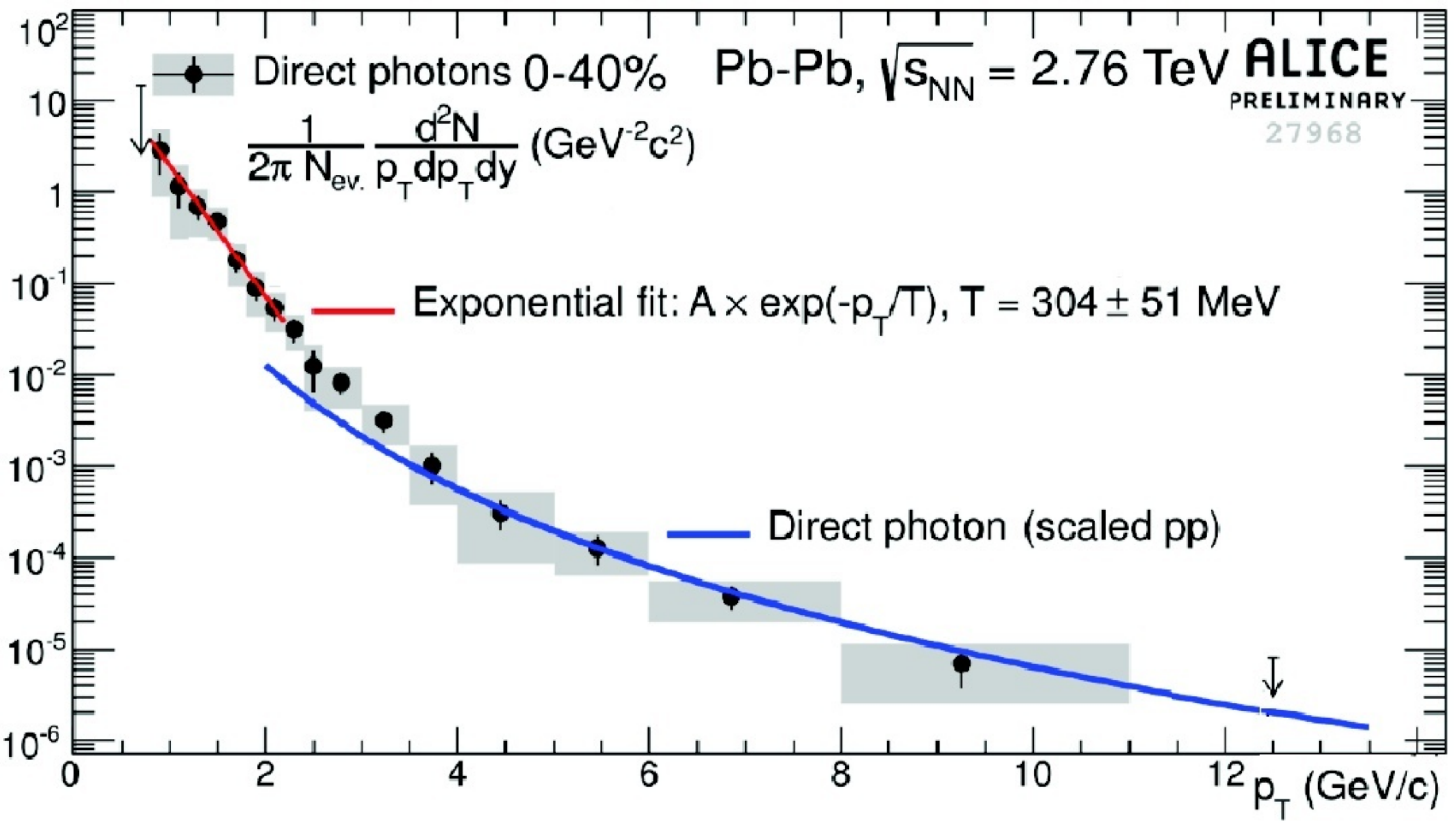}
}
\caption[]{Direct photon LHC-\hAA yield; Adapted from: F. Antinori presentation July 2014.}\label{directGamma}
\end{figure*}

A virtual photon with $q^2\ne 0$  is upon materialization a dilepton  $e^+e^-,\ \mu^+\mu^-$ in the final state. The dilepton yields, compared to photons, are about  a factor 1000 smaller; this creates  measurement challenges \eg\ for large $p_\bot$. Backgrounds from vector meson intermediate states and decays  are very large and difficult to control. Despite many  efforts to improve detection capabilities and the understanding of the background, this author considers the situation as  fluid and inconclusive: dilepton  radiance not directly attributed to hadrons is often reported and even more often challenged. An observer view is presented in Ref.\cite{Petousis:2012yi}. \\[-0.15cm]

\noindent{\bf  $J/\Psi(c\bar c)$ yield modification}

This is the other cornerstone observable often quoted in the context of the early QGP search. The interest in the bound states of heavy charm quarks $c\bar c$  and in particular  $J/\Psi$ is due to their yield evolution in the deconfined state as first proposed by  Matsui  and Satz~\cite{Matsui:1986dk} just when first result $J/\Psi$ became available. Given that the  variations in yield are subtle,  and that there are many model interpretations of the effects based on different views of  interaction of  $J/\Psi$  in the dense matter -- both confined and deconfined --  this has  been for  a long time a livid topic which is beyond the scope of this review~\cite{Andronic:2014sga}.  

Modern theory addresses both \lq melting\rq\ and recombination in QGP as processes that modify the final $J/\Psi$ yield\,\cite{Thews:2000rj,Blaizot:2015hya}. Recent results obtained by the  Alice collaboration\,\cite{Abelev:2013ila} support, in my opinion, the notion of recombinant $c\bar c$ formation. Some features of these results allow   suggesting that  a yield equilibrium between melting and recombination has been reached for more central collisions. This is clearly a research  topic, not yet suitable for a review analysis.\\[-0.15cm]

\noindent{\bf Particle  correlations and HBT}

Measurement of two particle and in particular two pion and two kaon correlations allows within the framework of geometric source interpretation the exploration of the three dimensional source size and the emission lifespan of the fireball. For a recent review and update of PHENIX-RHIC results see Ref.\cite{Adare:2015bcj} and for ALICE-LHC see Ref.\cite{Adam:2015vja}. These reports are the basis for our tacit assumption that soft hadrons emerge from the hadronization fireball with transverse size as large as $R\simeq 9$\,fm for most central collisions. Aside of two particle correlation,  more complex multi-particle correlations can be and are explored -- their non vanishing strength reminds us that  the  QGP source can  have  color-charge confinement related multi-particle effects that  remain difficult to quantify. As an example of recent work on  long range rapidity correlations see Ref.\cite{Altinoluk:2015uaa}   \\[-0.15cm] 

\noindent{\bf Fluctuations}

Any physical system that at first sight appears homogeneous will under a magnifying glass show large fluctuations; the color of the  sky and for that matter of  our planet originate in  how the atmospheric density fluctuations scatter light. To see QGP fluctuation effects we need to study  each individual event forming QGP apart from another. The SHARE suite of SHM programs also computes statistical particle yield fluctuations, see Subsection~\ref{YieldFluct}. The search is for large, nonstatistical fluctuations that would signal competition between two different phases of matter, a phase transformation. This   topic is attracting attention~\cite{Csernai:2012zf}. To see the phase transformation in action smaller reaction systems may provide more opportunity.

\subsection{Survey of LHC-ion program July 2015}\label{Subsec:Experiments}
The Large Hadron Collider (LHC)  in years of operation  sets aside 4 weeks of run time a year to the heavy ion beam experiments, typically \hAA (Pb-Pb) collisions but also $p$-Pb. The \pp collision LHC  run which lasts considerably longer  addresses Higgs physics and beyond the  standard model searches for new physics. This long run  provides   heavy ion experimental groups an excellent opportunity to obtain relevant data from the smallest collision system, creating a precise baseline against which \hAA is  evaluated. Furthermore, at the  LHC energy, one can hope that in some measurable fraction of  events  conditions for QGP  could be  met in select, triggered  events (\ie\ collision class feature  selected).

When LHC reaches energy of 7 TeV+7 TeV for protons, for Pb-Pb collisions this magnet setting will correspond to   a center-of-mass energy  of up to   $\sqrt{s_\mathrm{NN}}=5.52$\,TeV per nucleon pair in Pb-Pb collisions. However, due  to magnet training considerations  the  scheduled heavy ion run starting in mid-November   2015 should be  at $\sqrt{s_\mathrm{NN}}=5.125$\, TeV and the maximum energy achieved in the following year. The results we discuss in this review, see Section~\ref{AnalysisHadronization}, were obtained at a lower magnet setting  in the  LHC run 1, corresponding to  $\sqrt{s_\mathrm{NN}}=2.76$\, TeV.

Several experiments at LHC take AA collision data: 
\begin{enumerate}
\item 
The ALICE (A Large Ion Collider Experiment) was conceived specifically for the exploration of the QGP formed in nucleus-nucleus collisions at the LHC. Within the central rapidity domain $0.5\le y\le 0.5$, ALICE detectors are capable of precise tracking and identifying particles  over a large range of momentum. This permits the study of the production of strangeness, charm and resonances, but also multi-particle correlations, such as HBT and (moderate energy) jets. In addition, ALICE consists of a muon spectrometer allowing us to study at forward rapidities heavy-flavor and quarkonium production. The detector system also has the ability to trigger on different aspects of collisions, to select events on-line based on the particle multiplicity, or the presence of rare probes such as (di-)muons, and the electromagnetic energy from high-momentum electrons, photons and jets.

\item ATLAS (A Toroidal LHC ApparatuS) has made its name by being first to see jet quenching. It has high $p_\bot$ particle ID allowing the measurement of particle spectra in a domain inaccessible to other LHC experiments.

\item The CMS (Compact Muon Solenoid) offers high rate and high resolution calorimetry, charged particle tracking and muon identification over a wide acceptance, allowing detailed measurements of jets as well as heavy-quark open and bound states. The large solid angle coverage also provides unique opportunities in the study of global observables.
\end{enumerate}
The LHCb experiment has  at present no footprint in the study of \hAA collisions but has taken data in \pA trial run.

\begin{figure*}[tb]
\centering
\centering\resizebox{0.85\textwidth}{!}{%
\includegraphics{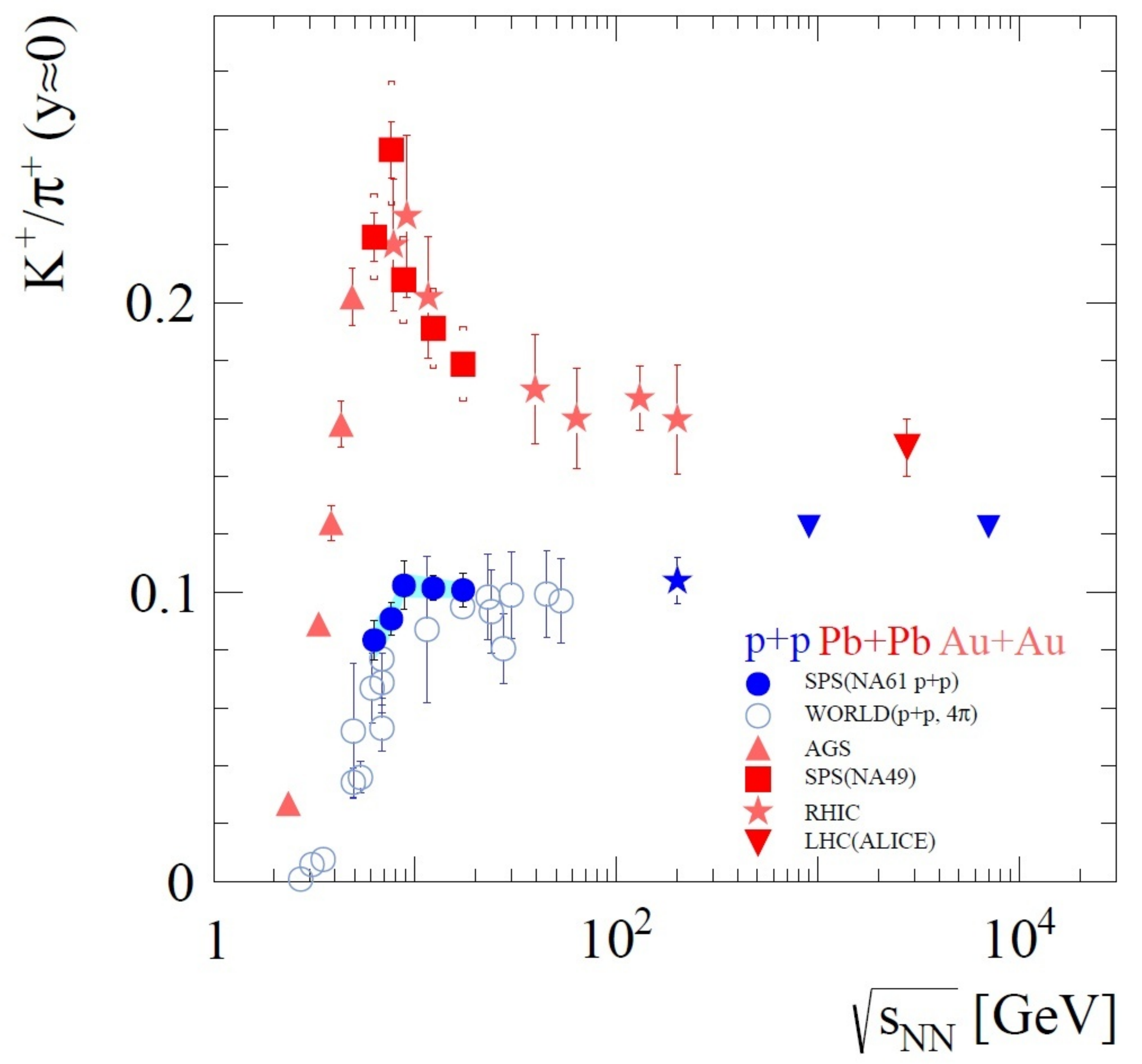}
\includegraphics{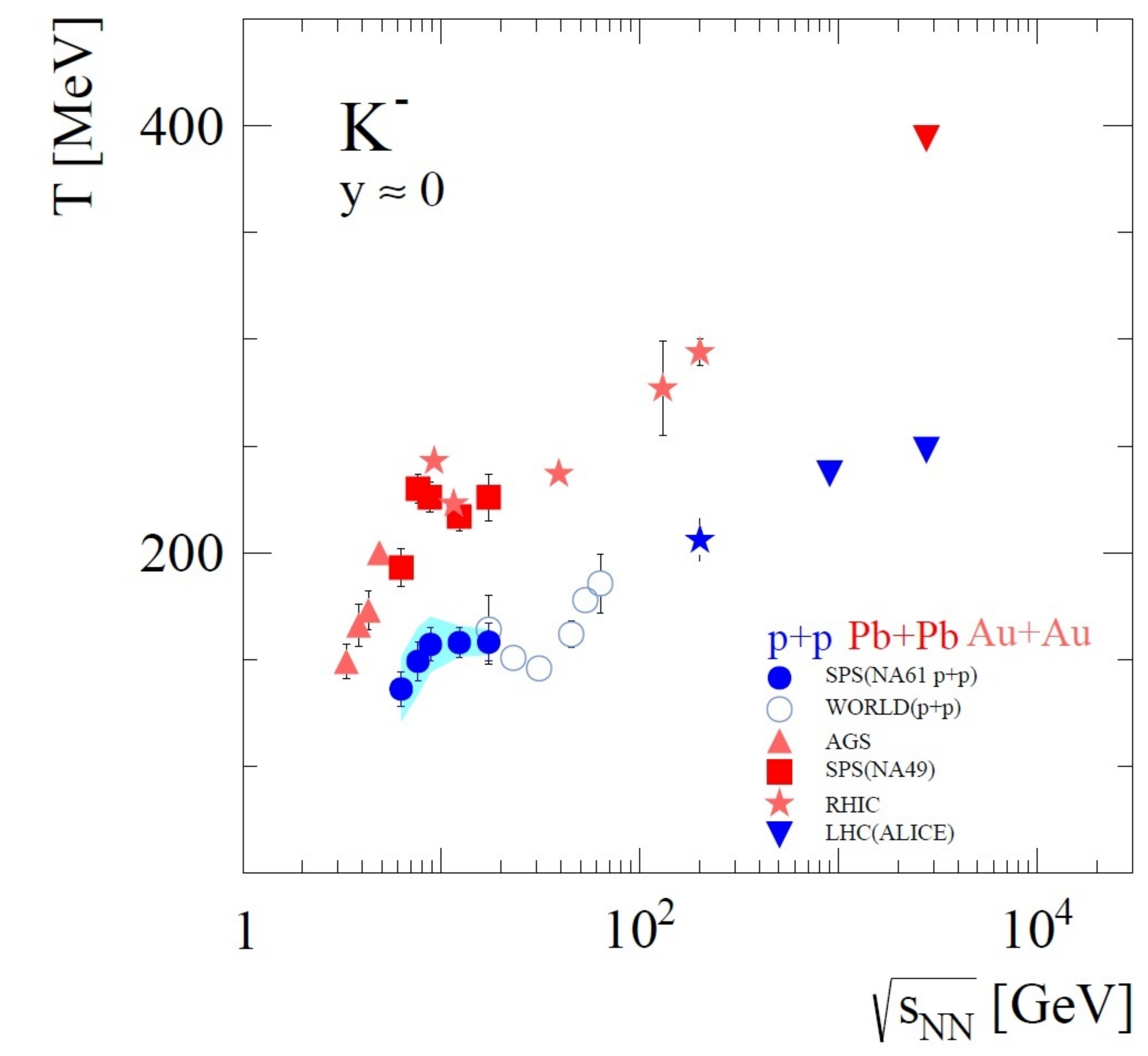}
}
  \caption{\label{fig:onset}
The so called horn (left) and step (right) structures
in energy dependence of the  K$^+/\pi^+$ ratio, and the inverse slope
parameter of  K$^-$ $m_\bot$ spectra, respectively.  signal
indicating threshold in strangeness to entropy yield 
in central Pb+Pb (Au+Au) collisions, from~\cite{Abgrall:sr2014}.
}
\end{figure*}

\subsection{Energy and $A$ scan}\label{HornThreshhold}
The smaller the size of colliding nuclei, the shorter is the collision time.  Thus in collisions of small sized objects such as \pp or light nuclei, one cannot presume, especially at a relatively low collision energy, that primordial and yet not well understood processes (compare Subsection~\ref{whatQGP}) will have time to generate the large amount of entropy leading to QGP formation  that would allow a statistical model to work well, and in particular would allow QGP formation. This than suggests that one should explore dependence on reaction volume size, both in terms of collision centrality and a scan of projectile ion $A$.

An important additional observation is that particle  production processes are more effective with increasing  collision energy. Therefore the chemical equilibration is achieved more rapidly at higher energy. It seems that just about everyone agrees to this even though one can easily argue the opposite, that more time is available at lower energy. In any  case, this urban legend that energy and time grow together is   the main reason why QGP search experiments started at the highest available accelerator energy. This said, the question about the threshold of QGP production as a function of  energy  is open.

Considered from  a theoretical  perspective one recognizes in an energy and $A$ scan the opportunity to explore qualitative features of the QCD phase diagram in the  $T,\muB$ plane. Of particular importance is the finding of the critical point where at a finite value of \muB\ the smooth transformation between quark-hadron phases turns into an expected 1st order transition, see Ref.\cite{Rajagopal:1999cp}.  There are other structure features of quark matter that  may become accessible, for a review see Ref.\cite{Alford:2007xm} and  comments at the end of   Subsection~\ref{Thdef}.

At CERN the multipurpose NA61 experiment surveys in its heavy-ion program tasks the domain in energy and collision system volume where threshold  of deconfinement is suspected in consideration of available data. This experiment responds to the results of a study of head-on Pb--Pb collisions as a function of energy  at SPS did produce by 2010 tantalizing hints of an energy threshold to new phenomena~\cite{Gazdzicki:2010iv,Grebieszkow:2011np,Abgrall:sr2014,Gazdzicki:2014bxa}.

There are significant discontinuities  as a function of collision energy in the  K$^+/\pi^+$ particle yield  ratio, see \rf{fig:onset} on left.   Similarly,  the inverse slope parameter of  the $m_\bot$ spectra of K$^-$, see \rf{fig:onset} on right, also displays  a local maximum near to 30\,$A$\,GeV, that is at 3.8+3.8 GeV, $\sqrt{s_\mathrm{NN}}=7.6$ GeV   collider energy collisions in both quantities. These behavior \lq thresholds\rq\ are to some degree   mirrored in the much smaller \pp reaction system also shown in  \rf{fig:onset}. These remarkable results  are interpreted as the onset of deconfinement as a function of collision energy.

Turning to comparable efforts at RHIC: in 2010 and 2011, RHIC ran the first phase of a  beam energy scan program (RHIC-BES) to probe  the nature of the phase boundary between hadrons and QGP as a function of \muB. With beam  energy settings $\sqrt{s_\mathrm{NN}}=7.7,$ $11.5,$ $19.6,$ $27,$ $39$ GeV, with $14.5$ GeV  included in year 2014, complementing the full energy of 200 GeV,  and the run at 62.4 GeV, a relatively wide domain of \muB\ can be probed, as the matter vs anti-matter excess increases when energy decreases. For a report on these result see Refs.~\cite{Sorensen:2013jka,Sahoo:2014bqa}.

Among the first phase of the beam energy scan  discoveries is the \muB\ dependence of azimuthal asymmetry of flow of matter, $v_2$. Particle yield ratio fluctuations show significant deviation from Poisson expectation within HRG model. This and other   results make it plausible that QGP is formed down to the lowest RHIC  beam energy of $\sqrt{s_\mathrm{NN}}=7.7$ GeV, corresponding to  fixed target collision experiments at 32 $A$ GeV. This is the collision energy where SPS energy scan also found  behavior characteristic of QGP, see \rf{fig:onset}. These interesting results  motivate   the second  RHIC-BES  phase after  detector upgrades are completed in 2018/19.

\section{What are the Conceptual Challenges\\ 
of the QGP/RHI Collisions Program?}\label{fullwhy}
In subsection~\ref{introwhy} we have   briefly  addressed the {\em Why?} of the RHI collision research program. Here  we return to  explore some of the  points raised, presenting a highly  subjective view of foundational opportunities that await us.  
\subsection{The origin of mass of matter}\label{MassMatter}
Confining quarks to a domain in space means that  the typical energy each of the  light  quarks will have  inside a hadron is $E_q\propto 1/R\gg m_q$, where $R$ is the size of the \lq hole\rq\ in the  vacuum -- a vacuole. Imposing a sharp  boundary and forbidding  a quark-leak results in a square-well-like  relativistic Dirac quantum waves. This model allows quantification of $E_q$. One further argues that the  size $R$ of  the  vacuole arises from the internal   Fermi and  Casimir  pressures balancing the  outside vacuum which presses to erase any vacuole comprising energy density that is  higher. 

In a nutshell this is the math known from within the context of  quark-bag model~\cite{Chodos:1974je,DeGrand:1975cf,Johnson:1975zp}, rounded off allowing color-magnetic hyperfine structure splitting. This   model explains how baryons and  mesons   have  a mass much greater than the sum of quark masses. It is also easy  to see that  a larger vacuole with hot quarks and  gluons would provide a good starting point to develop a dynamical model of expanding QGP fireball formed in RHI collisions.

The advent of lattice-QCD means we can address static time independent  properties of strongly interacting  particles. A test of bag models  ideas is  the  computation of the  hadron mass spectrum and demonstration that the  mass of hadrons is not determined by the mass of quarks bound inside. Indeed,  this has been shown~\cite{Durr:2008zz,Kronfeld:2012ym}; the confining  vacuum structure  contributes as much as 96\%  of the mass of the matter, the  Higgs field  the remaining few-\%. 

Based on both bag  model consideration and lattice-QCD we conclude  that  the quantum zero-point energy  of the localized, confined, light quarks  governs the mass of matter.  The ultimate word is, however, expected from an experiment. Most think that  setting quarks free in a large vacuole created in  RHI collision laboratory experiment is offering a decisive  opportunity to test this understanding of mass of matter. The same lattice-QCD that provided the  numerical evaluation of mass of matter, provides properties of the  hot QGP Universe. 

Others go even further to  argue  nothing needs to be  confirmed: given the QCD action,  the computer  provides hadron spectrum and other static properties of hadron structure. For a recent review of  \lq\lq Lattice results concerning low-energy particle physics,\rq\rq  see Ref.\cite{Aoki:2013ldr}. That is true: the relatively good agreement  of lattice-QCD theory with  low-energy particle physics proves that QCD is the theory of strong  interactions. In fact, many textbooks  argue that  this   has already been settled 20 years ago in accelerator experiments, so a counter question could be,  why bother to do lattice-QCD to prove QCD?  One  can present as example of a new insight the argument that the mass of matter is not due to the  Higgs field~\cite{Durr:2008zz,Kronfeld:2012ym}.  

However, the mass argument is not entirely complete. The vacuole size $R$ directly relates to QCD vacuum properties -- in bag models we relate it to the bag constant ${\cal B}$ describing the vacuum pressure acting on the  vacuole.  But is this hadron energy scale ${\cal B}^{1/4}\simeq 170$\,MeV fundamental?  The  understanding of the scale of the QCD vacuum structure  has not  been  part of  the present-day lattice-QCD. In lattice-QCD work one borrows  the energy scale from an observable. In my opinion hadron vacuum scale is due to the vacuum Higgs field, and thus  the  scale of hadron masses is after all due to Higgs field; it is just that the mechanism is not acting directly.

Let me explain this point of view: By the way of top interaction with Higgs there is a relation of the Higgs with the QCD vacuum scale;\\ 
a) The intersection between QCD and  the Higgs field  is provided by the top quark, given  the remarkable value of the minimal coupling $g_t$ 
\begin{equation}
g_t\equiv \frac{m_t}{\langle h\rangle}\simeq 1\;, \quad  \alpha^t_h\equiv  \frac{g_t^2}{ 4\pi}=0.08\simeq \alpha_s(m_t)=0.1\;.
\end{equation}
Note that the same strength of  interaction: top with gluons $\alpha_s(m_t)$, and with Higgs  field  fluctuations $\alpha^t_h$.\\
b) The size of QCD vacuum fluctuations  has been estimated at 0.3 fm~\cite{DiGiacomo:2000va}. This is large compared to the top quark Compton wavelength $\lambda_t=\hbar c/m_t=1.13\times 10^{-3}$\,fm. This means that  for the  top-field the QCD vacuum looks like a quasi-static mountainous random field driving large top-field fluctuations in the QCD vacuum.
 
The  possible relation of the QCD vacuum structure   via top quark with Higgs requires much more study, I hope that this will  keep some of us  busy  in coming years.

That something still needs improvement in our  understanding of strong interactions is in fact clear: Why  i) all hadrons we know have $qqq$ and $q\bar q$ structure states, and why ii) we do not observe internal excitations of quarks in bags appearing  as hadron resonances. These two questions  show that  how we interpret QCD within the bag model is  incomplete.

I hope to have  dented somewhat the  belief that  lattice-QCD is capable of replacing the  experimental study of vacuum structure. In a nutshell, lattice  neither explains scales of vacuum structure, nor  can it address any dynamical phenomena, by necessity present in any  laboratory recreation of the early Universe QGP conditions. In addition, the QCD vacuum structure paradigm needs an experimental confirmation.

\subsection{The quantum vacuum: Einstein's \ae ther}\label{EinsteinAether1}
The quantum vacuum state determines the prevailing form of the \lq fundamental\rq\ physics laws. Within the standard model, the nature of  particles and their  interactions is determined by the transport properties of the vacuum state. As just discussed above, the mass of matter is inherent in the scale of QCD, which itself relates in a way to be studied in the future with the Higgs vacuum structure.

The existence of a structured quantum vacuum as the carrier of the laws of physics was anticipated by Lorentz, and Einstein went further seeking to reconcile this with the principles of relativity. What we call quantum vacuum,  they called   \ae ther. The concluding paragraph from a lecture by  Albert Einstein is creating   the philosophical foundation of the quantum vacuum as carrier of laws of physics (translation by  author)~\cite{sidelights}  
\begin{quote}
\ldots space is endowed with physical qualities; in this sense the \ae ther exists. According to the general theory of relativity, space without \ae ther is unthinkable: without \ae ther light could not only not propagate,  but also there could be no measuring rods and clocks, resulting in  nonexistence of  space-time  distance as a physical concept. On the other hand, this \ae ther cannot be  thought to possess properties  characteristic of ponderable matter, such as having parts  trackable in time. Motion cannot be inherent to the \ae ther.
\end{quote}
A few months earlier, in November, 1919 Einstein announced the contents of this address in a letter to Lorentz: {\em It would have been more correct if I had limited myself, in my earlier publications, to  emphasizing only the non-existence of an \ae ther velocity, instead of arguing the total non-existence of the \ae ther \ldots}   

\subsection{The quantum vacuum: Natural constants}\label{EinsteinAether2}

In the quark--gluon plasma state of matter, we fuse and dissolve nucleons in the primordial \ae ther state, different in its structure and properties from the \ae ther of our experience. In Einstein's writings quoted above the case of transition between two coexistent \ae ther states was not foreseen, but properties such as the velocity of light were seen as being defined by the  \ae ther. One should thus  ask: Is velocity of light  the same out there (vacuole)  as it is around here? Such a  question seems on first sight empty as the velocity of light connects the definition of a unit of length  with the definition of a time increment.  However,  if $\bar c$ in the vacuole is the  same as $c$, it means that time `advances' at the same rate there as it does here. This assumption is not necessary.

{\it Is it possible, both in practical and in principle terms,  using  RHI collisions  to answer if $\bar c=c$, where the bar indicates the property in the  vacuole?\/}  

We can  for example study the relation between energy  and momentum of photons produced in QGP, and  the rate at which these  processes occur. The photon emitted is defined by its wavelength $k=1/\lambda$, the energy of the photon is $\hbar c k$. This  energy is different in the vacuole from what we observe in the laboratory -- energy conservation  for the photon is not maintained since the translation symmetry  in time would need to be violated to make time tick differently in different vacuum states.  However, global energy  conservation is assured. Transition radiation, Cherenkov radiation are more mundane examples of what happens when  a superluminal photon enters a dielectric medium. Thus we will need to differentiate with what would be  called medium  effect when considering  photon  propagation across the $\bar c\ne c$. boundary. That may be difficult.

Turning now to the  rate of photon production in the  vacuole:  we keep to  gauge invariance, thus charge cannot change between two quantum  vacuum states. The way the change from the  vacuole to the normal vacuum rate will be looked at  is that we assume the space size of the vacuole to be measured in units of length evaluated in the normal vacuum.  The rate of an electromagnetic process  in modified vacuum  should be, according to the Fermi golden rule proportional to   $\bar \hbar \bar \alpha^2=e^4\bar \hbar/(\bar\hbar^2\bar c^2)$.  This  expression reminds that we also can have $\bar \hbar\ne \hbar$, but the  result will involve the product $c\hbar$ only.  The rate per unit volume and time of an electromagnetic process is   in the  vacuole with $\Delta\bar t=\Delta L/\bar c$ is
\begin{equation}
\overline{\cal W}\propto \bar \hbar \frac{\bar \alpha^2}{\Delta^3L\,\Delta \bar t}\propto  \frac{1}{\bar \hbar \bar c}\frac{1}{\Delta^4L}\;.
\end{equation} 
The  number of events we observe is $\Delta L^4 \overline{\cal W}$. The  production of direct dileptons and direct photons is thus  predicted to scale with $(\bar \hbar \bar c)^{-1}$ in a space-time volume determined in our  vacuum by  for example the HBT method.

The above consideration cannot be applied to strong  interactions  since there is no meaning to $\alpha_s$ in the  normal vacuum; we always measure $\bar \alpha_s$. Similarly, the thermal properties of the vacuole, in particular addressing the  quark energy, are intrinsic properties. The direct connection of intrinsic to external properties occurs by  electromagnetic phenomena. The practical problem in using the rate of electromagnetic processes to compare in-out $ ( \hbar  c)^{-1}$   is that all production processes depend on scattering of electrically charged quanta (quarks) in QGP, and that in turn  depends  on a high  power of $T$. This means that small changes in $\hbar  c$ could be undetectable. However, it will be quite difficult to reconcile an order of magnitude $\hbar  c$  modification by pushing  $T$ and  HBT sizes.  We hope to see such studies in the near future, where one  tries to determine for electromagnetic processes an in-medium strength  of $\alpha$ as this is  how one  would reinterpret vacuole modified physical natural constants.

\subsection{The primordial Quark Universe in Laboratory}\label{QuarkUniverseLab} 
Relativistic heavy ion (RHI) collisions recreate  the extreme temperature   conditions prevailing in the early Universe: a) dominated by QGP;  b) in the era of evolution beginning at a few $\mu$s after the big-bang;  c) lasting through the time when  QGP froze into individual hadrons at about 20-30$\,\mu$s. We record especially at the LHC experiments the  initial  matter-antimatter symmetry a nearly net-baryon-free ($B=b-\bar b\to 0$) primordial QGP\footnote{Here $b,\bar b$  denotes baryons and antibaryons,  not  bottom quarks.} . The early Universe (but not the lab experiment) evolved through the matter-antimatter annihilation leaving behind the tiny $10^{-9}$ residual matter asymmetry fraction.   

The question in which era the present day net baryon  number of the Universe originates remains unresolved. Most believe that the net baryon asymmetry is not due to an  initial condition. For baryon  number to appear in the  Universe the   three Sakharov conditions have to be fulfilled:\\ 
1)  In terms of its evolution, the Universe cannot be  in the full   equilibrium  stage; or else whatever created the asymmetry will also undo it. This requirement is generally understood to mean that the asymmetry has to originate in the period of a phase transformation, and the focus of attention has been on electro-weak symmetry restoring condition at a temperature scale $1000\times \Th$. However the  time available for the  asymmetry to arise is in this condition on the scale of $10^{-8}$\,s  and not $ 10^{-5}$\,s   or longer if the  asymmetry  is related to QGP evolution, hadronization, and/or matter-antimatter annihilation period.\\
2) During this  period interactions must be able to differentiate between matter and antimatter, or else how could the residual asymmetry  be  matter dominated? This asymmetry requires CP-nonconservation, well known to be inherent in the SM as a complex phase of the Kobayashi-Maskawa flavor mixing,\\
3) If true global excess of baryons over antibaryons is to arise there must be a baryon number conservation violating process. This seems to be a requirement on fundamental interactions  which constrains most when and  how one must look for the asymmetry  formation. It would be hard to place this in the domain of physics today accessible to experiments as no such effect has come on the horizon.

A variant model of asymmetry could be a primordial  acoustical chemical potential wave inducing  an asymmetry  in the  local distribution of quantum numbers. It has been established that at the QGP hadronization $T=\Th$ temperature  a chemical potential amplitude at the level of 0.3 eV achieves the present day baryon  to photon number in our  domain of the Universe~\cite{Fromerth:2002wb}. Constrained by local, electrical charge neutrality, and $B=L$ (local net baryon density  equal to local net lepton density), this chemical potential amplitude is about $10^{-9} $ fraction of  \Th. 

This insight sets the  scale of energy we are looking for: the absence in the SM  of any force related to baryon number and operating at the  scale of eV is what  allows us to imagine local baryon number chemical fluctuation. This `random fluctuation' resolution of the  baryon asymmetry riddle  implies that  our matter domain in the Universe borders on an antimatter domain -- however a chemical potential wave means that this boundary  is where $\mu=0$ and thus where no asymmetry is present; today presumably a space domain void of any matter or antimatter. Therefore, a change from matter to antimatter across the boundary is impossible to detect by astronomical observations -- we have to look for antimatter dust straying into local particle detectors.  One of the declared objectives of the Alpha Magnetic Spectrometer (AMS) experiment  mounted on the International Space Station (ISS) is the search for antihelium, which is considered a characteristic signature of antimatter lurking in space~\cite{AMS}.

We recall that the acoustical density oscillation of matter is one of the results of the   precision  microwave background studies which explore the conditions in the Universe at temperatures near  the scale of $T=0.25$ eV where hydrogen recombines and photons begin to free-steam. This is the begin  of observational cosmology era. Another factor 30,000 into the primordial depth of the Universe expansion, we reach the big-bang nuclear synthesis stage  occurring at the scale of  $T\simeq 10$ keV. Abundance of helium compared to hydrogen constrains significantly the  timescale of the Universe expansion and hence the present day photon to baryon ratio.  A further  factor 30,000   increase of  temperature   is needed to reach the stage at which the hadronization of quark Universe occurs at Hagedorn temperature \Th. 

We have focused here on conservation, or not, of baryon  number in the  Universe. But another topic of current  interest is if the  hot QGP fireball in its visible energy component  conserves energy; the  blunt question to ask is: {\em What if the QGP radiates darkness, that is something we cannot see?\/}\cite{Birrell:2014cja}. I will return under separate cover to discuss the QGP in the early  Universe, connecting  these different  stages. For a preliminary  report see Ref.\cite{Rafelski:2013yka}. The understanding of the quark Universe deepens profoundly the reach of our understanding of our place in this world.

\section{Melting hadrons}\label{meltHad}
Two paths towards the quark phase of matter started  in parallel in 1964-65, when on one hand quarks were introduced triggering  the first quark matter paper~\cite{Ivanenko:1965dg}, and on another, Hagedorn recognized that the yields and spectra of hadrons were governed by new physics involving \Th and he proposed the SBM~\cite{Hagedorn:1965st}. This briefly addresses the events surrounding Hagedorn discovery and the resulting modern theory of hot hadronic matter.

\subsection{The tale of distinguishable particles}
 
In early 1978 Rolf Hagedorn shared with me a copy of his unpublished manuscript `Thermodynamics of Distinguishable Particles: A Key to High-Energy Strong Interactions?', a preprint CERN-TH-483 dated 12 October 1964. He said there were two copies; I was looking at one; another was in the CERN archives.  A quick glance sufficed to reveal that this was, actually, the  work  proposing a limiting temperature and the exponential mass spectrum.  Hagedorn explained that upon discussions of the contents of his paper  with L\'eon Van Hove, he evaluated in greater detail the requirements for the hadron mass spectrum and  recognized a needed fine-tuning.  Hagedorn concluded that his result was therefore too arbitrary to publish, and in the CERN archives one finds Hagedorn commenting on this shortcoming of the paper, see Chapter 18 in Ref.\protect\cite{HagedornBook}.

However,  Hagedorn's  `Distinguishable Particles'  is  a clear stepping stone on the road to modern understanding of strong interactions and particle production. The  insights gained in this work allowed Hagedorn  to rapidly invent  the Statistical Bootstrap Model (SBM). The SBM paper `Statistical Thermodynamics of Strong Interactions at High Energies', preprint CERN-TH-520 dated 25 January 1965, took more than a year  to appear in press\footnote{Publication was in  Nuovo Cim. Suppl. {\bf 3}, pp. 147--186 (1965) actually printed in April 1966.}~\cite{Hagedorn:1965st}. 

The beginning of a new idea in physics often seems to hang on a very fine thread: was  anything  lost when `Thermodynamics of Distinguishable Particles'  remained unpublished?  And  what would Hagedorn do after withdrawing his first limiting-temperature paper? My discussion of the matter with Hagedorn suggests that his vision at the time of how limiting temperature could be justified evolved very rapidly. Presenting his more complete  insight was   what interested Hagedorn and motivated his work. Therefore, he opted to work on the more complete theoretical model, and, publish it, rather than to deal with complications that  pressing  `Thermodynamics of Distinguishable Particles'  would generate.

While the withdrawal of the old, and the preparation of an entirely new paper seemed to be the right path to properly represent the evolving scientific understanding, today's perspective is different. In particular the insight   that the appearance of a large number of different hadronic states allows to effectively side-step the quantum physics nature of particles within statistical physics   became essentially invisible in the ensuing work. Few scientists realize  that this is a  key property in the  SBM, and the fundamental cause allowing the energy content to increase without an increase in temperature.

In the SBM model, a hadron  exponential mass spectrum with the required `fine-tuned' properties is a natural outcome. The absence of Hagedorn's  `Distinguishable Particles'  preprint  delayed the recognition of the importance of the invention of  the SBM model.   The SBM paper without its prequel looked like a mathematically esoteric work; the need for  exponential  mass spectrum was not immediately evident. 

Withdrawal of \lq Distinguishable Particles\rq\ also removed from view the fact that quantum physics in hot hadronic matter loses its relevance, as not even Boltzmann's $1/n!$ factor was needed, the  exponential mass spectrum effectively removes it. Normally, the greater the density of particles, the greater  the role of quantum physics.  To the best of my knowledge the dense,  strongly interacting  hadronic gas is the only physical system where the opposite happens.  Thus surfacing briefly in Hagedorn's withdrawn `Thermodynamics of Distinguishable Particles' paper, this original finding faded from view. Hagedorn presented  a  new idea that has set up his SBM model, and  for decades this new idea remained hidden in archives.

On the other hand, the  Hagedorn limiting temperature \Th got off the  ground. Within a span of only 90 days between the withdrawal of his manuscript, and the date of his new CERN-TH preprint, Hagedorn formulated the  SBM. Its salient feature is that the exponential mass spectrum arises from the principle that hadrons are clusters comprising lighter (already clustered) hadrons\footnote{In this  paper as is common today  we refer to all discovered hadron resonance states -- Hagedorn's clusters -- as resonances, and the undiscovered \lq heavy\rq\ resonances are called Hagedorn states}.  The key point of this second paper is   a theoretical model based on the very novel idea of hadrons made of other hadrons. Such a model bypasses the need to identify constituent content of all these particles. And, Hagedorn does not need to make explicit the phenomenon of Hadron distinguishability that  clearly was not easy to swallow just 30 years after quantum statistical distributions saw the light of day.

Clustering pions into new hadrons and then combining these new hadrons  with pions, and  with already preformed clusters, and so on, turned out to be a challenging but soluble mathematical exercise. The outcome   was that the number of states of a given mass was growing exponentially. Thus, in SBM,  the exponential mass spectrum required for the limiting temperature arose naturally ab-initio. Furthermore the model established a  relation between the limiting temperature, the exponential mass spectrum slope, and the pion mass, which provides the scale of energy in the model.

Models of the  clustering type are employed in other areas of physics. An example is the use of the $\alpha$-substructure in the description of   nuclei structure: atomic nuclei are made of individual nucleons, yet improvement of the understanding is achieved if we cluster  4 nucleons (two protons and two neutrons) into an $\alpha$-particle substructure.    

The difference between the SBM and  the nuclear $\alpha$-model is that the number of input building blocks in SBM \ie\ pions and more generally of all strongly interacting clusters is not conserved but is the result of constraints such  as  available energy. As result one finds  rapidly  growing  with energy size of phase space with undetermined number  of particles. This in turn   provides  justification for the use of the grand canonical statistical methods in the description of particle physics phenomena at a time when only a few particles were observed.

\subsection{Roots and contents of the SBM}
The development of SBM in 1964/65 had a few preceding pivotal milestones, see Chapter 17 in Ref.\cite{HagedornBook} for a fully  referenced list. One should know seeing point 1. below that it was Heisenberg who hired Hagedorn as a postdoc in June 1952 to work on cosmic emulsion \lq evaporation stars\rq, and soon after in 1954 sent him on to join the process of building CERN:
\begin{enumerate}
\item The realization and a first interpretation of 
many secondaries  produced in a {\em single} hadron--hadron collision   (Heisenberg 1936~\cite{Heisenberg:1936Schauer}),
\item The concept of the compound nucleus and its thermal behavior
(1936--1937).
\item The construction of simple statistical/thermodynamical models for particle production in analogy to compound nuclei (1948--1950) (Koppe 1949~\cite{Koppe:1949zza,Koppe:1949zzb}, Fermi 1950~\cite{Fermi:1950jd}).
\item[] {\bf Enter Hagedorn:}
\item The inclusion of resonances to represent interactions recognized via phase
shifts   (Belenky 1956~\cite{Belenky:1956}).
\item The discovery of limited $\langle p_{\perp}\rangle$ (1956).
\item The discovery that fireballs exist and that a typical \pp
collision seems to produce just two of them, projectile and target fireball (1954--1958).
\item The discovery that large-angle elastic cross-sections decrease
exponentially with CM energy (1963).
\item The discovery of the  parameter-free and numerically correct
description of this exponential decrease  buried in Hagedorn's
archived Monte Carlo phase-space results obtained earlier at CERN (1963).
\end{enumerate}

Hagedorn introduced a model   based on an unlimited sequence of heavier and heavier bound and resonance states he called clusters\footnote{In the older literature Hagedorn and others initially called   decaying clusters  fireballs, this is another example of how a  physics term  is recycled in a new setting.}, each being a possible constituent of a still heavier resonance, while at the same time being itself composed of lighter ones.  The pion is the lightest `one-particle-cluster'.   Hadron resonance states {\em are due to} strong interactions; if introduced as new, independent particles in a statistical model, they {\em express} the strong interactions  to which they owe their existence. To account in full for strong interactions effects   we need all resonances; that is, we need the complete mass spectrum $\rho (m)$.

In order to obtain the   mass spectrum $\rho (m)$, we will implement in mathematical terms the self-consistent requirement that a  cluster is composed of clusters. This leads to the \lq bootstrap condition and/or bootstrap equation\rq\ for the mass spectrum $\rho (m)$. The integral  bootstrap equation (BE) can be solved analytically with the result that the mass spectrum $\rho (m)$ has to grow exponentially. Consequently, any thermodynamics employing this mass spectrum has a singular temperature \Th  generated by the asymptotic mass spectrum $\rho (m) \sim \exp (m/T_0)$. Today this singular temperature is interpreted as the temperature where (for baryon chemical potential $\muB=0$) the phase conversion of hadron gas $\longleftrightarrow$ quark--gluon plasma occurs.

\subsection{Implementation of the model}\label{SBMexample}
Let us look at a simple toy  model proposed by  Hagedorn to illustrate the Frautschi-Yellin reformulation~\cite{Frautschi:1971ij,Yellin:1973nj} of the original model which we find in comparable detail in  Ref.\cite{appenA}, also shown in Subsection~\ref{SBMdef}, \req{2chap2eq3.11}. Like in SBM, in the toy model  particle clusters are composed of clusters; however we ignore   kinetic energy. Thus
\begin{align}
\rho (m) = &\delta (m-m_0)+ \label{311}\\
  &\sum^{\infty}_{n=2}\frac{1}{n!}\int\! \delta\!
\left (m -\sum^n_{i=1}\! m_i\right )\!\prod^n_{i=1} \rho (m_i)\D m_i\;.
\notag
\end{align}
In words, the cluster with mass $m$ is either the \lq input particle\rq\ with
mass $m_0$ or else it is composed of any number of clusters of any
masses $m_i$ such that $\Sigma m_i = m$. We Laplace-transform \req{311}:
\begin{align}
\int\!\!\rho (m)\E^{-\beta m}\D m = &\E^{-\beta m_0}+\label{312}\\
&\sum^{\infty}_{n=2}\frac{1}{n!}\prod^n_{i=1} \int\!\E^{-\beta m_i}\rho (m_i)\D m_i\;.
\notag
\end{align}
Define
\begin{equation}
z(\beta )\equiv \E^{-\beta m_0}\;,\quad G(z)\equiv \int\!\E^{-\beta m}\rho (m) \D m\;.
\label{313}
\end{equation}
Thus \req{312} becomes $G(z) = z + \exp[G(z)] -G(z) -1$ or
\begin{equation}
z = 2G(z) - \E^{G(z)}+1\;,
\label{314}
\end{equation}
which provides implicitly the function $G(z)$, the Laplace transform of the mass spectrum. 

A graphic solution is obtained  drawing $z(G)$ in \rf{F5} top frame a)  and transiting in  \rf{F5} from top a) to bottom frame  b) by exchanging the axis. The parabola-like maximum of $z(G)$ implies a square root singularity of
$G(z)$ at $z_0$, first remarked  by Nahm \cite{Nahm:1972zc}  
\[
z_{\rm max}(G)\equiv z_0 = \ln 4-1 = 0.3863 \ldots\;,\  G_0 = G(z_0) = \ln 2\;.
\]
as also shown in \rf{F5}.
\begin{figure}[t]\vspace{-3mm}
\centering\resizebox{0.4\textwidth}{!}{%
\rotatebox{-90}{%
\includegraphics{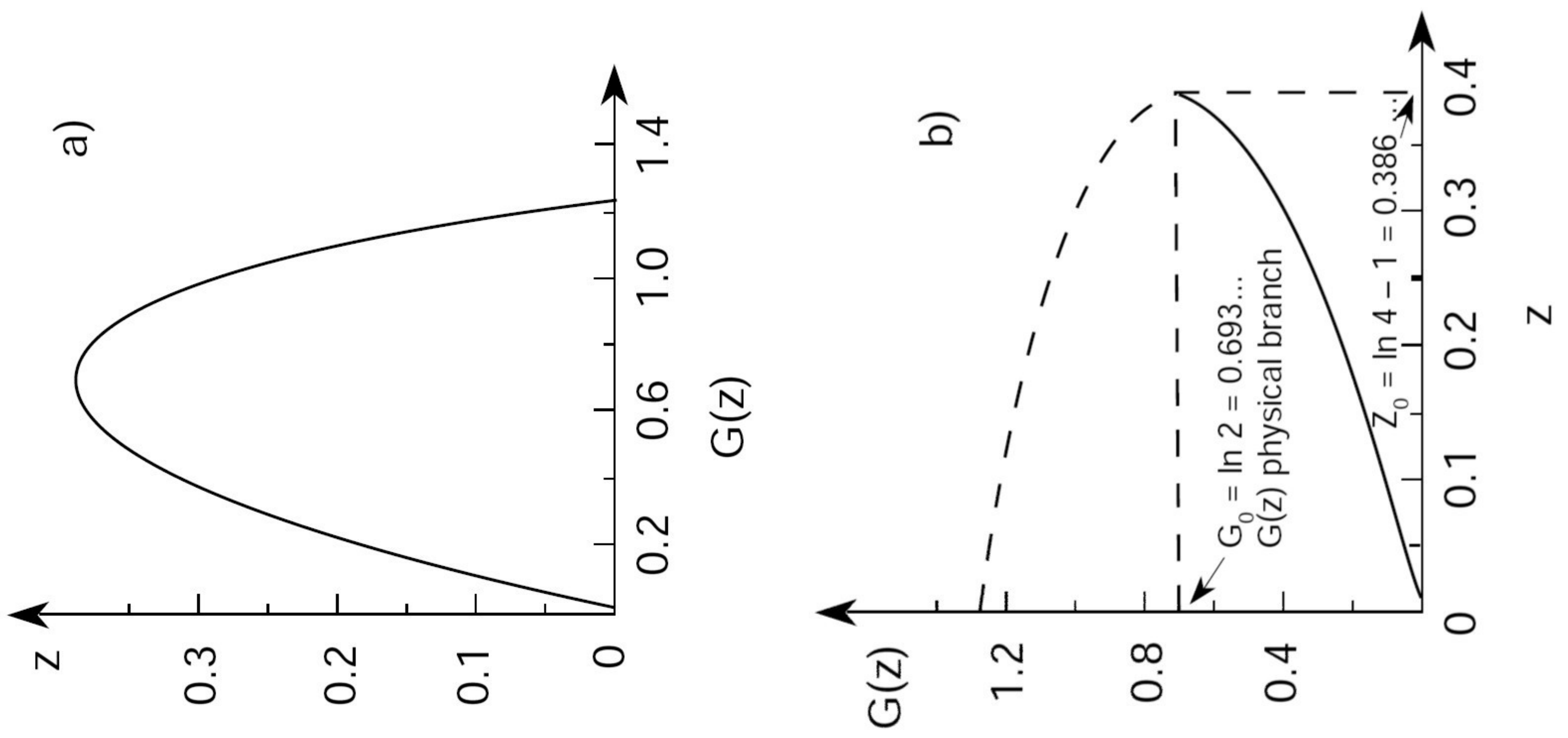}
}}
\caption[]{({\bf a}) $z (G)$ according to \req{314}. ({\bf b}) Bootstrap function $G(z)$, the
graphical solution of \req{314}. The {\it dashed line\/} represents the unphysical branch. The root singularity is at $\varphi_0=\ln(4/\E)=0.3863$}\label{F5}
\end{figure}

It is  remarkable that the \lq Laplace-transformed BE\rq\ \req{314}
is \lq universal\rq\ in the sense that it is not restricted to the above
toy model, but turns out to be the same in all (non-cutoff) realistic
SBM cases \cite{Frautschi:1971ij,Yellin:1973nj}. Moreover, it is independent of:
\begin{itemize}
\item the number of space-time dimensions \cite{Satz:1979ka}; the \lq toy model\rq\ extends this to the cae of \lq zero\rq-space dimensions,
\item the number of \lq input particles' ($z$ becomes a sum over
modified Hankel functions of input masses),
\item Abelian or non-Abelian symmetry constraints \cite{Redlich:1979bf}.
\end{itemize}

Upon inverse Laplace asymptotic transformation of the  Bootstrap function $G(z)$ one obtains
\begin{equation}\label{massStomod}
\rho (m) \sim m^{-3}\E^{ m/\Th}\;,
\end{equation}
 where in the present case  (not universally):
\begin{equation}
\Th = -\frac { m_0} {\ln z_0} = \frac{ m_0}{0.95}\;.
\label{315}
\end{equation}
Using the natural choice $m_0 = m_{\pi}$ we obtain:
\begin{equation}
\Th (\mbox{toy model}) = 145\,{\rm MeV}\;.
\label{316}
\end{equation}
The simple toy model  already yields all essential features of SBM: the  exponential mass spectrum with  $a=3$ and the  right magnitude of  \Th.

\subsection{Constituents of finite size}
The original point-volume bootstrap model was adapted   to be applicable to collisions of heavy ions where the reaction volume was relevant. This work began in 1977 and was in essence complete by  1980 \cite{Hagedorn:1978kc,Rafelski:1979cia,Hagedorn:1980kb}, see the details presented in Ref.\cite{appenA}.  The new physics is  that    cluster volumes   are introduced.

For overview of   work that followed see for example Ref.\cite{Zalewski:2015yea}. However, \lq in principle\rq\ we are today where the subject was when the initial model was completed in 1979. While many  refinements were proposed, these were in physical terms of marginal impact.  A new well-posed  question is how the van der Waals excluded volume extension of Hagedorn SBM  connects to present day lattice-QCD~\cite{Vovchenko:2014pka}, and we address this in the next Subsection.  

Before we discuss that, here follows one point of principle. Current work takes for  granted the ability to work in a context similar to non-relativistic gas including relativistic phase space. This is  not at all self-evident. To get there, see also Ref.\cite{appenA}, we argued that particle rest-frame volumes had to be proportional to particle masses. Following Touschek \cite{kkii}, we defined a \lq four-volume\rq; arbitrary  observer would attribute to each particle the 4-volume $V^{\mu}_i$ moving with particle four-velocity $u^{\mu}_i$
\begin{equation}
V^{\mu}_i = V_i u^{\mu}_i\;,\quad u^{\mu}_i\equiv \frac{p^{\mu}_i}{ m_i} \;.
\label{317}
\end{equation}
 
The entire  volume of all particles is comoving with the  four-velocity of the entire particle assembly of mass $m$
\begin{align}\label{318}
 &V^{\mu} = Vu^{\mu} \;,\quad u^{\mu}= \frac{p^{\mu}}{ m};\\ \notag
 & p^{\mu}=\sum^n_{i=1}p^{\mu}_i\;,\quad m=\sqrt{p_{\mu}p^{\mu}}
 \end{align}
We explored  a simple additive model applicable when all hadrons have the  same energy  density, see Ref.\cite{appenA}
\begin{equation}
\frac{V}{ m} = \frac{V_i}{ m_i} = Const = 4{\cal B}\;,
\label{319}
\end{equation}
where the proportionality constant is written $4{\cal B}$ in order to emphasize the similarity to MIT bags \cite{Chodos:1974je,DeGrand:1975cf,Johnson:1975zp}, which have the same mass--volume relation in absence of any other energy scale. However, in QCD two relevant scales enter in higher order: that of strange quark mass, and parameters characterizing the running of QCD parameters; the coupling constant $\alpha_s$ and mass of strange quark $m_s$ are here relevant.

Given that the assembly  of particles of mass $m$  occupies a comoving volume  $V$ and the same applies to the constituent particles and their volumes one can henceforth ignore the Lorentz covariance challenges associated with introduction of particle proper volume. However, this has been shown only  if all hadrons  have the same  energy  density, nobody extended this argument to a more general case.

In an independent consideration the energy spectrum of such SBM clusters we obtained and that of MIT bags was found to be the same \cite{Kapusta:1981ay,Kapusta:1981ue,Redlich:1982va}. This suggests   these two models are two different  views of the  same object, a snapshot taken once from the hadron side, and another time from the quark side.  MIT bags \lq consist of\rq\ quarks and gluons, SBM clusters of hadrons. This leads on to  a phase transition to connect these two aspects of the  same, as is further developed in  Ref.\cite{appenA}; the model of the phase boundary defined by MIT bags was continued by Gorenstein and collaborators, see Refs.\cite{Gorenstein:1981fa,Gorenstein:2005rc}.

\subsection{Connection with lattice-QCD}\label{connectLQCDSBM}
Today the transformation between  hadrons  and QGP is characterized within the lattice-QCD evaluation of the  thermal properties of the  quark-hadron Universe.  In the context of our introductory remarks we have addressed the close relation of the HRG with the  lattice results, see Subsection~\ref{HRGsec} and in particular \rf{P_HRG_BorsanyLQCD}. But what  does this agreement  between lattice-QCD and  HRG have to say  about  SBM of hadrons of finite size? That is an important question, it decides also the  fate of the 1979  effort described in  Ref.\cite{appenA}.

Vovchenko, Anchishkin, and Gorenstein~\cite{Vovchenko:2014pka} analyzed the  lattice-QCD for the pressure and energy density at   $T < 155,\muB=0$~MeV  within the hadron resonance gas model allowing for effects of both the excluded volume and the undiscovered part of Hagedorn mass spectrum. That work is within a specific model of finite sized hadron gas:   particles occupy a volume defined by $v=16\pi r^3/3$ where $r$ is a parameter in range $0<r<0.4$\,fm, and  it is the density of particles that characterizes the  size of excluded volume.

The shape  of their exponentially  extended mass spectrum  
\begin{equation}
\label{gorenrho}
\rho(m) = C\,\frac{\operatorname\theta(m-M_0)}{(m^{2}+m_0^2)^{a/2}}~\E^{ {m}/\Th}\;,
\end{equation}
where  the authors assumed with Hagedorn $\Th=160$~MeV, $m_0$=0.5~GeV, and placed the cutoff  at $M_0=2$~GeV. Especially the assumed   $a=5/2$ is in conflict with prior art, see Subsection~\ref{Thdef}, and the sharp cutoff leaves an unfilled \lq hole\rq\ in the intermediate mass domain. The authors report in a side remark that their results are insensitive to a change  $a=5/2\to a=3 $ with appropriate other changes but this does not resolve the above sharp cutoff matter. Note that the normalization parameter $C$ in \req{gorenrho} is the only free parameter and for $C=0$ the complement states are excluded (dashed lines in \rf{Gorenstein}), the model reverts to be HRG with finite size particle volume, but  only for $r\ne 0$, for $r=0$ we have point HRG. How this model modifies energy density $\varepsilon/T^4$  and  pressure $P/T^4$ is seen in  \rf{Gorenstein}.

\begin{figure}
\centering\resizebox{0.47\textwidth}{!}{%
\includegraphics{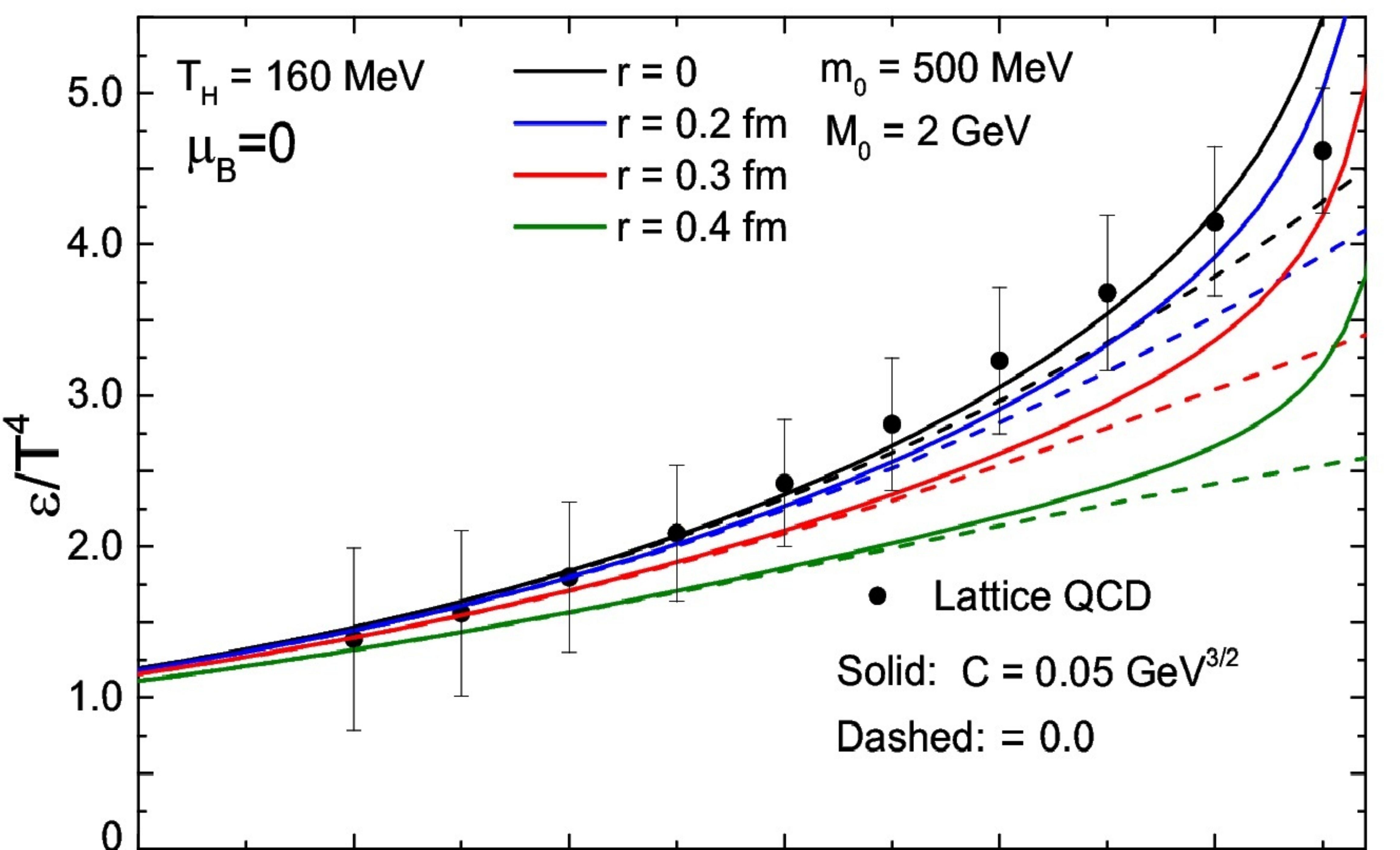}
}\\
\centering\resizebox{0.47\textwidth}{!}{%
\includegraphics{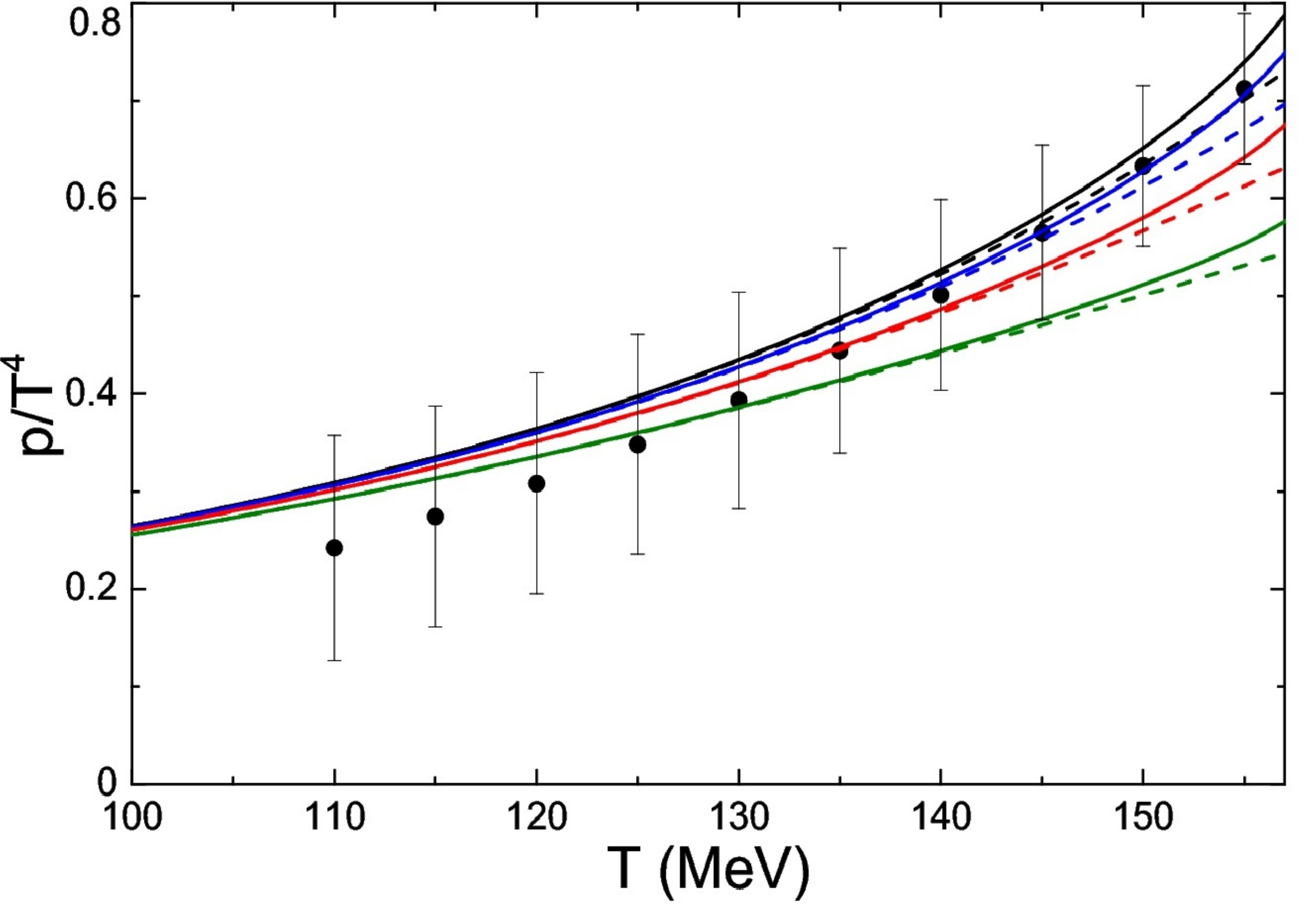}
}
\caption[]{Lattice-QCD~\cite{Borsanyi:2013bia}   energy density $\varepsilon/T^4$ and Pressure $P/T^4$ for $T < 155,\muB=0$~MeV, in RHG in a model with excluded volume parameter $r=0,0.2,.0.3,0.4$ (dashed lines) and allowing for extension of the HRG with exponential mass spectrum (solid lines) for assumed \Th=160\,MeV. See text. After Ref.\cite{Vovchenko:2014pka} }\label{Gorenstein}
\end{figure}

The authors conclude that lattice data exclude taking the two effects apart, \ie\ consideration of each of these individually.  This is so since for $C=0$  the fit of  pressure \rf{P_HRG_BorsanyLQCD} favors finite hadron volume parameter  $r\lesssim 0.4$~fm; however the best fit of energy density shows $r\cong 0$.  When both: excluded volume $r\ne 0$, and heavy resonances $C\ne 0$ are considered simultaneously the model works better: the effect of finite volume and the possibly yet undiscovered high mass Hagedorn mass spectrum thus complement each other   when considered simultaneously: the suppression effects for pressure $P/T^4$ and energy density $\varepsilon/T^4$  due to the excluded volume effects and the enhancement due to the Hagedorn mass spectrum make the data fit marginally  better as we can see in  \rf{Gorenstein}.

In effect Ref.\cite{Vovchenko:2014pka} tests in quantitative fashion the sensitivity of the  lattice-QCD results to physics interpretation: it seems that even if and when the lattice results should be a factor 5 more precise, the  correlation between the  contribution  of undiscovered states and the  van der Waals effect will compensate within error margin. 

As disappointing as these results may seem to some, it is a triumph for the physics developed in 1979. Namely, results of Ref.\cite{Vovchenko:2014pka}  also mean  that upon a reasonable choice of the energy density in  Hadrons  $4{\cal B}$,   the  model presented in  Ref.\cite{appenA} will fit well the present day  lattice-QCD data $T < 155,\muB=0$~MeV, since it has both the  correct mass spectrum, and the correct van der Waals repulsion effects due to finite hadron size. Therefore, this model is  bound to be  accurate as a function of \muB\ as well. 

Reading in  Ref.\cite{appenA}  it seems that  the perturbative QCD phase has  properties that do not match well to the SBM, requiring a strong 1st order phase transition matching to SBM. This was a result obtained with a fixed value of $\alpha_s$ in thermal-QCD. The results of lattice-QCD teach us that a more refined model with either  running  $\alpha_s$ and/or thermal quarks and gluon masses~\cite{Letessier:2002gp,Letessier:2003uj} is needed.   Contemporary investigation of the latest lattice-QCD results in  such terms is promising~\cite{Andersen:2014dua} as we already mentioned in Subsection~\ref{QGPstory}.

\section{Hadronization of QGP fireball}\label{HadronizationModel}
In this section the method and  implementation of the fireball hadronization model will be presented allowing us to address the task defined in Subsection~\ref{subsec:nonequilWhy}. This model was already described in Subsection~\ref{SHMsec} and thus we can proceed rapidly to develop the technical details.
\subsection{A large parameter set}
Our task is to describe precisely a multitude of hadrons by a relatively small set of parameters. This then allows us to characterize the drop of QGP at the time of hadronization. In our view, the key objective is to characterize the source of hadrons rather than to argue about the meaning of parameter values in a religious fashion. For this  procedure to succeed, it is necessary to allow for the greatest possible flexibility in the  characterization of the particle phase space, consistent with conservation  laws and related physical constraints at the time of QGP hadronization. For example,  the  number yield of strange and light quark pairs has to be   nearly preserved during  QGP hadronization. Such an analysis of experimental hadron yield results  requires a significant bookkeeping and fitting  effort, in  order to allow for resonances, particle widths, full decay trees and isospin multiplet sub-states. We use  SHARE (Statistical HAdronization with REsonances), a  data analysis program available   in three evolution stages for public  use~\cite{Torrieri:2004zz,Torrieri:2006xi,Petran:2013dva}.

The important parameters of the SHM, which control the relative yields of particles, are the particle specific fugacity factors ${\lambda\equiv \E^{\mu/T}}$ and the space occupancy factors $ \gamma $. The fugacity is related to particle chemical potential ${\mu} = T{\ln{\lambda}}$. $\mu$ follows a conserved quantity  and  senses the sign of \lq a charge\rq. Thus it flips sign between particles and  antiparticles. 

The resultant shape of the Fermi-Dirac distribution is seen in \req{fermiDist}. The occupancy   $  \gamma $ is in Boltzmann approximation  the ratio of produced   particles to the number of particles expected in chemical equilibrium. Since there is one quark and one antiquark in each meson, yield is proportional to  $  \gamma_q^2$ and accordingly   the baryon yield to $  \gamma_q^3$. When necessary we will distinguish the flavor of the valance quark content   $q=u,\,d,\,s,\ldots$

The occupancy parameters describing the abundance of valance quarks counted in  hadrons emerge in a complex evolution process described in  Subsection~\ref{subsec:nonequilWhy}. In general,  we expect    a nonequilibrium value $\gamma_i\neq 1$. A much simplified argument to that used in Subsection~\ref{subsec:nonequilWhy} is to  to assume that we have a completely equilibrated QGP with all quantum charges zero (baryon number, etc) and thus, in QGP all $\lambda_i=1, \gamma_i=1$. Just two parameters  describe the QGP under these conditions: temperature      $T$ and volume $V$. 

This state  hadronizes  preserving energy, and increasing or preserving entropy and essentially the  number of pairs of strange quarks.  On the hadron side  temperature      $T$ and volume $V$ would not suffice   to satisfy these  constraints, and thus  we must at least introduce $ \gamma_s>1$. The value is in general above unity because near to chemical equilibrium the QGP state contains a greater number of strange quark pairs compared to the hadron phase space.

\begin{table}
\caption{\label{tab:thermodata}Thermal parameters and their SHARE name. The values are to be presented in  units   GeV and fm$^3$, where applicable.}
\begin{center}
\begin{tabular}{p{6ex}p{9ex}p{37ex}}
\hline
Symbol & Parameter &    Parameter description\\\hline
$V$ & \texttt{norm} &   absolute normalization in fm$^3$\\
$T$ & \texttt{temp} &   chemical freeze-out temperature $T$ \\
$\lambda_q$ & \texttt{lamq}&  light quark fugacity factor\\
$\lambda_s$ & \texttt{lams}&   strangeness fugacity factor\\
$\gamma_q$ & \texttt{gamq }&    light quark phase space occupancy\\
$\gamma_s$ & \texttt{gams} &   strangeness phase space occupancy\\
$\lambda_3$ &\texttt{lmi3} &  $I_3$ fugacity factor (\req{eq:lambdas})\\
$\gamma_3$ & \texttt{gam3} &    $I_3$ phase space occupancy (\req{eq:gammas})\\\hline
$\lambda_c$ & \texttt{lamc} &     charm fugacity factor e.g. $\lambda_c=1$ \\
$N_{c+\bar c}$ & \texttt{Ncbc}&   number of $c+\bar{c}$ quarks\\
$T_c/T$ & \texttt{tc2t}&     ratio of charm to the light quark hadronization temperature \\\hline
\end{tabular}
\end{center}
\end{table}


Table~\ref{tab:thermodata} presents the here relevant parameters which must be input with their guessed values or assumed conditions, in order to run the SHARE with CHARM program as input in file  \texttt{thermo.data}.  When and if we allow $  \gamma_s$ to account for excess of strangeness content, we must also introduce  $  \gamma_q$ to account for a similar excess of QGP light quark content as already discussed in depth  in Subsection~\ref{subsec:nonequilWhy}. 

In regard to the  parameters $\gamma_q,  \gamma_s\ne 1$ we note:\\
(a) We do not know all hadronic particles,  and the incomplete hadron spectrum used in SHM can be  to some degree absorbed into values of $  \gamma_s, \tilde\gamma_q$;\\
(b) In our analysis of hadron production results we do not fit spectra but yields of particles. This is so since the dynamics of outflow of matter in an exploding fireball  is hard to control; integrated spectra (i.e., yields) are not affected by collective flow of hot matter.\\ 
(c) $\gamma_q,  \gamma_s\ne 1$  complement  $  \gamma_c,  \gamma_b$ to form a set of nonequilibrium parameters.  

Among the arguments advanced against use of chemical nonequilibrium parameters  is the urban legend that it is hard, indeed impossible, to find in the enlarged parameter space a stable fit to the hadron yield data. A large set of parameters often allows spurious local minima which cloud the physical minimum -- when  there are several fit minima,  a random search   can oscillate between such non-physics minima rendering the fit neither reproducible, nor physically relevant.

This problem is solved as follows using the  SHARE suite of programs:  we recall that SHARE   allows us to  use  any of the QGP  bulk properties to constrain fits to particle yield. In extreme, one can reverse the process: given a prescribed fireball bulk properties one can fit a   statistical parameter set, provided that the information that is introduced is sufficient. 

To find a physics best fit, what a practitioner of SHARE will do is to loosely constrain the physical bulk properties at hadronization. One speeds up considerably the convergence  by  requiring that fits satisfy some ballpark value such as $\epsilon =0.45\pm0.15$\,GeV/fm$^3$. Once a  good physics minimum is obtained, a constraint can be removed. If the minimum is very sharp, one must repeat this process recursively;  when imposing a value such as a favorite value of  freeze-out  $T$, the  convergence improvement constraint has to be adjusted. 
 
\subsection{Rapidity density yields $dN/dy$}\label{secdNdy}
In fitting the particle produced at RHIC and LHC energies we rarely have full information available about  the yields. The detectors are typically designed to either  cover the center of momentum domain (central rapidity) or  the  forward/backward \lq projectile/target\rq\ domains. Thus  practically always -- with the exception of results in SPS range of energies -- we do need to focus our analysis on particle yields emerging from a domain, typically characterized by rapidity $y$ of a particle. 

As a reminder, the rapidity of a particle $y$ replaces in set of kinematic variables the momentum component  parallel to the axis RHI  motion. For a particle of mass $m$ with momentum vector $\vec p=\vec p_\parallel+\vec p_\bot$ split into components parallel   and perpendicular   to the axis RHI  motion, the  relation is
\begin{equation}
p_\parallel=E_\bot \sinh y\;,\quad E_\bot=\sqrt{m^2+\vec p_\bot^{\,2}}
\end{equation}
which implies the useful relation $E=E_\bot\cosh y$. 

Rapidity is popular   due to the   additivity of the  value of $y$ under a change of reference frame in  $\parallel$-direction characterized by the Lorentz transformation where $\cosh y_\mathrm{L}=\gamma_\mathrm{L}, \sinh_\mathrm{L}=\beta_\mathrm{L}\gamma_\mathrm{L}$. In this restricted sense rapidity replaces velocity in the context of relativistic motion. The value of  $y$ is recognized realizing that a fireball emitting particles will have  some specific value of $y_\mathrm{ f}$ which we recognize displaying particle yields as function of rapidity, integrated with respect to $\vec p_\bot$.

The meaning of an analysis of particle  data multiplicities $dN/dy_p$ is that we look at the particles that emerged from $dV/dy_p$: in the fireball incremental volume $dV$ per unit of rapidity of emitted particles $dy_p$, 
\begin{equation}\label{dVdy}
\frac{dN}{dy} \propto \frac{dV}{dy}\equiv D_\bot(L) \frac{dL}{dy},
\end{equation}  
where $D_\bot$ is the transverse surface at hadronization  of the fireball. Considering the case of sufficiently high energy  where one  expects that  particle yields ${dN}/{dy_p}$ are flat as function of rapidity,  we can expect that $D_\bot(L)\simeq D_\bot(L=0)$ and thus ${dL}/{dy_p}=Const$, where $L=0$ corresponds to the CM-location of the hadron-hadron collision.  

The quantity ${dL}/{dy_p}$ relates to the dynamics of each of the positions $L$ from which measured particles emerge with a measured rapidity $y_p$. Each such location has its proper time $\tau$ which applies to both  the dynamics of the longitudinal volume element $dL$ and the dynamics of particle production from this volume element. We thus can write 
\begin{equation}\label{dLdy}
\frac{dL}{dy_p} \equiv \frac{{dL}/{d\tau}}{{dy_p}/{d\tau}}=
\frac{f(y_L)}{d(y_{th}+y_L)/d\tau}=Const.
\end{equation}  
In the last step we recognize the longitudinal dynamics introducing the local flow rapidity $y_L$ in the  numerator where   ${dL}/{d\tau}=f(y_L)$, and in the denominator given the additivity of rapidity we can break up the  particle rapidity into the longitudinal dynamics $y_L$ and  the  thermal component $y_{\rm th}$, describing the statistical thermal  production of particles.
We have  so obtained
\begin{equation}\label{dLdifeq0}
 f( y_L) =R\left(\frac{d y_{\rm th}}{d\tau}+\frac{d y_L}{d\tau}\right)\;.
\end{equation} 

Since we form $dN/dy$   observing many particles emitted forward ($y_{th}>0$) or backward  ($y_{th}<0$) in rapidity with respect to local rest frame, the statistical term averages out and thus we obtain as the  requirement for a flat $dN/dy$ that the local longitudinal flow satisfies  
\begin{equation}\label{dLdifeq}
 \frac{dL} {d\tau} = f(y_L) =R \frac{d y_L}{d\tau} \;,
\end{equation}  
that is a linear relation
\begin{equation}\label{Lofy}
L-L_0 =R(y_L-y_0)\;.
\end{equation}  

It is tempting to view  $ f( y_L)\equiv dL/d\tau = \sinh y_L$ as we would expect if $L$ were a coordinate of a material particle. The  implicit system of equations  allows  us then to determine the  dependence of $y_L$ and thus $L$ on $\tau$ and thus of time evolution in \req{dVdy} and the relation of $dV/dy$ with geometric (HBT) volume,  a connection that is at present not understood. This will be a topic for further study.

\subsection{Centrality classes}\label{ssec:particip}
When two atomic nuclei collide at relativistic speed, only matter in the collision path, see \rf{SpectatorFig2}, participates in the reaction.  Two fraction of nuclei are shaved of and fly by along collisions axis -- we call these nucleons spectators. 

\begin{figure}
\centering\resizebox{0.43\textwidth}{!}{%
\includegraphics{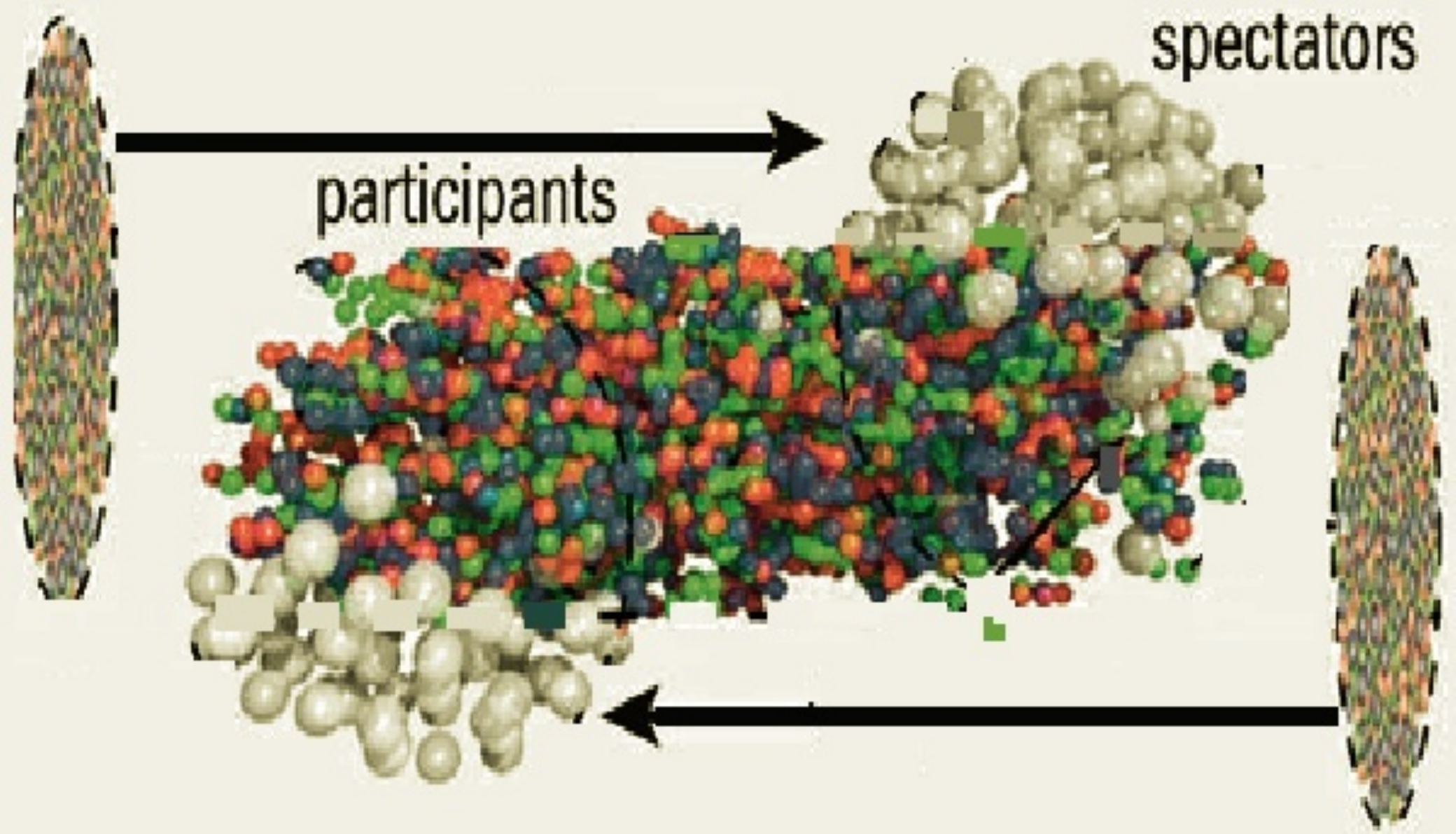}
}
\caption[]{Illustration of relativistic heavy ion collision: two Lorentz-contracted nuclei impact with offset, with some of the  nucleons  participating, and some remaining spectators, i.e. nucleons that miss the other nucleus, based on Ref.\cite{Toia:2013}}\label{SpectatorFig2}
\end{figure}

The sum of the number of participants and spectators must be exactly the number of nucleons introduced into the reaction: for Pb-Pb this number  is $2A=416$ or perhaps better said, there are $N_q=1248$ valance $u, d$-quarks, and for Au-Au we have $2A=358$ or $N_q=1074$. How many of these quarks actually have interacted in each reaction is hard to know or directly measure. One applies a \lq trigger\rq\ to accept a {\em class of collision events\/} which then is characterized in terms of some macroscopic observable relating to a nearly forward flying spectator. A numerical model connects the artificially created reaction classes with the mean number of participants $N_\mathrm{part}$  that contributed. For further details for the LHC work we refer to the recent ALICE review of their approach~\cite{Abelev:2013qoq}. 

In \rf{SpectatorFig} we see how this works. All inelastic collision classes are divided into groups related to how big a fraction of all inelastic events the trigger selects. So 0-5\% means that we are addressing the 5\% most central collisions, nearly head-on. How head-on this is we can see by considering the distribution in $N_\mathrm{part}$  one obtains in the Monte-Carlo Glauber model as shown in  \rf{SpectatorFig}.  

How do we know that  such classification, that is a characterization of events in terms of some forward observable which is model-converted into participant distribution,  is meaningful? Experimental work provides direct confirmation by connecting different observables~\cite{Abelev:2013qoq}.  I will in the analysis of other experimental results evaluate specific properties of the fireball of matter in terms of the number of participants. Some of these properties turn out to be very flat across many of the  collision classes as a function of $N_\mathrm{part}$ which entered into the discussion. This shows that the expected extensivity of the property holds: as more participants participate  the system expands accordingly. Moreover, this finding also validates the  analysis method, a point which  will be raised in due time.

\begin{figure*}
\centering\resizebox{0.7\textwidth}{!}{%
\includegraphics{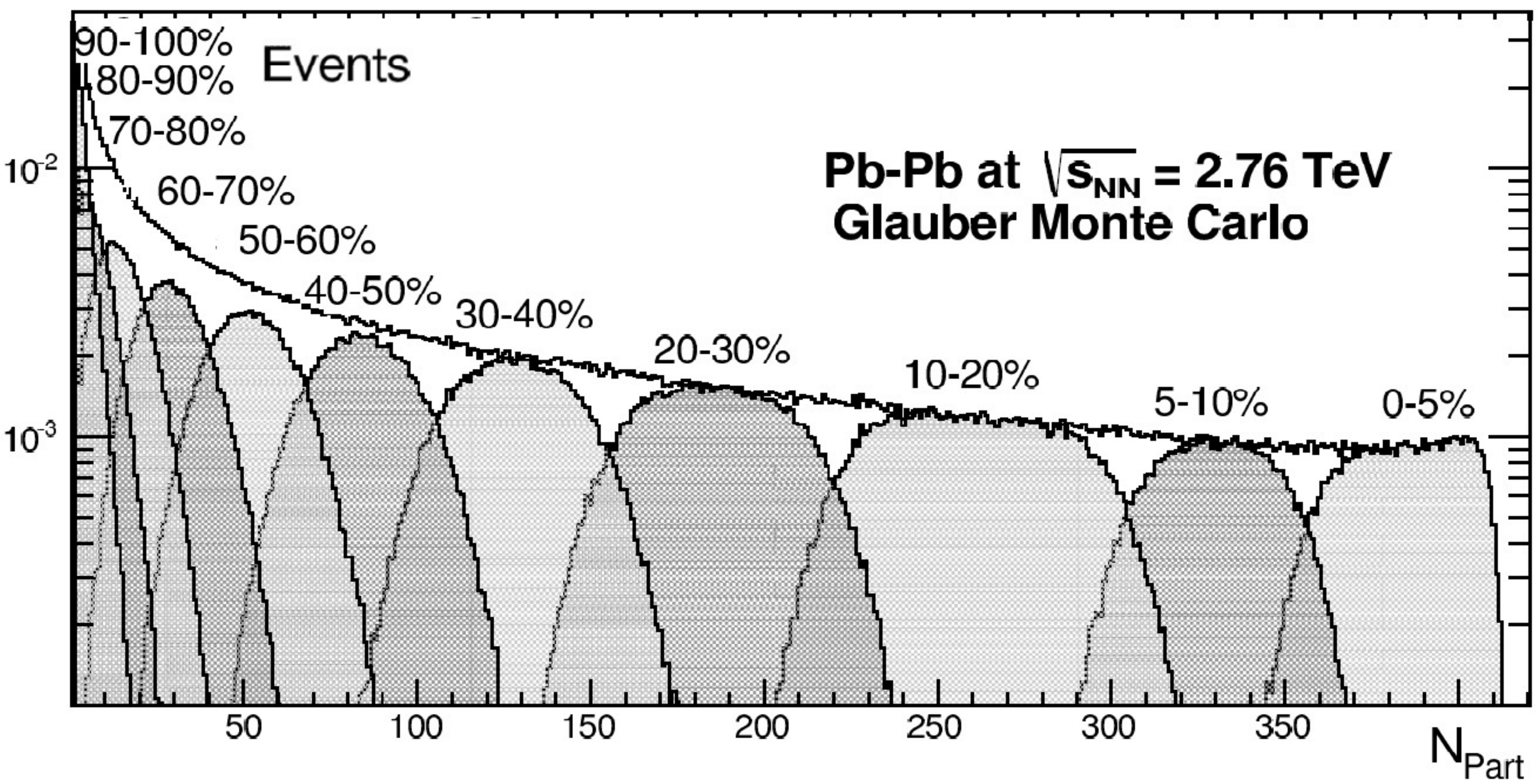}
}
\caption[]{Distribution in $N_\mathrm{part}$ for each of experimental trigger classes called a-b\% based on a MC Glauber model, data from Ref.\cite{Abelev:2013qoq}}\label{SpectatorFig}
\end{figure*}


\subsection{Particle yields and fluctuations}\label{YieldFluct}
\label{sec:yieldANDfluctuations}
For full and correct evaluation of the final hadron state in the LHC  era, one has to calculate 
\begin{enumerate}
\item Primary particle yields at chemical freeze-out,
\item Charm hadron decays for a given charm quark abundance, followed by
\item Decays of all hadron resonances.
\end{enumerate}
The point 2. is the new module that rounds of SHARE for LHC energies~\cite{Petran:2013dva}

Every hadron of species $i$ with energy $E_i=\sqrt{m_i^2+p_i^2}$ populates the energy states according to Fermi-Dirac or Bose-Einstein distribution function:
\begin{equation}
\label{eq:distribution}
n_i\equiv n_i\left( E_i \right) =\frac{1}{\Upsilon_i^{-1}\exp\left( E_i/T \right)\pm 1},
\end{equation}
where the upper sign corresponds to Fermions and the lower one to Bosons. The fugacity $\Upsilon_i$ of the $i$-th hadron species is described and reduced to the valence quark properties  in Subsection~\ref{sec:chemistry} below. Then the hadron species $i$ yield will correspond to the integral of the distribution function (Eq.\ref{eq:distribution}) over the phase space multiplied by the hadron spin degeneracy $g_i = (2J_i+1)$ and volume $V$
\begin{equation}
\label{eq:yield}
\langle N_i \rangle \equiv \langle N_i  (m_i,g_i,V,T,\Upsilon_i)\rangle  = g_i V \int\frac{\mathrm{d}^3p}{(2\pi)^3} \, n_i.
\end{equation}
The fluctuation of the yield \req{eq:yield} is:
\begin{equation}
\label{eq:fluctuation}
\left\langle (\Delta N_i)^2 \right\rangle 
= \Upsilon_i \left.\frac{\partial\langle N_i \rangle}{\partial\Upsilon_i}\right|_{T,V} = g_i V
\int \frac{\mathrm{d}^3p}{(2\pi)^3} \, n_i  \left( 1 \mp n_i  \right).
\end{equation}
It is more practical for numerical computation to express the above yields and fluctuations  as an expansion in modified Bessel functions  
\begin{align}
\langle N_i \rangle &= \frac{g_i V T^3}{2\pi^2}\sum\limits_{n=1}^\infty\frac{(\pm 1)^{n-1}\Upsilon_i^n}{n^3}{W}\left(\frac{nm_i}{T}\right),\label{eq:yieldexpansion} \\
\left\langle (\Delta N_i)^2 \right\rangle &= \frac{g_i V T^3}{2\pi^2}\!\sum\limits_{n=1}^\infty\frac{(\pm 1)^{n-1}\Upsilon_i^n}{n^3}\binom{2\!+\!n\!-\!1}{n}{W}\!\left(\!\frac{nm_i}{T}\!\right)\;,\label{eq:fluctuationexpansion}\\
 W(x)&\equiv x^2 K_2(x)\;.\label{eq:Wfunct}
\end{align}
These expansions can be calculated to any desired accuracy; for Bosons convergence requires  $\Upsilon_i\exp(-m_i/T)<1$, otherwise  the expansion makes no sense. For heavy ($m \gg T$) particles, such as charm hadrons, the Boltzmann distribution is a good approximation, i.e., it is sufficient to evaluate the first term of the expansion in \req{eq:yieldexpansion}, which is indeed implemented in the CHARM module of SHARE to reduce computation time at no observable loss of precision.

To evaluate the yield of hadron resonance with finite width $\Gamma_i$, one has to weigh the yield (Eq.\ref{eq:yield}) by the resonance mass using the Breit-Wigner distribution:
\begin{align}
\label{eq:yieldwithwidth1}
 \langle \tilde{N}_i^\Gamma \rangle &=\! \int\! \mathrm{d}M\, 
\frac{ \Gamma_i}{2\pi}\dfrac{\langle N_i(M,g_i,T,V,\Upsilon_i) \rangle  }{(M-m_i)^2+\Gamma_i^2/4}\\ \notag
  &\longrightarrow \langle N_i \rangle \text{ for } \Gamma_i\rightarrow 0.
\end{align}
For   low energy states with a large width one has to use the energy dependent resonance width, since an energy independent width implies a way  too large  probability of the resonance being formed with unrealistically small mass. 
The partial width of a decay channel $i\to j$ can be well approximated by 
\begin{equation}
\label{eq:partialwidth}
\Gamma_{i \to j}(M)=b_{i\to j}\Gamma_i \left[1-\left(\frac{m_{ij}}{M}\right)^2 \right]^{l_{ij}+1/2}
\  \text{ for } M>m_{ij},
\end{equation}
where $b_{i\to j}$ is the decay channel branching ratio, $m_{ij}$ is the decay threshold (i.e., sum of the decay product masses) and $l_{ij}$ is the angular momentum released in the decay. The total energy dependent width $\Gamma_i(M)$ is obtained using the partial widths \req{eq:partialwidth}  for all decay channels of the resonance in question as
\begin{equation}
\Gamma_i(M) = \sum\limits_j\Gamma_{i\to j}(M).
\end{equation}
For a resonance with a finite width, we can then replace \req{eq:yieldwithwidth1} by
\begin{equation}
\label{eq:yieldwithwidth2}
\langle N_i^\Gamma \rangle = \!\sum\limits_j\!\int\limits_{m_{ij}}^\infty \mathrm{d}M\,  \frac{\Gamma_{i\to j}(M)}{A_i}\dfrac{\langle N_i(M,g_i,T,V,\Upsilon_i) \rangle}{(M-m_i)^2+\Gamma_i(M)^2/4},
\end{equation}
where $A_i$ is a normalization constant 
\begin{equation}
A_i = \sum\limits_j\int\limits_{m_{ij}}^\infty \mathrm{d}M\,\dfrac{\Gamma_{i\to j}(M)}{(M-m_i)^2+\Gamma_i(M)^2/4}.
\end{equation}
Equation\,(\ref{eq:yieldwithwidth2}) is the form used in the program to evaluate hadron resonance yield, whenever calculation with finite width is required. Note that yield evaluation with finite width is implemented only for hadrons with no charm constituent quark; zero width($\Gamma_{c_i} = 0$) is used for all charm hadrons.

\subsection{Hadron fugacity $\Upsilon_i$ and  quark chemistry}
\label{sec:chemistry}
The fugacity of hadron states defines the yields of different hadrons based on their quark content. It can be calculated from the individual constituent quark fugacities. In the most general case, for a hadron consisting of $N_u^i, N_d^i ,N_s^i$ and $N_c^i$ up, down, strange and charm 
quarks respectively and $N_{\bar{u}}^i,N_{\bar{d}}^i,N_{\bar{s}}^i$ and $N_{\bar{c}}^i$ anti-quarks, the fugacity can be expressed as
\begin{align}
\label{eq:fugacity}
\Upsilon_i = &(\lambda_u\gamma_u)^{N_u^i}(\lambda_d\gamma_d)^{N_d^i}(\lambda_s\gamma_s)^{N_s^i}(\lambda_c\gamma_c)^{N_c^i}\times \\ \notag
&(\lambda_{\bar{u}}\gamma_{\bar{u}})^{N_{\bar{u}}^i}(\lambda_{\bar{d}}\gamma_{\bar{d}})^{N_{\bar{d}}^i}(\lambda_{\bar{s}}\gamma_{\bar{s}})^{N_{\bar{s}}^i}(\lambda_{\bar{c}}\gamma_{\bar{c}})^{N_{\bar{c}}^i},
\end{align}
where $\gamma_f$ is the phase space occupancy of flavor $f$ and $\lambda_f$ is the fugacity factor of flavor $f$. Note that we allow for non-integer quark content to account for states like $\eta$ meson, which is implemented as $\eta=0.55(u{\bar{u}}+d{\bar{d}}) + 0.45s{\bar{s}}$ in agreement with~\cite{Li:2007xf}.
It can be shown that for quarks and anti-quarks of the same flavor
\begin{equation}
\gamma_f = \gamma_{\bar{f}}\qquad\text{ and }\qquad \lambda_f = \lambda_{\bar{f}}^{-1},
\end{equation}
which reduces the number of variables necessary to evaluate the fugacity by  half.

It is a common practice to take advantage of the isospin symmetry and to treat the two lightest quarks ($q = u,d$) using light quark and isospin phase space occupancy and fugacity factors which are obtained via a transformation of parameters:
\begin{equation}
\gamma_q = \sqrt{\gamma_u\gamma_d},\qquad \gamma_3=\sqrt{\frac{\gamma_u}{\gamma_d}}\label{eq:gammas},
\end{equation}
with straightforward backwards transformation
\begin{equation}
\gamma_u = \gamma_q\gamma_3,\qquad \gamma_d = \gamma_q/\gamma_3,
\end{equation}
and similarly for the fugacity factors
\begin{align}
\lambda_q = \sqrt{\lambda_u\lambda_d},\qquad \lambda_3=\sqrt{\frac{\lambda_u}{\lambda_d}}\label{eq:lambdas},\\
\lambda_u = \lambda_q\lambda_3,\qquad \lambda_d = \lambda_q/\lambda_3.
\end{align}

Chemical potentials are closely related to fugacity; one can express an associated chemical potential $\mu_i$ for each hadron species $i$ via
\begin{equation}
\Upsilon_i = e^{\mu_i/T}.
\end{equation}
It is more common to express chemical potentials related to conserved quantum numbers of the system, such as baryon number $B$,  strangeness $s$, third component of isospin $I_3$ and charm $c$: 
\begin{align}
\mu_B &= 3T \log \lambda_q\;, \label{eq:mub}\\
\mu_S &= T \log \lambda_q/\lambda_s\;,\label{eq:mus}\\
\mu_{I_3} &= T \log \lambda_3\;, \label{eq:mu3}\\
\mu_C &= T \log \lambda_c\lambda_q\;.\label{eq:muc} 
\end{align}
Notice the inverse, compared to intuitive definition of $\mu_S$, which has a historical origin and is a source of frequent mistakes.

\subsection{Resonance decays}
\label{sec:decays}
The hadron yields observed include the post-hadronization decays of in general free streaming hadron states -- only a  few are stable enough to reach detectors. In fact heavier   resonances decay rapidly after the freeze-out and feed lighter resonances and `stable' particle yields. The final stable particle yields are obtained by allowing all resonances to decay sequentially from the heaviest to the lightest and thus correctly accounting for resonance cascades. 

The observable yield of each hadron $i$ including into the study the resonances populated by more massive resonances, is then a combination of primary production and feed from resonance decays
\begin{equation}
\label{eq:decayfeed}
\langle N_i \rangle = \langle N_i \rangle_\mathrm{primary} + \sum\limits_{j\neq i}B_{j\to i}\langle N_j \rangle,
\end{equation}
where $B_{j\to i}$ is the probability (branching ratio) that particle $j$ will decay into particle $i$. Applied recursively, \req{eq:decayfeed}  generated the  model result that corresponds to the experimentally observable yields of all hadrons, \lq stable\rq\ and unstable resonances, which are often of interest.

The SHARE program includes for non-charm hadrons  all decay channels with branching ratio $\geq 10^{-2}$  in data tables. To attain the parallel level of precision for the higher number of charm hadron decays (a few hundred(!) in some cases) with small branching ratios required to set the acceptance  for decay channels at a branching ratio $\geq 10^{-4}$. Since charm hadrons in many cases decay into more than three particles, a more complex  approach in implementing them had to be used~\cite{Petran:2013dva}. 

There is still a lot of uncertainty regarding charm decay channels. Some of them are experimentally difficult to confirm, but required and had to be estimated based on symmetries. For example, a measured $\Lambda_c^+$ decay channel 
\begin{equation}
\label{eq:LambdaCdecay}
\Lambda_c^+ \to p\overline{K}^0\pi^0 \qquad (3.3\pm1.0)\%,
\end{equation}
is complemented by the unobserved isospin symmetric channel 
\begin{equation}
\label{eq:LambdaCdecaySymmetric}
\Lambda_c^+ \to n\overline{K}^0\pi^+ \qquad (3.3\pm1.0)\%,
\end{equation}
with the same branching ratio.

The influence of resonance feed-down on fluctuations is the following:
\begin{equation}
\label{eq:decayfluct}
\langle (\Delta N_{j\to i})^2 \rangle = B_{j\to i}(\mathcal{N}_{j\to i} - B_{j\to i})\langle N_j \rangle \,+\, B_{j\to i}^2\langle(\Delta N_j)^2\rangle.
\end{equation}
The first term corresponds to the fluctuations of the mother particle $j$, which decays into particle $i$ with branching ratio $B_{j\to i}$. $\mathcal{N}_{j\to i}$ is the number of particles $i$ produced in the decay of $i$ (inclusive production) so that $\sum_iB_{j\to i} = \mathcal{N}_{j\to i}$. For nearly all decays of almost all resonances $\mathcal{N}_{j\to i}=1$, however, there are significant exceptions to this, including the production of multiple $\pi^0$, such as $\eta\to 3\pi^0$. The second term in \req{eq:decayfluct} corresponds to the fluctuation in the yield of the mother particle (resonance).

\section{Hadrons from QGP: What do we learn?}\label{AnalysisHadronization}
A comparison of lattice results with freeze-out conditions were shown in \rf{fig:phasediagram}.  The  band   near to the temperature axis displays the lattice estimate for \Th presented in Ref.~\cite{Borsanyi:2012rr},  $\Th=147\pm5\,\mathrm{MeV}$. As  \rf{fig:phasediagram} demonstrates, many SHM are in more or less severe conflict with this value of \Th. The model SHARE we detailed in previous Section~\ref{HadronizationModel}  is, however, in excellent agreement. One of the   reasons to write this review is to highlight how the change in understanding of \Th impacts the resultant choice that emerges in terms of  SHM applicability.

The SHARE toolbox permits a complete analysis of any sufficiently large family set of particle yields that is consistently presented in terms of a given reaction energy and  participant number  class  $N_\mathrm{part}$. Especially  as a function of $N_\mathrm{part}$ this is not always the case, whence some interpolation of data is a part of the analysis. We do not discuss this practical issue further here. The material selected for presentation  is not comprehensive and it is only representative of the work manifestly consistent with \rf{fig:phasediagram}.

Another criterion that we use is to focus on particle yields only.  Doing this, we need to mention upfront the work of Begun, Florkowski and Rybczynski~\cite{Begun:2013nga,Begun:2014rsa} which applies the  same nonequilibrium methods in an ambitious effort to describe all LHC particle spectra and does this with good success.  These results are directly  relevant to our study of LHC data presenting complementing information that  confirms our statistical parameter determination.

We  will also show, by an example, some of  the issues that have affected the SHM analysis carried out by  another group.
\subsection{Hadron source bulk properties before LHC}\label{subsec:Bulk}
Among the important features built into the SHARE program is the capability to fully describe the properties of the fireball that produces the particles analyzed. This is not done  in terms   of  produced particles:   each  carries away `content', such as the energy of the fireball.  We evaluate and sum  all fractional contributions to the fireball bulk properties from the observed and, importantly, unobserved particles, predicted by the fit in their  abundance. The energy content is only thermal, as we eliminate using  yields the  effect of expansion flow on the  spectra, \ie\ the dynamical collective flow energy of matter. Thus  the energy content we compute is the  \lq comoving\rq\ total thermal energy.

Given the large set of  parameters that SHARE makes available we fit all particles well and thus the physical properties that  we report are rather precise  images  of the  observed particle  yields. The question what the SHM parameters mean does not enter the discussion at all. If  a measurement  error has crept in then our results would look anomalous when inspected as a function of  collision energy or  collision centrality. 

The fit of SHM parameters then  provides an extrapolation from the  measured particle abundances to unmeasured yields of all particles known and listed in SHARE tables. Most of these are of no great individual relevance, being too massive. The bulk properties we report here are, for the most part, defined by particles directly observed. We expect smooth lines describing the fireball properties as a function  of $\sqrt{s_\mathrm{NN}}$, the CM energy per pair of nucleons or/and as function of collision centrality class $N_\mathrm{npart}$. Appearance of discontinuous behavior as a function of $\sqrt{s_\mathrm{NN}}$ can indicate a change related to QGP formation.

In \rf{EdepBulk} we see in the SPS and RHIC energy domain for most head-on collisions, from top to bottom, the   pressure  $P$, energy density $\varepsilon$, the entropy density $\sigma$, and  the net baryon density $\rho_{\mathrm{B}-\bar{\mathrm{B}}}$.  SPS and RHIC $4\pi$ data were used, for RHIC range also results obtained fitting $dN/dy$ are shown by the dashed line, particles originating in a volume   $dV/dy$, $y\in \{-0.5,0.5\}$. Only for the  baryon density can we recognize a serious difference; the  baryon density in the central rapidity region seems to be  a factor 5 below average  baryon density. Not shown is the change in the fitted volume, which is the one changing quantity (aside of $\rho_{\mathrm{B}-\bar{\mathrm{B}}}$). Volume grows to accommodate the  rapid rise in particle multiplicity with the available energy.

\begin{figure} 
\centering\resizebox{0.40\textwidth}{!}{%
\includegraphics{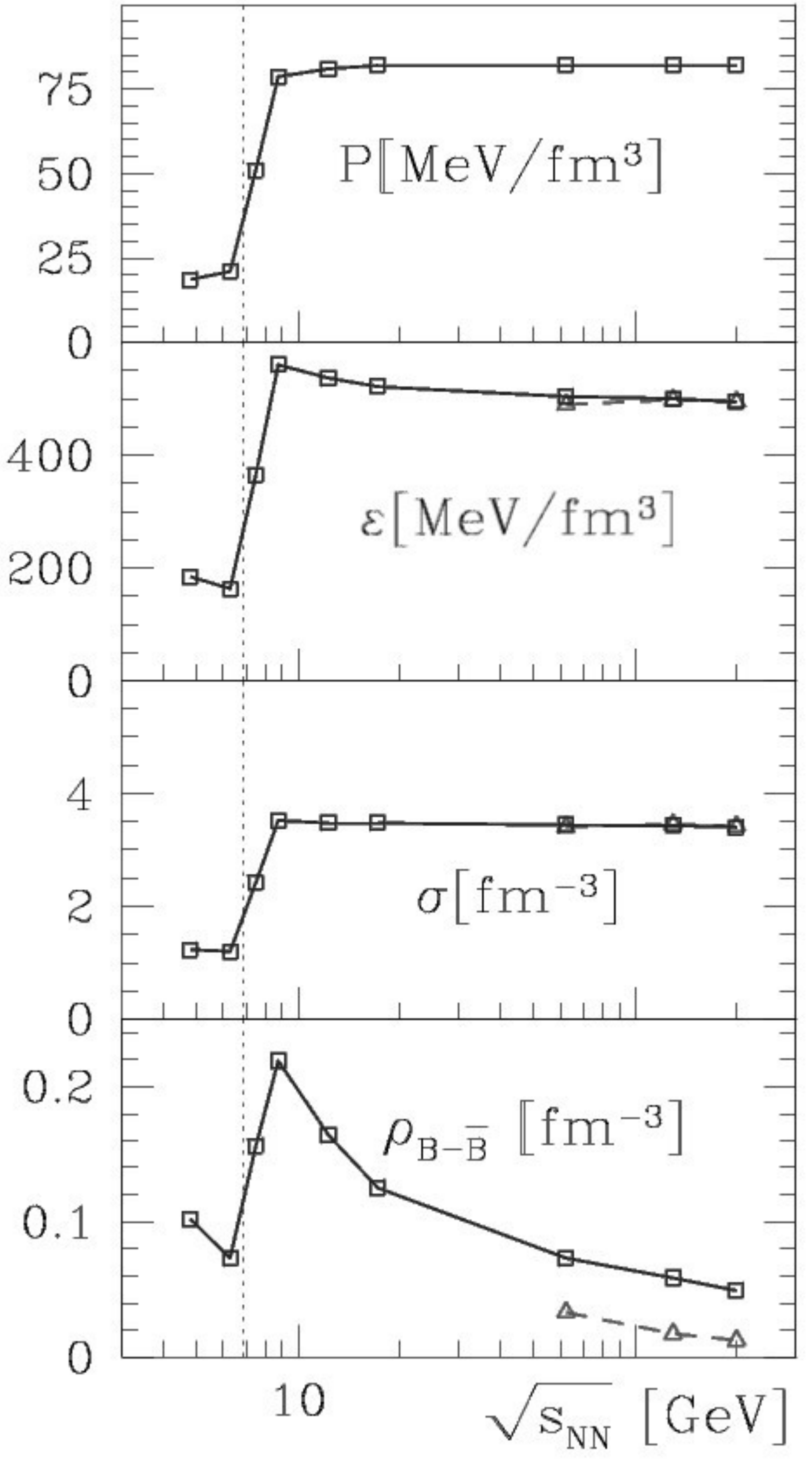}
}
\caption{\label{EdepBulk} 
Fireball bulk properties  in the SPS and RHIC energy domain, see text. Update of results published in Ref.\cite{Letessier:2005qe}.}
\end{figure}

Figure~\ref{EdepBulk} shows exciting features worth further  discussion. There can be no doubt that over a relatively small domain of collision energy -- in laboratory  frame, between   20 and 40 $A$\,GeV (SPS projectile per nucleon energies) and in CM frame  $\sqrt{s_\mathrm{NN}}\in \{6.5,7.5\}$ GeV  per nucleon pair -- the properties of the fireball change entirely. {\em Is this a signal of the onset of new physics? And if so why, is this happening at this energy?\/} Though this experimental result has been recognized  for nearly 10  years   now, Ref.\cite{Letessier:2005qe} and private communication by  M. Ga\'zdzicki, I have  no  clear answer to offer to these simple questions. 

We find a peak in the net baryon density, bottom frame of the   \rf{EdepBulk}. The $K^+/\pi^+$ peaking,  \rf{fig:onset},  discussed in Subsection~\ref{HornThreshhold} seems to be related to the effect of baryon stopping, perhaps a rise as function of  $\sqrt{s_\mathrm{NN}}$ in stopping power at first, when color bonds are broken, and a more gradual decline with increasing energy. But what makes quarks  stop just then? And why do they decide to stop less at higher energy, instead `shooting through'?  Note that a possible  argument that a decrease in baryon density is due to volume growth  is not right considering that  the thermal energy density $\epsilon$, and the entropy density $\sigma$  remain constant above the threshold in collision energy. 

I would   argue that when first color bonds are melted, gluons are stopped while quarks are more likely to run out. That  would agree with our finding in context of strangeness production, see Ref.\cite{appenB}, that  despite similar looking matrix elements   in perturbative QCD, gluons are much more effective in making things happen due to their \lq high\rq\ adjoint representation color charge; the best analogy would be  to say that  gluons have  double-color charge. The  high gluon  density  at first manages to stop some quarks but the probability decreases with increasing energy. It is remarkable how fast the  dimensionless $\rho_{\mathrm{B}-\bar{\mathrm{B}}}/\sigma\equiv b/S$ drops. This expresses the  ability  to stop quarks normalized to the ability  to produce entropy.

Seeing all these results, one cannot but ask what the  total  abundance of strange quark pairs will do. Before the discussion of results seen in \rf{EdepStrange}  it is wise to read the conclusions in Ref.\cite{appenB}   where in 1983 the overall strangeness yield enhancement alone was not predicted to be  a striking signature. In \rf{EdepStrange}  ratios are shown, in the top frame: the pair strangeness abundance $s$ per net baryon abundance $b$;  per entropy in the middle frame; and in the bottom frame we see the energy cost in GeV to make a strange quark pair, $E/s$; mind you that this  energy is the final state thermal fireball energy.

\begin{figure} 
\centering\resizebox{0.42\textwidth}{!}{%
\includegraphics{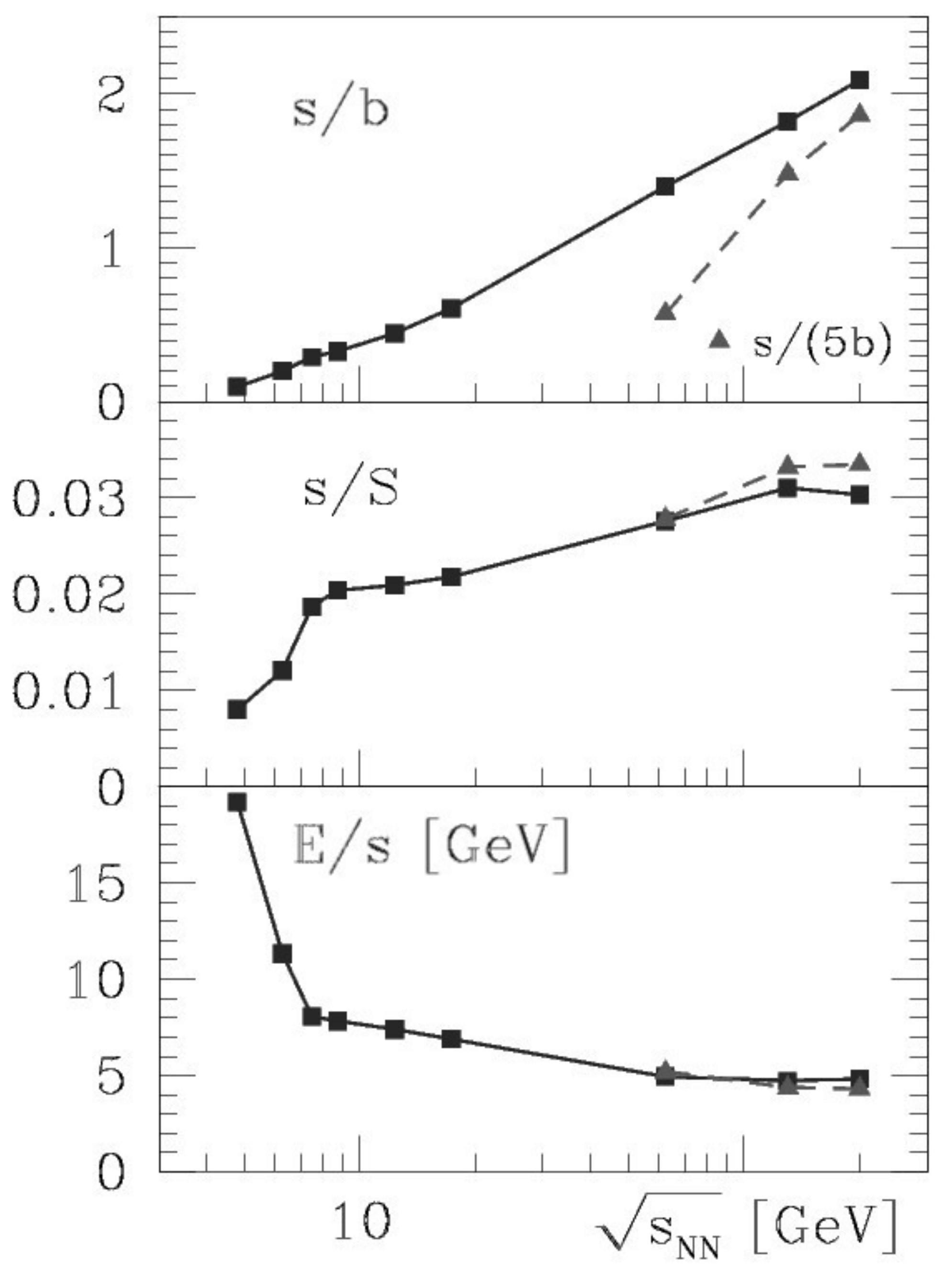}
}
\caption{\label{EdepStrange} 
Strangeness pair production $s$ yield from SPS and RHIC as a function $\sqrt{s_\mathrm{NN}}$: yield normalized by  net baryon abundance $b$ in top frame, entropy $S$ in middle frame. At bottom the energy cost to  produce  strangeness. Total particle yields, except for $dN/dy$ results   shown as dashed lines in the RHIC energy range. Update of results published in Ref.\cite{Letessier:2005qe}}
\end{figure}

We see in \rf{EdepStrange} that the $s/b$ ratio is smooth. This means that strangeness production takes off where baryon stopping takes off, being in the QGP attributed range of $\sqrt{s_\mathrm{NN}}$ faster than rise in entropy  production. And, the energy  cost of a pair seems to be very low at high energy: only  5--6 times the energy that the pair actually carries by itself, and  this factor reflects accurately on how abundant strangeness is in comparison to all the other  constituents of the  QGP fireball. This by itself clearly indicates that the yield converges to chemical QGP equilibrium. The clear break in the  cost of making  a strange quark pair near 30 GeV  energy shows the threshold above which strangeness, as compared to other  components, becomes an equal fireball partner.

Our analysis thus shows: (a) There is an onset of baryon transparency and entropy production at a very narrowly defined collision energy range. (b) Beyond this threshold in  collision energy  the hadronization proceeds more effectively into strange antibaryons. (c) The universality of hadronization source properties, such as energy density, or entropy density above the same energy threshold, suggest as explanation that a new phase of matter hadronizes. 

There is little doubt considering these cornerstone analysis results  that at SPS at and above the projectile energy of 30 $A$\,GeV  we produced a rapidly evaporating (hadronizing) drop of QGP. The analysis results we presented for the properties of the  fireball leave very  little space for other  interpretation. The properties of the  QGP fireball created in the energy range of  30--156 $A$ GeV
Pb--Pb collisions at CERN are just the same as those  obtained for RHIC beam energy scan, see end of Subsection~\ref{HornThreshhold}.

\subsection{LHC SHM analysis}\label{subsec:BulkLHC}
We consider now LHC results obtained at  $\sqrt{s_{NN}}=2760$ GeV as a function of participant number $N_\mathrm{part}$, Section~\ref{ssec:particip} and compare   with an earlier similar analysis  of  STAR results available at $\sqrt{s_\mathrm{NN}}= 62$ GeV~\cite{Petran:2011aa}. In comparison, there is a nearly a factor 50 difference in collision energy. The results presented here  for  LHC are from the ALICE experiment as analyzed in Refs.\cite{Petran:2013lja,Petran:2013qla,Petran:2013dea}. The experimental data inputs  were discussed extensively  in these references, the data source includes Refs.\cite{Abelev:2012wca,ABELEV:2013zaa,Abelev:2012vx,Abelev:2013xaa}.
The  analysis of hadron production as a function of participant number $N_\mathrm{part}$ at RHIC and  LHC proceeds in essentially the same way as already described. The results here presented were obtained without the contribution of charmed hadrons.

Given the  large set of available SHARE parameters  all particles are described very well, a non-complete example of the data included is seen in \rf{fig:LHCparticles}. Note that the central rapidity yields are divided by $N_\mathrm{part}/2$; that is they are per nucleon pair as in \pp collisions. This also means that our fit spans a range of a yield of $dN_{\uOmega}\simeq  10^{-4}$ for the most peripheral collisions to $dN_{\upi}\simeq 2000$ for the most central collisions, thus  more than 7 orders of magnitude alone of particles shown in \rf{fig:LHCparticles}.

\begin{figure}
\centering\resizebox{0.43\textwidth}{!}{%
\includegraphics{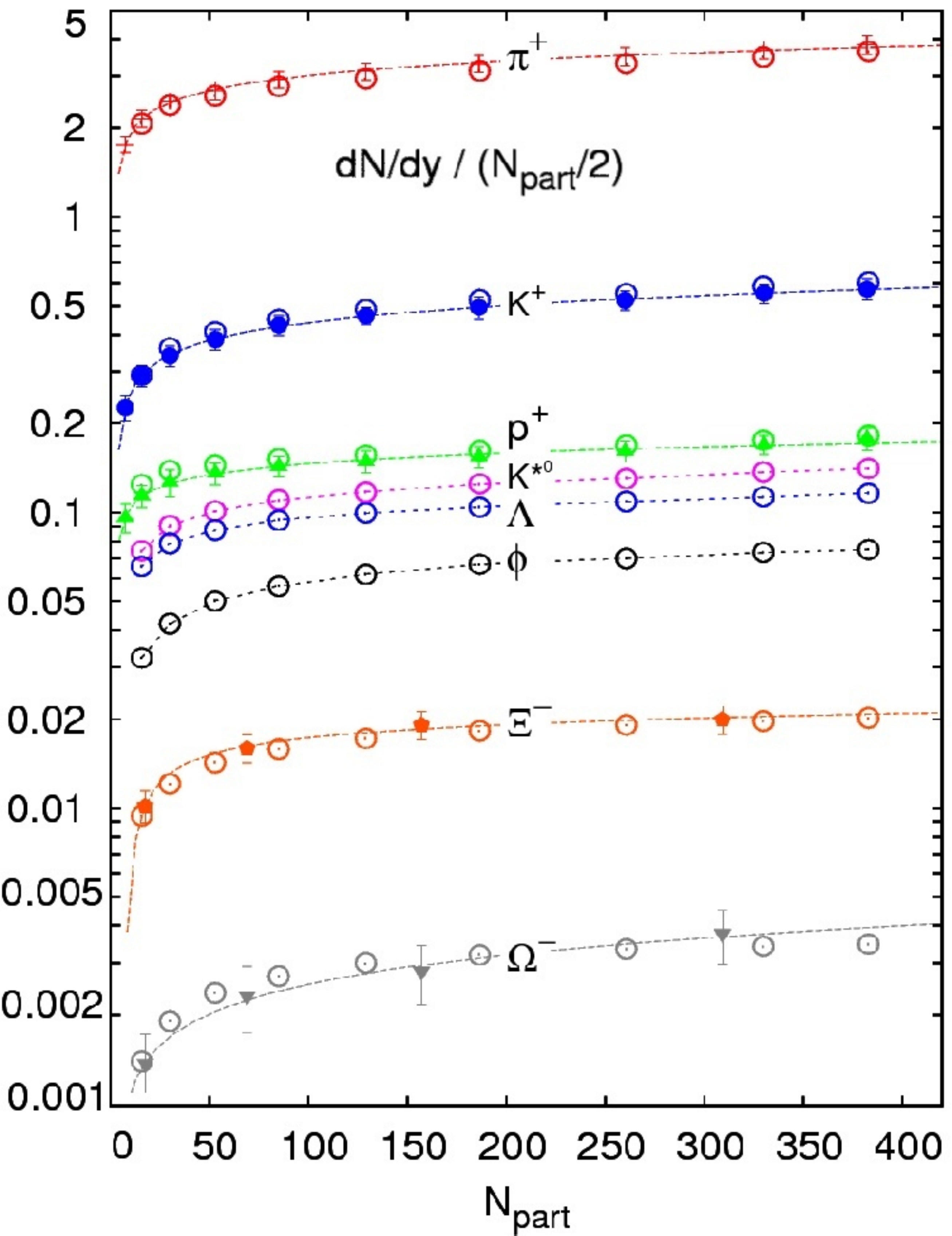}
}
\caption{\label{fig:LHCparticles}  LHC experimental data measured by the ALICE  experiment  in Pb--Pb collisions at $\sqrt{s_{NN}}=2.76$ TeV  as function of centrality described by $N_\mathrm{part}$, normalized by   $N_\mathrm{part}/2$. Results  adapted from  Refs.\cite{Petran:2013lja,Petran:2013qla,Petran:2013dea}}
\end{figure}

About three orders of magnitude of the large   range of yields $dN/dy$ that are fitted are absorbed  into the rapidly  changing volume $dV/dy$ from which these particles emerge, see \rf{fig:LHCRHICvolumespec}. Note that this result is already reduced by the  factor  $N_\mathrm{part}/2$; thus this is volume per colliding nucleon pair. For RHIC we see that this is a rather  constant value to which the  LHC results seem to converge for small value of  $N_\mathrm{part}$. However for large $N_\mathrm{part}$  at LHC the specific volume keeps growing. Keep in mind that the interpretation of $dV/dy$ is  difficult and a priori is  not geometric, see Subsection~\ref{secdNdy}.

\begin{figure}
\centering\resizebox{0.40\textwidth}{!}{%
\includegraphics{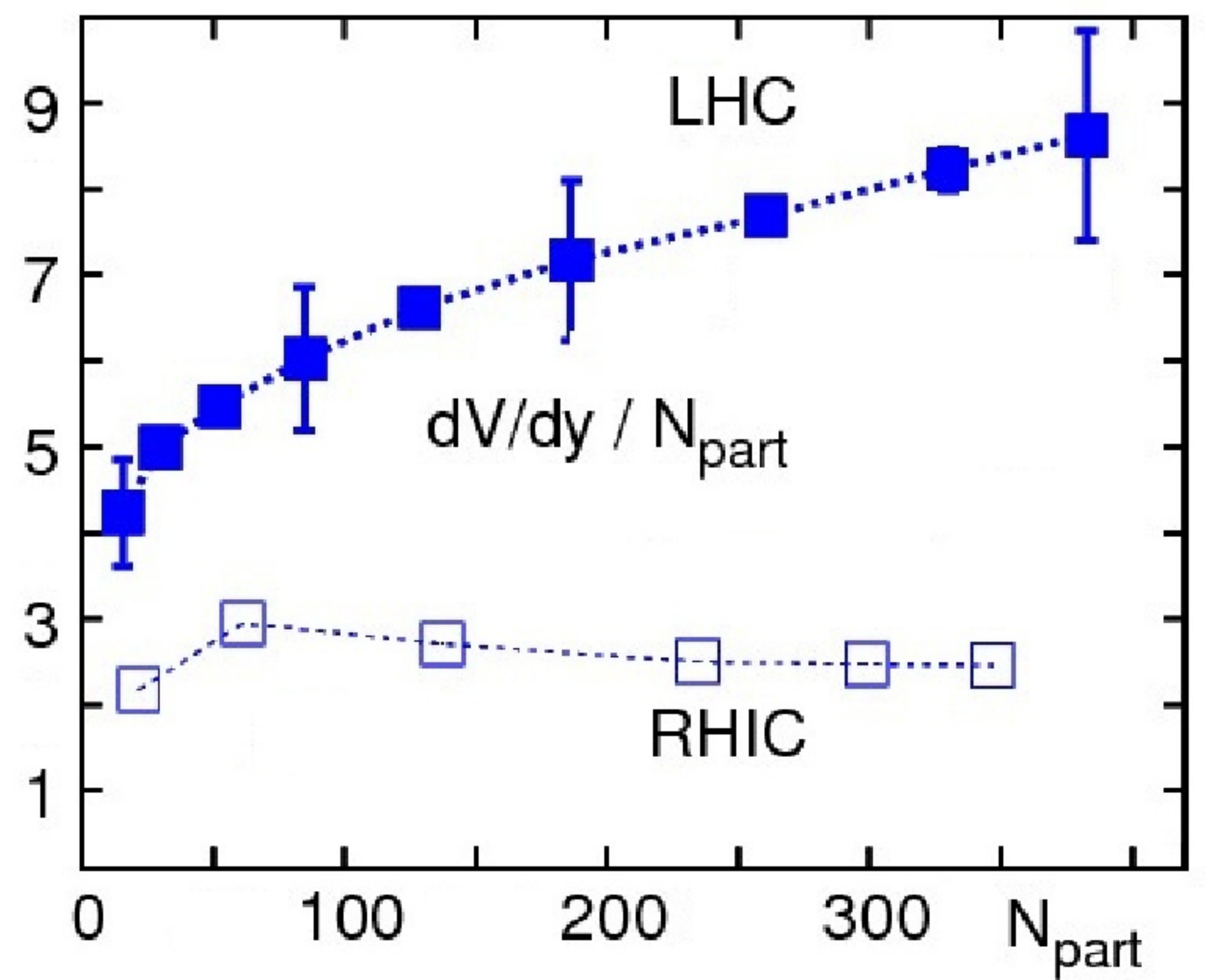}
}
\caption{\label{fig:LHCRHICvolumespec} The source volume $dV/dy$ at $\sqrt{s_{NN}}=2.76$ TeV, normalized by number of nucleon pairs  $N_\mathrm{part}/2$, as a function of the number of participants $N_{\rm part}$. For comparison, a similar  STAR $\sqrt{s_\mathrm{NN}}= 62$ GeV data analysis is shown. Results adapted from   Refs.\cite{Petran:2013lja,Petran:2013qla,Petran:2013dea}}
\end{figure}

The  corresponding   LHC  and RHIC chemical freeze-out temperature $T$, \rf{fig:Tchem}, varies both at RHIC and LHC in the same fashion with larger values found for smaller hadronization volumes. This  is natural, as scattering length for decoupling must be larger than the size of the  system and thus  the more dense hotter condition is possible for the smaller fireball. One  can also argue with the same outcome that the rapid expansion of the larger fireball can lead to stronger supercooling  of QGP which directly transforms into free-streaming hadrons.
The possibility  of direct QGP hadronization is supported by the strong chemical non-equilibrium with  $\gamma_q>1,\gamma_s>1$ for all collision centralities. These results are seen in \rf{fig:gammas}.

\begin{figure}
\centering\resizebox{0.40\textwidth}{!}{%
\includegraphics{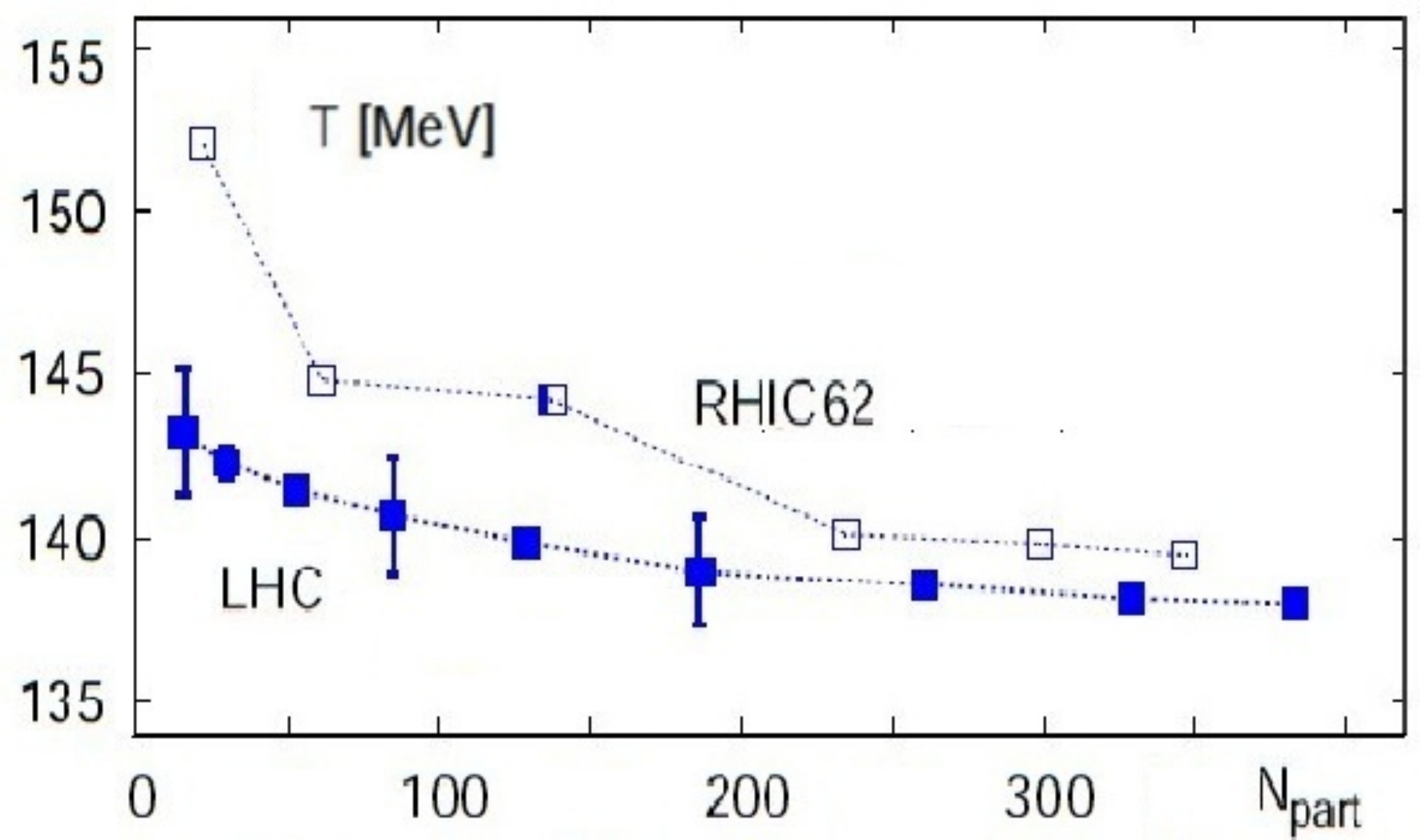}
}
\caption{\label{fig:Tchem} The chemical freeze-out temperature $T$ at$\sqrt{s_{NN}}=2.76$ TeV, as a function of the number of participants $N_{\rm part}$, lines guide the eye. Results  adapted from  Refs.\cite{Petran:2013lja,Petran:2013qla,Petran:2013dea}.}
\end{figure}

\begin{figure}
\centering\resizebox{0.45\textwidth}{!}{%
\includegraphics{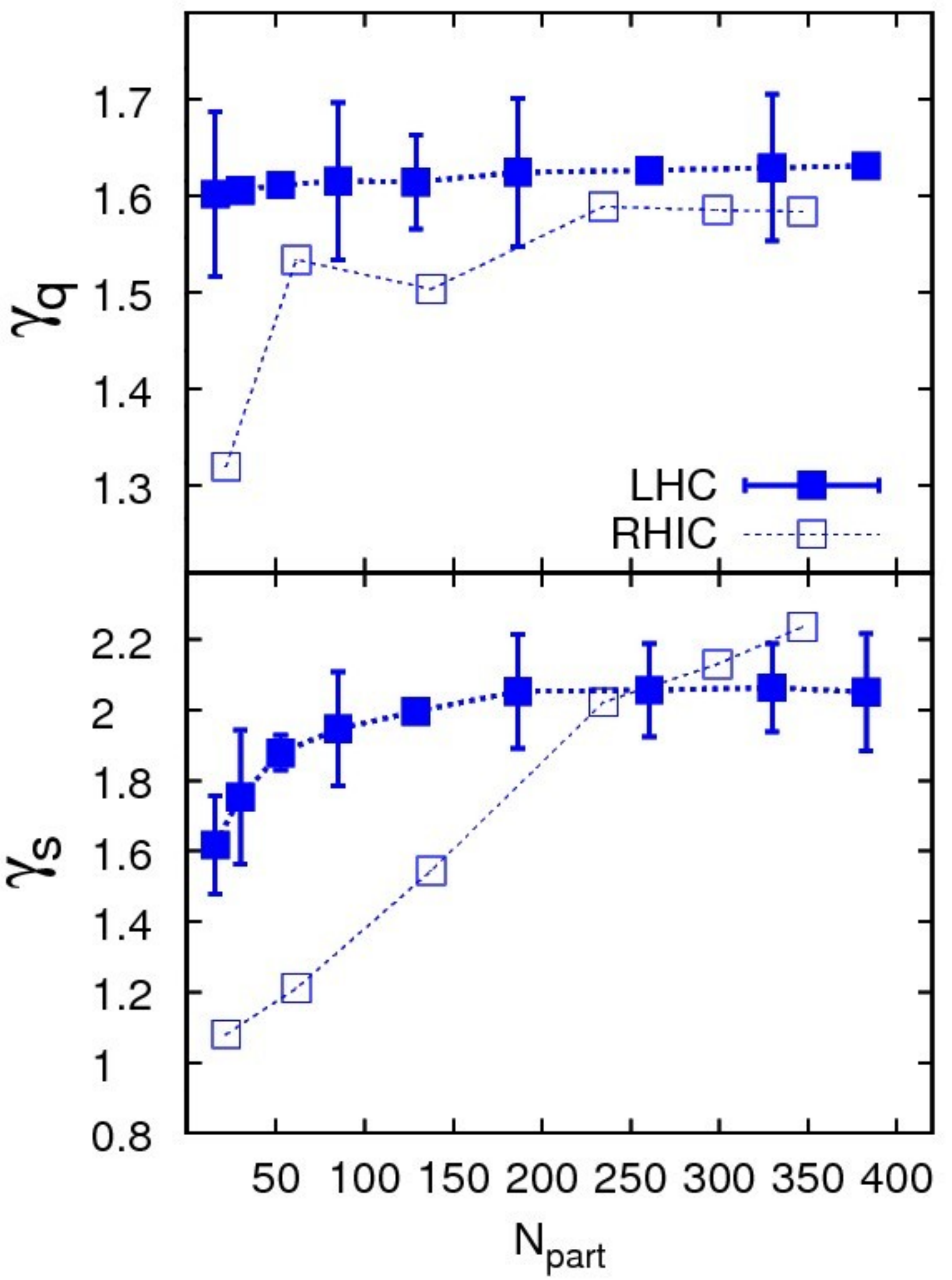}
}
\caption{\label{fig:gammas} Light  quark $\gamma_q$  and strange quark $\gamma_s$  fugacities at $\sqrt{s_{NN}}=2.76$ TeV as a function of the number of participants $N_{\rm part}$, lines guide the eye. Results  adapted from  Refs.\cite{Petran:2013lja,Petran:2013qla,Petran:2013dea}.}
\end{figure}

In \rf{fig:LHCproperties} we see the physical properties of the fireball as obtained by the  same procedure as discussed in Subsection~\ref{subsec:Bulk}. With increasing participant number all these bulk properties decrease steadily. This is the  most marked difference to the  RHIC results. We should here remember  that the hadronization volume at LHC  given the greater total energy content of the fireball is  much greater and thus  the  dynamics of fireball expansion should be  different.

\begin{figure}
\centering\resizebox{0.45\textwidth}{!}{%
\includegraphics{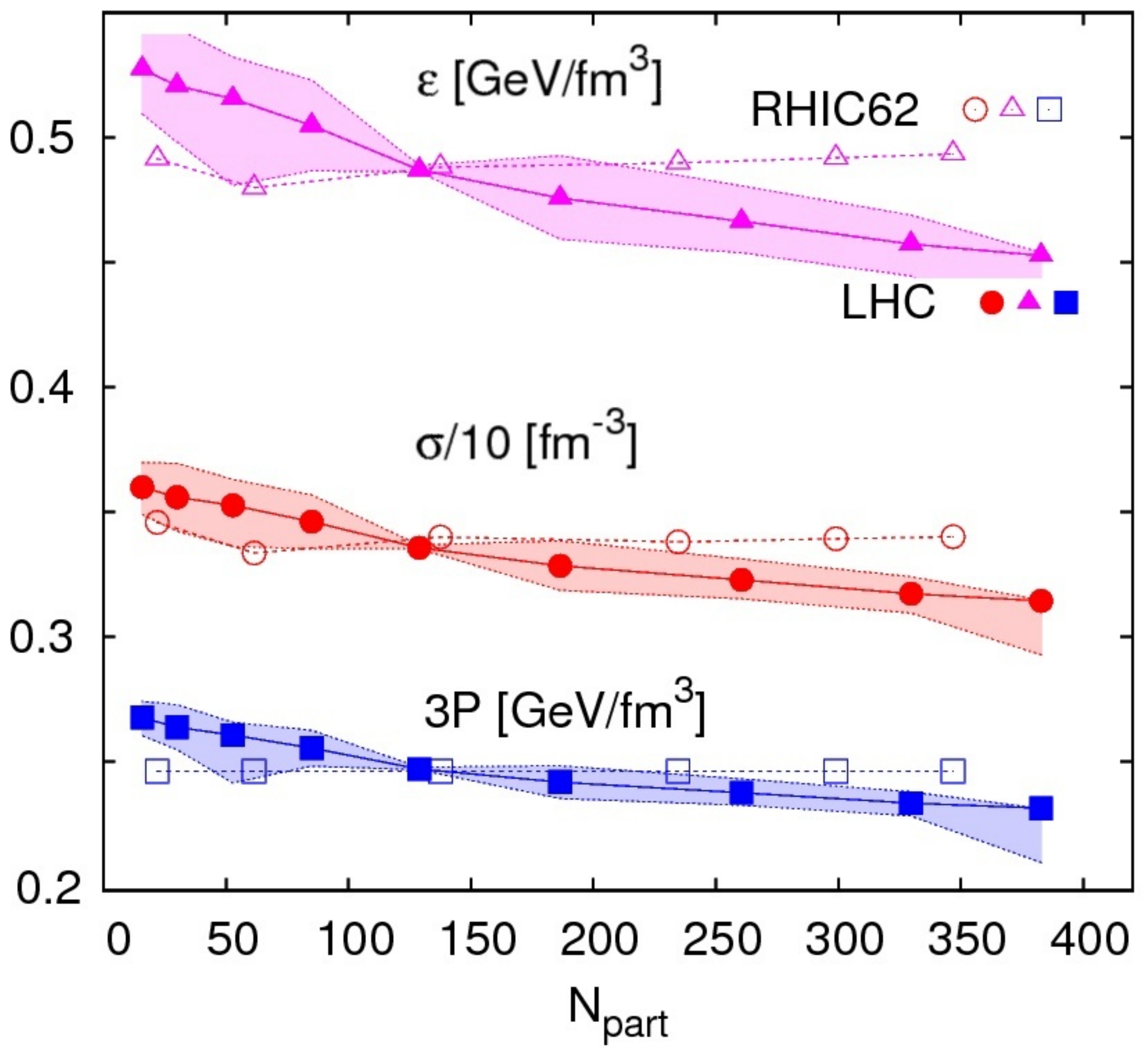}
}
\caption{\label{fig:LHCproperties} From top to bottom   as function of centrality described by $N_\mathrm{part}$: energy density  $\varepsilon$, entropy  density  $\sigma $ reduced by  a factor 10 to fit in figure, and  $3P$ at$\sqrt{s_{NN}}=2.76$ TeV. The dotted line are RHIC $\sqrt{s_\mathrm{NN}}= 62$ GeV analysis results not showing the  (larger) error band. Results  adapted from  Refs.\cite{Petran:2013lja,Petran:2013qla,Petran:2013dea}.}
\end{figure}

Results  seen in \rf{fig:LHCproperties} show  a remarkable universality,  both when LHC is compared to RHIC, and as a function of centrality; variation as a function of $N_\mathrm{part}$  is much smaller than that  seen in particle yields in \rf{fig:LHCparticles} (keep in mind that these results are divided by  $N_\mathrm{part}/2$). The universality of the hadronization condition is even more pronounced when we study, see \rf{fig:TraceHad},  $(\varepsilon-3P)/T^4$, the interaction measure $I_\mathrm{m}$ \req{interact_measure} (compare Subsection~\ref{hadronMatter}, \rf{BiroIm}).

\begin{figure}
\centering\resizebox{0.45\textwidth}{!}{%
\includegraphics{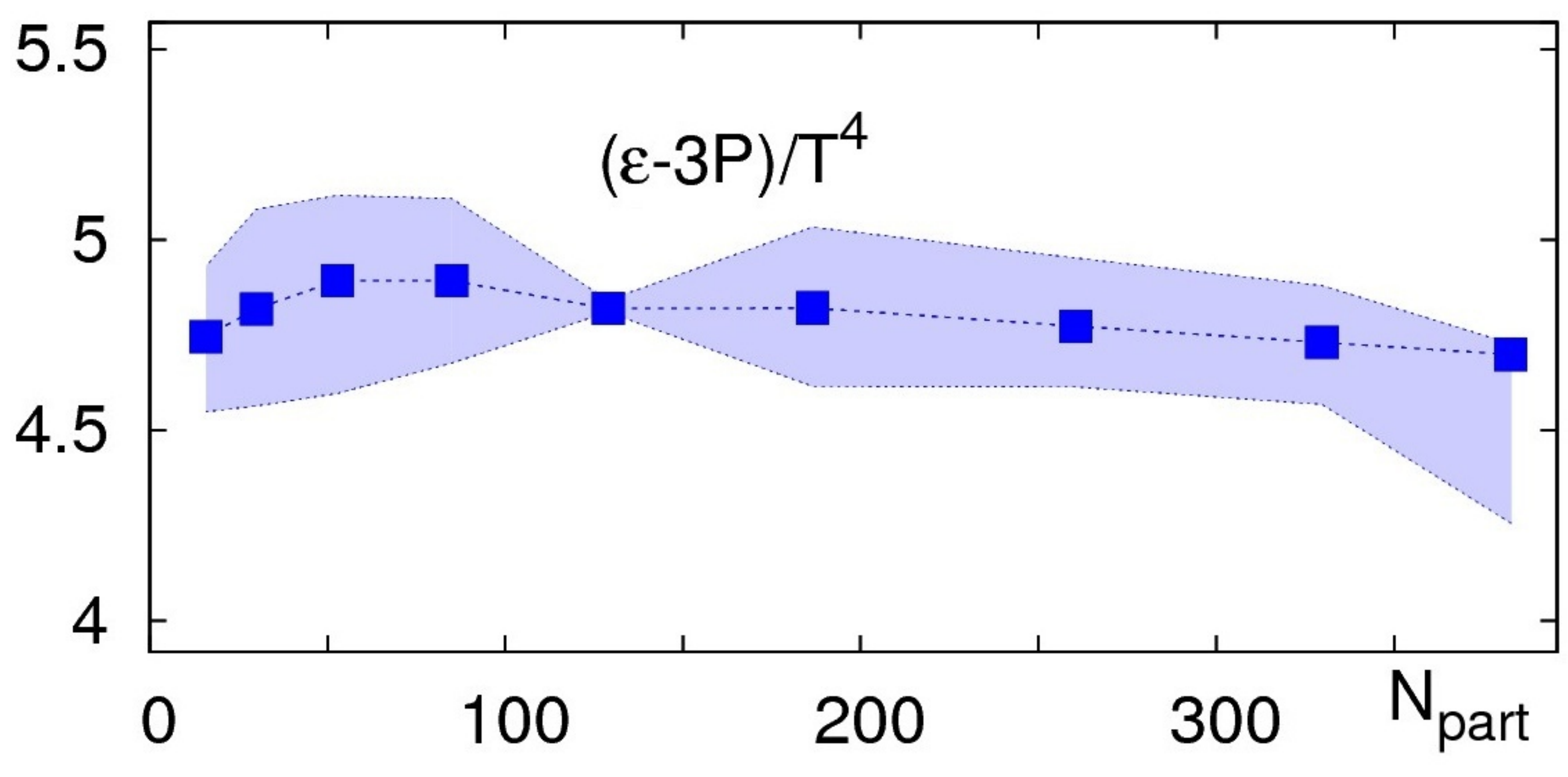}
}
\caption{\label{fig:TraceHad}  Hadronization universality: the  interaction measure $(\varepsilon-3P)/T^4$ evaluated at hadronization condition of the hadron fireball created in  $\sqrt{s_{NN}}=2.76$ TeV Pb--Pb collisions  as a function of centrality described by $N_\mathrm{part}$. Results  adapted from  Refs.\cite{Petran:2013lja,Petran:2013qla,Petran:2013dea}}
\end{figure}

We observe  that the lattice-QCD  maximum  from \rf{BiroIm} $(\varepsilon-3P)/T^4$ falls right into the uncertainty  band of this result. Only for $\gamma_q\simeq 1.6$  and $\gamma_s\simeq 2$ a high value for $I_\mathrm{m}$ shown in \rf{fig:TraceHad} can be obtained.  The equilibrium hadron gas results are about a factor 3 smaller in the relevant domain of temperature.

Turning our attention now to strangeness:  In the most  central  5\% Pb--Pb collisions at the LHC2760, a total of $dN_{s\bar{s}}/dy\simeq 600$ strange and anti-strange quarks per unit of rapidity is produced. For the more peripheral collisions the rise of the total strangeness yield is very rapid, as both the size of the reaction volume and within the small fireball the approach to saturation of strangeness production in the larger QGP fireball combine. 

It is of considerable interest to understand  the  magnitude of  strangeness QGP density at hadronization. We form a sum of all (strange) hadron multiplicities  $dN_h/dy$ weighting the sum with the strange content  $-3\le n_s^h\ge 3$ of any hadron $h$ and include  hidden strangeness, to obtain the  result shown in  \rf{fig:strangeness}. Within the error bar the  result is a constant; strange quarks and antiquarks in the fireball are 20\% more dense than are nucleons bound  in nuclei.
\begin{figure}
\centering\resizebox{0.49\textwidth}{!}{%
\includegraphics{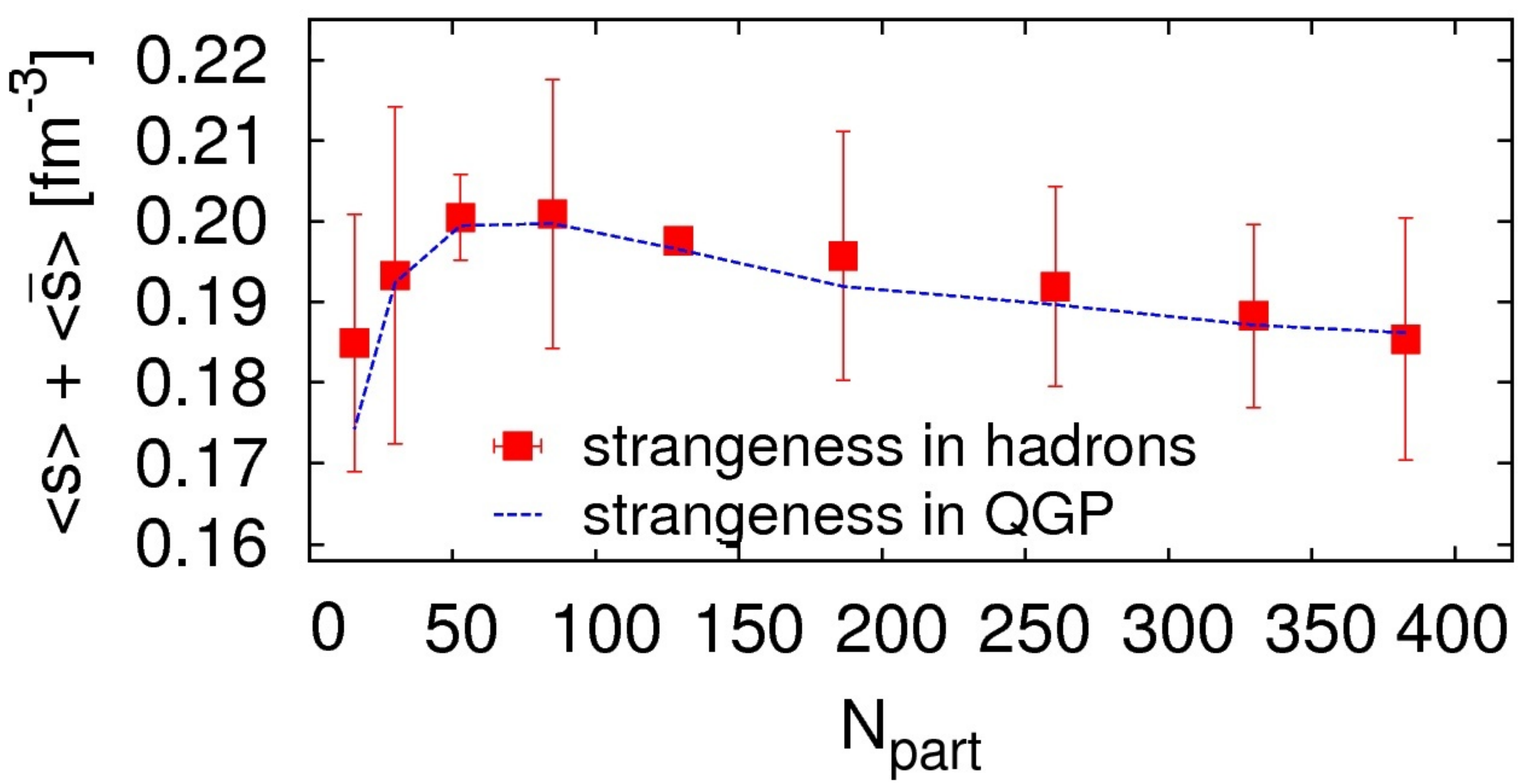}
}
\caption{\label{fig:strangeness} $\langle s \rangle +\langle \bar s\rangle$ strangeness density measured in the hadron phase (red squares)  as a function of centrality described by $N_\mathrm{part}$. The dashed (blue) line is a fit  with strangeness in the QGP phase, see text.   Results  adapted from  Refs.\cite{Petran:2013lja,Petran:2013qla,Petran:2013dea}}
\end{figure}

However,  is this  $\langle s \rangle +\langle \bar s\rangle$ strangeness density  shown by error bars in \rf{fig:strangeness} a density related to QGP?  To give this result a quantitative QGP meaning we evaluate QGP  phase strangeness density at a given $T$, see \req{eq:yieldexpansion}
\begin{align}
\label{eq:strangeness}
s(m_s,T; \gamma_s^{\scriptscriptstyle\mathrm{QGP}}) = -&\frac{gT^3}{2\pi^2} 
\sum\limits_{n=1}^\infty \left(-\gamma_s^{\scriptscriptstyle\mathrm{QGP}}\right)^{n}\times\\ \notag &\ \frac{1}{n^3} \left(\frac{nm_s}{T} \right)^2 K_2\left(\frac{nm_s}{T} \right),
\end{align}
where  $m_s$ is the (thermal) strange quark mass, $\gamma_s^{\scriptscriptstyle\mathrm{QGP}}$ is the phase space occupancy: here the superscript  QGP helps to distinguish  from that measured in the hadronization analysis as  $\gamma_s$, used without a superscript. The degeneracy is $g=12=2_\mathrm{spin} 3_\mathrm{color}  2_\mathrm{p}$  where the last factor accounts for the presence of both quarks and antiquarks. 

In central LHC  collisions, the large volume (longer lifespan) also means that strangeness approaches saturated yield  in the QGP. In peripheral collisions, the short lifespan of the fireball may not be sufficient to reach chemical equilibrium. Therefore we introduce a centrality dependent strangeness phase space occupancy $\gamma_s^{\scriptscriptstyle\mathrm{QGP}}(N_{part})$ which is to be used in \req{eq:strangeness}. 

A model of the centrality dependence of $\gamma_s^{\scriptscriptstyle\mathrm{QGP}}(N_{part})$ is not an important consideration, as the yield for $N_{part}> 30$ is nearly constant.  The   value of strangeness density  requires $m_s=299\,\mathrm{MeV}$ in a  QGP fireball at hadronization. For   $m_s \simeq 140\,\mathrm{MeV} $ (mass at a scale of $\mu\simeq 2\pi T \simeq 0.9\,\mathrm{GeV}$).   $\gamma_{s\,\mathrm{final}}^{\scriptscriptstyle\mathrm{QGP}}\simeq 0.77$ is found. The  higher value of $m_s$ makes more sense in view of the need to account for the thermal effects. Thus we conclude that for  $N_{part}> 30$ the fireball contains QGP chemical equilibrium strangeness abundance, with strangeness thermal mass  $m_s=299\,\mathrm{MeV}$~\cite{Petran:2013dea}.

The  ratio of strangeness to entropy  is  easily recognized to be, for  QGP, a measure of the relative number of strange   to  all particles - adding a factor $\simeq 4$ describing the  amount of entropy that each particle carries. Thus  a QGP source will weigh in with ratio $s/S\simeq 0.03$~\cite{Kuznetsova:2006bh}. This is about factor 1.4 larger than one computes for hadron phase at the same $T$, and this factor describes the strangeness enhancement effect in abundance which was predicted to be that small, see  Ref.\cite{appenB}. However, if a QGP fireball was formed we do expect a rather  constant $s/S$ as a function of $N_{\mathrm{part}}$.

\subsection{\label{sec:earlier}Earlier work}
Results of SHM that  provide freeze-out $T$ well above \Th seen in \rf{fig:phasediagram} should today be considered obsolete. As an example let us enlarge here on the results of  Ref.\cite{BraunMunzinger:2001ip}  which would be marked in \rf{fig:phasediagram} GSI-RHIC at  $T\simeq174\pm 7,\ \muB\simeq 46\pm 5 $\,MeV  corresponding to a fit of  $\sqrt{s_\mathrm{NN}}=130$\,GeV RHIC results (but  the point is not shown above the upper $T$ margin). This reference assumes full chemical equilibrium. They draw attention to agreements with other results and expectations, both in their conclusions, as well as in the body of their text, verbatim:   
\begin{quote}
\lq\lq The chemical freeze-out temperature $T_f\simeq 168\pm 2.4$ MeV found from a thermal analysis   of experimental data in Pb–Pb collisions at SPS is remarkably consistent within error with the critical temperature $T_c\simeq 170\pm 8$ MeV obtained from lattice Monte Carlo simulations of QCD at vanishing baryon density [15] and [16]\rq\rq
\end{quote}
 Their  lattice references are   [15,  from the year  2001]~\cite{Karsch:2001vs}  and  [16, from the year 1999]~\cite{Karsch:1999vy}. The   two references  disagree  in regard to value  of \Th, verbatim: 
\begin{quote}
(1999)\lq\lq If the quark mass dependence does not change drastically closer to the chiral limit the current data suggest $T_c \simeq (170$--$ 190)$\,MeV    for 2-flavor QCD in the chiral limit.   In fact, this estimate also holds for 3-flavor QCD.\rq\rq. \\
(2001) \lq\lq The 3-flavor theory, on the other hand, leads to consistently smaller values of the critical temperature,\ldots  3 flavor QCD: $T_c =  (154\pm 8)$\,MeV\rq\rq
\end{quote}
While the authors of Ref.\cite{BraunMunzinger:2001ip} were clearly encouraged by the 1999 side remark in Ref.\cite{Karsch:1999vy} about 3 flavors,   they also cite in the same breath the correction~\cite{Karsch:2001vs} which renders their RHIC SHM fit invalid:   for a lattice result   $\Th=154\pm 8$ MeV chemical freeze-out at $T\simeq 174\pm 7$\,MeV  seems  inconsistent  since  $T<\Th$ strictly.

When reading Ref.\cite{BraunMunzinger:2001ip}  in Spring 2002 I further spotted that it is technically  wanting. Namely, the  experimental $\overline{\uXi}/\uXi$ ratio used in the paper predicts a value $\mu_{\uXi}=\muB-2\mu_\mathrm{S}=18.8$\,MeV while the  paper   determines from this ratio a value   $\mu_{\uXi}=\muB-2\mu_\mathrm{S} =9.75$\,MeV. In conclusion: the cornerstone manuscript of the GSI group is at the time of publication  inconsistent with the lattice used as justification showing chemical freeze-out $T>\Th$ by 20 MeV, and its computational part contains a technical mistake. But, this paper had a  \lq good\rq\ confidence level.

The key  argument of the paper is that   $\chi^2/\mathrm{dof}\simeq 1$.   However,  $\chi^2$ depends in that case on large error bars in the  initial 130 GeV RHIC results. Trusting $\chi^2$ alone  is not appropriate to judge a fit result\footnote{Hagedorn explained the abuse  of $\chi^2$ as follows: he carried an   elephant and mouse transparency set, showing how both transparencies are fitted by a third one   comprising a partial picture of something. Both mouse and elephant fitted the something very well. In order  to distinguish mouse from elephant one needs external scientific understanding; in his example,  the scale, was required.}. A  way to say this is to argue that  a fit   must be \lq confirmed\rq\ by theory, and indeed that is what Ref.\cite{BraunMunzinger:2001ip}  claimed, citing Ref.\cite{Karsch:2001vs} which however, provided a result  in direct disagreement. 

Thus we can conclude that Ref.\cite{BraunMunzinger:2001ip} at time of publication  had already proved itself wrong. And while \lq humanum errare est\rq, students lack the experience to capture theirs effectively. Today this work  is cited more than 500 times -- meaning that despite the obvious  errors and omissions it has   entered   into the contemporary knowledge base. Its  results confuse  the uninitiated deeply. These results could only  be erased by a direct withdrawal note by  the  authors.

\subsection{Evaluation of LHC SHM fit results}\label{subsec:EvalSHM}
The chemical non-equilibrium   SHM  describes very well all available   LHC-2760 hadron production data obtained in a wide range of centralities $N_\mathrm{part}$, measured  in the CM  within the rapidity interval $-0.5\le 0\le 0.5$.  A value of freeze-out  temperature that is clearly below the range for \Th reported in \rf{fig:phasediagram}  for lattice-QCD  arises only  when accepting a full  chemical  nonequilibrium outcome. Chemical nonequilibrium is expected for the hadron phase space if QGP fireball was in chemical equilibrium. In that sense, theory  supports the finding, and this  result also has a very good $\chi^2/\mathrm{ndf} < 1$  for all collision centralities. 

The value  of the ratio $p/\pi|_{\rm experiment}\!\! = 0.046\pm0.003$~\cite{Abelev:2012vx,Abelev:2012wca} is a LHC result  that any model of particle production in RHI collisions must  agree with. The value  $p/\pi \simeq 0.05$  is a natural outcome of the chemical non-equilibrium fit with $\gamma_q\simeq 1.6$.  This result was   predicted in Ref.\cite{Rafelski:2010cw}: $p/\pi|_{\rm prediction} = 0.047\pm 0.002$ for the hadronization pressure  seen at RHIC and SPS  $P=82\pm 5$\,MeV/fm$^3$. Chemical equilibrium model predicts and fits a very much larger result. This is  the so-called proton anomaly; there is no anomaly if one does not dogmatically prescribe chemical equilibrium conditions.

A recent study of the proton spectra within the  freeze-out model developed in Krakow  confirms the chemical nonequilibrium~\cite{Begun:2013nga}:  In \rf{fig:pikpspectra} we show a comparison between a spectral fit of pions, kaons and protons within the equilibrium and nonequilibrium approaches. These results show   the strong overprediction of soft protons and some overprediction of kaons that one finds in the equilibrium model. The chemical nonequilibrium model provides an excellent description of this key data. 

\begin{figure}
\centering\resizebox{0.40\textwidth}{!}{%
\includegraphics{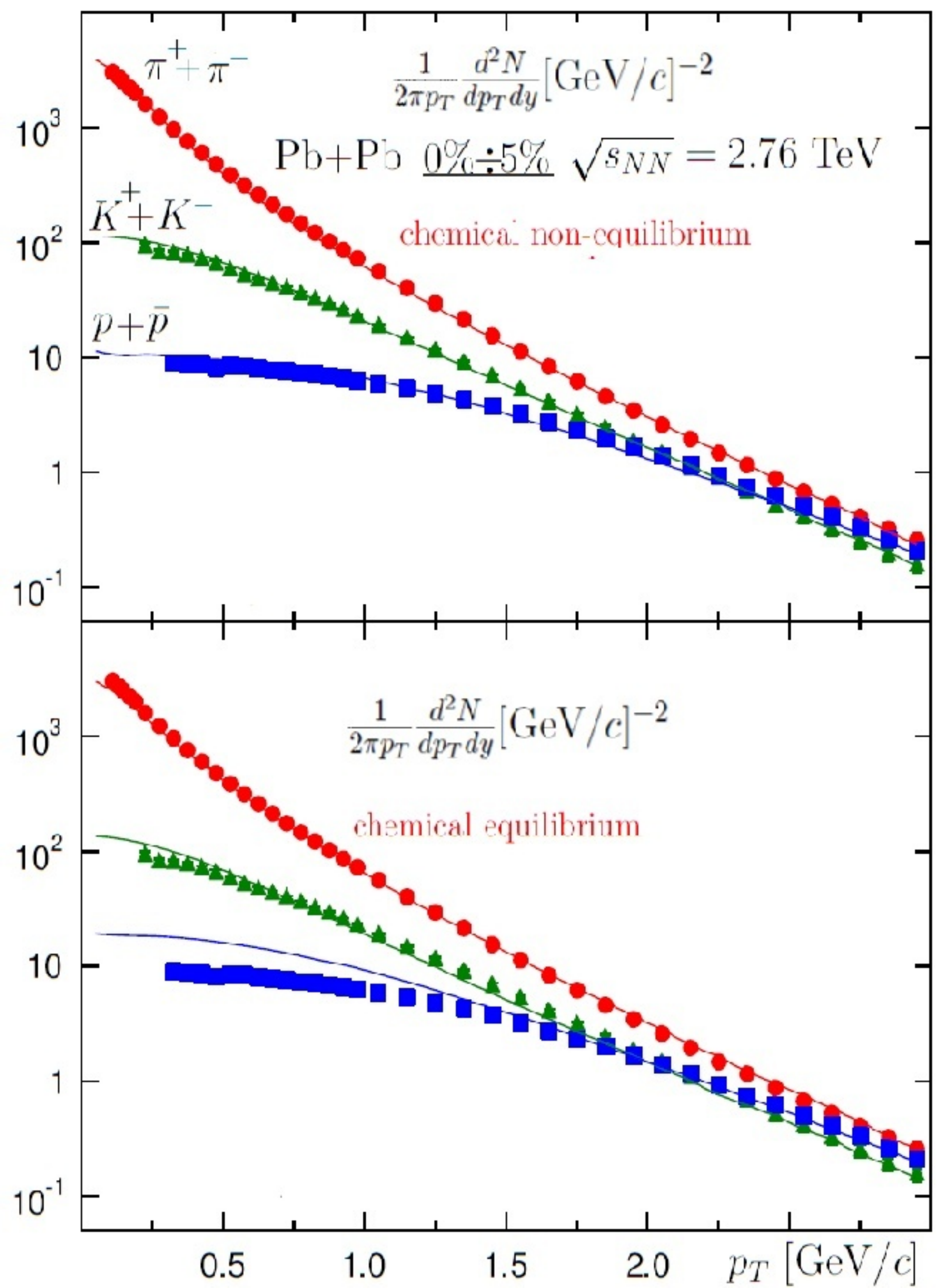}
}
\caption{\label{fig:pikpspectra} Data: most central spectra of pions, kaons and protons from ALICE experiment~\cite{Abelev:2012wca,Abelev:2013vea}  as a function of $p_\bot$. Lines: Top the nonequilibrium model for  parameters of this review; bottom the outcome with equilibrium constraint. Figure  adapted from  Ref.\cite{Begun:2013nga}}
\end{figure}

Further  evidence for the chemical non-equilibrium  outcome of  SHM analysis arises from the universality of hadronization at  LHC, RHIC and SPS: the  bulk properties of the fireball that we determine are all very similar to each  other. This  can be seen by comparing RHIC-SPS results  presented in  \rf{EdepBulk}  with those shown in \rf{fig:LHCproperties} for LHC-RHIC.

This universality includes the strangeness content of the fireball. The LHC particle multiplicity  data has relatively small errors, allowing establishment  of relatively precise results. The strong nonequilibrium result   $\gamma_s \to 2$ seen in \rf{fig:gammas} allows the description of the large abundance of multi-strange hadrons despite the relatively small value of freeze-out temperature. The value of light quark fugacity,   $\gamma_q \to 1.6 $ allows a match in the high entropy content of the  QGP fireball with the  $\gamma_q $ enhanced  phase space of hadrons, especially mesons. As noted above, this effect naturally provides the  correct  $p/\pi$ ratio at small $T$.

There are two noticeable differences that appear in comparing RHIC62 to LHC2760 results; in \rf{fig:LHCRHICvolumespec}  we see as a function of centrality  the specific  volume parameter $(dV/dy)/N_\mathrm{part}$.  The noticeable difference is that at RHIC this  value is  essentially constant, while at LHC there is clearly a visible  increase. One can  associate this with a corresponding   increase in entropy per participant, implying that a novel component in entropy production must have opened up in the LHC energy regime. This additional entropy production also explains why at LHC the maximum value of specific strangeness pair yield per entropy is smaller when compared to  RHIC62 for most central collisions, see \rf{fig:sSratio}.

\begin{figure}
\centering\resizebox{0.45\textwidth}{!}{%
\includegraphics{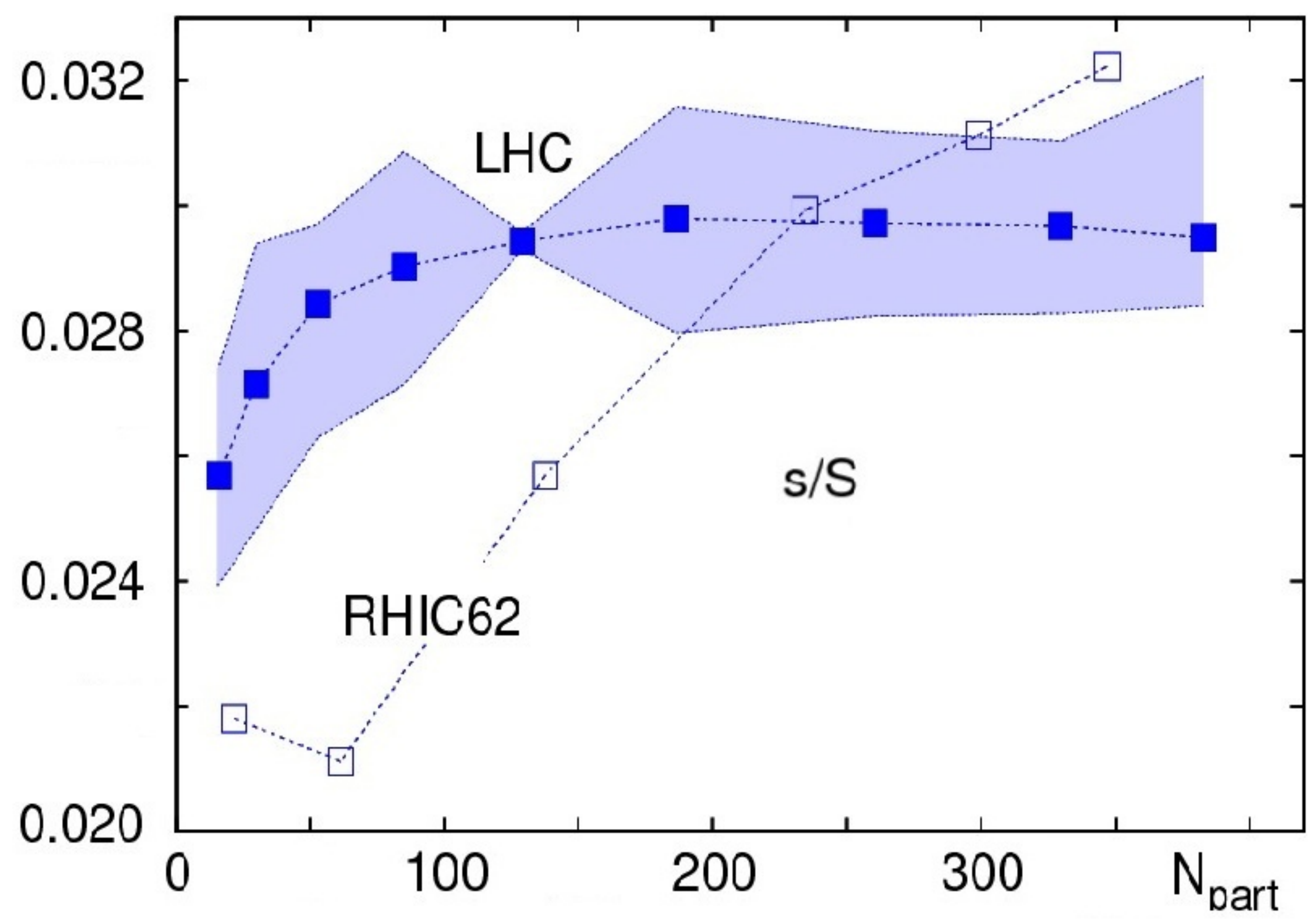}
}
\caption{\label{fig:sSratio} Ratio of strangeness pair yield  to entropy $s/S$ as function of  collision centrality described by $N_{\mathrm{part}}$. Results  adapted from  Refs.\cite{Petran:2013lja,Petran:2013qla,Petran:2013dea}}
\end{figure}

The results found in the LHC-SHM analysis characterize a fireball that  has properties  which can be directly compared with results of lattice-QCD, and which  have  not been  as yet reported;  thus this analysis offers a prediction which can be used to verify the consistency of SHM results with lattice. For example, note  the dimensionless ratio of the number  of strange quark pairs  with entropy $s/S\to 0.03$. Since strange quarks have a mass scale, this ratio can be expected to be a function of temperature  in  lattice-QCD evaluation. The interesting question is,  at what $T$ will lattice obtain this strangeness hadronization condition $s/S= 0.03$? 

Further, there is a variation as a function of centrality seen in \rf{fig:sSratio}; $s/S$ decreases with decreasing $N_\mathrm{part}$. Seeing that freeze-out $T$ increases, compare \rf{fig:Tchem}, we obtain the prediction that as freeze-out $T$ increases, $s/S$ decreases. On first sight this  is counterintuitive as we would think that  at higher  $T$ there is more strangeness. This is an interesting behavior that may provide an opportunity to better understand the relation of  the  freeze-out analysis with lattice-QCD results.

\section{Comments and Conclusions}\label{Comments}
To best of my knowledge nobody has attempted a synthesis of the theory of hot hadronic matter, lattice-QCD results, and the statistical hadronization model. This is done here  against the  backdrop of the rich volume of soft hadron production results  that have emerged in the  past 20 years in the field of RHI collisions, covering the entire range of SPS, RHIC and now LHC  energy, ranging a factor 1000 in $\sqrt{s_\mathrm{NN}}$.  

This is certainly  not  the ultimate word  since we expect  new and important experimental results in the next few years: LHC-ion operations will reach near  maximum energy by the end of 2015; further decisive  energy increases could take another lifespan. On the other energy range end, we are reaching out to the domain where we expect the QGP formation energy threshold,  both at RHIC-BES, and at SPS-NA61.  The future will show if new experimental facilities today in construction and/or  advanced    planning will come online within this  decade and join in the study of QGP formation threshold. Such plans have  been made both at GSI and at DUBNA laboratories. 

In order for this report to be also a readable  RHI collision introduction, I provided   pages distributed across the manuscript, suitable both  for students  starting in the  field, and readers from other  areas of science who are interested in the topic. I realize that most of technical material is not accessible to these two groups, but it is better to build a bridge  of understanding than to do nothing. Moreover, some historical considerations may be welcome in these circles.

I did set many of the insights into their historical context which  I have  witnessed personally. In my eyes understanding the  history, how topics came to be looked at the  way they are, helps both the present generation to learn what we know, and the future generation in resolving the misunderstandings that block progress. 

For  this reason I felt that many insights that I needed to develop could be presented here  equally well in the format of work done many years ago. Therefore, in the jointly published   Refs.\cite{appenA,appenB} I present two unpublished reports from conference proceedings, long gone from library shelves, which I think provide quite appropriate background material for this report. Perhaps I  should have abridged  this bonus  material to omit a few obsolete developments,  and/or to avoid duplication across these 80+ pages. However,   any contemporary change will modify in a damaging way the historical context of these presentations.

There was a very special reason to prepare this report now. I took up this task after I finished editing a book to honor 50  years of Hagedorn remarkable achievements, Ref.\cite{HagedornBook}. Given the historical context and the  target of interest in the book being also a person, I could not inject there  all the  results that  the reader sees in these pages. The overlap between this work and Ref.\cite{HagedornBook} is small, mostly when I describe how the field developed historically  before 1985. 

This background material as presented here is  more extensive compared to Ref.\cite{HagedornBook} as I can go into the  detail without concern  about the contents balance of an edited book.  For this reason a reader of Ref.\cite{HagedornBook}  should look at this text as an extension, and conversely,  the reader of this review should also obtain Ref.\cite{HagedornBook}  which is freely available on-line, published in open format by the publisher in order to access some of the hard to get references used in this volume. 

This immediately takes us to the question that experts in the field will pose: is   in this synthesis anything scientifically new? The answer is yes. The list is actually quite long, and the advice is: read, and fill the gaps where the developments stop. Let me  point here to  one result, the new item which is really not all that new:   1978~\cite{Rafelski:1979cia}, Hagedorn and I discovered that hadronic matter with exponential mass spectrum at the point of singularity   \Th  has a universally  vanishing speed of sound $c_s\to 0$ in leading order. 

Universally means that  this is true  for all   functions that I have  tried, including  in particular a variation on preexponential singularity index $a$. In \lq leading order\rq\ means that the  most singular parts of both  energy density and pressure are considered. This result is found in a  report that was submitted to a Bormio Winter Meeting proceedings, Chapter  23 in Ref.\cite{HagedornBook} which today is archived electronically at CERN~\cite{Rafelski:1979cia}. 

The reason that  this   important result $c_s\to 0$ was not preprinted is that Hagedorn and  I were working on  a larger manuscript from which this Bormio text was extracted. We did not want to preclude the publication of the large paper. However,  at the  same time we were developing a new field of physics. The  main manuscript was never to be finished and submitted, but the  field of physics took off. Thus the Bormio report is all that remains in public view.  


This  result,  $c_s|_{T\to\Th}\to 0$  has other remarkable consequences: Sound  velocity that goes to zero at the critical boundary implies that  the matter is  sticky there,  when pressed from inside many unusual things can happen, one  being filamenting  break-up we call sudden hadronization -- the  amazing thing is that while I was writing this, Giorgio Torrieri was just reminding me about  this insight he had shared before with me.  Could  $c_s|_{T\to\Th}\to 0$   actually be the  cause why  the  SHM study of the  fireball properties obtains such clean sets of results?  

Any universal hot hadron matter critical property  can   be  tested with lattice-QCD and the results available do  show a range where $c_s|_{T\to\Th} \simeq  0$. Thus, the  value of \Th may   become  available as the point of a minimum sound velocity. This criterion comes with  the  Hagedorn exponential mass spectrum \lq attached\rq: the  result is valid if and when there is an exponential growth in hadron mass spectrum.  

In this text I also  answer  simple questions which turn out to have complicated answers. For example, what is, and   when was, QGP discovered? It turns out that QGP as a phrase meant something else initially, and all kinds of variants such as: hot quark matter; hadron plasma, were in use. This makes literature  search difficult.

I also  tracked the  reporting  and  interviews  from the time when CERN decided in February 2000 to step forward with its announcement of the QGP discovery.  I  learned that the then director of BNL was highly skeptical of the CERN  results. And I was shocked to learn that one of the two authors of the CERN scientific consensus report declared a few months down the road that he was mistaken. 

Seeing these initial doubts, and  being  expert on strangeness I thought  I ought to take a late deep look at how the signature of CERN February 2000 announcement  held up in past 15 years. I am happy  to tell that it  is doing very well; the  case of QGP at CERN-SPS in terms of strange antibaryon signature is very convincing.  I hope the  reader will join me in this  evaluation, seeing  the results shown. These are not comprehensive (my apologies) but  sufficient  to make the point.

I have  spent a lot of time, ink, and  paper, to explain here why RHI collisions and QGP physics  is, was, and remains a frontier of our understanding of physics. It is true that for the trees we sometimes lose the  view of the  forest. Thus  at a few opportunities in this report  I went outside of the trees to tell how the forest looks today, after 35 years of healthy  growth. While some will see my  comments as speculative, others may choose to work out the consequences, both in theory and experiment. 
 
\begin{acknowledgement}
\indent {\bf Commendation:}
I am deeply indebted to Rolf Hagedorn of CERN-TH whose continued mentoring nearly 4 decades ago provided much of the guidance and motivation in my  long pursuit of  strangeness in quark--gluon plasma and hadronization mechanisms. Rolf Hagedorn  was the scientist whose dedicated, determined personal commitment formed the deep roots of  this novel area of physics. In 1964/65 Hagedorn proposed   the Hagedorn Temperature \Th and the Statistical Bootstrap Model (SBM). These  novel ideas opened up the physics of hot hadronic matter to at first, theoretical and later, experimental study, in relativistic heavy  ion collision experiments. 

{\bf Acknowledgments:}  I thank (alphabetically)  Michael Danos (deceased),  Wojciech Florkowski,   Marek Ga\'zdzicki, Rolf Hagedorn (deceased),   Peter Koch, Inga Kuznetsova, Jean Letessier, Berndt M\"uller,  Emanuele Quercigh,  Krzysztof Redlich, Helmut Satz, and  Giorgio Torrieri, who have all contributed in an essential way  to further   my  understanding of relativistic heavy ion collisions,  hot hadronic matter, statistical hadronization, and the strangeness signature of QGP. I thank Tamas Biro for critical comments of a draft manuscript helping to greatly improve the contents; and Victoria Grossack for her kind assistance with the manuscript presentation.

I thank the CERN-TH for hospitality in Summer 2015 while this project was created and completed. This work has been in part supported by the US Department of Energy, Office of Science, Office of Nuclear Physics  under  award number   DE-FG02-04ER41318 .
\end{acknowledgement}






\appendix
\renewcommand{\thesection}{\Alph{section}~}
\renewcommand{\thesubsection}{\arabic{subsection}}
\setcounter{table}{0}
\setcounter{figure}{0}
\setcounter{footnote}{0}
\setcounter{equation}{0}
\renewcommand{\theequation}{\thesection\arabic{equation}}
\renewcommand{\thefigure}{\thesection\arabic{figure}}
\makeatletter
\renewcommand\@biblabel[1]{[\thesection#1]}
\renewcommand\@cite[1]{[\thesection#1]}
\makeatother
\section{Extreme States of Nuclear Matter -- 1980} 
\noindent{\bf From: \lq\lq Workshop on Future Relativistic Heavy Ion Experiments\rq\rq\
held 7-10 October 1980  at: GSI, Darmstadt, Germany, R. Bock and R. Stock, editors. Printed in:  GSI81-6 Orange Report, pp. 282-324}\\ 

\noindent ABSTRACT: {\small The theory of hot nuclear fireballs consisting of all possible finite-size hadro\-nic constituents in chemical and thermal equilibrium is presented.  As a complement of this hadro\-nic gas phase characterized by maximal temperature and energy density, the quark bag description of the hadro\-nic fireball is considered. Preliminary calculations of temperatures and mean transverse momenta of particles emitted in high multiplicity relativistic nuclear collisions together with some considerations on the observability of quark matter are offered.}
\subsection{Overview}\label{3chap5sec1}

I wish to describe, as derived from known traits of strong interactions, the likely thermodynamic properties of hadro\-nic matter in two different phases: the hadro\-nic gas consisting of strongly interacting but individual baryons and mesons, and the dissolved phase of a relatively weakly interacting quark-gluon plasma. The equations of state of the hadro\-nic gas can be used to derive the particle temperatures and mean transverse momenta in relativistic heavy ion collisons, while those of the quark-gluon plasma are more difficult to observe experimentally. They may lead to recognizable effects for strange particle yields. Clearly, the ultimate aim is to understand the behavior of hadro\-nic matter in the region of the phase transition from gas to plasma and to find characteristic features which will allow its experimental observation. More work is still needed to reach this goal. This report is an account of my long and fruitful collaboration with R. Hagedorn \cite{3chap5bib1}.

The theoretical techniques required for the description of the two phases are quite different: in the case of hadro\-nic gas, a strongly attractive interaction has to be accounted for, which leads to the formation of the numerous hadro\-nic resonances -- which are in fact bound states of several (anti) quarks. if this is really the case, then our intuition demands that at sufficiently high particle (baryon) density the individuality of such a bound state will be lost. In relativistic physics in particular, meson production at high temperatures might already lead to such a transition at moderate baryon density. As is currently believed, the quark--quark interaction is of moderate strength, allowing a perturbative treatment of the quark-gluon plasma as relativistic Fermi and Bose gases. As this is a very well studied technique to be found in several reviews \cite{3chap5bib2a,3chap5bib20a,3chap5bib2b,3chap5bib2c,3chap5bib2d,3chap5bib2e,3chap5bib2f}, we shall present the relevant results for the relativistic Fermi gas and restrict the discussion to the interesting phenomeno\-logical consequences. Thus the theoretical part of this report will be devoted  mainly to the strongly interacting phase of hadro\-nic gas. We will also describe some experimental consequences for relativistic nuclear collisions such as particle temperatures, i.e., mean transverse momenta and entropy.

As we will deal with relativistic particles throughout this work, a suitable generalization of standard thermodynamics is necessary, and we follow the way described by Touschek \cite{3chap5bib3}. Not only is it the most elegant, but it is also by simple physical arguments the only {\it physical\/} generalization of the concepts of thermodynamics to relativistic particle kinematics. Our notation is such that $\hbar=c=k=1$. The inverse temperature $\beta$ and volume $V$ are generalized to become four-vectors:
\begin{eqnarray}
E &\;\longrightarrow\;& p^\mu = (p^0,\vec{p})= mu^\mu\;,\quad u_\mu u^\mu=1\;,\nonumber\\\noalign{\smallskip}
\frac{1}{T} &\;\longrightarrow\;& \beta^\mu = (\beta^0,\vec{\beta})= \frac{1}{T}v^\mu\;,\quad v_\mu v^\mu=1\;,\label{3chap5eq1}\\\noalign{\smallskip}
V &\;\longrightarrow\;& V^\mu = (V^0,\vec{V})= Vw^\mu\;,\quad w_\mu w^\mu=1\;,\nonumber
\end{eqnarray}
where $u^\mu$, $v^\mu$, and $w^\mu$ are the four-velocities of the total mass, the thermometer, and the volume, respectively. Usually, $\langle u^\mu\rangle=v^\mu=w^\mu$.

We will often work in the frame in which all velocities have a timelike component only. In that case we shall often drop the Lorentz index $\mu$, as we shall do for the arguments $V=V_\mu$, $\beta=\beta_\mu$ of different functions.

The attentive reader may already be wondering how the approach outlined here can be reconciled with the concept of quark confinement. We will now therefore explain why the occurrence of the high temperature phase of hadro\-nic matter -- the quark-gluon plasma -- is still consistent with our incapability to liberate quarks in high energy collisions. It is thus important to realize that the currently accepted theory of hadro\-nic structure and interactions, quantum chromodynamics \cite{3chap5bib4}, supplemented with its phenomeno\-logical extension, the MIT bag model \cite{3chap5bib5}, allows the formation of large space domains filled with (almost) free quarks. Such a state is expected to be unstable and to decay again into individual hadrons, following its free expansion. The mechanism of quark confinement requires that all quarks recombine to form hadrons again. Thus the quark-gluon plasma may be only a transitory form of hadro\-nic matter formed under special conditions and therefore quite difficult to detect experimentally.

We will recall now the relevant postulates and results that characterize the current understanding of strong interactions in quantum chromodynamics (QCD). The most important postulate is that the proper vacuum state in QCD is not the (trivial) perturbative state that we (naively) imagine to exist everywhere and which is little changed when the interactions are turned on/off. In QCD, the true vacuum state is believed to a have a complicated structure which originates in the glue (\lq photon') sector of the theory. The perturbative vacuum is an excited state with an energy density ${\cal B}$ above the true vacuum. It is to be found inside hadrons where perturbative quanta of the theory, in particular quarks, can therefore exist. The occurrence of the true vacuum state is intimately connected to the glue--glue interaction. Unlike QED, these massless quanta of QCD, also carry a charge -- color -- that is responsible for the quark--quark interaction.

In the above discussion, the confinement of quarks is a natural feature of the hypothetical structure of the true vacuum. If it is, for example, a color superconductor, then an isolated charge cannot occur. Another way to look at this is to realize that a single colored object would, according to Gauss' theorem, have an electric field that can only end on other color charges. In the region penetrated by this field, the true vacuum is displaced, thus effectively raising the mass of a quasi-isolated quark by the amount ${\cal B}V_{\rm field}$.

Another feature of the true vacuum is that it exercises a pressure on the surface of the region of the perturbative vacuum to which quarks are confined. Indeed, this is just the idea of the original MIT bag model \cite{3chap5bib6}. The Fermi pressure of almost massless light quarks is in equilibrium with the vacuum pressure ${\cal B}$. When many quarks are combined to form a giant quark bag, then their properties inside can be obtained using standard methods of many-body theory \cite{3chap5bib2a,3chap5bib20a,3chap5bib2b,3chap5bib2c,3chap5bib2d,3chap5bib2e,3chap5bib2f}. In particular, this also allows the inclusion of the effect of internal excitation through a finite temperature and through a change in the chemical composition.

A further effect that must be taken into consideration is the quark--quark interaction. We shall use here the first order contribution in the QCD running coupling constant $\alpha_{s}(q^2)=g^2/4\pi$. However, as $\alpha_{s}(q^2)$ increases when the average momentum exchanged between quarks decreases, this approach will have only limited validity at relatively low densities and/or temperatures. The collective screening effects in the plasma are of comparable order of magnitude and should reduce the importance of perturbative contributions as they seem to reduce the strength of the quark--quark interaction.

From this general description of the hadro\-nic plasma, it is immediately apparent that, at a certain value of temperature and baryon number density, the plasma must disintegrate into individual hadrons. Clearly, to treat this process and the ensuing further nucleonisation by perturbative QCD methods is impossible. It is necessary to find a semi-phenomeno\-logical method for the treatment of the thermodynamic system consisting of a gas of quark bags. The hadro\-nic gas phase is characterized by those reactions between individual hadrons that lead to the formation of new particles (quark bags) only. Thus one may view \cite{3chap5bib7,3chap5bib8,3chap5bib9} the hadro\-nic gas phase as being an assembly of many different hadro\-nic resonances, their number in the interval $(m^2,m^2+\D m^2)$ being given by the mass spectrum $\tau(m^2,b)\D m^2$. Here the baryon number $b$ is the only discrete quantum number to be considered at present. All bag--bag interaction is contained in the mutual transmutations from one state to another. Thus the gas phase has the characteristic of an infinite component ideal gas phase of extended objects. The quark bags having a finite size force us to formulate the theory of an extended, though otherwise ideal multicomponent gas.

It is a straightforward exercise, carried through in the beginning of the next section, to reduce the grand partition function $Z$ to an expression in terms of the mass spectrum $\tau(m^2,b)$. In principle, an experimental form of $\tau(m^2,b)$ could then be used as an input. However, the more natural way is to introduce the statistical bootstrap model \cite{3chap5bib7}, which will provide us with a theoretical $\tau$ that is consistent with assumptions and approximations made in determining $Z$.

In the statistical bootstrap, the essential step consists in the realization that a composite state of many quark bags is in itself an \lq elementary' bag \cite{3chap5bib1,3chap5bib10}. This leads directly to a nonlinear integral equation for $\tau$. The ideas of the statistical bootstrap have found a very successful application in the description of hadro\-nic reactions \cite{3chap5bib11} over the past decade. The present work is an extension \cite{3chap5bib1,3chap5bib9,3chap5bib12} and application \cite{3chap5bib1,3chap5bib13} of this method to the case of a system containing any number of finite size hadro\-nic clusters with their baryon numbers adding up to some fixed number. Among the most successful predictions of the statistical bootstrap, we record here the derivation of the limiting hadro\-nic temperature and the exponential growth of the mass spectrum.

We see that the theoretical description of the two hadro\-nic phases -- the individual hadron gas and the quark-gluon plasma -- is consistent with observations and with the present knowledge of elementary particles. What remains is the study of the possible phase transition between those phases as well as its observation. Unfortunately, we can argue that in the study of temperatures and mean transverse momenta of pions and nucleons produced in nuclear collisions, practically all information about the hot and dense phase of the collision is lost, as most of the emitted particles originate in the cooler and more dilute hadro\-nic {\it gas\/} phase of matter. In order to obtain reliable information on quark matter, we must presumably perform more specific experiments. We will briefly point out that the presence of numerous $\overline{s}$ quarks in the quark plasma suggest, as a characteristic experiment, the observation $\overline{\uLambda}$ hyperons.

We close this report by showing that, in nuclear collisions, unlike $pp$ reactions, we can use equilibrium thermodynamics in a large volume to compute the yield of strange and anti-strange particles. The latter, e.g., $\overline{\uLambda}$, might be significantly different from what one expects in $pp$ collisions and give a hint about the properties of the quark-gluon phase.

\subsection{Thermodynamics of the Gas Phase and the SBM}\label{3chap5sec2}

Given the grand partition function $Z(\beta,V,\lambda)$ of a many-body system, all thermodynamic quantities can be determined by differentiation of $\ln Z$ with respect to its arguments. Here, $\lambda$ is the fugacity introduced to conserve a discrete quantum number, here the baryon number. The conservation of strangeness can be carried through in a similar fashion leading then to a further argument $\lambda_{s}$ of $Z$. Whenever necessary, we will consider $Z$ to be implicitly dependent on $\lambda_{s}$.

The grand partition function is a Laplace transform of the level density $\sigma(p,V,b)$, where $p_\mu$ is the four-momentum and $b$ the baryon number of the many-body system enclosed in the volume $V\,$:
\begin{equation}\label{3chap5eq2}
Z(\beta,V,\lambda) = \sum_{b=-\infty}^\infty \lambda^b\int\sigma(p,V,b)\E^{-\beta_\mu p^\mu}\D^4 p\;.
\end{equation}
We recognize the usual relations for the thermodynamic expectation values of the baryon number,
\begin{subequations}
\begin{equation}\label{3chap5eq3a}
\langle b\rangle = \lambda\frac{\partial}{\partial\lambda} \ln Z(\beta,V,\lambda)\;,
\end{equation}
and the energy--momentum four-vector,
\begin{equation}\label{3chap5eq3b}
\langle p_\mu\rangle = -\frac{\partial}{\partial\beta_\mu}\ln Z(\beta,V,\lambda)\;,
\end{equation}
\end{subequations}
which follow from the definition in Eq.~(\ref{3chap5eq2}).

The theoretical problem is to determine $\sigma(p,V,b)$ in terms of known quantities. Let us suppose that the physical states of the hadro\-nic gas phase can be considered as being built up from an arbitrary number of massive objects, henceforth called clusters, characterized by a mass spectrum $\tau(m^2,b)$, where $\tau(m^2,b)\D m^2$ is the number of different elementary objects (existing in nature) in the mass interval $(m^2,m^2+\D m^2)$ and having the baryon number $b$. As particle creation must be permitted, the number $N$ of constituents is arbitrary, but constrained by four-momentum conservation and baryon conservation. Neglecting quantum statistics (it can be shown that, for $T\gtrsim 40$~MeV, Boltzmann statistics is sufficient), we have
\begin{equation}\label{3chap5eq4}
\begin{array}{ll}
\sigma(p,V,b)=&\displaystyle\sum_{N=0}^\infty\frac{1}{N!}\int\delta^4\big(p-\sum_{i=1}^Np_i\big)\times\\[0.2cm]
 &\displaystyle \sum_{\{b_i\}} \delta_k\big(b-\sum_{i=1}^Nb_i\big)\times\\[0.5cm]
 &\displaystyle \prod_{i=1}^N\displaystyle\frac{2\Delta_\mu p_i^\mu}{(2\pi)^3}\,\tau (p_i^2,b_i)\D^4p_i\;.
\end{array}
\end{equation}
The sum over all allowed partitions of $b$ into different $b_i$ is included and $\Delta$ is the volume available for the motion of the constituents, which differs from $V$ if the different clusters carry their proper volume $V_{{\rm c}i}\,$:
\begin{equation}\label{3chap5eq5}
\Delta^\mu = V^\mu-\sum_{i=1}^N V_{{\rm c}i}^\mu\;.
\end{equation}
The phase space volume used in Eq.~(\ref{3chap5eq4}) is best explained by considering what happens for one particle of mass $m_0$ in the rest frame of $\Delta_\mu$ and $\beta_\mu\,$:
\begin{equation}
\begin{array}{ll}
\displaystyle\!\!\int\!\!\D^4p_i\frac{2\Delta_\mu p_i^\mu}{(2\pi)^3}\,\E^{-\beta\cdot p}\delta_0(p_i^2-m^2)
&= \displaystyle\!\Delta_0\!\int\!\!\frac{\D^3p_i}{(2\pi)^3}\,\E^{-\beta_0\sqrt{p^2+m^2} }\\[0.5cm]
&=\displaystyle\Delta_0\displaystyle\frac{Tm^2}{2\pi^2} K_2(m/T)\;.\label{3chap5eq6}
\end{array}
\end{equation}
The density of states in Eq.~(\ref{3chap5eq4}) implies that the creation and absorption of particles in kinetic  and chemical equilibrium is limited only by four-momentum and baryon number conservation. These processes represent the strong hadro\-nic interactions which are dominated by particle productions. $\tau(m^2,b)$ contains all participating elementary particles and their resonances. Some remaining interaction is here neglected or, as we do not use the complete experimental $\tau$, it may be considered as being taken care of by a suitable choice of $\tau$. The short range repulsive forces are taken into account by the introduction of the proper volume $V$ of hadro\-nic clusters.

One more remark concerning the available volume $\Delta$ is in order here. If $V$ were considered to be given and an independent thermodynamic quantity, then in Eq.~(\ref{3chap5eq4}), a further built-in restriction limits the sum over $N$ to a certain $N_{\rm max}$, such that the available volume $\Delta$ in Eq.~(\ref{3chap5eq5}) remains positive. However, this more conventional assumption of $V$ as the independent variable would significantly obscure our mathematical formalism. It is important to realize that we are {\it free\/} to select the available volume $\Delta$ as the independent thermodynamic variable and to consider $V$ as a thermodynamic expectation value to be computed from Eq.~(\ref{3chap5eq5}):
\begin{equation}\label{3chap5eq7}
V^\mu\;\longrightarrow\;\langle V^\mu\rangle = \Delta^\mu+\big\langle V_{\rm c}^\mu(\beta,\Delta,\lambda)\big\rangle\;.
\end{equation}
Here $\langle V_{\rm c}^\mu\rangle$ is the average sum of proper volumes of all hadro\-nic clusters contained in the system considered. As already discussed, the standard quark bag leads to the proportionality between the cluster volume and hadron mass. Similar arguments within the bootstrap model \cite{3chap5bib9}, as for example discussed in the preceding   lecture  by R. Hagedorn~\cite{3chap5bib10}, also lead to
\begin{equation}\label{3chap5eq8}
\langle V_{\rm c}^\mu\rangle = \frac{\big\langle p^\mu(\beta,\Delta,\lambda)\big\rangle}{4{\cal B}}\;,
\end{equation}
where $4{\cal B}$ is the (at this point arbitrary) energy density of isolated hadrons in the quark bag model \cite{3chap5bib5}.

Since our hadrons are under pressure from neighbors in hadro\-nic matter, we have in principle to take instead of $4{\cal B}$ the energy density of a quark bag exposed to a pressure $P$ [see Eq.~(\ref{3chap5eq54}) below]
\[
{\varepsilon}_{\rm bag} = 4{\cal B}+3P\;.
\]
Combining Eqs.~(\ref{3chap5eq7})--(\ref{3chap5eq9}), we find, with ${\varepsilon}(\beta,\Delta,\lambda) = \langle p^\mu\rangle/\langle V^\mu\rangle = \langle E\rangle/\langle V\rangle$, that
\begin{equation}\label{3chap5eq9}
\frac{\Delta}{\langle V(\beta,\Delta,\lambda)\rangle} = 1-\frac{{\varepsilon}(\beta,\Delta,\lambda)}{4{\cal B}+3P(\beta,\Delta,\lambda)}\;.
\end{equation}
As we shall see, the pressure $P$ in the hadro\-nic matter never rises above $\simeq 0.4{\cal B}$, see \rf{3chap5fig5}a below, and arguments following \req{3chap5eq60}. Consequently, the inclusion of $P$ above -- the compression of free hadrons by the hadro\-nic matter by about 10\% -- may be omitted for now from further discussion. However, we note that both ${\varepsilon}$ and $P$ will be computed as $\ln Z$ becomes available, whence Eq.~(\ref{3chap5eq9}) is an implicit equation for $\Delta/\langle V\rangle$.

It is important to record that the expression in Eq.~(\ref{3chap5eq9}) can approach zero only when the energy density of the hadro\-nic gas approaches that of matter consisting of one big quark bag: ${\varepsilon}\rightarrow 4{\cal B}$, $P\rightarrow 0$. Thus the density of states in Eq.~(\ref{3chap5eq4}), together with the choice of $\Delta$ as a thermodynamic variable, is a consistent physical choice only up to this point. Beyond we assume that a description in terms of interacting quarks and gluons is the proper physical description. Bearing all these remarks in mind, we now consider the available volume $\Delta$ as a thermodynamic variable which by definition is positive. Inspecting Eq.~(\ref{3chap5eq4}) again, we recognize that the level density of the extended objects in volume $\langle V\rangle$ can be interpreted for the time being as the level density of point particles in a fictitious volume $\Delta\,$:
\begin{equation}\label{3chap5eq10}
\sigma(p,V,b)=\sigma_{\rm pt}(p,\Delta,b)\;,
\end{equation}
whence this is also true for the grand canonical partition function in Eq.~(\ref{3chap5eq2}):
\begin{equation}\label{3chap5eq11}
Z(\beta,V,\lambda) = Z_{\rm pt}(\beta,\Delta,\lambda)\;.
\end{equation}
Combining Eqs.~(\ref{3chap5eq2}) and (\ref{3chap5eq4}), we also find the important relation
\begin{equation}\label{3chap5eq12}
\ln Z_{\rm pt}(\beta,\Delta,\lambda)=\sum_{b=-\infty}^\infty\lambda^b\int\frac{2\Delta_\mu p^\mu}{(2\pi)^3}\tau(p^2,b)\E^{-\beta_\mu p^\mu}\D^4p\;.
\end{equation}
This result can only be derived when the sum over $N$ in Eq.~(\ref{3chap5eq4}) extends to infinity, thus as long as $\Delta/\langle V\rangle$ in Eq.~(\ref{3chap5eq9}) remains positive.

In order to continue with our description of hadro\-nic matter, we must now determine a suitable mass spectrum $\tau$ to be inserted into Eq.~(\ref{3chap5eq4}). For this we now introduce the statistical bootstrap model.
The basic idea is rather old, but has undergone some development more recently making it clearer, more consistent, and perhaps more convincing. The details may be found in \cite{3chap5bib9} and the references therein. Here a simplified naive presentation is given. We note, however, that our present interpretation is non-trivially different from that in \cite{3chap5bib9}.

The basic postulate of statistical bootstrap is that the mass spectrum $\tau(m^2,b)$ containing all the \lq particles', i.e., elementary, bound states, and resonances (clusters), is generated by the {\it same\/} interactions which we see at work if we consider our thermodynamical system. Therefore, if we were to compress this system until it reaches its natural volume $V_{\rm c}(m,b)$, then it would itself be almost a cluster appearing in the mass spectrum $\tau(m^2,b)$. Since $\sigma(p,\Delta,b)$ and $\tau(p^2,b)$ are both densities of states (with respect to the different parameters $\D^4p$ and $\D m^2$), we postulate that
\begin{equation}\label{3chap5eq13}
\sigma(p,\Delta,b)\Big|_{\langle V\rangle\underset{\Delta\rightarrow 0}{\;\longrightarrow\;}V_{\rm c}(m,b)} \;\;\widehat{=}\;\; {\rm const.}\times\tau(p^2,b)\;,
\end{equation}
where $\widehat{=}$ means \lq corresponds to' (in some way to be specified). As $\sigma(p,\Delta,b)$ is [see Eq.~(\ref{3chap5eq4})] the sum over $N$ of $N$-fold convolutions of $\tau$, the above \lq bootstrap postulate' will yield a highly nonlinear integral equation for $\tau$.

The bootstrap postulate (\ref{3chap5eq13}) requires that $\tau$ should obey the equation resulting from replacing $\sigma$ in Eq.~(\ref{3chap5eq4}) by some expression containing $\tau$ linearly and by taking into account the volume condition expressed in Eqs.~(\ref{3chap5eq7}) and (\ref{3chap5eq8}).

We cannot simply put $V=V_{\rm c}$ and $\Delta=0$, because now, when each cluster carries its own dynamically determined volume, $\Delta$ loses its original meaning and must be redefined more precisely. Therefore, in Eq.~(\ref{3chap5eq4}), we tentatively replace
\begin{equation}\label{3chap5eq14}
\begin{array}{rl}
\displaystyle \sigma(p,V_{\rm c},b)\;\longrightarrow\;& \displaystyle\frac{2V_{\rm c}(m,b)\cdot p}{(2\pi)^3} \tau(p^2,b)\\[6mm]
&= \displaystyle\frac{2m^2}{(2\pi)^34{\cal B}}\tau(p^2,b)\;,\\[6mm]
\displaystyle \frac{2\Delta\cdot p_i}{(2\pi)^3}\tau(p_i^2,b_i)\;\longrightarrow\;& \displaystyle\frac{2V_{\rm c}(m_i,b_i)\cdot p_i}{(2\pi)^3} \tau(p_i^2,b_i)\\[6mm]
&=\displaystyle \frac{2m_i^2}{(2\pi)^34{\cal B}}\tau(p_i^2,b_i)\;.
\end{array}
\end{equation}
Next we argue that the explicit factors $m^2$ and $m_i^2$ arise from the dynamics and therefore must be absorbed into $\tau(p_i^2,b_i)$ as dimensionless factors\footnote{Here is the essential difference with \cite{3chap5bib9}, where another choice was made.} $m_i^2/m_0^2$. Thus,
\begin{equation}\label{3chap5eq15}
\begin{array}{rl}
\displaystyle \sigma(p,V_{\rm c},b)\;\longrightarrow\;&\displaystyle  \frac{2m_0^2}{(2\pi)^3 4{\cal B}} \tau(p^2,b) = H\tau(p^2,b)\;,\\[6mm]
\displaystyle \frac{2\Delta\cdot p_i}{(2\pi)^3}\tau(p_i^2,b_i)\;\longrightarrow\;& \displaystyle \frac{2m_0^2}{(2\pi)^3 4{\cal B}} \tau(p_i^2,b_i) = H\tau(p_i^2,b_i)\;,
\end{array}
\end{equation}
with 
\[
H:=\frac{2m_0^2}{(2\pi)^34{\cal B}}\;,
\]
where either $H$ or $m_0$ may be taken as a new free parameter of the model, to be fixed later. (If $m_0$ is taken, then it should be of the order of the \lq elementary masses' appearing in the system, e.g., somwhere between $m_\pi$ and $m_{\rm N}$ in a model using pions and nucleons as elementary input.) Finally, if clusters consist of clusters which consist of clusters, and so on, this should end at some \lq elementary' particles (where what we consider as elementary is fixed by convention). Inserting Eq.~(\ref{3chap5eq15}) into Eq.~(\ref{3chap5eq4}), the bootstrap equation (BE) then reads
\begin{eqnarray}\label{3chap5eq16}
&&H\tau(p^2,b) = Hg_b\delta_0(p^2-\overline{m}{}_b^2)+\\\noalign{\smallskip} \nonumber
  && +\sum_{N=2}^\infty\frac{1}{N!}\int\!\delta^4\!\bigg(p-\sum_{i=1}^Np_i\bigg)\sum_{\{b_i\}}\delta_k\bigg(b-\sum_{i=1}^Nb_i\bigg)\times \\\noalign{\smallskip}
 && \hspace*{4.5cm} \times \prod_{i=1}^N H\tau(p_i^2,b_i)\D^4p_i\;.\nonumber
\end{eqnarray}
Clearly, the bootstrap equation (\ref{3chap5eq16}) has not been derived. We have made it more or less plausible and state it as a postulate. For more motivation, see \cite{3chap5bib9}. In other words, the bootstrap equation means that the cluster with mass $\sqrt{p^2}$ and baryon number $b$ is either elementary (mass $\overline{m}{}_b$, spin isospin multiplicity $g_b$), or it is composed of any number $N\geq 2$ of subclusters having the same internal composite structure described by this equation. The bar over $\overline{m}{}_b$ indicates that one has to take the mass which the \lq elementary particle' will have effectively when present in a large cluster, e.g., in nuclear matter, $\overline{m}=m-\langle E_{\rm bind}\rangle$, and $\overline{m}{}_{\rm N}\approx 925$~MeV. That this must be so becomes obvious if one imagines Eq.~(\ref{3chap5eq16}) solved by iteration (the iteration solution exists and is the physical solution). Then $H\tau(p^2,b)$ becomes in the end a complicated function of $p^2$, $b$, all $\overline{m}{}_b$, and all $g_b$. In other words, in the end a single cluster consists of the \lq elementary particles'. As these are all bound into the cluster, their mass $\overline{m}$ should be the effective mass, not the free mass $m$. This way we may include a small correction for the long-range attractive meson exchange by choosing $\overline{m}{}_{\rm N}= m-15$~MeV.

Let us make a brief excursion to the bag model at this point. There the mass of a hadron is computed from the assumption of an {\it isolated\/} particle (= bag) with its size and mass being determined from the equilibrium between the vacuum pressure ${\cal B}$ and the internal Fermi pressure of the (valence) quarks. In a hadron gas, this is not true as a finite pressure is exerted on hadrons in matter. After a short calculation, we find the pressure dependence of the bag model hadro\-nic mass:
\begin{equation}\label{3chap5eq17}
\frac{M(P)}{M(0)}=\frac{1+3P/4{\cal B}}{(1+P/{\cal B})^{3/4}} 
=  \left[1+\frac{3}{32}\left(\frac{P}{\cal B}\right)^2+\cdots\right]\;.
\end{equation}
We have already noted that the pressure never exceeds $0.4{\cal B}$ in the hadro\-nic gas phase, see \rf{3chap5fig5}a below, and arguments following \req{3chap5eq60}. Hence we see that the increase in mass of constituents (quark bags) in the hadro\-nic gas never exceeds 1.5\% and is at most comparable with the 15~MeV binding in $\overline{m}$. In general, $P$ is about $0.1B$ and the pressure effect may be neglected.

Thus we can consider the \lq input' first term in Eq.~(\ref{3chap5eq16}) as being fixed by pions, nucleons, and whenever necessary by the usual strange members of meson and baryon multiplets. Furthermore, we note that the bootstrap equation (\ref{3chap5eq16}) makes use of practically all the same approximations as our description of the level density in Eq.~(\ref{3chap5eq4}). Thus the solution of Eq.~(\ref{3chap5eq16}) is particularly suitable for our use.

We solve the BE by the same double Laplace transformation which we used before Eq.~(\ref{3chap5eq2}). We define
\begin{eqnarray}
\varphi(\beta,\lambda) &:=& 
\!\int\E^{-\beta_\mu p^\mu}\!\!\sum_{b=-\infty}^\infty\!\!\lambda^b Hg_b\delta_0(p^2-\overline{m}{}_b^2)\D^4p\nonumber \\ \nonumber
&=&2\pi HT\!\!\sum_{b=-\infty}^\infty\!\!\lambda^b g_b\overline{m}{}_b K_1(\overline{m}{}_b/T)\;,\nonumber\\\noalign{\smallskip}
\Phi(\beta,\lambda) &:=& 
\!\int\E^{-\beta_\mu p^\mu}\!\!\sum_{b=-\infty}^\infty\!\!\lambda^bH\tau(p^2,b)\D^4p\;.
\label{3chap5eq18}
\end{eqnarray}
Once the set of input particles $\{\overline{m}{}_b,g_b\}$ is given, $\varphi(\beta,\lambda)$ is a known function, while $\Phi(\beta,\lambda)$ is unknown. Applying the double Laplace transformation to the BE, we obtain
\begin{equation}\label{3chap5eq19}
\Phi(\beta,\lambda)=\varphi(\beta,\lambda)+\exp\Phi(\beta,\lambda)-\Phi(\beta,\lambda)-1\;.
\end{equation}
This implicit equation for $\Phi$ in terms of $\varphi$ can be solved without regard for the actual $\beta,\lambda$ dependence. Writing
\begin{equation}\label{3chap5eq20}
G(\varphi) := \Phi(\beta,\lambda)\;,\qquad \varphi = 2G-\E^G+1\;,
\end{equation}
we can draw the curve $\varphi(G)$  and then invert it graphically  (see \rf{3chap5fig1})  to obtain $G(\varphi)=\Phi(\beta,\lambda)$. $G(\varphi)$ has a square root singularity at $\varphi=\varphi_0=\ln(4/\E)=0.3863$. Beyond this value, $G(\varphi)$ becomes complex. 
Apart from this graphical solution, other forms of solution are known:
\begin{equation}\label{3chap5eq21}
G(\varphi) = \sum_{n=1}^\infty s_n\varphi^n = \sum_{n=0}^\infty w_n(\varphi_0-\varphi)^{n/2} = \stackrel{\mbox{integral}}{\mbox{representation}}\;.
\end{equation}
The expansion in terms of $(\varphi_0-\varphi)^{n/2}$ has been used in our numerical work (12 terms yield a solution within computer accuracy) and the integral representation will be published elsewhere\footnote{Extensive discussion of the analytical properties of the bootstrap function was publisched in: R.~Hagedorn and J.~Rafelski:
Analytic Structure and Explicit Solution of an Important Implicit Equation, 
  Commun.\ Math.\ Phys.\  {\bf 83}, (1982) 563}.
\begin{figure}
\centering\resizebox{0.42\textwidth}{!}{%
\includegraphics{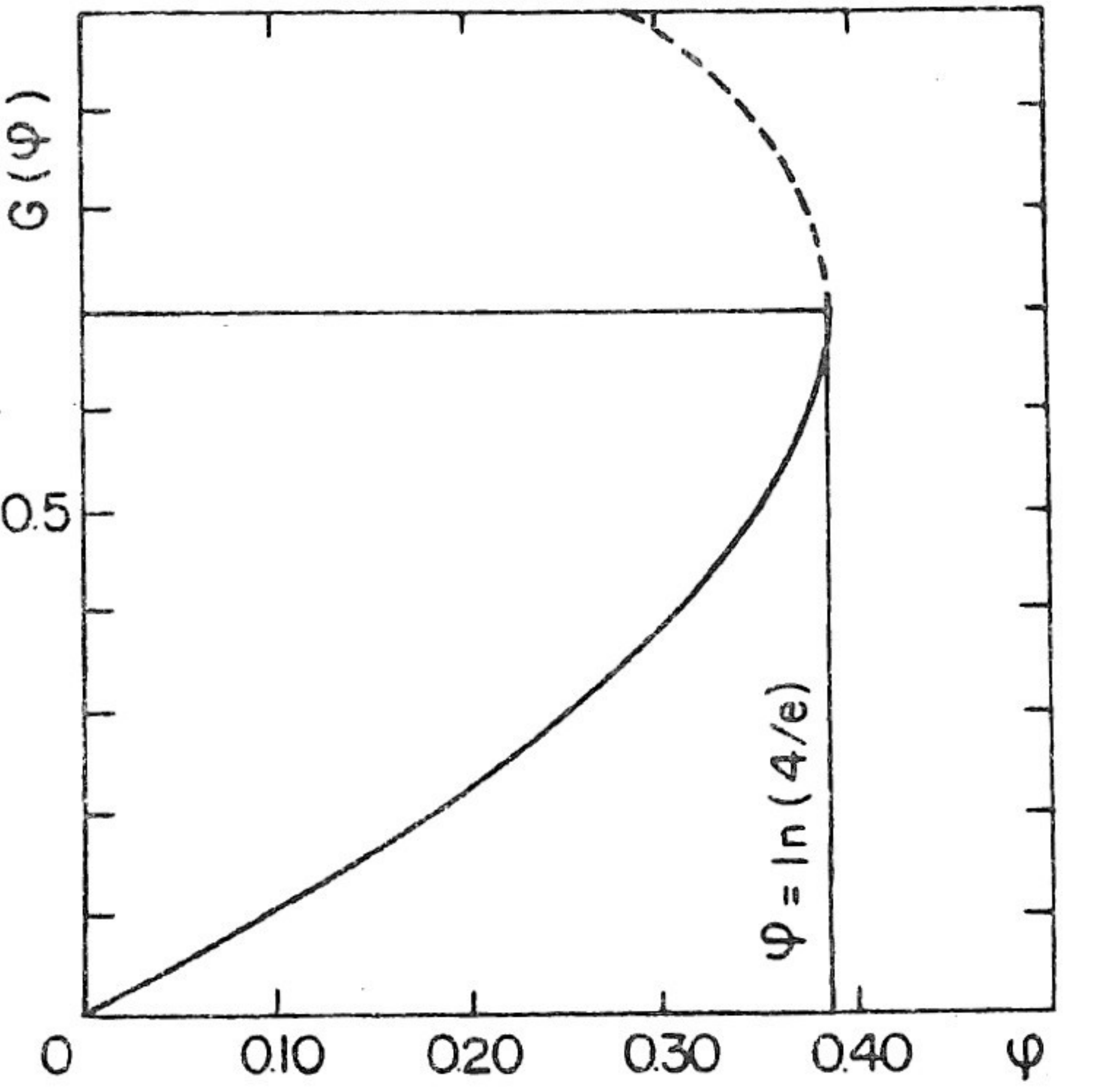}
}
\caption[]{Bootstrap function $G(\varphi)$. The {\it dashed line\/} represents the unphysical branch. The root singularity is at $\varphi_0=\ln(4/\E)=0.3863$}\label{3chap5fig1}
\end{figure}
Henceforth, we consider $\Phi(\beta,\lambda)=G(\varphi)$ to be a known function of $\varphi(\beta,\lambda)$. Consequently, $\tau(m^2,b)$ is also in principle known. From the singularity at $\varphi=\varphi_0$, it follows \cite{3chap5bib1} that $\tau(m^2,b)$ grows, for $m\gg m_{\rm N}b$, exponentially $\sim m^{-3}\exp(m/T_0)$. In some weaker form, this has been known for a long time \cite{3chap5bib7,3chap5bib15,3chap5bib16}.

\subsection{The Hot Hadronic Gas}\label{3chap5sec3}

The definition of $\Phi(\beta,\lambda)$ in Eq.~(\ref{3chap5eq18}) in terms of the mass spectrum allows us to write a very simple expression for $\ln Z$ in the gas phase (passing now to the rest frame of the gas):
\begin{equation}\label{3chap5eq22}
\ln Z(\beta,V,\lambda)=\ln Z_{\rm pt}(\beta,\Delta,\lambda) = -\frac{2\Delta}{(2\pi)^3H}\frac{\partial}{\partial\beta}\Phi(\beta,\lambda)\;.
\end{equation}
We recall that Eqs.~(\ref{3chap5eq9}) and (\ref{3chap5eq19}) define (implicitly) the quantities $\Delta$ and $\Phi$ in terms of the physical variables $V$, $\beta$, and $\lambda$.

Let us now introduce the energy density ${\varepsilon}_{\rm pt}$ of the hypothetical pointlike particles as
\begin{equation}\label{3chap5eq23}
{\varepsilon}_{\rm pt}(\beta,\lambda) =  -\frac{\partial}{\Delta\partial\beta}\ln Z_{\rm pt}(\beta,\Delta, \lambda)  = \frac{2}{(2\pi)^3H}\frac{\partial^2}{\partial\beta^2}\Phi(\beta,\lambda)\;,
\end{equation}
which will turn out to be quite helpful as it is independent of $\Delta$. The proper energy density is
\begin{equation}\label{3chap5eq24}
{\varepsilon}(\beta,\lambda)=\frac{1}{\langle V\rangle}\left(-\frac{\partial}{\partial\beta}\ln Z\right)=\frac{\Delta}{\langle V\rangle}{\varepsilon}_{\rm pt}\;,
\end{equation}
while the pressure follows from
\begin{equation}\label{3chap5eq25}
P(\beta,\lambda)\langle V\rangle = T\ln Z(\beta,V,\lambda)=T\ln Z_{\rm pt}(\beta,\Delta,\lambda)\;,
\end{equation}
\begin{equation}\label{3chap5eq26}
P(\beta,\lambda)= \frac{\Delta}{\langle V\rangle}\left[-\frac{2T}{(2\pi)^3H}\frac{\partial}{\partial\beta} \Phi(\beta,\lambda)\right] =:\frac{\Delta}{\langle V\rangle}P_{\rm pt}\;.
\end{equation}
Similarly, for the baryon number density, we find
\begin{equation}\label{3chap5eq27}
\nu(\beta,\lambda)= \frac{\langle b\rangle}{\langle V\rangle} =: \frac{\Delta}{\langle V\rangle} \nu_{\rm pt}(\beta,\lambda)\;,
\end{equation}
with 
\begin{equation}\label{3chap5eq28}
\nu_{\rm pt}(\beta,\lambda)= \frac{1}{\Delta} \lambda\frac{\partial}{\partial\lambda}\ln Z_{\rm pt} = -\frac{2}{(2\pi)^3H}\lambda\frac{\partial}{\partial\lambda} \frac{\partial}{\partial\beta}\Phi(\beta,\lambda)\;.
\end{equation}
From Eqs.~(\ref{3chap5eq23})--(\ref{3chap5eq23}), the crucial role played by the factor $\Delta/\langle V\rangle$ becomes apparent. We note that it is quite straightforward to insert Eqs.~(\ref{3chap5eq24}) and (\ref{3chap5eq25}) into Eq.~(\ref{3chap5eq9}) and solve the resulting quadratic equation to obtain $\Delta/\langle V\rangle$ as an explicit function of ${\varepsilon}_{\rm pt}$ and $P_{\rm pt}$. First we record the limit $P\ll B\,$:
\begin{equation}\label{3chap5eq29}
\frac{\Delta}{\langle V\rangle} = 1-\frac{{\varepsilon}(\beta,\lambda)}{4{\cal B}} = \left[1+\frac{{\varepsilon}_{\rm pt}(\beta,\lambda)}{4{\cal B}} \right]^{-1}\;,
\end{equation}
while the correct expression is
\begin{equation}\label{3chap5eq30}
\frac{\Delta}{\langle V\rangle} = \frac{1}{2} -\frac{{\varepsilon}_{\rm pt}}{6P_{\rm pt}} -\frac{2{\cal B}}{3P_{\rm pt}} + \sqrt{ \frac{4{\cal B}}{3P_{\rm pt}} + \left(\frac{1}{2} -\frac{{\varepsilon}_{\rm pt}}{6P_{\rm pt}} -\frac{2{\cal B}}{3P_{\rm pt}}  \right)^2}\;.
\end{equation}
The last of the important thermodynamic quantities is the entropy $S$. By differentiating Eq.~(\ref{3chap5eq25}), we find
\begin{equation}\label{3chap5eq31}
\frac{\partial}{\partial\beta}\ln Z = \frac{\partial}{\partial\beta}\beta P\langle V\rangle = P\langle V\rangle - T\frac{\partial}{\partial T}(P\langle V\rangle)\;.
\end{equation}
Considering $Z$ as a function of the chemical potential, viz.,
\begin{equation}\label{3chap5eq32}
Z(\beta,V,\lambda) = Z(\beta,V,\E^{\mu\beta}) = \tilde{Z}(\beta,V,\mu)= \tilde{Z}_{\rm pt}(\beta,\Delta,\mu)\;,
\end{equation}
we find
\begin{equation}\label{3chap5eq33}
\frac{\partial}{\partial\beta}\ln Z\bigg|_{\mu,\Delta} = \frac{\partial}{\partial\beta}\ln \tilde{Z}_{\rm pt}(\beta,\Delta,\mu)=-E+\mu\langle b\rangle\;,
\end{equation}
with $E$ being the total energy. From Eqs.~(\ref{3chap5eq31}) and (\ref{3chap5eq33}), we find the \lq first law' of thermodynamics to be
\begin{subequations}
\begin{equation}\label{3chap5eq34a}
E=-P\langle V\rangle +T\frac{\partial}{\partial T}(P\langle V\rangle) + \mu\langle b\rangle\;.
\end{equation}
Now quite generally,
\begin{equation}\label{3chap5eq34b}
E=-P\langle V\rangle +TS+ \mu\langle b\rangle\;,
\end{equation}
\end{subequations}
so that
\begin{equation}\label{3chap5eq35}
S= \frac{\partial}{\partial T}\Big[P(\beta,\Delta,\mu)\langle V(\beta,\Delta,\mu)\rangle \Big]\Big|_{\mu,\Delta}\;.
\end{equation}
Equations (\ref{3chap5eq25}) and (\ref{3chap5eq33}) now allow us to write
\begin{equation}\label{3chap5eq36}
S= \frac{\partial}{\partial T}(P\langle V\rangle) = \ln\tilde{Z}_{\rm pt}(T,\Delta,\mu)+\frac{E-\mu b}{T}\;.
\end{equation}
The entropy density in terms of the already defined quantities is therefore
\begin{equation}\label{3chap5eq37}
{\cal S}=\frac{S}{\langle V\rangle}=\frac{P+{\varepsilon}-\mu \nu}{T}\;.
\end{equation}
We shall now take a brief look at the quantities $P$, ${\varepsilon}$, $\nu$, $\Delta/\langle V\rangle$. They can be written in terms of $\partial \Phi(\beta,\lambda)/\partial\beta$ and its derivatives. We note that [see Eq.~(\ref{3chap5eq20})]
\begin{equation}\label{3chap5eq38}
\frac{\partial}{\partial\beta}\Phi(\beta,\lambda)=\frac{\partial G(\varphi)}{\partial\varphi}\frac{\partial\varphi}{\partial\beta}\;,
\end{equation}
and that $\partial G/\partial\varphi\sim (\varphi_0-\varphi)^{-1/2}$ near to $\varphi=\varphi_0=\ln (4/\E)$ (see  \rf{3chap5fig1}).
 Hence at $\varphi=\varphi_0$, we find a singularity in the point particle quantities ${\varepsilon}_{\rm pt}$, $\nu_{\rm pt}$, and $P_{\rm pt}$. This implies that all hadrons have coalesced into one large cluster. Indeed, from Eqs.~(\ref{3chap5eq24}), (\ref{3chap5eq26}), (\ref{3chap5eq27}), and (\ref{3chap5eq29}), we find
\begin{equation}\label{3chap5eq39}
\begin{array}{r c l}
{\varepsilon} & \longrightarrow & 4{\cal B}\;,\\\noalign{\smallskip}
P & \longrightarrow & 0\;,\\\noalign{\smallskip}
\Delta/\langle V\rangle & \longrightarrow & 0\;.
\end{array}
\end{equation}
We can easily verify that this is correct by establishing the average number of clusters present in the hadro\-nic gas. This is done by introducing an artificial fugacity $\xi^N$ in Eq.~(\ref{3chap5eq4}) in the sum over $N$, where $N$ is the number of clusters. Denoting by $Z(\xi)$ the associated grand canonical partition functions in Eq.~(\ref{3chap5eq22}), we find
\begin{equation}\label{3chap5eq40}
\langle N\rangle = \xi\frac{\partial}{\partial\xi}\ln Z_{\rm pt}^\xi(\beta,\Delta,\lambda;\xi)\bigg|_{\xi=1} = -\frac{2\Delta}{(2\pi)^3H}\frac{\partial}{\partial\beta}\Phi(\beta,\lambda)\;,
\end{equation}
which leads to the useful relation 
\begin{equation}\label{3chap5eq41}
P\langle V\rangle = \langle N\rangle T\;.
\end{equation}
Thus as $P\langle V\rangle\rightarrow 0$, so must $\langle N\rangle$, the number of clusters, for finite $T$. We record the astonishing fact that the hadron gas phase obeys an \lq ideal' gas equation, although of course $\langle N\rangle$ is not constant as for a real ideal gas but a function of the thermodynamic variables.

The boundary given by 
\begin{equation}\label{3chap5eq42}
\varphi(\beta,\lambda)=\varphi_0=\ln(4/\E)
\end{equation}
thus defines a critical curve in the $\beta,\lambda$ plane. Its position depends, of course, on the actually given form of $\varphi(\beta,\lambda)$, i.e., on the set of \lq input' particles $\{\overline{m}{}_b,g_b\}$ assumed and the value of the constant $H$ in Eq.~(\ref{3chap5eq15}). In the case of three elementary pions $\pi^+$, $\pi^0$, and $\pi^-$ and four elementary nucleons (spin $\otimes$ isospin) and four antinucleons, we have from Eq.~(\ref{3chap5eq18})

\begin{figure}
\centering\resizebox{0.45\textwidth}{!}{%
\includegraphics{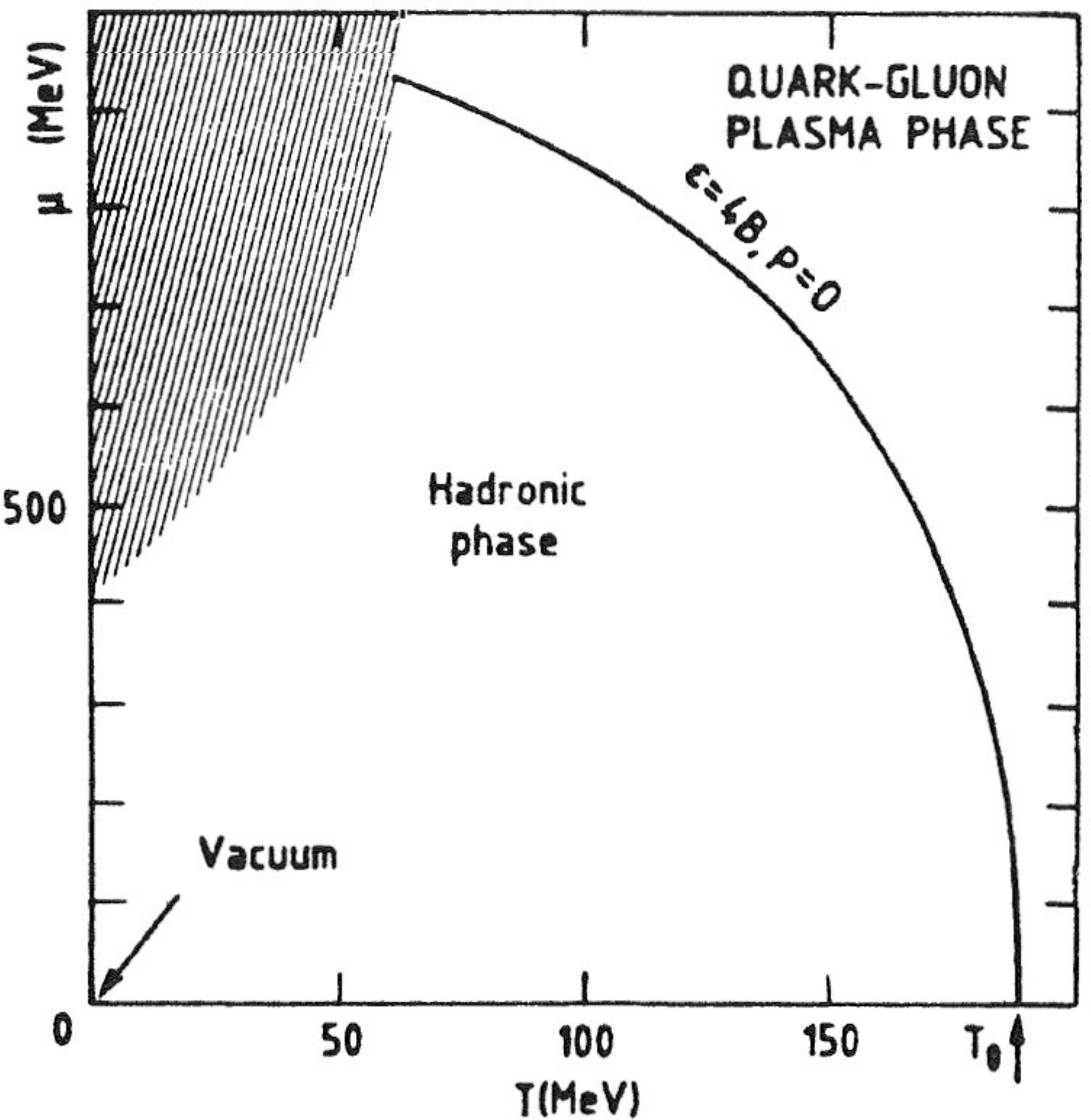}
}
\caption[]{The critical curve corresponding to $\varphi(T,\mu)=\varphi_0$ in the $\mu,T$ plane. Beyond it, the usual hadro\-nic world ceases to exist. In the {\it shaded region\/}, our theory is not valid, because we neglected Bose-Einstein and Fermi-Dirac statistics}\label{3chap5fig2}
\end{figure}
\begin{subequations}
\begin{equation}\label{3chap5eq43a}
\begin{array}{rl}
\varphi(\beta,\lambda)=2\pi HT &\left[3m_\pi K_1(m_\pi/T)+\right.\\[0.3cm]
&\left.+4\left(\lambda+\frac{1}{\lambda}\right) \overline{m}{}_{\rm N} 
K_1(\overline{m}{}_{\rm N}/T)\right]\;,
\end{array}
\end{equation}
and the condition (\ref{3chap5eq42}), written in terms of $T$ and $\mu=T\ln \lambda$, yields the curve shown in  \rf{3chap5fig2},  i.e., the \lq critical curve'.
For $\mu=0$, the curve ends at $T=T_0$, where $T_0$, the \lq limiting temperature of hadro\-nic matter', is the same as that appearing in the mass spectrum \cite{3chap5bib7,3chap5bib9,3chap5bib15,3chap5bib16} $\tau(m^2,b)\sim m^{-3}\exp(m/T_0)$ (for $b\gg bm_{\rm N}$).

The value of the constant $H$ in Eq.~(\ref{3chap5eq15}) has been chosen \cite{3chap5bib13} to yield $T_0=190$~MeV. This apparently large value of $T_0$ seemed necessary to yield a maximal average decay temperature of the order of 145~MeV, as required by \cite{3chap5bib17}. (However, a new value of the bag constant then induces a change \cite{3chap5bib1} to a lower value of $T_0=180$~MeV.) Here we use
\begin{equation}\label{3chap5eq43b}
\begin{array}{r c l l}
H &=& 0.724~{\rm GeV}^{-2}\;,\quad \quad   T_0&=0.19~{\rm GeV}\;,\\
\noalign{\smallskip}
m_0 &=& 0.398~{\rm GeV}\quad   [\mathrm{when\ } {\cal B}&=(145~\mathrm{MeV})^4 ]\;,
\end{array}
\end{equation}
\end{subequations}
where the value of $m_0$ lies as expected between $m_\pi$ and $m_{\rm N}$ [$(m_\pi m_{\rm N})^{1/2}=0.36$~GeV].

The critical curve limits the hadron gas phase. By approaching it, all hadrons dissolve into a giant cluster, which is not in our opinion a hadron solid \cite{3chap5bib14}. We would prefer to identify it with a quark-gluon plasma. Indeed, as the energy density along the critical curve is constant ($=4{\cal B}$), the critical curve can be attained and, if the energy density becomes $>4{\cal B}$, we enter into a region which cannot be described without making assumptions about the inner structure and dynamics of the \lq elementary particles' $\{\overline{m}{}_b,g_b\}$ -- here pions and nucleons -- entering into the input function $\varphi(\beta,\lambda)$. Considering pions and nucleons as quark-gluon bags leads naturally to this interpretation.

\subsection{The Quark--Gluon Phase}\label{3chap5sec4}

We now turn to the discussion of the region of the strongly interacting matter in which the energy density would be equal to or higher than $4{\cal B}$. As a basic postulate, we will assume that it consists of -- relatively weakly -- interacting quarks. To begin with, only u and d flavors will be considered as they can easily be copiously produced at $T\gtrsim 50$~MeV. Again the aim is to derive the grand partition function $Z$. This is a standard exercise. For the massless quark Fermi gas up to first order in the interaction \cite{3chap5bib1,3chap5bib2a,3chap5bib20a,3chap5bib2b,3chap5bib2c,3chap5bib2d,3chap5bib2e,3chap5bib2f,3chap5bib12}, the result is
\begin{equation}\label{3chap5eq44}
\begin{array}{rl}
\ln Z_{q}(\beta,\lambda) =\displaystyle \frac{8V}{6\pi^2\beta^{3}}& \displaystyle \!\!\Bigg[\!\left(1-\frac{2\alpha_{s}}{\pi}\right)\! \left(\frac{1}{4}\ln^4\!\!\lambda_{q} + \frac{\pi^2}{2}\ln^2\!\!\lambda_{q}\!\right)\! +\\[0.5cm]
&\ +\displaystyle \left(1-\frac{50}{21}\frac{\alpha_{s}}{\pi}\right)\frac{7\pi^4}{60}\Bigg]\;,
\end{array}
\end{equation}
valid in the limit $m_{q}<T\ln\lambda_{q}$.

Here $g=(2s+1)(2I+1)C=12$ counts the number of the components of the quark gas, and $\lambda_{q}$ is the fugacity related to the quark number. As each quark has baryon number 1/3, we find
\begin{equation}\label{3chap5eq45}
\lambda_{q}^3=\lambda = \E^{\mu/T}\;,
\end{equation}
where as before $\lambda$ allows for conservation of the baryon number. Consequently,
\begin{equation}\label{3chap5eq46}
3\mu_{q} = \mu\;.
\end{equation}
The glue contribution is
\begin{equation}\label{3chap5eq47}
\ln Z_{\rm g}(\beta,\lambda)=V\frac{8\pi^2}{45}\beta^{-3}\left(1-\frac{15}{4}\frac{\alpha_{s}}{\pi}\right)\;.
\end{equation}
We notice the two relevant differences with the photon gas:
\begin{itemize}
\item The occurence of the factor eight associated with the number of gluons.
\item The glue--glue interaction as gluons carry color charge.
\end{itemize}
Finally, let us introduce the vacuum term, which accounts for the fact that the perturbative vacuum is an excited state of the \lq true' vacuum which has been renormalized to have a vanishing thermodynamic potential, $\Omega=-\beta^{-1}\ln Z$. Hence in the perturbative vacuum,
\begin{equation}\label{3chap5eq48}
\ln Z_{\rm vac}=-\beta {\cal B}V\;.
\end{equation}
This leads to the required positive energy density ${\cal B}$ within the volume occupied by the colored quarks and gluons and to a negative pressure on the surface of this region. At this stage, this term is entirely phenomeno\-logical, as discussed above. The equations of state for the quark-gluon plasma are easily obtained by differentiating
\begin{equation}\label{3chap5eq49}
\ln Z = \ln Z_{q} + \ln Z_{\rm g} + \ln Z_{\rm vac}
\end{equation}
with respect to $\beta$, $\lambda$, and $V$. The baryon number density,  energy, and pressure are respectively:
\begin{equation}\label{3chap5eq51}
\nu = \frac{1}{V}\lambda\frac{\partial}{\partial \lambda}\ln Z = \frac{2T^3}{\pi^2}\left(1-\frac{2\alpha_{s}}{\pi}\right)\left(\frac{1}{3^4}\ln^3\lambda + \frac{\pi^2}{9}\ln\lambda\right)\;,
\end{equation}
\begin{eqnarray}
{\varepsilon} &=& -\frac{1}{V}\frac{\partial}{\partial \beta}\ln Z= {\cal B}\nonumber\\\noalign{\smallskip}
  &+& \frac{6}{\pi^2}T^4\Bigg[\left(1-\frac{2\alpha_{s}}{\pi}\right)\left(\frac{1}{4\cdot 3^4}\ln^4\lambda + \frac{\pi^2}{2\cdot 3^2}\ln^2\lambda\right)\nonumber\\\noalign{\smallskip}
    &+& \left(\!1-\frac{50}{21}\frac{\alpha_{s}}{\pi}\!\right)\frac{7\pi^4}{60}\Bigg] + \frac{8\pi^2}{15}T^4\left(\!1-\frac{15}{4}\frac{\alpha_{s}}{\pi}\!\right)\;,\label{3chap5eq50}
\end{eqnarray}
\begin{eqnarray}
P &=& T\frac{\partial}{\partial V}\ln Z=-{\cal B}\nonumber\\\noalign{\smallskip}
  &+& \frac{2T^4}{\pi^2}\Bigg[\left(1-\frac{2\alpha_{s}}{\pi}\right)\left(\frac{1}{4\cdot 3^4}\ln^4\lambda + \frac{\pi^2}{2\cdot 3^2}\ln^2\lambda\right)\nonumber\\\noalign{\smallskip}
    &+&\left(\!1-\frac{50}{21}\frac{\alpha_{s}}{\pi}\!\right)\frac{7\pi^4}{60}\Bigg] + \frac{8\pi^2}{45}T^4\left(\!1-\frac{15}{4}\frac{\alpha_{s}}{\pi}\!\right)\;.\label{3chap5eq52}
\end{eqnarray}
Let us first note that, for $T\ll \mu$ and $P=0$, the baryon chemical potential tends to
\begin{equation}\label{3chap5eq53}
\begin{array}{rl}
\mu_{\rm B}=3\mu_{q}\;\longrightarrow\;&\displaystyle 3{\cal B}^{1/4}\left[\frac{2\pi^2}{(1-2\alpha_{s}/\pi)}\right]^{1/4} \hspace{-8pt}= 1010~{\rm MeV}\,,\\[0.4cm]
& [\alpha_{s}=1/2\,,\;\; {\cal B}^{1/4}=145~{\rm MeV}]\;,
\end{array}
\end{equation}
which assures us that interacting cold quark matter is an excited state of nuclear matter. We have assumed that, except for $T$, there is no relevant dimensional parameter, e.g., quark mass $m_{q}$ or the quantity $\Lambda$ which enters into the running coupling constant $\alpha_{s}(q^2)$. Therefore the relativistic relation between the energy density and pressure, viz., ${\varepsilon}-{\cal B}=3(P+{\cal B})$, is preserved, which leads to 
\begin{equation}\label{3chap5eq54}
P=\frac{1}{3}({\varepsilon}-4{\cal B})\;,
\end{equation}
a relation we have used occasionally before [see Eq.~(\ref{3chap5eq9})].

\begin{figure*} 
\centering\resizebox{0.49\textwidth}{!}{%
\includegraphics{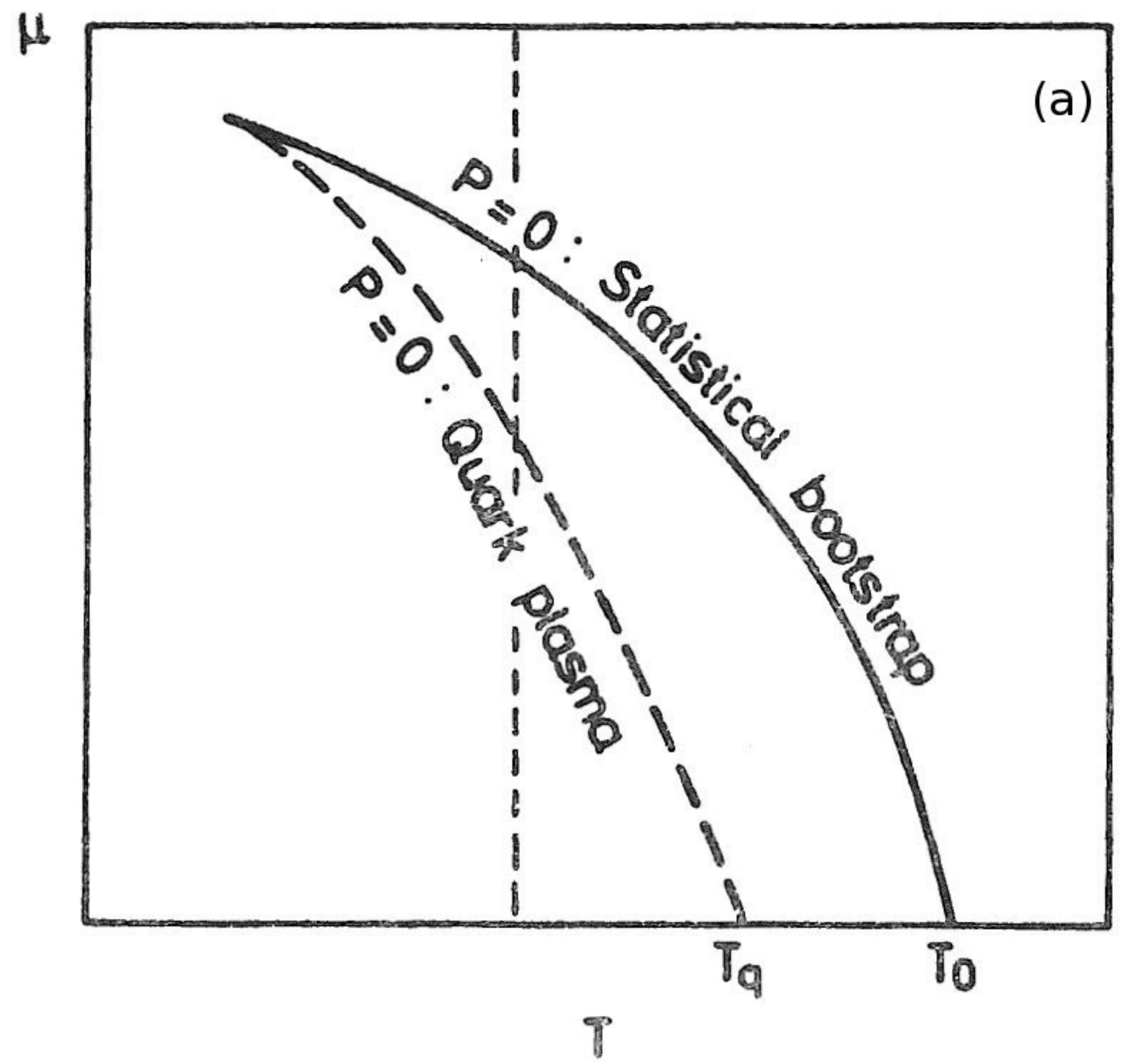}
}
\centering\resizebox{0.49\textwidth}{!}{%
\includegraphics{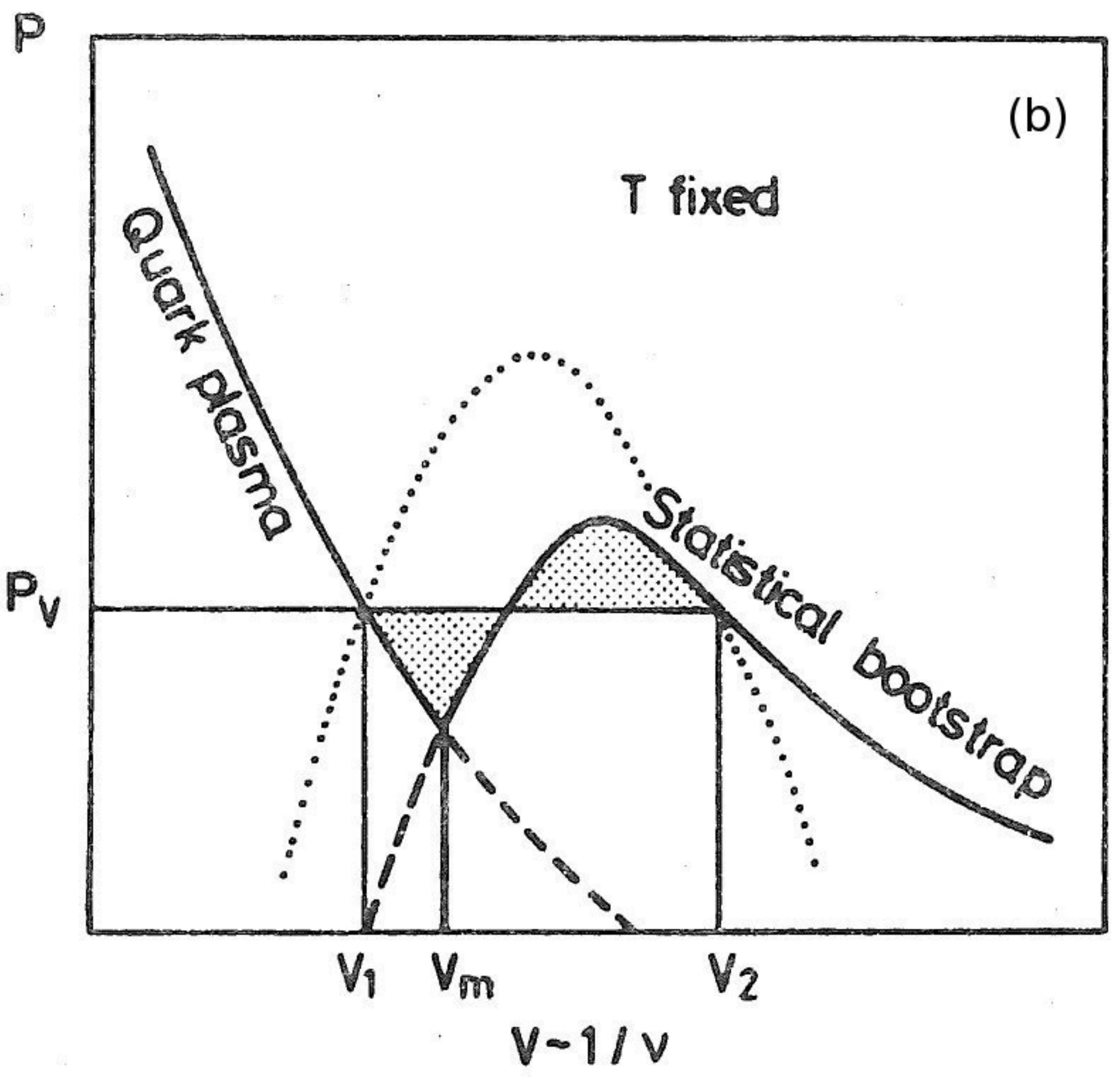}
}
\caption[]{{\bf a)} The critical curves ($P=0$) of the two models in the $T,\mu$ plane (qualitatively). The region below the {\it full line\/} is described by the statistical bootstrap model and the region above the {\it broken line\/} by the quark-gluon plasma. The critical curves can be made to coincide. {\bf  b)} $P,V$ diagram (qualitative) of the phase transition (hadron gas to quark-gluon plasma) along the {\it broken line\/} $T=$ const. of \rf{3chap5fig3}a. The coexistence region is found from the usual Maxwell construction (the {\it shaded areas\/} being equal)}\label{3chap5fig3}
\end{figure*}

From Eq.~(\ref{3chap5eq54}), it follows that, when the pressure vanishes, the energy density is $4{\cal B}$, independent of the values of $\mu$ and $T$ which fix the line $P=0$. This behavior is consistent with the hadro\-nic gas phase. This may be used as a reason to choose the parameters of both phases in such a way that the two lines $P=0$ coincide. We will return to this point again below. For $P>0$, we have ${\varepsilon}>4{\cal B}$. Recall that, in the hadro\-nic gas, we had $0<{\varepsilon}<4{\cal B}$. Thus, above the critical curve of the $\mu,T$ plane, we have the quark-gluon plasma exposed to an external force.

In order to obtain an idea of the form of the $P=0$ critical curve in the $\mu,T$ plane for the quark-gluon plasma, we rewrite Eq.~(\ref{3chap5eq52}) using Eqs.~(\ref{3chap5eq45}) and (\ref{3chap5eq46}) for $P=0$:
\begin{equation}\label{3chap5eq55}
\begin{array}{rl}
{\cal B}=&\displaystyle\frac{1-2\alpha_{s}/\pi}{162\pi^2}\big[\mu^2+(3\pi T)^2\big]^2\\[0.4cm]
+&\displaystyle\frac{T^4\pi^2}{45}\left[ 12\left(1-\frac{5}{3}\frac{\alpha_{s}}{\pi}\right) + 8\left(1-\frac{15}{4}\frac{\alpha_{s}}{\pi}\right)\right]\;.
\end{array}
\end{equation}
Here, the last term is the glue pressure contribution. (If the true vacuum structure is determined by the glue--glue interaction, then this term could be modified significantly.) We find that the greatest lower bound on temperature $T_{q}$ at $\mu=0$ is about 
\begin{equation}\label{3chap5eq56}
T_{q}\sim {\cal B}^{1/4}\approx \mbox{145--190~MeV}\,.
\end{equation}
This result can be considered to be correct to within 20\%. Its order of magnitude is as expected. Taking Eq.~(\ref{3chap5eq55}) as it is, we find for $\alpha_{s}=1/2$, $T_{q}=0.88{\cal B}^{1/4}$. Omitting the gluon contribution to the pressure, we find $T_{q}=0.9{\cal B}^{1/4}$. It is quite likely that, with the proper treatment of the glue field and the plasma corrections, and with larger ${\cal B}^{1/4}\sim 190$~MeV, the desired value of $T_{q}=T_0$ corresponding to the statistical bootstrap choice will follow. Furthermore, allowing some reasonable $T,\mu$ dependence of $\alpha_{s}$, we can then easily obtain an agreement between the critical curves.

\begin{figure*}
\centering\resizebox{0.98\textwidth}{!}{%
\includegraphics{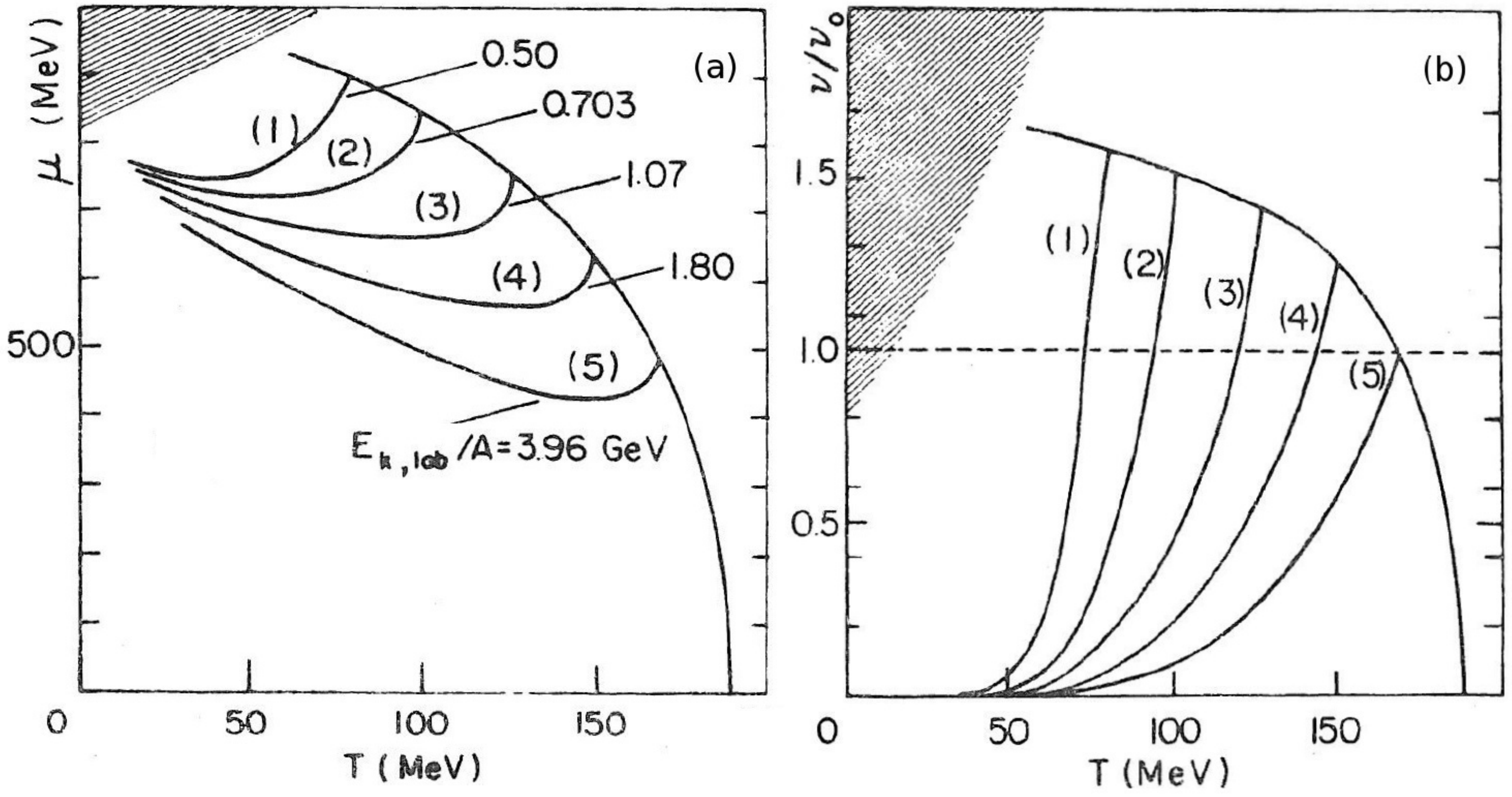} 
}
\caption[]{{\bf a)} The critical curve of hadron matter (bootstrap), together with some \lq cooling curves' in the $T,\mu$ plane. While the system cools down along these lines, it emits particles. When all particles have become free, it comes to rest on some point on these curves (\lq freeze out'). In the {\it shaded region\/}, our approach may be invalid  {\bf b)} The critical curve of hadron matter (bootstrap), together with some \lq cooling curves' [same energy as in \rf{3chap5fig4}a] in the variables $T$ and $\nu/\nu_0 =$ ratio of baryon number density to normal nuclear baryon number density. In the {\it shaded region\/}, our approach may be invalid}\label{3chap5fig4} 
\end{figure*}

However, it is not necessary for the two critical curves to coincide, even though this would be preferable. As the quark plasma is the phase into which individual hadrons dissolve, it is sufficient if the quark plasma pressure vanishes within the boundary set for non-vanishing positive pressure of the hadro\-nic gas. It is quite satisfactory for the theoretical development that this is the case. In \rf{3chap5fig3}a, a qualitative picture of the two $P=0$ lines is shown in the $\mu,T$ plane. Along the dotted straight line at constant temperature, we show in \rf{3chap5fig3}b the pressure as a function of the volume (a $P,V$ diagram). The volume is obtained by inverting the baryon density at constant fixed baryon number:
\begin{equation}\label{3chap5eq57}
V=\frac{\langle b\rangle}{\nu}\;.
\end{equation}
The behavior of $P$ ($V,T=$ const.) for the hadro\-nic gas phase is as described before in the statistical bootstrap model. For large volumes, we see that $P$ falls with rising $V$. However, when hadrons get close to each other so that they form larger and larger lumps, the pressure drops rapidly to zero. The hadro\-nic gas becomes a state of few composite clusters (internally already consisting of the quark plasma). The second branch of the $P$ ($V,T=$ const.) line meets the first one at a certain volume $V=V_{\rm m}$.

The phase transition occurs for $T=$ const. in \rf{3chap5fig3}b   at a vapor pressure $P_{\rm v}$ obtained from the conventional Maxwell construction: the shaded regions in \rf{3chap5fig3}b   are equal. Between the volumes $V_1$ and $V_2$, matter coexists in the two phases with the relative fractions being determined by the magnitude of the actual volume. This leads to the occurrence of a third region, viz., the coexistence region of matter, in addition to the pure quark and hadron domains. For $V<V_1$, corresponding to $\nu>\nu_1\sim 1/V_1$, all matter has gone into the quark plasma phase.

The dotted line in \rf{3chap5fig3}b encloses (qualitatively) the domain in which the coexistence between the two phases of hadro\-nic matter seems possible. We further note that, at low temperatures $T\leq 50$~MeV, the plasma and hadro\-nic gas critical curves meet each other in \rf{3chap5fig3}a. This is just the domain where, at present, our description of the hadro\-nic gas fails, while the quark-gluon plasma also begins to suffer from infrared difficulties. Both approaches have a very limited validity in this domain.

The qualitative discussion presented above can be easily supplemented with quantitative results. But first we turn our attention to the modifications forced onto this simple picture by the experimental circumstances in high energy nuclear collisions.

\subsection{Nuclear collisions and  inclusive particle spectra}\label{3chap5sec5}

We assume that in relativistic collisions triggered to small impact parameters by high multiplicities and absence of projectile fragments \cite{3chap5bib18}, a hot central fireball of hadro\-nic matter can be produced. We are aware of the whole problematic connected with such an idealization. A proper treatment should include collective motions and distribution of collective velocities, local temperatures, and so on \cite{3chap5bib19a,3chap5bib19b,3chap5bib19c,3chap5bib19d}, as explained in the lecture  by R. Hagedorn~\cite{3chap5bib10}. Triggering for high multiplicities hopefully eliminates some of the complications. In nearly symmetric collisions (projectile and target nuclei are similar), we can argue that the numbers of participants in the center of mass of the fireball originating in the projectile or target are the same. Therefore, it is irrelevant how many nucleons do form the fireball -- and the above symmetry argument leads, in a straightforward way, to a formula for the center of mass energy per participating nucleon:
\begin{equation}\label{3chap5eq58}
U:=\frac{E_{\rm c.m.}}{A} = m_{\rm N}\sqrt{1+\frac{E_{\rm k,lab}/A}{2m_{\rm N}}}\;,
\end{equation}
where $E_{\rm k,lab}/A$ is the projectile kinetic energy per nucleon in the laboratory frame. While the fireball changes its baryon density and chemical composition ($\pi+{p} \leftrightarrow \Delta$, etc.) during its lifetime through a change in temperature and chemical potential, the conservation of energy and baryon number assures us that $U$ in Eq.~(\ref{3chap5eq58}) remains constant, assuming that the influence on $U$ of pre-equilibrium emission of hadrons from the fireball is negligible. As $U$ is the total energy per baryon available, we can, supposing that kinetic and chemical equilibrium have been reached, set it equal to the ratio of thermodynamic expectation values of the total energy and baryon number:
\begin{equation}\label{3chap5eq59}
U = \frac{\langle E\rangle}{\langle b\rangle} = \frac{{\rm E}(\beta,\lambda)}{\nu(\beta,\lambda)}\;.
\end{equation}
Thus we see that, through Eq.~(\ref{3chap5eq59}), the experimental value of $U$ in Eq.~(\ref{3chap5eq58}) fixes a relation between allowable values of $\beta,\lambda\,$: the available excitation energy defines the temperature and the chemical composition of hadro\-nic fireballs. In \rf{3chap5fig4}a and \rf{3chap5fig4}b, these paths are shown for a choice of kinetic energies $E_{\rm k,lab}/A$ in the $\mu,T$ plane and in the $\nu,T$ plane, respectively. In both cases, only the hadro\-nic gas domain is shown. 

\begin{figure*}
\centering\resizebox{0.92\textwidth}{!}{%
\includegraphics{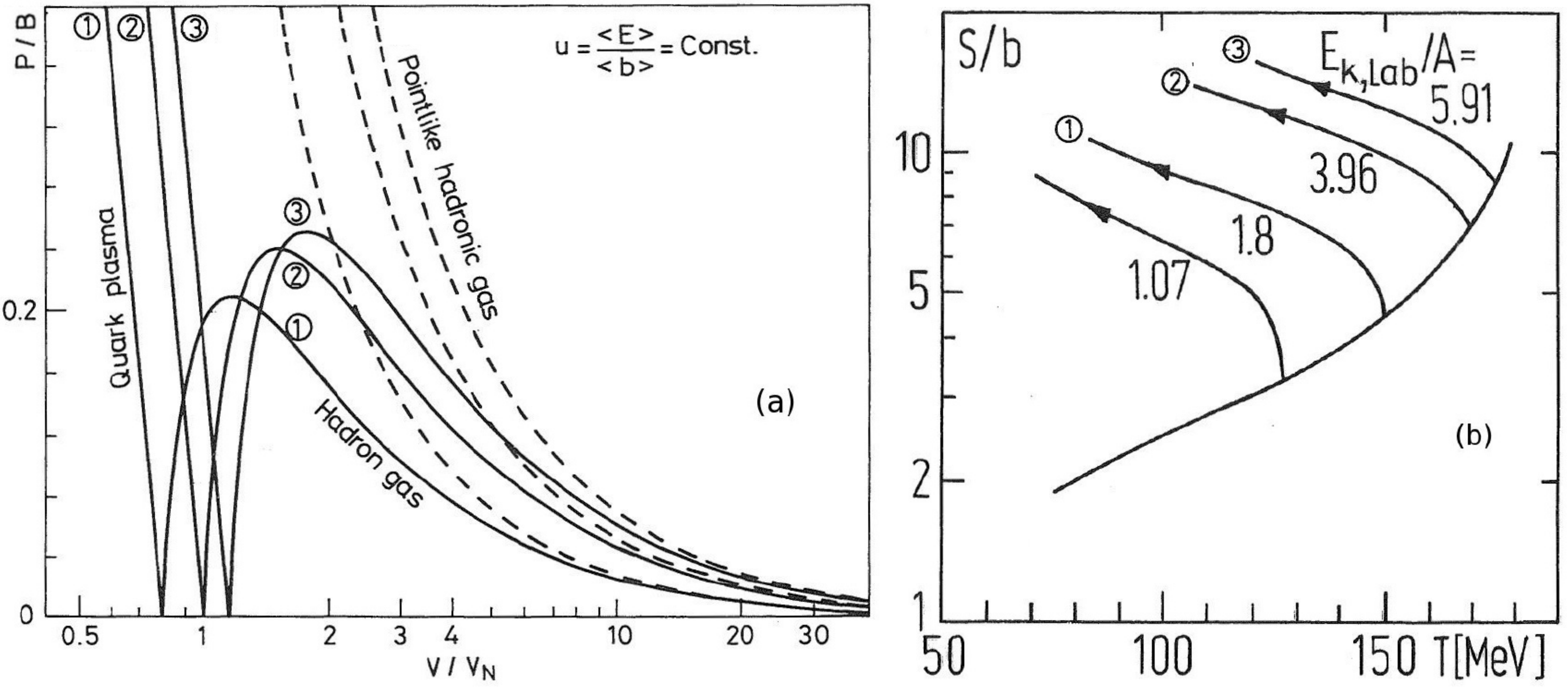}
}
\caption[]{{\bf a)} $P,V$ diagram of \lq cooling curves' belonging to different kinetic laboratory energies per nucleon: (1) 1.8~GeV, (2) 3.965~GeV, (3) 5.914~GeV. In the history of a collision, the system comes down the quark lines and jumps somewhere over to the hadron curves (Maxwell). {\it Broken lines\/} show the diverging pressure of pointlike bootstrap hadrons {\bf b)}The total specific entropy per baryon in the hadro\-nic gas phase. Same energies per nucleon as in  \rf{3chap5fig5}a, and a fourth value 1.07 GeV}\label{3chap5fig5} 
\end{figure*}

We wish to note several features of the curves shown in \rf{3chap5fig4}a and \rf{3chap5fig4}b that will be relevant in later considerations:
\begin{enumerate}
\item Beginning at the critical curve, the chemical potential first drops rapidly when $T$ decreases and then rises slowly as $T$ decreases further (\rf{3chap5fig4}a). This corresponds to a monotonically falling baryon density with decreasing temperature (\rf{3chap5fig4}b, but implies that, in the initial expansion phase of the fireball, the chemical composition changes more rapidly than the temperature.
\item The baryon density in \rf{3chap5fig4}b is of the order of 1--1.5 of normal nuclear density. This is a consequence of the choice of ${\cal B}^{1/4}=145$~MeV. Were ${\cal B}$ three times as large, i.e., ${\cal B}^{1/4}=190$~MeV, which is so far not excluded, then the baryon densities in this figure would triple to 3--$5\nu_0$. Furthermore, we observe that, along the critical curve of the hadro\-nic gas, the baryon density falls with rising temperature. This is easily understood as, at higher temperature, more volume is taken up by the numerous mesons.
\item Inspecting \rf{3chap5fig4}b, we see that, at given $U$, the temperatures at the critical curve and those at about $\nu_0/2$ differ little (10\%) for low $U$, but more significantly for large $U$. Thus, highly excited fireballs cool down more before dissociation (\lq freeze out'). As particles are emitted all the time while the fireball cools down along the lines of \rf{3chap5fig4}a and \rf{3chap5fig4}b, they carry kinetic energies related to various different temperatures. The inclusive single particle momentum distribution will yield only averages along these cooling lines.
\end{enumerate}
Another remark which does not follow from the curves shown is:
\begin{enumerate}\setcounter{enumi}{3}
\item Below about 1.8 GeV, an important portion of the total energy is in the collective (hydrodynamical) motion of hadro\-nic matter, hence the cooling curves at constant excitation energy do not properly describe the evolution of the fireball.
\end{enumerate}
Calculations of this kind can also be carried out for the quark plasma. They are, at present, uncertain due to the unknown values of $\alpha_{s}$ and ${\cal B}^{1/4}$. Fortunately, there is one particular property of the equation of state of the quark-gluon plasma that we can easily exploit. 

Combining Eq.~(\ref{3chap5eq54}) with Eq.~(\ref{3chap5eq59}), we obtain
\begin{equation}\label{3chap5eq60}
P=\frac{1}{3}(U\nu-4{\cal B})\;.
\end{equation}
Thus, for a given $U$ (the available energy per baryon in a heavy ion collision), Eq.~(\ref{3chap5eq60}) describes the pressure--volume ($\sim 1/\nu$) relation. By choosing to measure $P$ in units of ${\cal B}$ and $\nu$ in units of normal nuclear density $\nu_0=0.14/$fm$^3$, we find
\begin{equation}\label{3chap5eq61}
\frac{P}{\cal B} = \frac{4}{3}\left(\gamma\frac{U}{m_{\rm N}}\frac{\nu}{\nu_0} - 1\right)\;,
\end{equation}
with
\[\gamma := \frac{m_{\rm N}\nu_0}{4{\cal B}} = 0.56\;,\ \mathrm{for:}\ {\cal B}^{1/4}=145~{\rm MeV}\;,\  \nu_0=0.14/{\rm fm}^3\;.
\]
Here, $\gamma$ is the ratio of the energy density of normal nuclei (${\varepsilon}_{\rm N}=m_{\rm N}\nu_0$) and of quark matter or of a quark bag (${\varepsilon}_{q}=4{\cal B}$). In \rf{3chap5fig5}a, this relation is shown for three projectile energies: $E_{\rm k,lab}/A=1.80$~GeV, 3.965~GeV, and 5.914~GeV, corresponding to $U=1.314$~GeV, 1.656~GeV, and 1.913~GeV, respectively. We observe that, even at the lowest energy shown, the quark pressure is zero near the baryon density corresponding to 1.3 normal nuclear density, given the current value of ${\cal B}$.

Before discussing this point further, we note that the hadro\-nic gas branches of the curves in \rf{3chap5fig5}a and \rf{3chap5fig5}b show a quite similar behavior to that shown at constant temperature in \rf{3chap5fig3}b. Remarkably, the two branches meet each other at $P=0$, since both have the same energy density ${\varepsilon}=4{\cal B}$ and therefore $V(P=0)\sim 1/\nu=U/{\varepsilon}=U/4{\cal B}$. However, what we cannot see by inspecting \rf{3chap5fig5}a and \rf{3chap5fig5}b is that there will be a discontinuity in the variables $\mu$ and $T$ at this point, except if parameters are chosen so that the critical curves of the two phases coincide. Indeed, near to $P=0$, the results shown in \rf{3chap5fig5}a should be replaced by points obtained from the Maxwell construction. The pressure in a nuclear collision will never fall to zero. It will correspond to the momentary vapor pressure of the order of $0.2{\cal B}$ as the phase change occurs.

A further aspect of the equations of state for the hadro\-nic gas is also illustrated in \rf{3chap5fig5}a. Had we ignored the finite size of hadrons (one of the van der Waals effects) in the hadron gas phase then, as shown by the dash-dotted lines, the phase change could never occur because the point particle pressure would diverge where the quark pressure vanishes. In our opinion, one cannot say it often enough: inclusion of the finite hadro\-nic size and of the finite temperature when considering the phase transition to quark plasma lowers the relevant baryon density (from 8--$14\nu_0$ for cold point-nucleon matter) to 1--$5\nu_0$ (depending on the choice of ${\cal B}$) in 2--5~GeV/$A$ nuclear collisions. The possible formation of quark-gluon plasma in nuclear collisions was first discussed quantitatively in Ref.\cite{3chap5bib20a}, see also Ref.\cite{3chap5bib20b}.

The physical picture underlying our discussion is an explosion of the fireball into vacuum with little energy being converted into collective motion, e.g., hydrodynamical flow, or being taken away by fast pre-hadronization particle emission. Thus the conserved internal excitation energy can only be shifted between thermal (kinetic) and chemical excitations of matter. \lq Cooling' thus really means that, during the explosion, the thermal energy is mostly convered into chemical energy, e.g., {\it pions are produced\/}.

While it is at present hard to judge the precise amount of expected deviation from the cooling curves shown in \rf{3chap5fig2},  it is possible to show that they are  entirely inconsistent with the notion of reversible adiabatic, i.e., entropy conserving, expansion. As the expansion proceeds along $U=$ const. lines, we can compute the entropy per participating baryon using Eqs.~(\ref{3chap5eq36}) and (\ref{3chap5eq37}), and we find a significant growth of total entropy. As shown in \rf{3chap5fig5}b, the entropy rises initially in the dense phase of the matter by as much as 50--100\% due to the pion production and resonance decay. Amusingly enough, as the newly produced entropy is carried mostly by pions, one will find that the entropy carried by protons remains constant. With this remarkable behavior of the entropy, we are in a certain sense, victims of our elaborate theory. Had we used, e.g., an ideal gas of Fermi nucleons, then the expansion would seem to be entropy conserving, as pion production and other chemistry were forgotten. Our fireballs have no tendency to expand reversibly and adiabatically, as many reaction channels are open. A more complete discussion of the entropy puzzle can be found in \cite{3chap5bib1}.

Inspecting \rf{3chap5fig4}a and \rf{3chap5fig4}b again, it seems that a possible test of the equations of state for the hadro\-nic gas consists in measuring the temperature in the hot fireball zone, and doing this as a function of the nuclear collision energy. The plausible assumption made is that the fireball follows the \lq cooling' lines shown in \rf{3chap5fig4}a  and \rf{3chap5fig4}b until final dissociation into hadrons. This presupposes that the surface emission of hadrons during the expansion of the fireball does not significantly alter the available energy per baryon. This is more likely true for sufficiently large fireballs. For small ones, pion emission by the surface may influence the energy balance. As the fireball expands, the temperature falls and the chemical composition changes. The hadro\-nic clusters dissociate and more and more hadrons are to be found in the \lq elementary' form of a nucleon or a pion. Their kinetic energies are reminiscent of the temperature found at each phase of the expansion.

To compute the experimentally observable final temperature \cite{3chap5bib1,3chap5bib13}, we shall argue that a time average must be performed along the cooling curves. Not knowing the reaction mechanisms too well, we assume that the temperature decreases approximately linearly with the time in the significant expansion phase. We further have to allow that a fraction of particles emitted can be reabsorbed in the hadro\-nic cluster. This is a geometric problem and, in a first approximation, the ratio of the available volume $\Delta$ to the external volume $V_{\rm ex}$ is the probability that an emitted particle not be reabsorbed, i.e., that it can escape:
\begin{equation}\label{3chap5eq62}
R_{\rm esc} =\frac{\Delta}{V_{\rm ex}} = 1-\frac{{\varepsilon}(\beta,\lambda)}{4{\cal B}}\;.
\end{equation}
The relative emission rate is just the integrated momentum spectrum
\begin{equation}\label{3chap5eq63}
R_{\rm emis} =\int\!\!\frac{\D^3p}{(2\pi)^3}\E^{-\sqrt{p^2+m^2}/T+\mu/T} = \frac{m^2T}{2\pi^2}K_2(m/T)\,\E^{\mu/T}\;.
\end{equation}
The chemical potential acts only for nucleons. In the case of pions, it has to be dropped from the above expression. For the mean temperature, we thus find
\begin{equation}\label{3chap5eq64}
\langle T\rangle = \frac{\displaystyle\int_{\rm c} R_{\rm esc}R_{\rm emis}T\D T}{\displaystyle\int_{\rm c} R_{\rm esc}R_{\rm emis}\D T}\;,
\end{equation}
where the subscript c on the integral indicates here a line integral along that particular cooling curve in \rf{3chap5fig4}a  and \rf{3chap5fig4}b which belongs to the energy per baryon fixed by the experimentalist.

In practice, the temperature is most reliably measured through the measurement of mean transverse momenta of the particles. It may be more practical therefore to calculate the average transverse momentum of the emitted particles. In principle, to obtain this result we have to perform a similar averaging to the one above. For the average transverse momentum at given $T,\mu$, we find \cite{3chap5bib8}
\begin{equation}\label{3chap5eq65}
\begin{array}{rl}
\langle p_\perp(m,T,\mu)\rangle_p =&\displaystyle \frac{\displaystyle \int p_\perp \E^{-\sqrt{p^2+m^2}-\mu)/T} \D^3p}{\displaystyle \int \E^{-\sqrt{p^2+m^2}-\mu)/T} \D^3p} \\[0.9cm]
=& \displaystyle\frac{\sqrt{\pi mT/2}\,K_{\frac 5 2}\!\left(\frac{m}{T}\right)\E^{\mu/T}}{K_2\!\left(\frac{m}{T}\right)\E^{\mu/T}}\;.
\end{array}
\end{equation}
The average over the cooling curve is then
\begin{equation}\label{3chap5eq66}
\big\langle\langle p_\perp(m,T,\mu)\rangle_p\big\rangle_{\rm c} = \frac
{\displaystyle \int_{\rm c}\!\frac{\Delta}{V_{\rm ex}}T^{3/2} \sqrt{\frac{\pi m}{2}}\,K_{\frac 5 2}\!\left(\frac{m}{T}\right)\E^{\mu/T}\D T}
{\displaystyle \int_{\rm c}\!\frac{\Delta}{V_{\rm ex}}TK_2\!\left(\frac{m}{T}\right)\E^{\mu/T}\D T}\;.
\end{equation}
We did verify numerically that the order of averages does not matter:
\begin{equation}\label{3chap5eq67}
\big\langle p_\perp(m,\langle T\rangle_{\rm c},\mu)\big\rangle_p \approx \big\langle\langle p_\perp(m,T,\mu)\rangle_p\big\rangle_{\rm c}\;,
\end{equation}
which shows that the mean transverse momentum is also the simplest (and safest) method of determining the average temperature (indeed better than fitting ad hoc exponential type functions to $p_\perp$ distributions).

In the presented calculations, we chose the bag constant ${\cal B}=(145~{\rm MeV})^4$, but we now believe that a larger ${\cal B}$ should be used. As a consequence of our choice and the measured pion temperature of $\langle T\rangle_\pi^{\rm ex}=140$~MeV at highest ISR energies, we have to choose the constant $H$ such that $T_0=190$~MeV [see Eq.~(\ref{3chap5eq43b})].

The average temperature, as a function of the range of integration over $T$, reaches different limiting values for different particles. The limiting value obtained thus is the observable \lq average temperature' of the debris of the interaction, while the initial temperature $T_{\rm cr}$ at given $E_{\rm k,lab}$ (full line in \rf{3chap5fig6}) is difficult to observe. When integrating along the cooling line as in Eq.~(\ref{3chap5eq64}), we can easily, at each point, determine the average hadro\-nic cluster mass. The integration for protons is interrupted (protons are \lq frozen out') when the average cluster mass is about half the nucleon isobar mass. We have also considered baryon density dependent freeze-out, but such a procedure depends strongly on the unreliable value of ${\cal B}$.

\begin{figure}
\centering\resizebox{0.38\textwidth}{!}{%
\includegraphics{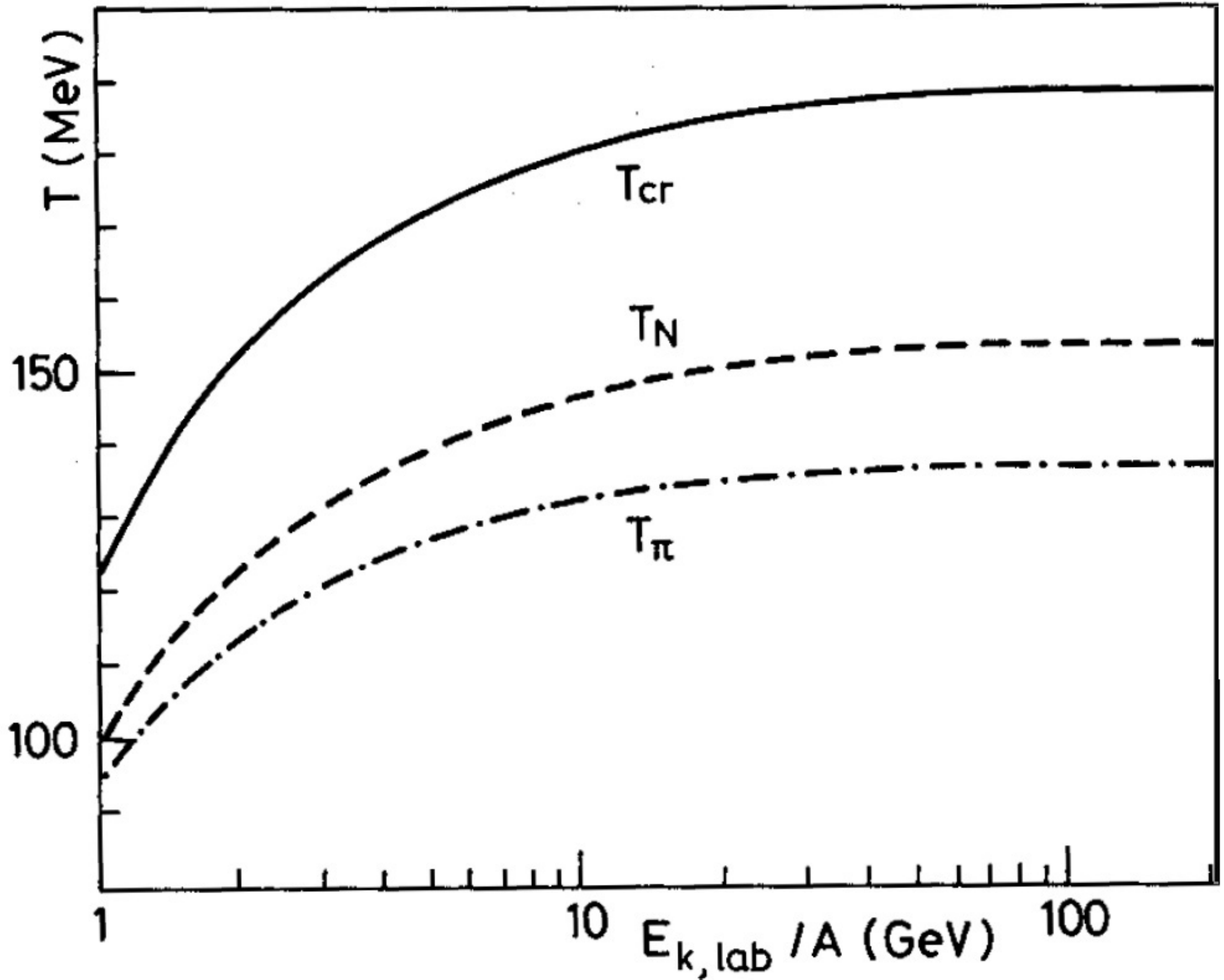}
}
\caption[]{Mean temperatures for nucleons and pions together with the critical temperature belonging to the point where the \lq cooling curves' start off the critical curve (see \rf{3chap5fig4}a). The mean temperatures are obtained by integrating along the cooling curves. Note that $T_{\rm N}$ is always greater than $T_\pi$}.\label{3chap5fig6}
\end{figure}

Our choice of the freeze-out condition was made in such a way that the nucleon temperature at $E_{\rm k,lab}/A=1.8$~GeV is about 120~MeV. The model dependence of our freeze-out introduces an uncertainty of several MeV in the average temperature. In \rf{3chap5fig6}, the pion and nucleon average temperatures are shown as a function of the heavy ion kinetic energy. Two effects contributed to the difference between the $\pi$ and N temperatures:
\begin{enumerate}
\item The particular shape of the cooling curves (\rf{3chap5fig4}a). The chemical potential drops rapidly from the critical curve, thereby damping relative baryon emission at lower $T$. Pions, which do not feel the baryon chemical potential, continue being created also at lower temperatures.
\item The freeze-out of baryons occurs earlier than the freeze-out of pions.
\end{enumerate}
A third effect has been so far omitted -- the emission of pions from two-body decay of long-lived resonances \cite{3chap5bib1} would lead to an effective temperature which is lower in nuclear collisions.

\begin{figure}
\centering\resizebox{0.38\textwidth}{!}{%
\includegraphics{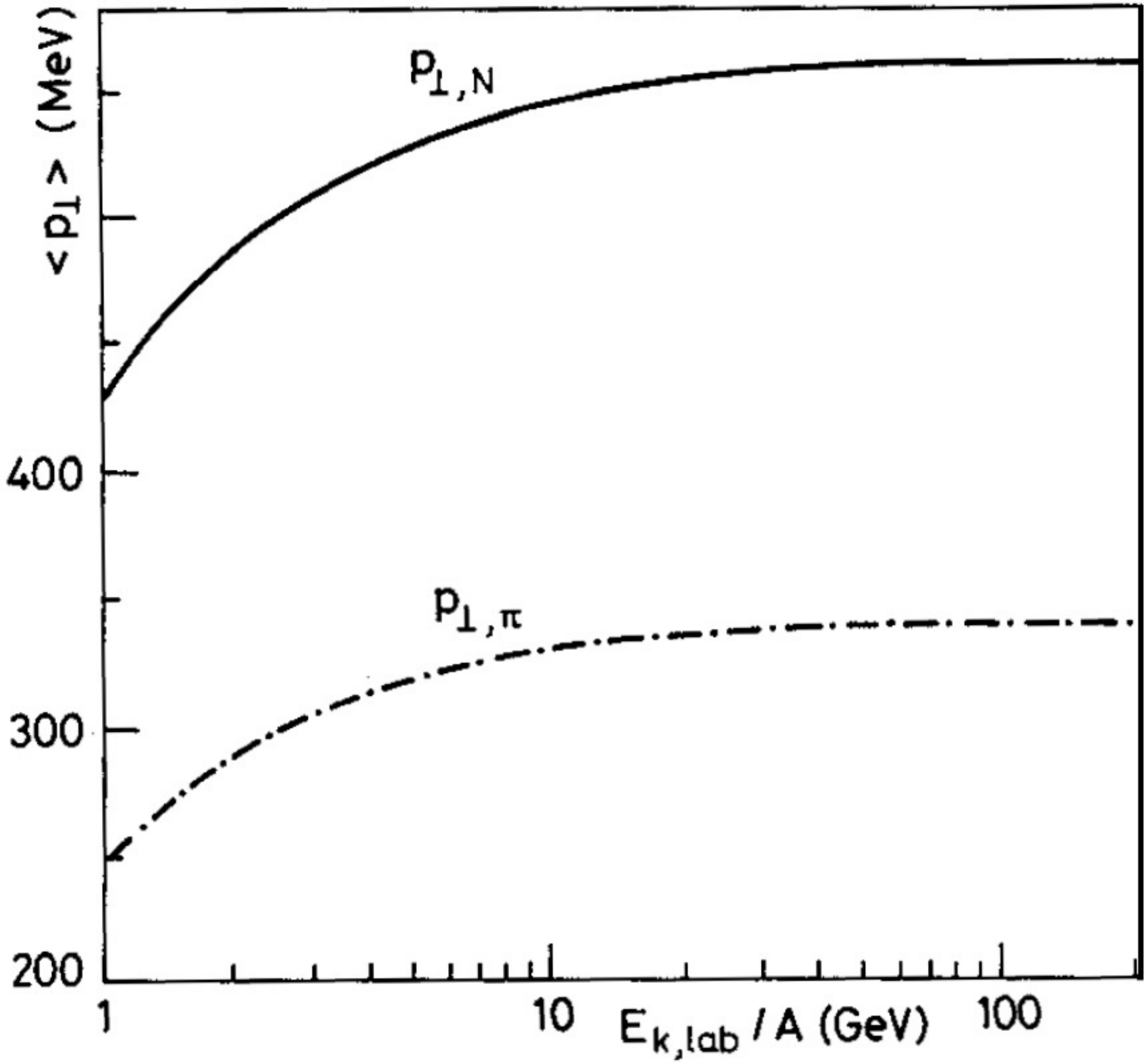}
}
\caption[]{Mean transverse momenta of nucleons and pions found by integrating along the \lq cooling curves'.}\label{3chap5fig7}
\end{figure}

In \rf{3chap5fig7}, we show the dependence of the average transverse momenta of pions and nucleons on the kinetic energy of the heavy ion projectiles.

\subsection{Strangeness in Heavy Ion Collisions}\label{3chap5sec6}

From the averaging process described here, we have learned that the temperatures and transverse momenta of particles originating in the hot fireballs are more reminiscent of the entire history of the fireball expansion than of the initial hot compressed state, perhaps present in the form of quark matter. We may generalize this result and then claim that most properties of inclusive spectra are reminiscent of the equations of state of the hadro\-nic gas phase and that the memory of the initial dense state is lost during the expansion of the fireballs as the hadro\-nic gas rescatters many times while it evolves into the final kinetic and chemical equilibrium state.

In order to observe properties of quark-gluon plasma, we must design a thermometer, an isolated degree of freedom weakly coupled to the hadro\-nic matter. Nature has, in principle (but not in practice) provided several such thermometers: leptons and heavy flavors of quarks. We would like to point here to a particular phenomenon perhaps quite uniquely characteristic of quark matter. First we note that, at a given temperature, the quark-gluon plasma will contain an equal number of strange ($s$) quarks and antistrange ($\overline{s}$) quarks, naturally assuming that the hadro\-nic collision time is much too short to allow for light flavor weak interaction conversion to strangeness. Thus, assuming equilibrium in the quark plasma, we find the density of the strange quarks to be (two spins and three colors)
\begin{equation}\label{3chap5eq68}
\frac{s}{V}=\frac{\overline{s}}{V} = 6\int\frac{\D^3p}{(2\pi)^3}\E^{-\sqrt{p^2+m_{s}^2}/T} = 3\frac{Tm_{s}^2}{\pi^2}K_2(m_{s}/T)\;,
\end{equation}
neglecting for the time being the perturbative corrections and, of course, ignoring weak decays. As the mass $m_{s}$ of the strange quarks in the perturbative vacuum is believed to be of the order of 280--300~MeV, the assumption of equilibrium for $m_{s}/T\sim 2$ may indeed be correct. In Eq.~(\ref{3chap5eq68}), we were able to use the Boltzmann distribution again, as the density of strangeness is relatively low. Similarly, there is a certain light antiquark density ($\overline{q}$ stands for either $\overline{u}$ or $\overline{d}$):
\begin{equation}\label{3chap5eq69}
\frac{\overline{q}}{V} = 6\int\frac{\D^3p}{(2\pi)^3}\E^{-|p|/T-\mu_{q}/T} = \E^{-\mu_{q}/T} T^3\frac{6}{\pi^2}\;,
\end{equation}
where the quark chemical potential is $\mu_{q}=\mu/3$, as given by Eq.~(\ref{3chap5eq46}). This exponent suppresses the q$\overline{q}$ pair production.

What we intend to show is that there are many more $\overline{s}$ quarks than antiquarks of each light flavor. Indeed,
\begin{equation}\label{3chap5eq70}
\frac{\overline{s}}{\overline{q}}  = \frac{1}{2}\left(\frac{m_{s}}{T}\right)^2 K_2\left(\frac{m_{s}}{T}\right)\E^{\mu/3T}\;.
\end{equation}
The function $x^2K_2(x)$ is, for example, tabulated in \cite{3chap5bib21}. For $x=m_{s}/T$ between 1.5 and 2, it varies between 1.3 and 1. Thus, we almost always have more $\overline{s}$ than $\overline{q}$ quarks and, in many cases of interest, $\overline{s}/\overline{q}\sim 5$. As $\mu\rightarrow 0$, there are about as many $\overline{\rm u}$ and $\overline{q}$ quarks as there are $\overline{s}$ quarks.

When the quark matter dissociates into hadrons, some of the numerous $\overline{s}$ may, instead of being bound in a q$\overline{s}$ kaon, enter into a $\overline{q}\,\overline{q}\,\overline{s}$ antibaryon and, in particular\footnote{$\overline{\uSigma}{}^0$  decays into $\overline{\uLambda}$ by emitting a photon and is always counted within the $\overline{\uLambda}$ abundance.}, a $\overline{\uLambda}$ or $\overline{\uSigma}{}^0$. The probability for this process seems to be comparable to the similar one for the production of antinucleons by the antiquarks present in the plasma. What is particularly noteworthy about the $\overline{s}$-carrying antibaryons is that they can conventionally only be produced in direct pair production reactions. Up to about $E_{\rm k,lab}/A=3.5$~GeV, this process is very strongly suppressed by energy--momen\-tum conservation because, for free $pp$ collisions, the threshold is at about 7~GeV. We would thus like to argue that a study of the $\overline{\uLambda}$ and $\overline{\uSigma}{}^0$ in nuclear collisions for $2<E_{\rm k,lab}/A<4$~GeV could shed light on the early stages of the nuclear collisions in which quark matter may be formed.

Let us mention here another effect of importance in this context: the production rate of a pair of particles with a conserved quantum number like strangeness will usually be suppressed by the Boltzmann factor $\E^{-2m/T}$, rather than a factor $\E^{-m/T}$ as is the case in thermomechanical equilibrium (see, for example, the addendum in \cite{3chap5bib8}). As relativistic nuclear collisions are just on the borderline between those two limiting cases, it is important when considering the yield of strange particles to understand the transition between them. We will now show how one can describe these different cases in a unified statistical description \cite{3chap5bib22}.

As we have already implicitly discussed [see Eq.~(\ref{3chap5eq12})], the logarithm of the grand partition function $Z$ is a sum over all different particle configurations, e.g., expressed with the help of the mass spectrum. Hence, we can now concentrate in particular on that part of $\ln Z$ which is exclusively associated with the strangeness.

As the temperatures of interest to us and which allow appreciable strangeness production are at the same time high enough to prevent the strange particles from being thermodynamically degenerate, we can restrict ourselves again to the discussion of Boltzmann statistics only.

The contribution to $Z$ of a state with $k$ strange particles is
\begin{equation}\label{3chap5eq71}
Z_k=\frac{1}{k!}\bigg[\sum_s Z_I^s(T,V)\bigg]^k\;,
\end{equation}
where the one-particle function $Z_1$ for a particle of mass $m_{s}$ is given in Eq.~(\ref{3chap5eq16}). To include both particles {\it and\/} antiparticles as two thermodynamically independent phases in Eq.~(\ref{3chap5eq71}), the sum over $s$ in Eq.~(\ref{3chap5eq71}) must include them both. As the quantum numbers of particles (p) and antiparticles (a) must always be present with {\it exactly\/} the same total number, not each term in Eq.~(\ref{3chap5eq71}) can contribute. Only when $n=k/2=$\,number of particles\,=\,number of antiparticles is exactly fulfilled do we have a physical state. Hence,
\begin{equation}\label{3chap5eq72}
Z_{2n}^{\rm pair} = \frac{1}{(2n)!}\left(\begin{array}{c} 2n\\ n\end{array}\right)\bigg(\sum_{s_{\rm p} }Z^{s_{\rm p} }_1\bigg)^n \bigg(\sum_{s_{\rm a}}Z^{s_{\rm a}}_1\bigg)^n\;.
\end{equation}
We now introduce the fugacity factor $f^n$ to be able to count the number of strange pairs present. Allowing an arbitrary number of pairs to be produced, we obtain
\begin{equation}\label{3chap5eq73}
\begin{array}{rl}
Z_s(\beta,V;f)=&\displaystyle\sum_{n=0}^\infty \frac{f^n}{n!n!}\bigg(\sum_{s_{\rm p} }Z^{s_{\rm p} }_1\bigg)^n \bigg(\sum_{s_{\rm a}}Z^{s_{\rm a}}_1\bigg)^n\\[0.6cm]
 =&\displaystyle I_0(\sqrt{4y})\;,
\end{array}
\end{equation}
where $I_0$ is the modified Bessel function and 
\begin{equation}\label{3chap5eq74}
y=f\bigg(\sum_{s_{\rm p} }Z^{s_{\rm p} }_1\bigg) \bigg(\sum_{s_{\rm a}}Z^{s_{\rm a}}_1\bigg)\;.
\end{equation}
We have to maintain the difference between the particles (p) and antiparticles (a), as in nuclear collisions the symmetry is broken by the presence of baryons and there is an associated need for a baryon fugacity (chemical potential $\mu$) that controls the baryon number. We obtain
\begin{equation}\label{3chap5eq75}
\begin{array}{rl}
Z_1^{\rm p,a}:=&\displaystyle\sum_{s_{\rm p,a}}Z_1^{s_{\rm p,a}} \\[0.6cm]
=&\displaystyle\frac{VT^3}{2\pi^2}\Big\{2W(x_{\rm K})+2\E^{\pm\mu/T}\big[W(x_{\sLambda}) + 3W(x_{\sSigma})\big]\Big\}\;,
\end{array}
\end{equation}
for particles ($+\mu$) and antiparticles ($-\mu$), where $W(x)=x^2K_2(x)$, $x_i=m_i/T$, and all kaons and hyperons are counted. In the quark phase, we have
\begin{equation}\label{3chap5eq76}
Z^{\rm p,a}_{1,{q}} = \frac{VT^3}{2\pi^2}\Big[6\,\E^{\pm\mu/3T}W(x_{s})\Big]\;,
\end{equation}
with $Tx_{s}=m_{s} \sim 280$~MeV. We note in passing that the baryon chemical potential cancels out in $y$ of Eq.~(\ref{3chap5eq74}) when Eq.~(\ref{3chap5eq76}) is inserted in the quark phase [compare with Eq.~(\ref{3chap5eq68})].

By differentiating $\ln Z_{s}$ of Eq.~(\ref{3chap5eq73}) with respect to $f$, we find the strangeness number present at given $T$ and $V\,$:
\begin{equation}\label{3chap5eq77}
\langle n\rangle_{s} = f\frac{\partial }{\partial f}\ln Z_{s}\Big|_{f=1} = \frac{I_1(\sqrt{4y})}{I_0(\sqrt{4y})}\sqrt{y}\;.
\end{equation}

For large $y$, that is, at given $T$ for large volume $V$, we find $\langle n\rangle_{s}=\sqrt{y}\sim\E^{-m/T}$, as expected. For small $y$, we find $\langle n\rangle_{s}=y\sim\E^{-2m/T}$. In \rf{3chap5fig8}, we show the dependence of the quenching factor $I_1/I_0=\eta$ as a function of the volume $V$ measured in units of $V_{\rm h}=4\pi/3$ fm$^3$ for a typical set of parameters: $T=150$, $\mu=550$~MeV (hadro\-nic gas phase).

\begin{figure}
\centering\resizebox{0.45\textwidth}{!}{%
\includegraphics{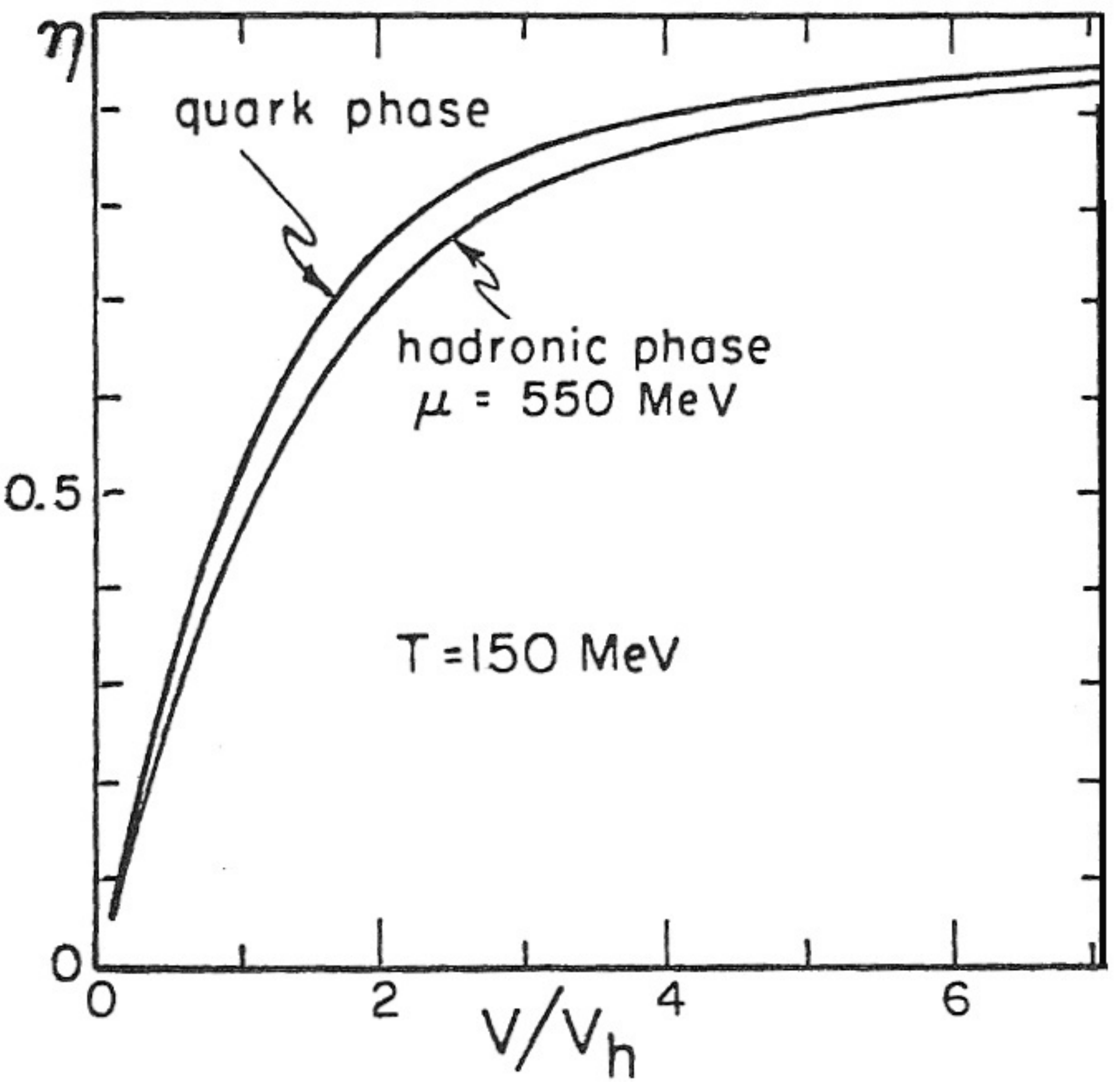}
}
\caption[]{The quenching factor for strangeness production as a function of the active volume $V/V_{\rm h}$, where $V_{\rm h}=4\pi/3$ fm$^3$}\label{3chap5fig8}
\end{figure}

\noindent The following observations follow from inspection of \rf{3chap5fig8}:
\begin{enumerate}
\item The strangeness yield is a qualitative measure of the hadro\-nic volume in thermodynamic equilibrium.
\item Total strangeness yield is not an indicator of the phase transition to quark plasma, as the enhancement ($\sqrt{\eta_{q}/\eta}$ $=1.25$) in yield can be reinterpreted as being due to a change in hadro\-nic volume.
\item We can expect that, in nuclear collisions, the active volume will be sufficiently large to allow the strangeness yield to correspond to that of \lq infinite' volume for reactions triggered on \lq central collisions'. Hence, e.g., $\uLambda$ production rate will significantly exceed that found in $pp$ collisions.
\end{enumerate}
Our conclusions about the significance of $\overline{\uLambda}$ as an indicator of the phase transition to quark plasma remain valid as the production of $\overline{\uLambda}$ in the hadro\-nic gas phase will only be possible in the very first stages of the nuclear collisions, if sufficient center of mass energy is available.

\subsection{Summary}\label{3chap5sec7}

Our aim has been to obtain a description of hadro\-nic matter valid for high internal excitations. By postulating the kinetic and chemical equilibrium, we have been able to develop a thermodynamic description valid for high temperatures and different chemical compositions. In our work we have found two physically different domains: firstly, the hadro\-nic gas phase, in which individual hadrons can exist as separate entities, but are sometimes combined into larger hadro\-nic clusters, while in the second domain, individual hadrons dissolve into one large cluster consisting of hadro\-nic constituents, viz., the quark-gluon plasma.

In order to obtain a theoretical description of both phases, we have used some `common' knowledge and plausible interpretations of currently available experimental observations. In particular, in the case of hadro\-nic gas, we have completely abandoned a more conventional Lagrangian approach in favour of a semi-phenomeno\-logical statistical bootstrap model of hadro\-nic matter that incorporates those properties of hadro\-nic interaction that are, in our opinion, most important in nuclear collisions.

In particular, the attractive interactions are included through the rich, exponentially growing hadro\-nic mass spectrum $\tau(m^2,b)$, while the introduction of the finite volume of each hadron is responsible for an effective short-range repulsion. Aside from these manifestations of strong interactions, we only satisfy the usual conservation laws of energy, momentum, and baryon number. We neglect quantum statistics since quantitative study has revealed that this is allowed above $T\approx 50$~MeV. But we allow particle production, which introduces a quantum physical aspect into the otherwise `classical' theory of Boltzmann particles.

Our approach leads us to the equations of state of hadro\-nic matter which reflect what we have included in our considerations. It is the {\it quantitative\/} nature of our work that allows a detailed comparison with experiment. This work has just begun and it is too early to say if the features of strong interactions that we have chosen to include in our considerations are the most relevant ones. It is important to observe that the currently predicted pion and nucleon mean transverse momenta and temperatures show the required substantial rise (see \rf{3chap5fig7}) as required by the experimental results available at $E_{\rm k,lab}/A=2$~GeV (BEVALAC, see \cite{3chap5bib18}) and at 1000~GeV (ISR, see \cite{3chap5bib17}). Further comparisons involving, in particular, particle multiplicities and strangeness production are under consideration.

We also mention the internal theoretical consistency of our two-fold approach. With the proper interpretation, the statistical bootstrap leads us, in a straightforward fashion, to the postulate of a phase transition to the quark-gluon plasma. This second phase is treated by a quite different method. In addition to the standard Lagrangian quantum field theory of weakly interacting particles at finite temperature and density,\,we also introduce the\,phenomeno\-logical vacuum pressure and energy density\,$\cal B$.

Perhaps the most interesting aspect of our work is the realization that the transition to quark matter will occur at much lower baryon density for highly excited hadro\-nic matter than for matter in the ground state ($T=0$). The precise baryon density of the phase transition depends somewhat on the bag constant, but we estimate it to be at about 2--$4\nu_0$ at $T=150$~MeV. The detailed study of the different aspects of this phase transition, as well as of possible characteristic signatures of quark matter, must still be carried out. We have given here only a very preliminary report on the status of our present understanding.

We believe that the occurrence of the quark plasma phase is observable and we have proposed therefore a measurement of the $\bar{\uLambda}/\overline{p}$ relative yield between 2 and 10~GeV/$N$ kinetic energies. In the quark plasma phase, we expect a significant enhancement of $\bar{\uLambda}$ production which will most likely be visible in the $\bar{\uLambda}/\overline{p}$ relative rate.\\

\noindent{\bf 1980 Acknowledgements}
Many fruitful discussions with the GSI/LBL Relativistic Heavy Ion group stimulated the ideas presented here. I would like to thank R. Bock and R. Stock for their hospitality at GSI during this workshop. As emphasized before, this work was performed in collaboration with R. Hagedorn. Supported by Deutsche Forschungsgemeinschaft.



\renewcommand{\theequation}{\thesection\thesubsection.\arabic{equation}}
\setcounter{table}{0}
\setcounter{figure}{0}
\setcounter{footnote}{0}
\setcounter{equation}{0}
\section{Strangeness and Phase Changes in Hot Hadronic Matter -- 1983}
\noindent{\bf From: \lq\lq Sixth High Energy Heavy Ion Study\rq\rq\\
held 28 June -- 1 July 1983 at: LBNL, Berkeley, CA, USA;
Printed in:  LBL-16281  pp. 489-510; 
Also: report number UC-34C; DOE CONF-830675; 
Also: preprint  CERN-TH-3685 available at\\
     {\small https://cds.cern.ch/record/147343/files/198311019.pdf}}\\

\noindent ABSTRACT: {\small Two phases of hot hadronic matter are described with emphasis put on their distinction. Here the role of strange particles as a characteristic observable of the quark-gluon plasma phase is particularly explored.}
\setcounter{equation}{0}
\subsection{Phase transition or perhaps transformation:\protect\newline Hadronic gas and the quark-gluon plasma}\label{3chap8Csec1}

I explore here consequences of the hypothesis that the energy available in the collision of two relativistic heavy nuclei, at least in part of the system, is equally divided among the accessible degrees of freedom. This means that there exists a domain in space in which, in a suitable Lorentz frame, the energy of the longitudinal motion has been largely transformed to transverse degrees of freedom. The physical variables characterizing such a `fireball' are energy density, baryon number density, and total volume. The {\it basic\/} question concerns the {\it internal structure\/} of the fireball. It can consist either of individual hadrons, or instead, of quarks and gluons in a new physical phase, the plasma, in which they are deconfined and can move freely over the volume of the fireball. It appears that the phase transition from the hadronic gas phase to the quark-gluon plasma is controlled mainly by the energy density of the fireball. Several estimates\footnote{An incomplete list of quark-gluon plasma papers includes:\cite{3chap8Cbib1a,3chap8Cbib1b,3chap8Cbib1c,3chap8Cbib1d,3chap8Cbib1e,3chap8Cbib1f,3chap8Cbib1ff,3chap8Cbib1g,3chap8Cbib1h,3chap8Cbib1i}.} lead to 0.6--1~GeV/fm$^3$ for the critical energy density, to be compared with nuclear matter 0.16~GeV/fm$^3$.

We first recall that the unhandy extensive variables, viz., energy, baryon number, etc., are replaced by intensive quantities. To wit, the temperature $T$ is a measure of energy per degree of freedom; the baryon chemical potential $\mu$ controls the mean baryon density. The statistical quantities such as entropy (= measure of the number of available states), pressure, heat capacity, etc., will also be functions of $T$ and $\mu$, and will have to be determined. The theoretical techniques required for the description of the two quite different phases, viz., the hadronic gas and the quark-gluon plasma, must allow for the formulation of numerous hadronic resonances on the one side, which then at sufficiently high energy density dissolve into the state consisting of their constituents\footnote{These ideas originate in Hagedorn's statistical bootstrap theory~\cite{3chap8Cbib2a,3chap8Cbib2b}.}. At this point, we must appreciate the importance and help by a finite, i.e., nonzero temperature in reaching the transition to the quark-gluon plasma: to obtain a high particle density, instead of only compressing the matter (which as it turns out is quite difficult), we also heat it up; many pions are generated in a collision, allowing the transition to occur at moderate, even vanishing baryon density \cite{3chap8Cbib3}.

\begin{figure}[b]
\centering\resizebox{0.45\textwidth}{!}{%
\includegraphics{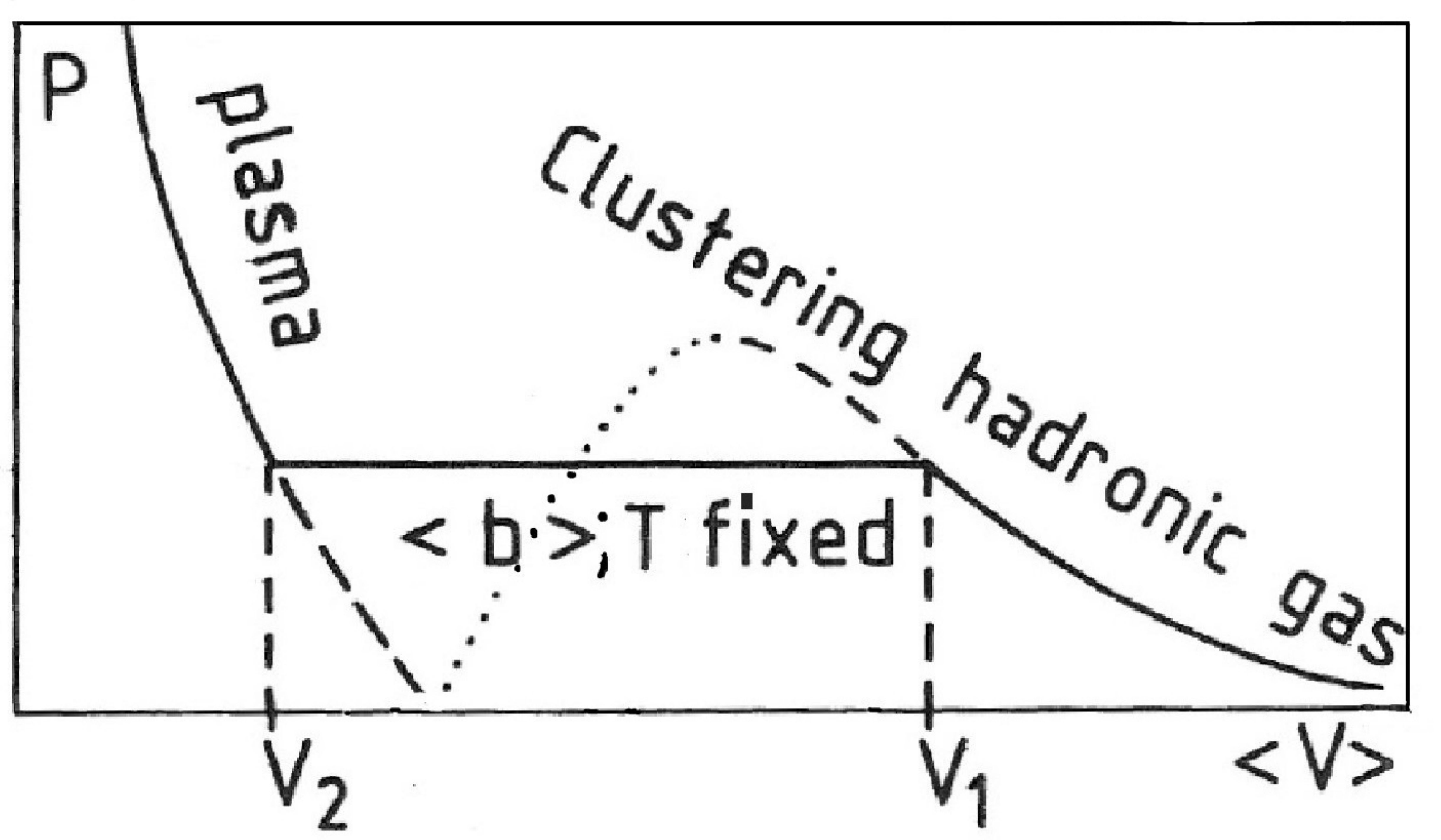}
}
\caption[]{$p$--$V$ diagram for the gas--plasma first order transition, with the {\it dotted curve\/} indicating a model-dependent, unstable domain between overheated and undercooled phases.}\label{3chap8Cfig1}
\end{figure}

\begin{table*}[t]
\centering
\caption[]{Phase transition of hot hadronic matter in theoretical physics}\label{3chap8Ctab1}
\setlength\tabcolsep{5pt}
\begin{tabular}{@{}|l c l c l|}
\noalign{\smallskip}\hline\noalign{\smallskip}
Object & $\longrightarrow$ & Observational hypothesis & $\longrightarrow$ & Theoretical consequence\\\noalign{\smallskip}\hline\noalign{\smallskip}
Nature & $\longrightarrow$ & Internal SU(3) symmetry & $\longrightarrow$ & First order phase transition\\
 &  &  &  & (on a lattice)\\\noalign{\smallskip}\hline\noalign{\smallskip}
Nature & $\longrightarrow$ & Bootstrap $\widehat{=}$ resonance & $\longrightarrow$ & First order phase transition\\
 &  & dominance of hadronic &  & in a phenomenological\\
  &  & interactions &  & bootstrap approach\\\noalign{\smallskip}\hline\noalign{\smallskip}
? & $\longrightarrow$ & Internal SU(2) symmetry & $\longrightarrow$ & Second order phase transition\\
 &  &  &  & (on a lattice)\\\noalign{\smallskip}\hline
\end{tabular}
\end{table*}

Consider, as an illustration of what is happening, the $p,V$ diagram shown in \rf{3chap8Cfig1}. Here we distinguish three domains. The hadronic gas region is approximately a Boltzmann gas where the pressure rises with reduction of the volume. When the internal excitation rises, the individual hadrons begin to cluster. This reduces the increase in the Boltzmann pressure, since a smaller number of particles exercises a smaller pressure. In a complete description of the different phases, we have to allow for a coexistence of hadrons with the plasma state in the sense that the internal degrees of freedom of each cluster, i.e., quarks and gluons, contribute to the total pressure even before the dissolution of individual hadrons. This does indeed become necessary when the clustering overtakes the compressive effects and the hadronic gas pressure falls to zero as $V$ reaches the proper volume of hadronic matter. At this point the pressure rises again very quickly, since in the absence of individual hadrons, we now compress only the hadronic constituents. By performing the Maxwell construction between volumes $V_1$ and $V_2$, we can in part account for the complex process of hadronic compressibility alluded to above. 

As this discussion shows, and detailed investigations confirm \cite{3chap8Cbib4a,3chap8Cbib4b,3chap8Cbib4c,3chap8Cbib4d}, we cannot escape the conjecture of a first order phase transition in our approach. This conjecture of \cite{3chap8Cbib1g} has been criticized, and only more recent lattice gauge theory calculations have led to the widespread acceptance of this phenomenon, provided that an internal SU(3) (color) symmetry is used -- SU(2) internal symmetry leads to a second order phase transition \cite{3chap8Cbib1i}. It is difficult to assess how such hypothetical changes in actual internal particle symmetry would influence phenomenological descriptions based on an observed picture of nature. For example, it is difficult to argue that, were the color symmetry SU(2) and not SU(3), we would still observe the resonance dominance of hadronic spectra and could therefore use the bootstrap model. {\it All\/} present understanding of phases of hadronic matter is based on approximate models, which requires that Table~\ref{3chap8Ctab1} be read from left to right.

I believe that the description of hadrons in terms of bound quark states on the one hand, and the statistical bootstrap for hadrons on the other hand, have many common properties and are quite complementary. Both the statistical bootstrap and the bag model of quarks are based on quite equivalent phenomenological observations. While it would be most interesting to derive the phenomenological models quantitatively from the accepted fundamental basis -- the Lagrangian quantum field theory of a non-Abelian SU(3) `glue' gauge field coupled to colored quarks -- we will have to content ourselves in this report with a qualitative understanding only. Already this will allow us to study the properties of hadronic matter in both aggregate states: the hadronic gas and the state in which individual hadrons have dissolved into the plasma consisting of quarks and of the gauge field quanta, the gluons.

It is interesting to follow the path taken by an isolated quark-gluon plasma fireball in the $\mu,T$ plane, or equivalently in the $\nu,T$ plane. Several cases are depicted in \rf{3chap8Cfig2}. In the Big Bang expansion, the cooling shown by the dashed line occurs in a Universe in which most of the energy is in the radiation. Hence, the baryon density $\nu$ is quite small. In normal stellar collapse leading to cold neutron stars, we follow the dash-dotted line parallel to the $\nu$ axis. The compression is accompanied by little heating.

\begin{figure}[b]
\centering\resizebox{0.45\textwidth}{!}{%
\includegraphics{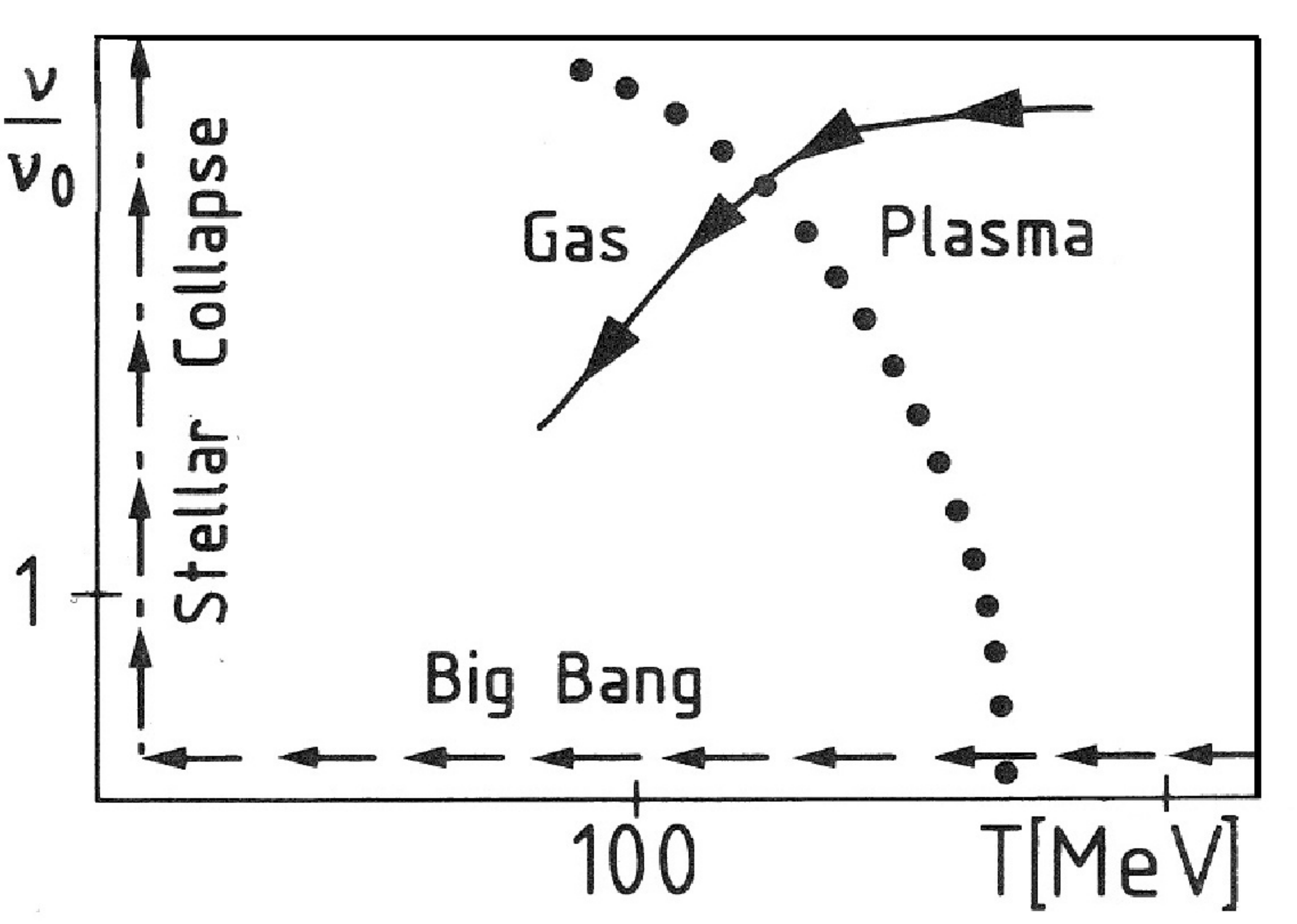}
}
\caption[]{Paths taken in the $\nu,T$ plane by different physical events.}\label{3chap8Cfig2}
\end{figure}

In contrast, in nuclear collisions, almost the entire $\nu,T$ plane can be explored by varying the parameters of the colliding nuclei. We show an example by the full line, and we show only the path corresponding to the cooling of the plasma, i.e., the part of the time evolution after the termination of the nuclear collision, assuming a plasma formation. The figure reflects the circumstance that, in the beginning of the cooling phase, i.e., for 1--$1.5\times 10^{-23}$~s, the cooling happens almost exclusively by the mechanism of pion radiation \cite{3chap8Cbib5a,3chap8Cbib5b}. In typical circumstances, about half of the available energy has been radiated away before the expansion, which brings the surface temperature close to the temperature of the transition to the hadronic phase. Hence a possible, perhaps even likely, scenario is that in which the freezing out and the expansion happen simultaneously. These highly speculative remarks are obviously made in the absence of experimental guidance. A careful study of the hadronization process most certainly remains to be performed.

In closing this section, let me emphasize that the question whether the transition hadronic gas $\longleftrightarrow$ quark-gluon plasma is a phase transition (i.e., discontinuous) or continuous phase transformation will probably only be answered in actual experimental work; as all theoretical approaches suffer from approximations unknown in their effect. For example, in lattice gauge computer calculations, we establish the properties of the {\it lattice\/} and not those of the continuous space in which we live. 

The remainder of this report is therefore devoted to the study of strange particles in different nuclear phases and their relevance to the observation of the quark-gluon plasma.

\setcounter{equation}{0}
\subsection{Strange particles in hot nuclear gas}\label{3chap8Csec2}

My intention in this section is to establish quantitatively the different channels in which the strangeness, however created in nuclear collisions, will be found. In our following analysis (see Ref.\cite{3chap8Cbib8}) a tacit assumption is made that the hadronic gas phase is practically a superposition of an infinity of different hadronic gases, and all information about the interaction is hidden in the mass spectrum $\tau(m^2,b)$ which describes the number of hadrons of baryon number $b$ in a mass interval $\D m^2$ and volume $V\sim m$. When considering strangeness-carrying particles, all we then need to include is the influence of the non-strange hadrons on the {\it baryon chemical potential\/} established by the non-strange particles. 

The total partition function is approximately multiplicative in these degrees of freedom:
\begin{equation}\label{3chap8Ceq2.1}
\ln Z = \ln Z^{\mbox{\footnotesize non-strange}} + \ln Z^{\rm strange}\;.
\end{equation}
For our purposes, i.e., in order to determine the particle abundances, it is sufficient to list the strange particles separately, and we find
\begin{equation}
\begin{array}{rl} 
   \displaystyle \ln Z^{\rm strange}(T,V,\lambda_{s},\lambda_{q}) = C\Big\{2W(x_{\rm K})
   &\displaystyle(\lambda_{s}\lambda_{q}^{-1} + \lambda_{s}^{-1}\lambda_{q})\\[0.4cm]
  \displaystyle \phantom{C\Big\{} + 2\big[W(x_{\sLambda}) + 3W(x_{\sSigma})\big]
  & \displaystyle \label{3chap8Ceq2.2}
(\lambda_{s}\lambda_{q}^{2} + \lambda_{s}^{-1}\lambda_{q}^{-2})\Big\}\;, 
\end{array}
\end{equation}
where
\begin{equation}\label{3chap8Ceq2.3}
W(x_i) = \left(\frac{m_i}{T}\right)^2K_2\left(\frac{m_i}{T}\right)\;.
\end{equation}
We have $C=VT^3/2\pi^2$ for a fully equilibrated state. However, strangeness-creating ($x\rightarrow {s}+{\bar s}$) processes in hot hadronic gas may be too slow (see below) and the total abundance of strange particles may fall short of this value of $C$ expected in {\it absolute strangeness chemical equilibrium\/}. On the other hand, strangeness exchange cross-sections are very large (e.g., the K$^-p$ cross-section is $\sim 100$~mb in the momentum range of interest), and therefore any momentarily available strangeness will always be distributed among all particles in \req{3chap8Ceq2.2} according to the values of the fugacities $\lambda_{q}=\lambda_{\rm B}^{1/3}$ and $\lambda_{s}$. Hence we can speak of a {\it relative strangeness chemical equilibrium\/}. 

We neglected to write down quantum statistics corrections as well as the multistrange particles $\uXi$ and $\uOmega^-$, as our considerations remain valid in this simple approximation \cite{3chap8Cbib6}. Interactions are effectively included through explicit reference to the baryon number content of the strange particles, as just discussed. Non-strange hadrons influence the strange faction by establishing the value of $\lambda_{q}$ at the given temperature and baryon density.

The fugacities $\lambda_{s}$ and $\lambda_{q}$ as introduced here control the strangeness and the baryon number, respectively. While $\lambda_{s}$ counts the strange quark content, the up and down quark content is counted by $\lambda_{q}=\lambda_{\rm B}^{1/3}$.

Using the partition function Eq.~(\ref{3chap8Ceq2.2}), we calculate for given $\mu$, $T$, and $V$ the mean strangeness by evaluating 
\begin{equation}\label{3chap8Ceq2.4}
\langle n_{s}-n_{{\bar s}}\rangle = \lambda_{s}\frac{\partial}{\partial\lambda_{s}} \ln Z^{\rm strange}(T,V,\lambda_{s},\lambda_{q})\;,
\end{equation}
which is the difference between strange and antistrange components. This expression must be equal to zero due to the fact that the strangeness is a conserved quantum number with respect to strong interactions. From this condition, we get\footnote{Notation has been changed $\gamma\to F$ in order to avoid confusion with phase space occupancy $\gamma$.}
\begin{equation}\label{3chap8Ceq2.5}
\lambda_{s} = \lambda_{q}\left|\frac{W(x_{\rm K})+\lambda_{\rm B}^{-1} \big[W(x_{\sLambda}) + 3W(x_{\sSigma})\big]}{W(x_{\rm K})+\lambda_{\rm B}\big[W(x_{\sLambda}) + 3W(x_{\sSigma})\big]}\right|^{1/2} \equiv \lambda_{q}F\;,
\end{equation}
a result contrary to intuition: $\lambda_{s}\neq 1$ for a gas with total $\langle s\rangle =0$. We notice a strong dependence of $F$ on the baryon number. For large $\mu$, the term with $\lambda_{\rm B}^{-1}$ will tend to zero and the term with $\lambda_{\rm B}$ will dominate the expression for $\lambda_{s}$ and $F$. As a consequence, the particles with fugacity $\lambda_{s}$ and strangeness $S=-1$ (note that by convention strange quarks $s$ carry $S=-1$, while strange antiquarks ${\bar s}$ carry $S=1$) are suppressed by a factor $F$ which is always smaller than unity. Conversely, the production of particles which carry the strangeness $S=+1$ will be favored by $F^{-1}$. This is a consequence of the presence of nuclear matter: for $\mu=0$, we find $F=1$.

In nuclear collisions, the mutual chemical equilibrium, that is, a proper distribution of strangeness among the strange hadrons, will most likely be  achieved. By studying the relative yields, we can exploit this fact and eliminate the absolute normalization $C$ [see \req{3chap8Ceq2.2}] from our considerations. We recall that the value of $C$ is uncertain for several reasons: 
\begin{itemize}
\item[i] $V$ is unknown.
\item[ii] $C$ is strongly $(t,r)$-dependent, through the space-time dependence of $T$.
\item[iii] Most importantly, the value $C=VT^3/2\pi^2$ assumes absolute chemical equilibrium,  which is not achieved owing to the shortness of the collision. 
\end{itemize}
Indeed, we have [see  \req{3chap8Ceq4.3} for in plasma strangeness formation and further details and solutions]
\begin{equation}\label{3chap8Ceq2.6}
\frac{\D C}{\D t} = A_{\rm H}\left[1-\frac{C(t)^2}{C(\infty)^2}\right]\;,
\end{equation}
and the time constant for strangeness production in nuclear matter can be estimated to be~\cite{3chap8Cbib7} 
$$\tau_{\rm H}=C(\infty)/2A_{\rm H}\sim 10^{-21}\,{\rm s}\,.$$
Thus $C$ does not reach $C(\infty)$ in plasmaless nuclear collisions. If the plasma state is formed, then the relevant $C>C(\infty)$ (since strangeness yield in plasma is above strangeness yield in hadron gas (see  below).

Now, why should we expect {\it relative\/} strangeness equilibrium to be reached faster than {\it absolute\/} strangeness equilibrium \cite{3chap8Cbib8}? Consider the strangeness {\it exchange\/} interaction
\begin{equation}\label{3chap8Ceq2.7}
{\rm K}^-{p} \;\longrightarrow\;\uLambda\upi^0
\end{equation}
\vskip -0.25cm\centerline{\resizebox{0.25\textwidth}{!}{%
\includegraphics{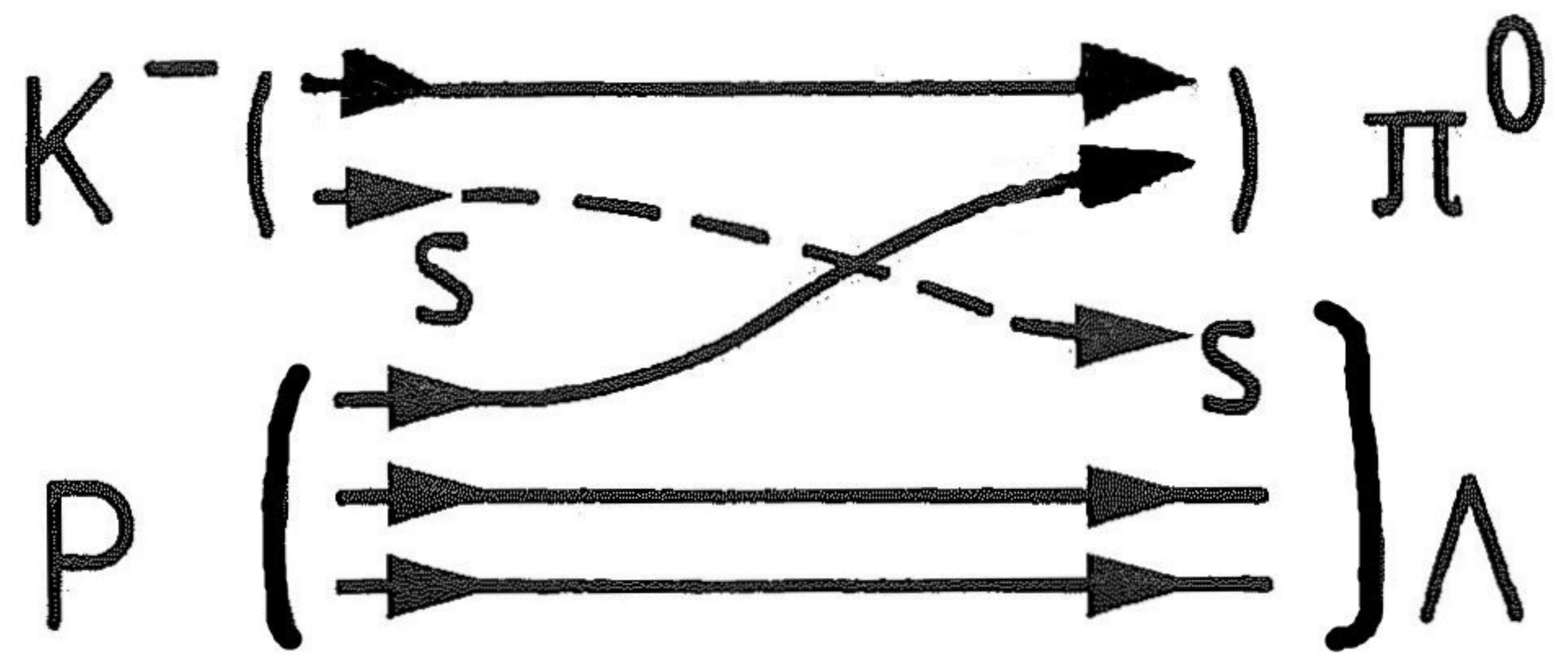}
}}
\vskip 0.3cm
\noindent which has a cross-section of about 10~mb at low energies, while the $s{\bar s}$ \lq strangeness creating' associate production
\begin{equation}\label{3chap8Ceq2.8}
{p} {p} \;\longrightarrow\;{p} \uLambda{\rm K}^+
\end{equation}
\vskip -0.35cm\centerline{\resizebox{0.25\textwidth}{!}{%
\includegraphics{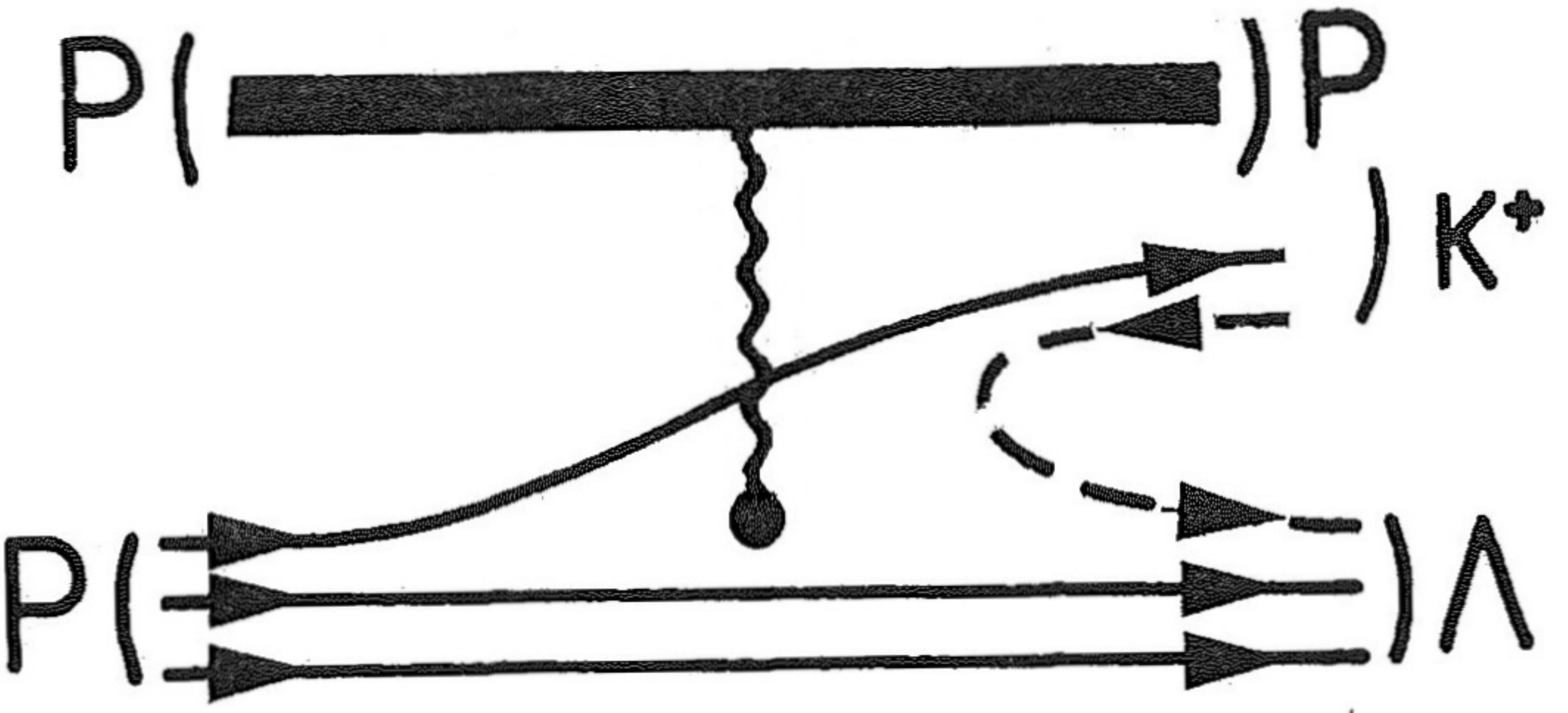}
}}
\vskip 0.3cm
\noindent has a cross-section of less than 0.06~mb, i.e., 150 times smaller. Since the latter reaction is somewhat disfavored by phase space, consider further the reaction 
\begin{equation}\label{3chap8Ceq2.9}
\upi{p} \;\longrightarrow\;{\rm Y}{\rm K}
\end{equation}
\vskip -0.3cm\centerline{\resizebox{0.25\textwidth}{!}{%
\includegraphics{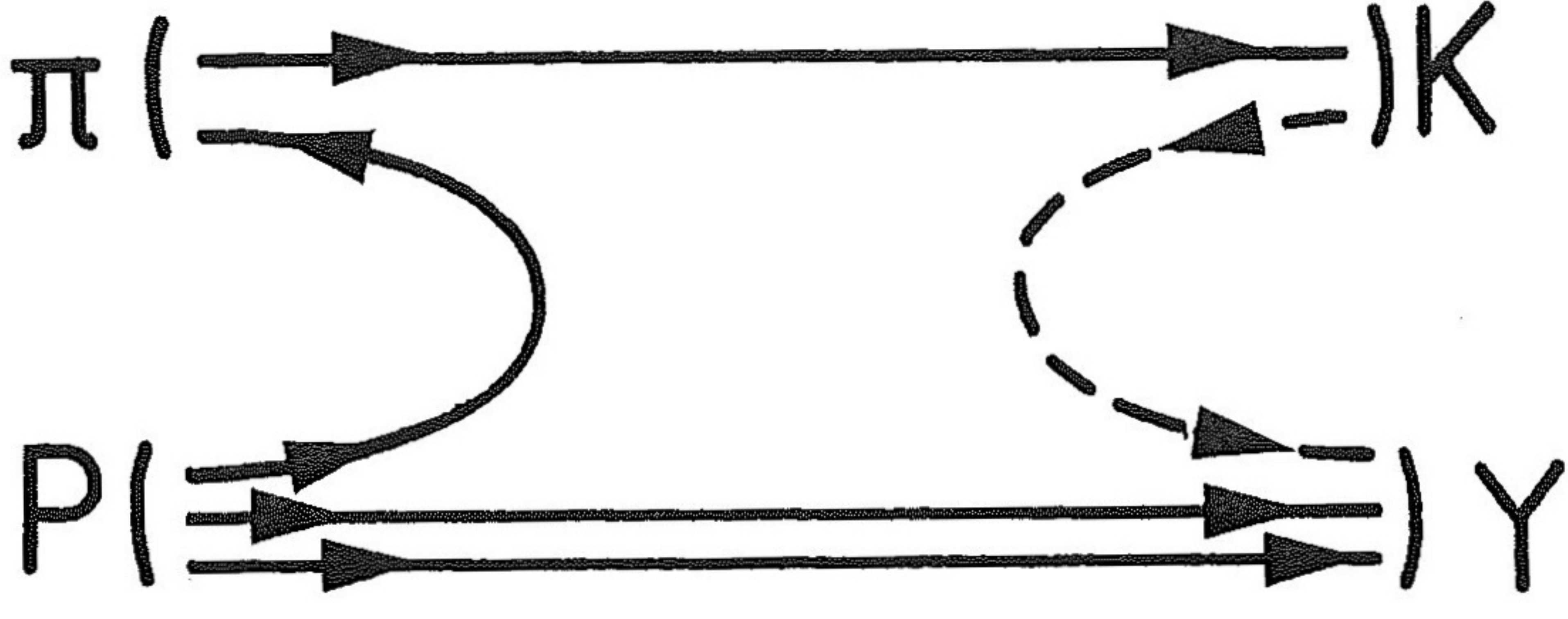}
}}
\vskip 0.3cm
\noindent where Y is any hyperon (strange baryon). This has a cross-section of less than 1~mb, still 10 times weaker than {\it one\/} of the $s$-exchange channels in  \req{3chap8Ceq2.7}. Consequently, I expect the relative strangeness equilibration time to be about ten times shorter than the absolute strangeness equilibration time, namely $10^{-23}$~s, in hadronic matter of about twice nuclear density.

We now compute the relative strangeness abundances expected from nuclear collisions. Using \req{3chap8Ceq2.5}, we find from Eq.~(\ref{3chap8Ceq2.2}) the grand canonical partition sum for zero average strangeness:
\begin{equation}
\begin{array}{rl}
\displaystyle \ln Z_0^{\rm strange} C\Big[
 & \displaystyle  2W(x_{\rm K})\big(F\lambda_{\rm K} + F^{-1}\lambda_{\overline{\rm K}}\big)\\[0.2cm]
 + &\displaystyle  2W(x_{\sLambda})\big(F\lambda_{\rm B}\lambda_{\sLambda} + 
         F^{-1}\lambda_{\rm B}^{-1}\lambda_{\overline{\sLambda}}\big)\\[0.2cm]
  + & \displaystyle  6W(x_{\sSigma})\big(F\lambda_{\rm B}\lambda_{\sSigma} +
          F^{-1}\lambda_{\rm B}^{-1}\lambda_{\overline{\sSigma}}\big)\Big]\;,\label{3chap8Ceq2.10}
\end{array}
\end{equation}
where, in order to distinguish different hadrons, dummy fugacities $\lambda_i$, $i={\rm K}$, $\overline{\rm K}$, $\uLambda$, $\overline{\uLambda}$, $\uSigma$, $\overline{\uSigma}$ have been written. The strange particle multiplicities then follow from
\begin{equation}\label{3chap8Ceq2.11}
\langle n_i\rangle = \lambda_i\frac{\partial}{\partial\lambda_i}\ln Z_0^{\rm strange}\Big|_{\lambda_i=1}\;.
\end{equation}
Explicitly, we find (notice that the power of $F$ follows the $s$-quark content):
\begin{equation}\label{3chap8Ceq2.12}
\langle n_{{\rm K}^\pm}\rangle = CF^\mp W(x_{\rm K})\;,
\end{equation}
\begin{equation}\label{3chap8Ceq2.13}
\langle n_{\sLambda/\sSigma^0}\rangle = CF^{+1} W(x_{\sLambda/\sSigma^0}) \E^{+\mu_{\rm B}/T}\;,
\end{equation}
\begin{equation}\label{3chap8Ceq2.14}
\langle n_{\overline{\sLambda}/\overline{\sSigma}{}^0}\rangle = CF^{-1} W(x_{\overline{\sLambda}/\overline{\sSigma}{}^0}) \E^{-\mu_{\rm B}/T}\;.
\end{equation}
In \req{3chap8Ceq2.14}  we have indicated that the multiplicity of antihyperons can only be built up if antibaryons are present according to their (small) phase space. This still seems an unlikely proposition, and the statistical approach may be viewed as providing an upper limit on their multiplicity.

From the above equations, we can derive several very instructive conclusions. In \rf{3chap8Cfig3} we show the ratio 
$$\langle n_{{\rm K}^+}\rangle/\langle n_{{\rm K}^-}\rangle=F^{-2}$$ 
as a function of the baryon chemical potential $\mu$ for several temperatures that can be expected and which are seen experimentally. We see that this particular ratio is a good measure of the baryon chemical potential in the hadronic gas phase, provided that the temperatures are approximately known. The mechanism for this process is as follows: the strangeness exchange reaction of \req{3chap8Ceq2.7} tilts to the left (K$^-$) or to the right (abundance $F\sim{\rm K}^+$), depending on the value of the baryon chemical potential.

\begin{figure}
\centering\resizebox{0.45\textwidth}{!}{%
\includegraphics{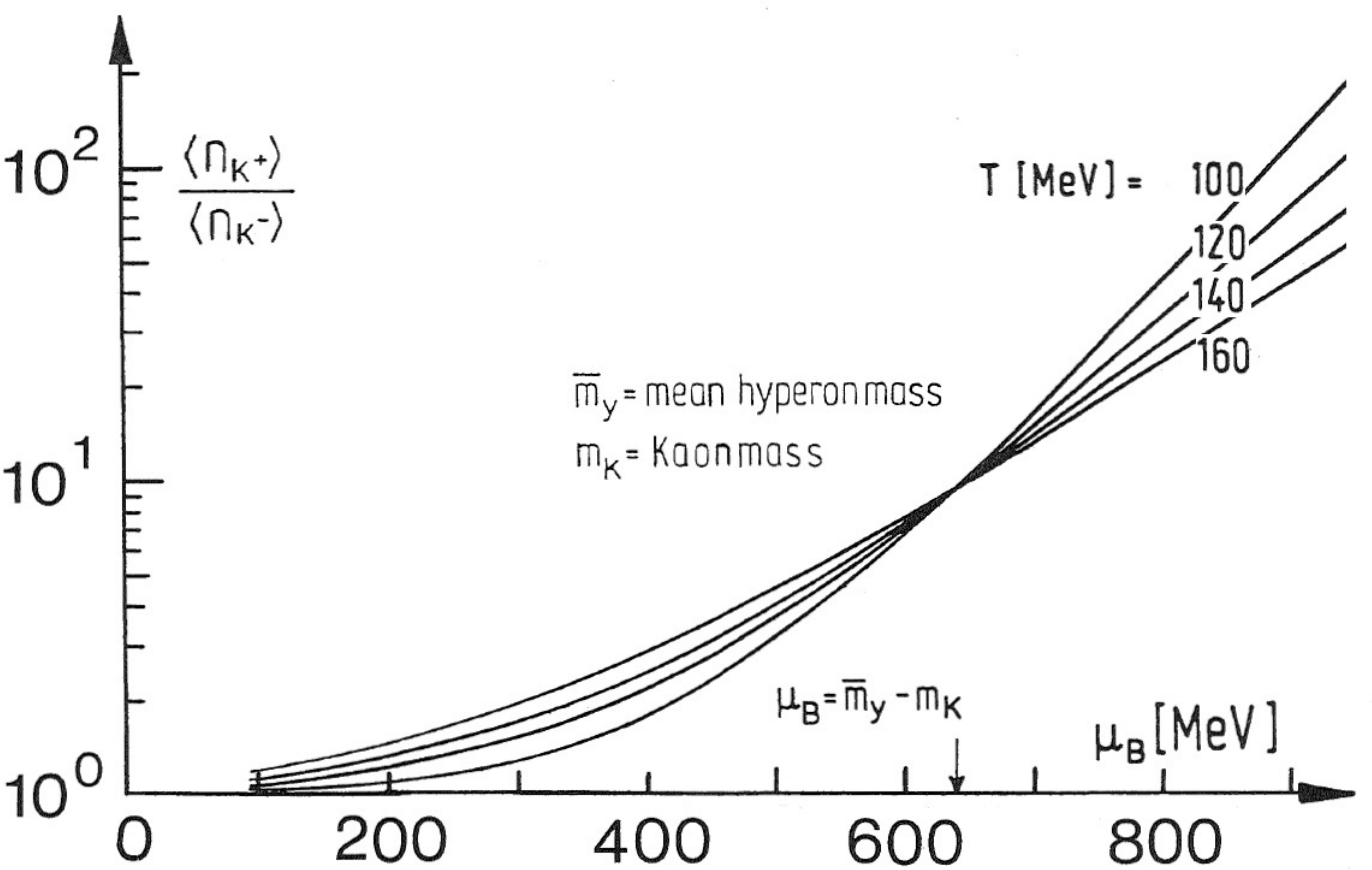}
}
\caption[]{The ratio $\langle n_{{\rm K}^+}\rangle/\langle n_{{\rm K}^-}\rangle\equiv F^{-2}$ as a function of the baryon chemical potential $\mu$, for $T=100,(20),160$\,MeV. The lines cross where   $\mu=\overline{m}_\mathrm{Y}-m_\mathrm{K}$; $\overline{m}_\mathrm{Y}$ is the mean hyperon mass.}\label{3chap8Cfig3}
\end{figure}

In  \rf{3chap8Cfig4} the long dashed  line shows the upper limit for the abundance of $\overline{\uLambda}$ as measured in terms of $\uLambda$ abundances. Clearly visible is the substantial relative suppression of $\overline{\uLambda}$, in part caused by the baryon chemical potential factor of \req{3chap8Ceq2.14}, but also by the strangeness chemistry (factor $F^2$), as in K$^+$K$^-$ above. Indeed, the actual relative number of $\overline{\uLambda}$ will be even smaller, since $\uLambda$ are in relative chemical equilibrium and $\overline{\uLambda}$ in hadron gas are not: the reaction ${\rm K}^+\overline{p} \rightarrow\overline{\uLambda}\upi^0$, analogue to \req{3chap8Ceq2.7}, will be suppressed by {\it low\/} $\overline{p} $ abundance. Also indicated in \rf{3chap8Cfig4} by shading is a rough estimate for the $\overline{\uLambda}$ production in the plasma phase, which suggests that anomalous $\overline{\uLambda}$ abundance may be an interesting feature of highly energetic nuclear collisions \cite{3chap8Cbib16}, for further discussion see Section~\ref{3chap8Csec5} below.
\begin{figure}[b]
\centering\resizebox{0.45\textwidth}{!}{%
\includegraphics{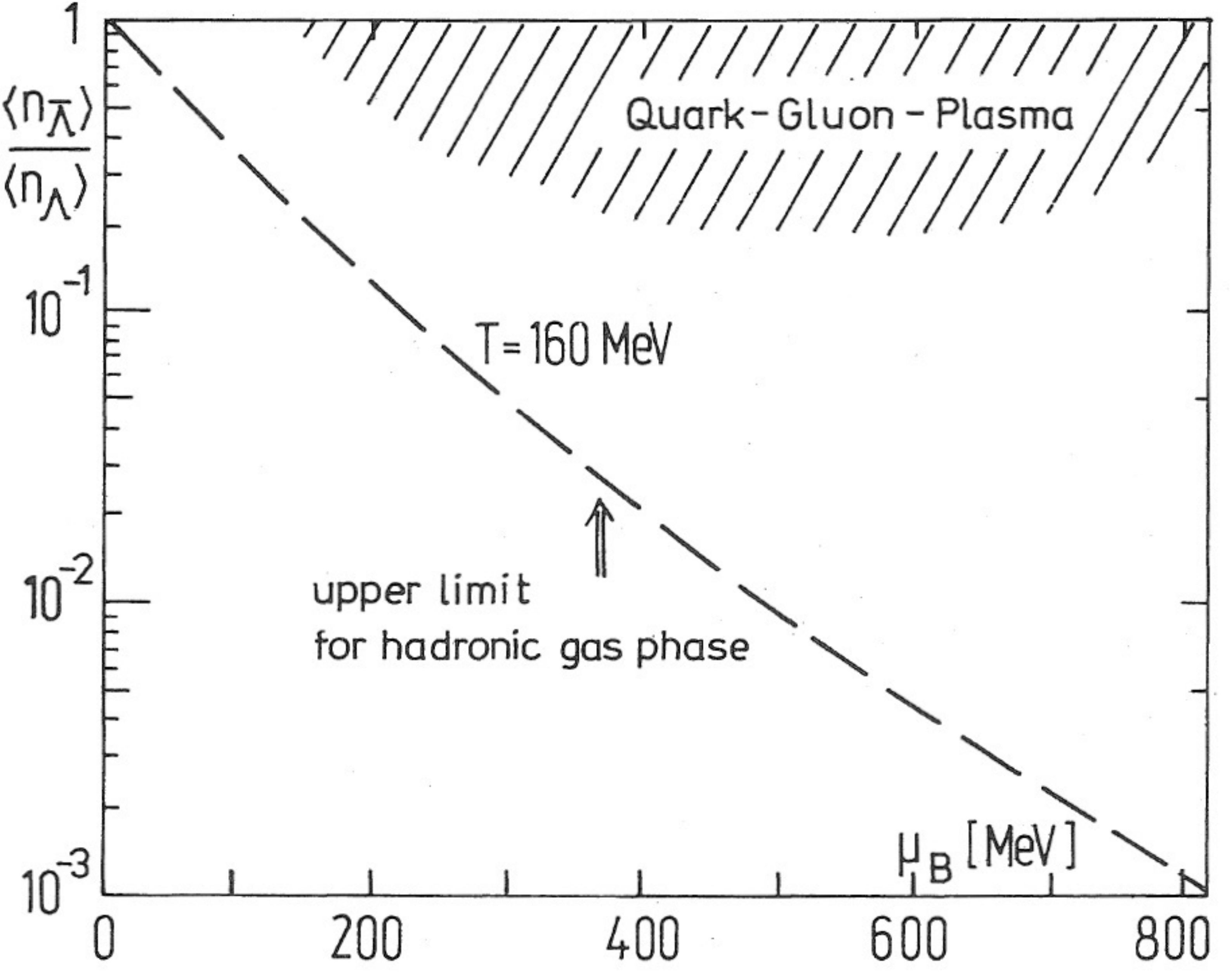}
}
\caption[]{Relative abundance of $\overline{\pLambda}/\pLambda$. The actual yield from the hadronic gas limit may still be 10--100 times smaller than the statistical value shown.}\label{3chap8Cfig4}
\end{figure}

\setcounter{equation}{0}
\subsection{Quark-Gluon Plasma}\label{3chap8Csec3}

From the study of hadronic spectra, as well as from hadron--hadron and hadron--lepton interactions, there has emerged convincing evidence for the description of hadronic structure in terms of quarks \cite{3chap8Cbib9}. For many purposes it is entirely satisfactory to consider baryons as bound states of three fractionally charged particles, while mesons are quark--antiquark bound states. The Lagrangian of quarks and gluons is very similar to that of electrons and photons, except for the required summations over flavour and color:
\begin{equation}\label{3chap8Ceq3.1}
L=\overline{\psi}\big[F\cdot(p-gA)-m\big]\psi - \frac{1}{4}F_{\mu\nu}F^{\mu\nu}\;.
\end{equation}
The flavour-dependent masses $m$ of the quarks are small. For u, d flavours, one estimates $m_{\rm u,d}\sim5$--20~MeV. The strange quark mass is usually chosen at about 150~MeV \cite{3chap8Cbib10a,3chap8Cbib10b}. The essential new feature of QCD, not easily visible in \req{3chap8Ceq3.1}, is the non-linearity of the field strength $F$ in terms of the potentials $A$. This leads to an attractive glue--glue interaction in select channels {\it and\/}, as is believed, requires an improved (non-perturbative) vacuum state in which this interaction is partially diagonalized, providing for a possible perturbative approach.

The energy density of the perturbative vacuum state, defined with respect to the true vacuum state, is by definition a positive quantity, denoted by $\cal B$. This notion has been introduced originally in the MIT bag model \cite{3chap8Cbib11a,3chap8Cbib11b,3chap8Cbib11c}, logically, e.g., from a fit to the hadronic spectrum, which gives
\begin{equation}\label{3chap8Ceq3.2}
{\cal B}=\big[\mbox{(140--210)~MeV}\big]^4 = \mbox{(50--250)~MeV/fm}^3\;.
\end{equation}
The central assumption of the quark bag approach is that, inside a hadron where quarks are found, the true vacuum structure is displaced or destroyed. One can turn this point around: quarks can only propagate in domains of space in which the true vacuum is absent. This statement is a reformulation of the quark confinement problem. Now the remaining difficult problem is to show the incompatibility of quarks with the true vacuum structure. Examples of such behavior in ordinary physics are easily found; e.g., a light wave is reflected from a mirror surface, magnetic field lines are expelled from superconductors, etc. In this picture of hadronic structure and quark confinement, all {\it colorless\/} assemblies of quarks, antiquarks, and gluons can form stationary states, called a quark bag. In particular, all higher combinations of the three-quark baryons $(qqq)$ and quark--antiquark mesons ($q{\bar q}$) form a permitted state.

As the $u$ and $d$ quarks are almost massless inside a bag, they can be produced in pairs, and at moderate internal excitations, i.e., temperatures, many  $q{\bar q}$ pairs will be present. Similarly, $s{\bar s}$ pairs will also be produced. We will return to this point at length below. Furthermore, real gluons can be excited and will be included here in our considerations.

Thus, what we are considering here is a {\it large\/} quark bag with substantial, equilibrated internal excitation, in which the interactions can be handled (hopefully) perturbatively. In the large volume limit, which as can be shown is valid for baryon number $b\gtrsim 10$, we simply have for the light quarks the partition function of a Fermi gas which, for practically massless u and d quarks can be given analytically (see ref.\cite{3chap8Cbib1b} and \cite{3chap8Cbib12a,3chap8Cbib12b}), even including the effects of interactions through first order in $\alpha_{s}=g^2/4\pi\,$:
\begin{equation}\label{3chap8Ceq3.3}
\begin{array}{rl}
\displaystyle \ln Z_{q}(\beta,\mu) = \frac{gV}{6\pi^2}\beta^{-3}
&\displaystyle\left\{ \hspace{-2pt}\left(1-\frac{2\alpha_{s}}{\pi}\right)\hspace{-2pt}\left[\frac{1}{4}(\mu \beta)^4 + \frac{\pi^2}{2}(\mu\beta)^2\right]\right. \\[0.3cm]
 +& \displaystyle\left.\left(1-\frac{50}{21}\frac{\alpha_{s}}{\pi}\right)\frac{7\pi^4}{60}\right\}\;.
\end{array}
\end{equation}
Similarly, the glue is a Bose gas:
\begin{equation}\label{3chap8Ceq3.4}
\ln Z_{g}(\beta,\lambda) = V\frac{8\pi^2}{45}\beta^{-3}\left(1-\frac{15}{4}\frac{\alpha_{s}}{\pi}\right)\;,
\end{equation}
while the term associated with the difference to the true vacuum, the bag term, is
\begin{equation}\label{3chap8Ceq3.5}
\ln Z_{\rm bag}=-{\cal B}V\beta\;.
\end{equation}
It leads to the required positive energy density $\cal B$ within the volume occupied by the colored quarks and gluons and to a negative pressure on the surface of this region. At this stage, this term is entirely phenomenological, as discussed above. The equations of state for the quark-gluon plasma are easily obtained by differentiating
\begin{equation}\label{3chap8Ceq3.6}
\ln Z=\ln Z_{q}+\ln Z_{g}+\ln Z_{\rm vac}\;,
\end{equation}
with respect to $\beta$, $\mu$, and $V$.

An assembly of quarks in a bag will assume a geometric shape and size such as to make the total energy $E(V,b,S)$ as small as possible at fixed given baryon number and fixed total entropy $S$. Instead of just considering one bag we may, in order to be able to use the methods of statistical physics, use the microcanonical ensemble. We find from the first law of thermodynamics, viz.,
\begin{equation}\label{3chap8Ceq3.7}
\D E = -P\D V+T\D S+\mu\D b\;,
\end{equation}
that 
\begin{equation}\label{3chap8Ceq3.8}
P=-\frac{\partial E(V,b,S)}{\partial V}\;.
\end{equation}
We observe that the stable configuration of a single bag, viz., $\partial E/\partial V=0$, corresponds to the configuration with vanishing pressure $P$ in the microcanonical ensemble. Rather than work in the microcanonical ensemble with fixed $b$ and $S$, we exploit the advantages of the grand canonical ensemble and consider $P$ as a function of $\mu$ and $T\,$:
\begin{equation}\label{3chap8Ceq3.9}
P=-\frac{\partial}{\partial V} \big[T\ln Z(\mu,T,V)\big]\;,
\end{equation}
with the result 
\begin{equation}\label{3chap8Ceq3.10}
P=\frac{1}{3}(\varepsilon-4{\cal B})\;,
\end{equation}
where $\varepsilon$ is the energy density:
\begin{eqnarray}
\varepsilon &=& \frac{6}{\pi^2}\left\{\left(1-\frac{2\alpha_{s}}{\pi}\right)\left[\frac{1}{4}\left(\frac{\mu}{3}\right)^4 + \frac{1}{2}\left(\frac{\mu}{3}\right)^2(\pi T)^2\right]\right.\nonumber\\\noalign{\smallskip}
&&\phantom{\frac{6}{\pi^2}\Bigg\{}+\left. \left(1-\frac{50}{21}\frac{\alpha_{s}}{\pi}\right)\frac{7}{60}(\pi T)^4\right\}\nonumber\\\noalign{\smallskip}
  && \phantom{\frac{6}{\pi^2}\Bigg\{} + \left(1-\frac{15}{4}\frac{\alpha_{s}}{\pi}\right)\frac{8}{15\pi^2}(\pi T)^4 +{\cal B}\;.\label{3chap8Ceq3.11}
\end{eqnarray}
In Eq.~(\ref{3chap8Ceq3.10}), we have used the relativistic relation between the quark and gluon energy density and pressure:
\begin{equation}\label{3chap8Ceq3.12}
P_{q}=\frac{1}{3}\varepsilon_{q}\;,\qquad P_{g}=\frac{1}{3}\varepsilon_{g}\;.
\end{equation}
From  \req{3chap8Ceq3.10}, it follows that, when the pressure vanishes in a static configuration, the energy density is $4{\cal B}$, independently of the values of $\mu$ and $T$ which fix the line $P=0$. We note that, in both quarks and gluons, the interaction conspires to {\it reduce\/} the effective available number of degrees of freedom. At $\alpha_{s}=0$, $\mu=0$, we find the handy relation
\begin{equation}\label{3chap8Ceq3.13}
\varepsilon_{q}+\varepsilon_{g}=\left(\frac{T}{160~{\rm MeV}}\right)^4\;\left[\frac{\rm GeV}{{\rm fm}^3}\right]\;.
\end{equation}
It is important to appreciate how much entropy must be created to reach the plasma state. From Eq.~(\ref{3chap8Ceq3.6}), we find for the   entropy density ${\cal S}\,$  and the baryon density $\nu$:
\begin{equation}\label{3chap8Ceq3.14}
\begin{array}{rl}
{\cal S} =&\displaystyle \frac{2}{\pi}\hspace{-2pt}\left(\hspace{-2pt}1-\frac{2\alpha_{s}}{\pi}\hspace{-2pt}\right)\hspace{-2pt}\left(\frac{\mu}{3}\right)^2\pi T+
 \frac{14}{15\pi}\hspace{-2pt}\left(\hspace{-2pt}1-\frac{50}{21}\frac{\alpha_{s}}{\pi}\hspace{-2pt}\right)\hspace{-2pt}(\pi T)^3 \\[0.4cm]
+&\displaystyle \frac{32}{45\pi}\hspace{-2pt}\left(\hspace{-2pt}1-\frac{15}{4}\frac{\alpha_{s}}{\pi}\hspace{-2pt}\right)\hspace{-2pt}(\pi T)^3\;,
\end{array}
\end{equation}
\vskip 0.2cm
\begin{equation}\label{3chap8Ceq3.15}
\nu = \frac{2}{3\pi^2}\left\{\left(1-\frac{2\alpha_{s}}{\pi}\right)\left[\left(\frac{\mu}{3}\right)^3 + \frac{\mu}{3}(\pi T)^2\right]\right\}\;,
\end{equation}
which leads for $\mu/3=\mu_{q}<\pi T$ to the following expressions for the entropy per baryon [including the gluonic entropy second $T^3$ term in Eq.~(\ref{3chap8Ceq3.14})]:
\begin{equation}\label{3chap8Ceq3.16}
\frac{\cal S}{\nu}\approx \frac{37}{15}\pi^2\frac{T}{\mu_{q}}\quad\overset{T\sim \mu_{q}}{\longleftrightarrow}\quad 25\,!
\end{equation}
As this simple estimate shows, plasma events are extremely entropy-rich, i.e., they contain very high particle multiplicity. In order to estimate the particle multiplicity, one may simply divide the total entropy created in the collision by the entropy per particle for massless black body radiation, which is $S/n=4$. This suggests that, at $T\sim\mu_{q}$, there are roughly six pions per baryon. 

\setcounter{equation}{0}
\subsection{Strange Quarks in Plasma}\label{3chap8Csec4}

In lowest order in perturbative QCD, $s{\bar s}$ quark pairs can be created by gluon fusion processes, \rf{3chap8Bfig2}a,b,c; and by annihilation of light quark-antiquark pairs,  
see \rf{3chap8Bfig2}d.
The averaged total cross-sections for these processes were calculated by Brian Combridge \cite{3chap8Cbib13}. 

\begin{figure}
\centering\resizebox{0.45\textwidth}{!}{%
\includegraphics{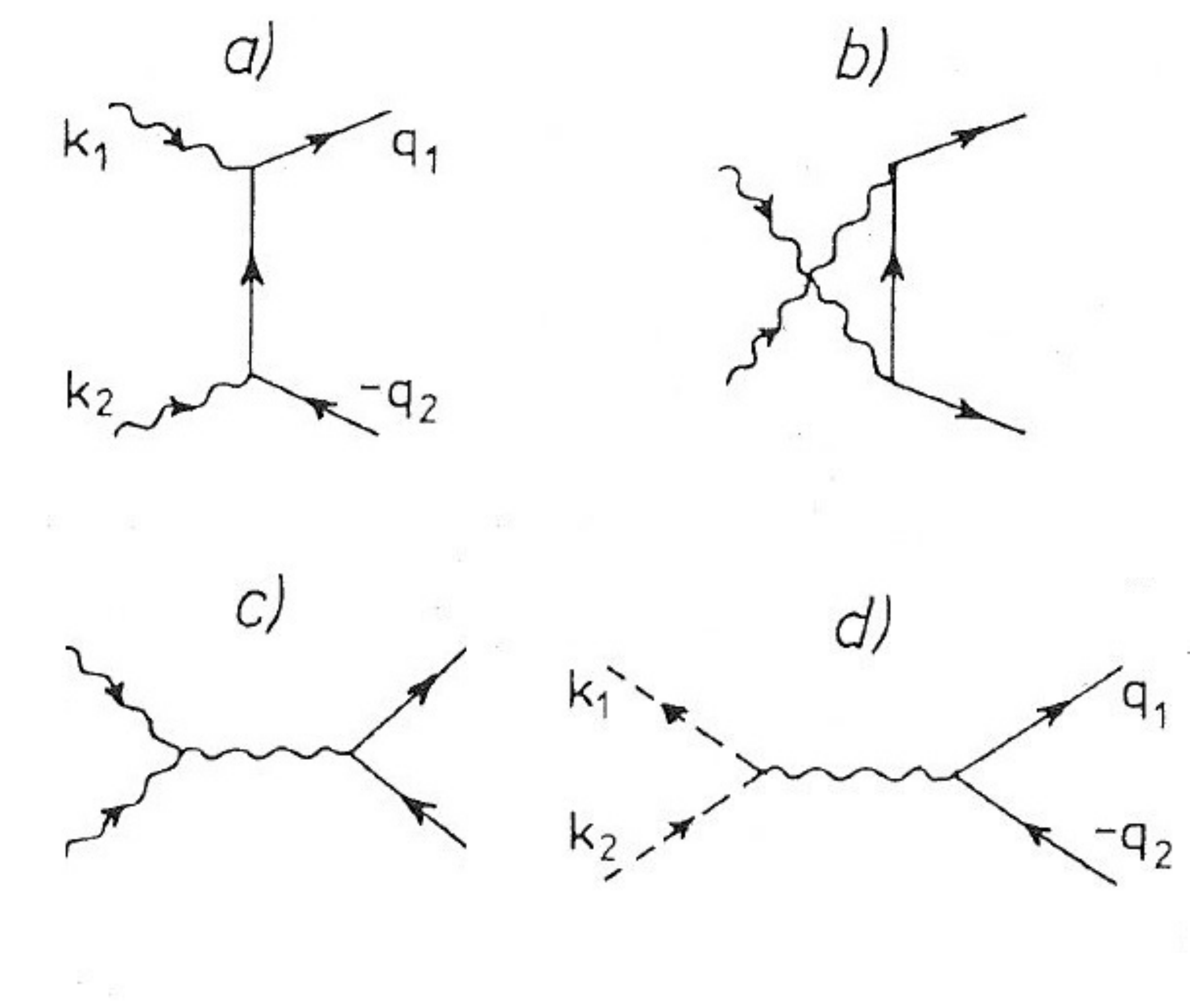}
}
\caption[]{Lowest order QCD diagrams for $s\bar s$  production:  {\bf a,b,c}) $ gg \rightarrow s\bar s  $, and
{\bf d}) $ q\bar q \rightarrow s\bar s $.}\label{3chap8Bfig2}
\end{figure}

Given the averaged cross-sections, it is easy to calculate the rate of events per unit time, summed over all final and averaged over initial states:
\begin{eqnarray}
\frac{\D N}{\D t} &=& 
\int\D^3x\! \sum_i\!\int\!\frac{\D^3k_1\D^3k_2}{(2\pi)^3|k_1|(2\pi)^3|k_2|}
\rho_{i,1}(k_1,x) \rho_{i,2}(k_2,x)\nonumber\\\noalign{\smallskip}
  && \times\int_{4M^2}^\infty\D s\,\delta\big(s-(k_1+k_2)^2\big)k_1^\mu k_{2\mu}\overline{\sigma}(s)\;.\label{3chap8Ceq4.1}
\end{eqnarray}
The factor $k_1\cdot k_2/|k_1||k_2|$ is the relative velocity for massless gluons or light quarks, and we have introduced a dummy integration over $s$ in order to facilitate the calculations. The phase space densities $\rho_i(k,x)$ can be approximated by assuming the $x$-independence of temperature $T(x)$ and the chemical potential $\mu(x)$, in the so-called local statistical equilibrium. Since $\rho$ then only depends on the absolute value of $\vec{k}$ in the rest frame of the equilibrated plasma, we can easily carry out the relevant integrals and obtain for the dominant process of the gluon fusion reaction \rf{3chap8Bfig2}a,b,c
the invariant rate per unit time and volume \cite{3chap8Cbib14}:
\begin{equation}\label{3chap8Ceq4.2}
{\cal A}=\frac{\D^4 N}{\D^3x\D t} \approx {\cal A}_{g} = \frac{7\alpha_{s}^2}{6\pi^2}MT^3\E^{-2M/T}\!\left(\!1+\frac{51}{14}\frac{T}{M}+\cdots\!\right)\!\;,
\end{equation}
where $M$ is the strange quark mass\footnote{In \req{3chap8Ceq4.2} a  factor 2 was included to reduce the invariant rate $\cal A$, see Erratum:
``Strangeness Production in the Quark-Gluon Plasma''
Johann Rafelski and Berndt M\"uller
Phys. Rev. Lett. {\bf 56}, 2334 (1986). This factor did not carry through to any  of the following results. However, additional definition factors `2' show up below in  Eqs.\ref{3chap8Ceq4.4}, \ref{3chap8Ceq4.5}.}.

The abundance of $s{\bar s}$ pairs cannot grow forever. At some point the $s{\bar s}$ annihilation reaction will restrict the strange quark population. It is important to appreciate that the $s{\bar s}$ pair annihilations may not proceed via the two-gluon channel, but instead occasionally through $\ugamma$G (photon-Gluon) final states \cite{3chap8Cbib15}. The noteworthy feature of such a reaction is the production of relatively high energy $\ugamma$'s at an energy of about 700--900~MeV ($T=160$~MeV) stimulated by coherent glue emission. These $\ugamma$'s will leave the plasma without further interactions and provide an independent confirmation of the $s$-abundance in the plasma.

\begin{figure}
\centering\resizebox{0.45\textwidth}{!}{%
\includegraphics{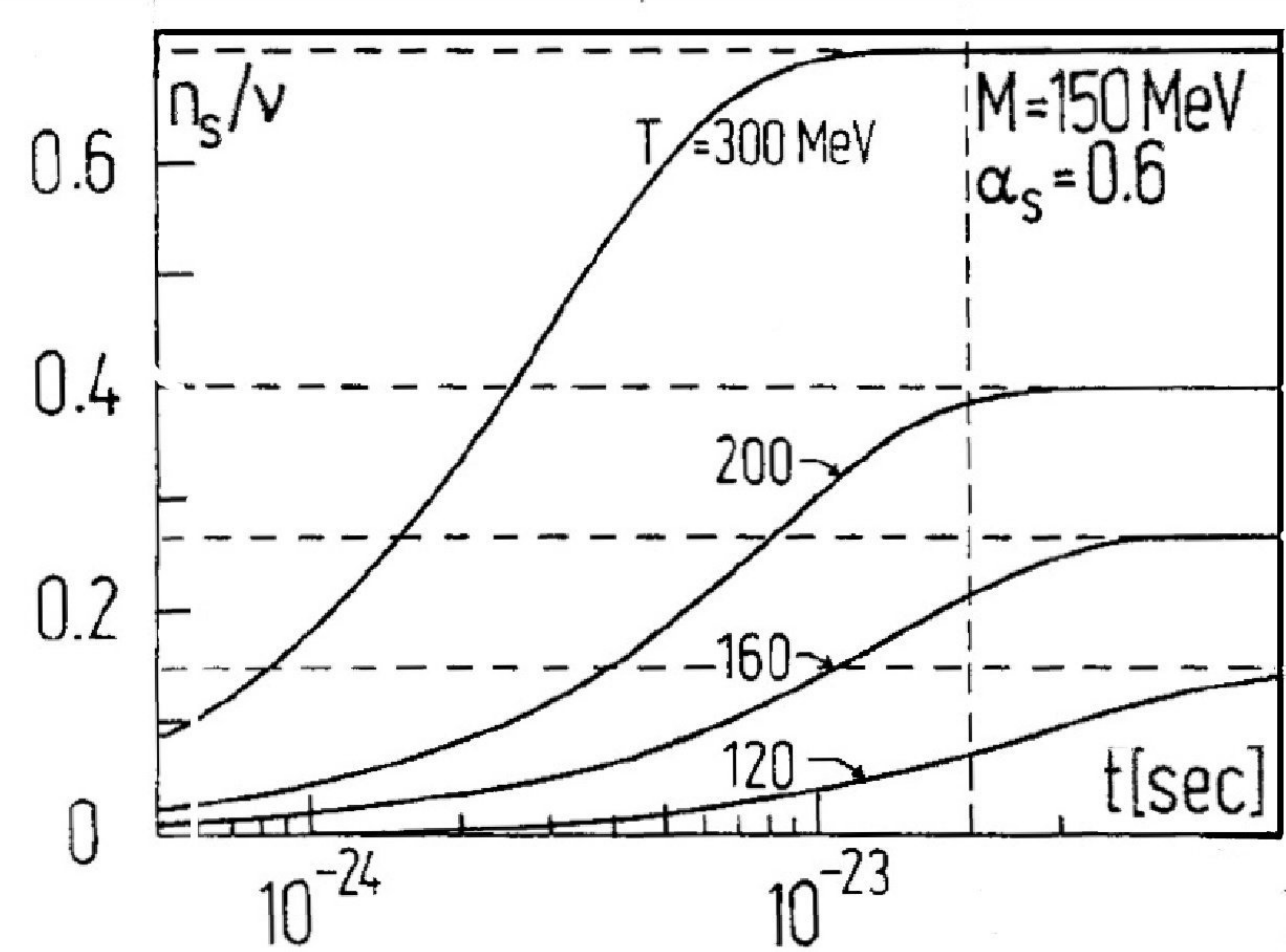}
}
\caption[]{Time evolution of the  strange quark to baryon number abundance in the plasma for various temperatures $T\sim\mu_{q}=\mu/3$. $M=150$~MeV, $\alpha_{s}=0.6$\,.}\label{3chap8Cfig6}
\end{figure}

The loss term of the strangeness population is proportional to the square of the density $n_{s}$ of strange and antistrange quarks. With $n_{s}(\infty)$ being the saturation density at large times, the following differential equation determines $n_{s}$ as a function of time\,\cite{3chap8Cbib2b}

\begin{equation}\label{3chap8Ceq4.3}
\frac{\D n_{s}}{\D t}\approx A\left\{1-\left[\frac{n_{s}(t)}{n_{s}(\infty)}\right]^2\right\}\;.
\end{equation}
Thus we find
\begin{equation}\label{3chap8Ceq4.4}
n_{s}(t) = n_{s}(\infty) 
{\displaystyle 
\frac{\tanh(t/2\tau)+\frac{n_{s}(0)}{n_{s}(\infty)}}   
      {1+\frac{n_{s}(0)}{n_{s}(\infty)}\tanh(t/2\tau)}
}\;,\qquad \tau=\frac{n_{s}(\infty)}{2{\cal A}}\;.
\end{equation}
where
\begin{equation}\label{3chap8Ceq4.5}
\tau=\frac{n_{s}(\infty)}{2{\cal A}}\;.
\end{equation}
The relaxation time $\tau$ of the strange quark density in Eq.~(\ref{3chap8Ceq4.5}) is  obtained using the saturated phase space in \req{3chap8Ceq4.5}. We have \cite{3chap8Cbib14}
\begin{equation}\label{3chap8Ceq4.6}
\tau\approx \tau_{g} =\left(\frac{\pi}{2}\right)^{1/2} \frac{9M^{1/2}}{7\alpha_{s}^{2}}
T^{-3/2}\E^{M/T}\left(1+\frac{99}{56}\frac{T}{M}+\cdots\right)^{-1}\;.
\end{equation}
For $\alpha_{s}\sim 0.6$ and $M\sim T$, we find from \req{3chap8Ceq4.6}  
that $\tau\sim 4\times 10^{-23}$~s. $\tau$ falls off rapidly with increasing temperature. Figure~\ref{3chap8Cfig6} shows the approach of $n_{s}(t)$, normalized with baryon density,  to the fully saturated phase space as a function of time.
For $M\lesssim T=160$~MeV, the saturation requires $4\times 10^{-23}$~s, while for $T=200$~MeV, we need $2\times 10^{-23}$~s, corresponding to the anticipated lifetime of the plasma. But it is important to observe that, even at $T=120$~MeV, the phase space is half-saturated in $2\times 10^{-23}$~s, a point to which we will return below. Another remarkable fact is 
the high abundance of strangeness relative to baryon number seen in  \rf{3chap8Cfig6} -- here, baryon number was computed assuming $T\sim\mu_{q}=\mu/3$ [see  \req{3chap8Ceq3.15}]. These two facts, namely:
\begin{enumerate}
\item high relative strangeness abundance in plasma,
\item practical saturation of available phase space, 
\end{enumerate}
have led me to suggest the observation of strangeness as a possible signal of quark-gluon plasma \cite{3chap8Cbib16}. 

There are two elements in point (1) above: firstly, strangeness in the quark-gluon phase is practically as abundant as the anti-light quarks $\overline{\rm u}=\overline{\rm d}={\bar q}$, since both phase spaces have similar suppression factors: for $\overline{\rm u}$, $\overline{\rm d}$ it is the baryon chemical potential, for  $s,{\bar s}$ the mass ($M\approx \mu_{q}$): 
\begin{subeqnarray}
\frac{s}{V} &=& \frac{\overline{s}}{V} = 6\int \frac{\D^3p}{(2\pi)^3}\frac{1}{\E^{\sqrt{p^2+M^2}/T}+1}\;,\label{3chap8Ceq4.7a}\\\noalign{\smallskip}
\frac{\overline{q}}{V} &=& 6\int \frac{\D^3p}{(2\pi)^3}\frac{1}{\E^{|p|/T+\mu_{q}/T}+1}\;.\label{3chap8Ceq4.7b}
\end{subeqnarray}
Note that the chemical potential of quarks suppresses the ${\bar q}$ density. This phenomenon reflects on the chemical equilibrium between $q{\bar q}$ and the presence of a light quark density associated with the net baryon number. Secondly, strangeness in the plasma phase is more abundant than in the hadronic gas phase (even if the latter phase space is saturated) when compared at the {\em same\/} temperature and baryon chemical potential in the phase transition region. The rationale for the comparison at fixed thermodynamic variables, rather than at fixed values of microcanonical variables such as energy density and baryon density, is outlined in the next section. I record here only that the abundance of strangeness in the plasma is well above that in the hadronic gas phase space (by factors 1--6) and the two become equal only when the baryon chemical potential $\mu$ is so large that abundant production of hyperons becomes possible. This requires a hadronic phase at an energy density of 5--10~GeV/fm$^3$.

\setcounter{equation}{0}
\subsection{How to Discover the Quark--Gluon Plasma}\label{3chap8Csec5}

Here only the role of the strange particles in the anticipated discovery will be discussed. My intention is to show that, under different possible transition scenarios, characteristic anomalous strange particle patterns emerge. Examples presented are intended to provide some guidance to future experiments and are not presented here in order to imply any particular preference for a reaction channel. I begin with a discussion of the observable quantities.

The temperature and chemical potential associated with the hot and dense phase of nuclear collision can be connected with the observed particle spectra, and, as discussed here, particle abundances. The last grand canonical variable -- the volume -- can be estimated from particle interferences. Thus, it is possible to use these measured variables, even if their precise values are dependent on a particular interpretational model, to uncover possible rapid changes in a particular observable. In other words, instead of considering a particular particle multiplicity as a function of the collision energy $\sqrt{s}$, I would consider it as a function of, e.g., mean transverse momentum $\langle p_\perp\rangle$, which is a continuous function of the temperature (which is in turn continuous across any phase transition boundary).

To avoid possible misunderstanding of what I want to say, here I consider the (difficult) observation of the width of the K$^+$ two-particle correlation function in momentum space as a function of the average K$^+$ transverse momentum obtained at given $\sqrt{s}$. Most of K$^+$ would originate from the plasma region, which, when it is created, is relatively small, leading to a comparatively large width. (Here I have assumed a first order phase transition with substantial increase in volume as matter changes from plasma to gas.) If, however, the plasma state were not formed, K$^+$ originating from the entire hot hadronic gas domain would contribute a relatively large volume which would be seen; thus the width of the two-particle correlation function would be small. Thus, a first order phase transition implies a jump in the K$^+$ correlation width as a function of increasing $\langle p_\perp\rangle_{{\rm K}^+}$, as determined in the same experiment, varying $\sqrt{s}$.

From this example emerges the general strategy of my approach: search for possible discontinuities in observables derived from discontinuous quantities (such as volume, particle abundances, etc.) as a function of quantities measured experimentally and related to thermodynamic variables always continuous at the phase transition: temperature, chemical potentials, and pressure. This strategy, of course, can only be followed if, as stated in the first sentence of this report, approximate local thermodynamic equilibrium is also established.

Strangeness seems to be particularly useful for plasma diagnosis, because its characteristic time for chemical equilibration is of the same order of magnitude as the expected lifetime of the plasma: $\tau\sim 1$--$3\times 10^{-23}$~s. This means that we are dominantly creating strangeness in the zone where the plasma reaches its hottest stage -- freezing over the abundance somewhat as the plasma cools down. However, the essential effect is that the strangeness abundance in the plasma is greater, by a factor of about 30, than that expected in the hadronic gas phase at the same values of $\mu,T$. Before carrying this further, let us note that, in order for strangeness to disappear partially during the phase transition, we must have a {\it slow\/} evolution, with time constants of $\sim 10^{-22}$~s. But even so, we would end up with strangeness-saturated phase space in the hadronic gas phase, i.e., roughly ten times more strangeness than otherwise expected. For similar reasons, i.e., in view of the rather long strangeness production time constants in the hadronic gas phase, strangeness abundance survives practically unscathed in this final part of the hadronization as well. {\em  Facit:\/}
\begin{quote}
{\normalsize if a phase transition to the plasma state has occurred, then on return to the hadron phase, there will be most likely significantly more strange particles around than there would be (at this $T$ and $\mu$) if the hadron gas phase had never been left.}
\end{quote}
In my opinion, the simplest observable proportional to the strange particle multiplicity is the rate of V-events from the decay of strange baryons (e.g., $\uLambda$) and mesons (e.g., K$_{s}$) into two charged particles. Observations of this rate require a visual detector, e.g., a streamer chamber. To estimate the multiplicity of V-events, I reduce the total strangeness created in the collision by a factor 1/3 to select only neutral hadrons and another factor 1/2 for charged decay channels. We thus have
\begin{equation}\label{3chap8Ceq5.1}
\langle n_{\rm V}\rangle \approx \frac{1}{6}\frac{\langle s\rangle+\langle\overline{s}\rangle}{\langle b\rangle}\langle b\rangle \sim \frac{\langle b\rangle}{15}\;,
\end{equation}
where I have taken $\langle s\rangle/\langle b\rangle\sim 0.2$ (see Fig.~\ref{3chap8Cfig6}). Thus for events with a large baryon number participation, we can expect to have several V's per collision, which is 100--1000 times above current observation for Ar-KCl collision at 1.8~GeV/Nuc kinetic energy \cite{3chap8Cbib17}.

Due to the high ${\bar s}$ abundance, we may further expect an enrichment of strange antibaryon abundances \cite{3chap8Cbib16}. I would like to emphasize here ${\bar s}\,{\bar s}\,{\bar q}$ states (anticascades) created by the accidental coagulation of two ${\bar s}$ quarks helped by a gluon $\rightarrow{\bar q}$ reaction. Ultimately, the ${\bar s}\,{\bar s}\,{\bar q}$ states become ${\bar s}\,{\bar q}\,{\bar q}$, either through an ${\bar s}$ exchange reaction in the gas phase or via a weak interaction much, much later. However, half of the ${\bar s}\,{\bar q}\,{\bar q}$ states are then visible as $\overline{\uLambda}$ decays in a visual detector. This anomaly in the apparent $\overline{\uLambda}$ abundance is further enhanced by relating it to the decreased abundance of antiprotons, as described above.

Unexpected behavior of the plasma--gas phase transition can greatly influence the channels in which strangeness is found. For example, in an extremely particle-dense plasma, the produced $s{\bar s}$ pairs may stay near to each other -- if a transition occurs without any dilution of the density, then I would expect a large abundance of $\uphi(1020)$ $s{\bar s}$ mesons, easily detected through their partial decay mode (1/4\%) to a $\umu^+\umu^-$ pair.

Contrary behavior will be recorded if the plasma is cool at the phase transition, and the transition proceeds slowly -- major coagulation of strange quarks can then be expected with the formation of $sss$ and ${\bar s}\,{\bar s}\,{\bar s}$ baryons and in general $({s})^{3n}$ clusters. Carrying this even further, supercooled plasma may become \lq strange' nuclear (quark) matter \cite{3chap8Cbib18}. Again, visual detectors will be extremely successful here, showing substantial decay cascades of the same heavy fragment.

In closing this discussion, I would like to give warning about the pions. From the equations of state of the plasma, we have deduced in Sect.~\ref{3chap8Csec3} a very high specific entropy per baryon. This entropy can only increase in the phase transition and it leads to very high pion multiplicity in nuclear collisions, probably created through pion radiation from the plasma \cite{3chap8Cbib5a,3chap8Cbib5b} and sequential decays. Hence by relating anything to the pion multiplicity, e.g., considering ${\rm K}/\upi$ ratios, we dilute the signal from the plasma. Furthermore, pions are not at all characteristic for the plasma; they are simply indicating high entropy created in the collision. However, we note that the ${\rm K}/\upi$ ratio {\it can\/} show substantial deviations from values known in \pp collisions -- but the interpretations of this phenomenon will be difficult.

It is important to appreciate that the experiments discussed above would certainly be quite complementary to the measurements utilizing electromagnetically interacting probes, e.g., dileptons, direct photons. Strangeness-based measurements have the advantage that they have much higher counting rates than those recording electromagnetic particles.\\

\noindent {\bf 1983 Acknowledgements} 
I would like to thank R. Hagedorn, B. M\"uller, and P. Koch for fruitful and stimulating discussions, and R. Hagedorn for a thorough criticism of this manuscript. Thie work was in part supported by Deutsche Forschungsgemeinschaft.



 
\end{document}